\documentclass[fleqn,usenatbib]{mnras}

\usepackage[T1]{fontenc}

\DeclareRobustCommand{\VAN}[3]{#2}
\let\VANthebibliography\thebibliography
\def\thebibliography{\DeclareRobustCommand{\VAN}[3]{##3}\VANthebibliography}

\usepackage{graphicx}	
\usepackage{amsmath}	
\usepackage{amssymb}	
\usepackage{pdflscape}

\usepackage{newtxtext,newtxmath}

\title[Line luminosities of Wolf-Rayet stars]{Line Luminosities of Galactic and Magellanic Cloud Wolf-Rayet stars}

\author[Crowther et al.]{
Paul A. Crowther$^{1}$\thanks{paul.crowther@sheffield.ac.uk}, G. Rate$^{1}$, Joachim M. Bestenlehner$^{1}$\\
1. Department of Physics and Astronomy, University of Sheffield, Sheffield, S3 7RH, UK}

\date{Accepted 2022 July 5. Received 2022 July 5; in original form 2022 April 8}

\pubyear{2022}

\begin{document}
\label{firstpage}
\pagerange{\pageref{firstpage}--\pageref{lastpage}}
\maketitle

\begin{abstract}
We provide line luminosities and spectroscopic templates of prominent optical emission lines of 133 Galactic Wolf-Rayet stars by exploiting {\it Gaia} DR3 parallaxes and optical spectrophotometry, and provide comparisons with 112 counterparts in the Magellanic Clouds. Average line luminosities of the broad blue (He\,{\sc ii} $\lambda$4686, C\,{\sc iii} $\lambda\lambda$4647,51, N\,{\sc iii} $\lambda\lambda$4634,41, N\,{\sc v} $\lambda\lambda$4603,20) and yellow (C\,{\sc iv} $\lambda\lambda$5801,12)  emission features for WN, WN/C, WC and WO stars have application in characterising the Wolf-Rayet populations of star-forming regions of distant, unresolved galaxies. Early-type WN stars reveal lower line luminosities in more metal poor environments, but the situation is less clear for late-type WN stars. LMC WC4--5 line luminosities are higher than their Milky Way counterparts, with line luminosities of Magellanic Cloud WO stars higher than Galactic stars. We highlight other prominent optical emission lines, N\,{\sc iv} $\lambda\lambda$3478,85 for WN and WN/C stars, O\,{\sc iv} $\lambda\lambda$3403,13 for WC and WO stars and O\,{\sc vi} $\lambda\lambda$3811,34 for WO stars. We apply our calibrations to representative metal-poor and metal-rich WR galaxies, IC~4870 and NGC~3049, respectively, with spectral templates also applied based on a realistic mix of subtypes. Finally, the global blue and C\,{\sc iv} $\lambda\lambda$5801,12 line luminosities of the Large (Small) Magellanic Clouds are 2.6$\times 10^{38}$ erg\,s$^{-1}$ (9$\times 10^{36}$ erg\,s$^{-1}$) and 8.8$\times 10^{37}$ erg\,s$^{-1}$ (4$\times 10^{36}$ erg\,s$^{-1}$), respectively, with the cumulative WR line luminosity of the Milky Way estimated to be an order of magnitude higher than the LMC.
\end{abstract}





\section{Introduction}

Wolf-Rayet (WR) stars, the evolved descendants of (very) high mass stars, exhibit broad optical emission lines owing to their dense stellar winds \citep{2007ARA&A..45..177C}. There are three flavours of  WR stars known, nitrogen-sequence (WN) stars dominated by helium and nitrogen features, carbon-sequence (WC) stars with prominent carbon and helium lines, and rare oxygen-sequence (WO) stars with dominant oxygen and carbon features\footnote{The Wolf-Rayet phenomenon is common to evolved massive stars (classical WN, WC, WO), very massive main sequence stars (H-rich WN, Of/WN) and a subset of central stars of Planetary Nebulae ([WC], [WN])}. Their unusual spectroscopic signatures permit their use as  tracers of star-formation in galaxies \citep{1976MNRAS.177...91A, 1981A&A...101L...5K}. Indeed, a subset of starburst galaxies have been coined Wolf-Rayet galaxies owing to the presence of broad He\,{\sc ii} $\lambda$4686 emission from WN stars, and occasionally C\,{\sc iv} $\lambda$5808 emission from WC stars \citep{1992ApJ...401..543V}. The advent of highly multiplexed instrumentation (e.g. SDSS) plus large-field integral field spectroscopic capabilities (e.g MUSE) has identified large numbers of WR stars in external galaxies \citep{2008A&A...485..657B, 2016A&A...592A.105M, 2017A&A...603A.130M, 2021MNRAS.500.2076G, 2021MNRAS.503.6112S}.

\begin{table*}
\begin{center}
\caption{Source of optical spectrophotometry of Galactic and Magellanic Cloud WR stars for this study, including representative spectral resolutions at $\lambda$=5000\AA.}
\label{spectroscopy}
\begin{tabular}{l@{\hspace{2mm}}l@{\hspace{2mm}}l@{\hspace{2mm}}l@{\hspace{2mm}}l@{\hspace{2mm}}c@{\hspace{3mm}}l@{\hspace{3mm}}l@{\hspace{2mm}}l}
\hline
ID & Telescope & Instrument                     & Epoch        & Sp.& Wavelength & Flux & Ref & Notes\\
     &             &                                      &                    & Res (\AA)  & Coverage (\AA) & Calib. &  &    \\
\hline
AD  &ANU  2.3m            & DBS           & Dec 1997       &   5    & 3200--11000 & 10\% & 2  & Southern WR stars, 6070--6400\AA\ detector gap. \\ 
AR          & AAT                    & RGO    &  Mar 1992--Dec 1994 & 2 & 3680--6000 & 10\% & 1 & LMC/SMC WN and WN/C stars. \\ 
CS   & CTIO 1.5m           & SIT             & Nov 1981--Feb 1985 & 10 & 3400--7270 & 10\% & 16 & Southern Milky Way WR stars. Variable $\lambda_{\rm max}$.  \\ 
HF   & HST                    & FOS                       & Jan 1996--Jan 1997 &  3    & 3230--6820   & 10\% & 4--6 & LMC WN stars. G400 only ($\lambda_{\rm max}$=4780\AA) except for R136. \\ 
 HS    & HST                      & STIS      &  Mar 2014--Sep 2016 & 10 & 2900--10250 & 10\%  & 17 &  AB5 (HD~5980). \\ 
II91 & INT                      & IDS           & Sep 1991                & 2 & 3320--7300 & 20\% & 7  & Northern Milky Way WN and WN/C stars. Variable $\lambda_{\rm min/max}$. \\ 
II96 & INT                      & IDS          & Jul 1996                & 3 & 3620--6810   & 10\% & 8 & Northern Milky Way WN stars. \\ 
II13  & INT                     & IDS        & Sep 2013               & 5 & 3800--9350 & 20\% &  9 & Northern Milky Way WR stars. \\ 
KI     & KPNO 0.91m      & IRS          & Oct 1980--Feb 1983  & 9 & 3450--6900 & 10\% & 10 & Northern Milky Way WR stars. \\ 
 MM   & Magellan            & MagE   &  Sep 2014-Dec 2020  &  1.2           & 3170--9440 & 10\% & 13, 14 & LMC WR stars. \\ 
SC & Mt Stromlo 1.9m & Coud\'{e}     & Dec 1995     &      1     &4700--6700 & 20\% & 1 &   LMC late-type WN stars. \\ 
WI94 & WHT                  & ISIS        & Jun 1994   &        3        & 4450--6030 & 10\% & 11 &  Northern Milky Way WN and WN/C stars. \\ 
WI02  & WHT                 & ISIS       & Aug 2002   &        3.5       & 3400--9500 & 10\% & 12 & Northern Milky Way WC stars. \\ 
VM    & VLT                    & MUSE  & Aug 2014  &       3      & 4600--9350 & 10\% & 15 & LMC WN stars. Calibration via BAT99--100 (HST/FOS). \\  
VU    & VLT                       & UVES    & Jan 2002--Jan  2003                   & 0.1  & 3200--10240 & 10\% & 18, 19  & Southern Milky Way WR stars. \\ 
VX    & VLT                      & XShooter &  Nov 2011--Aug 2013    &   0.8  & 3100--24700 & 10\% & 20, 21 &Southern Milky Way and LMC WR stars. \\    
\hline
\end{tabular}
\begin{footnotesize}
1: \citet{1997A&A...320..500C}; 
2: \citet{2002A&A...392..653C}, 
3: \citet{2006A&A...449..711C}; 
4: \citet{1997ApJ...477..792D}; 
5: \citet{1998ApJ...493..180M};\\
6: \citet{1999AJ....118.1684W};
7: \citet{1995A&A...293..427C}; 
8: \citet{1997MNRAS.290L..59C}; 
9: This study;
10: \citet{1984ApJ...281..789M}; 
11: \citet{1995A&A...304..269C};\\
12: \citet{2006ApJ...636.1033C}; 
13: \citet{2017ApJ...841...20N}; 
14: \citet{2022ApJ...931..157A};
15: \citet{2018A&A...614A.147C}; 
16: \citet{1988AJ.....96.1076T}; \\
17: \citet{2019MNRAS.486..725H}; 
18: \citet{2003Msngr.114...10B};
19: \citet{2022arXiv221109130B};
20: \citet{2015A&A...581A.110T};
21: Rubin-Diez et al. (in prep.);
\end{footnotesize}
\end{center}
\end{table*}


Population synthesis models have incorporated WR stars via either empirical calibrations \citep{1998ApJ...497..618S} or synthetic spectra \citep{2002MNRAS.337.1309S, 2017PASA...34...58E}. The former approach has relied heavily on calibrations of WR stars in the metal-poor Magellanic Clouds \citep{1990ApJ...348..471S, 2006A&A...449..711C} while the latter depend on predictions from single or binary evolutionary models and suitable prescriptions for theoretical wind densities at a range of metallicities. Various  mass-loss prescriptions are available, and differ between very massive stars exhibiting WR spectral morphologies \citep{2020MNRAS.493.3938B} and classical WR stars \citep{2000A&A...360..227N, 2020MNRAS.491.4406S} although these are generally untested for a broad range of metallicities. By way of example, \citet{2017PASA...34...58E} utilise the grid of PoWR \citep{2002A&A...387..244G} model atmospheres\footnote{https://www.astro.physik.uni-potsdam.de/PoWR/} for spectral synthesis, yet predictions are highly dependent on the choice of WR models.

Empirical calibrations have been produced for WR stars in the Magellanic Clouds, but uncertain distances to Milky Way WR stars has severely hindered calibrations at high metallicity to date. The advent of reliable Galactic WR distances from {\it Gaia} \citep{2020MNRAS.493.1512R} has opened up the possibility of solar metallicity calibrations and spectral templates. In this study we exploit  {\it Gaia} DR3 parallaxes and archival spectrophotometry of large numbers of Galactic WR stars to provide calibrations at (near) solar composition, with which to complement synthetic spectroscopy, and revisit Magellanic Cloud calibrations in view of new discoveries and higher quality observations since \citet{2006A&A...449..711C}.

 This paper is structured as follows. Section~\ref{obs} discusses the observational datasets employed to construct WR line luminosities, Section~\ref{results} compares Milky Way WR line luminosities with Magellanic Cloud counterparts and includes templates for each environment, Section~\ref{WR-gal}  applies calibrations and templates to representative WR galaxies at low and high metallicity, with  conclusions drawn in Section~\ref{summary}.

\begin{figure}
\centering
	\includegraphics[width=0.9\linewidth,bb=50 180 505 635,angle=-90]{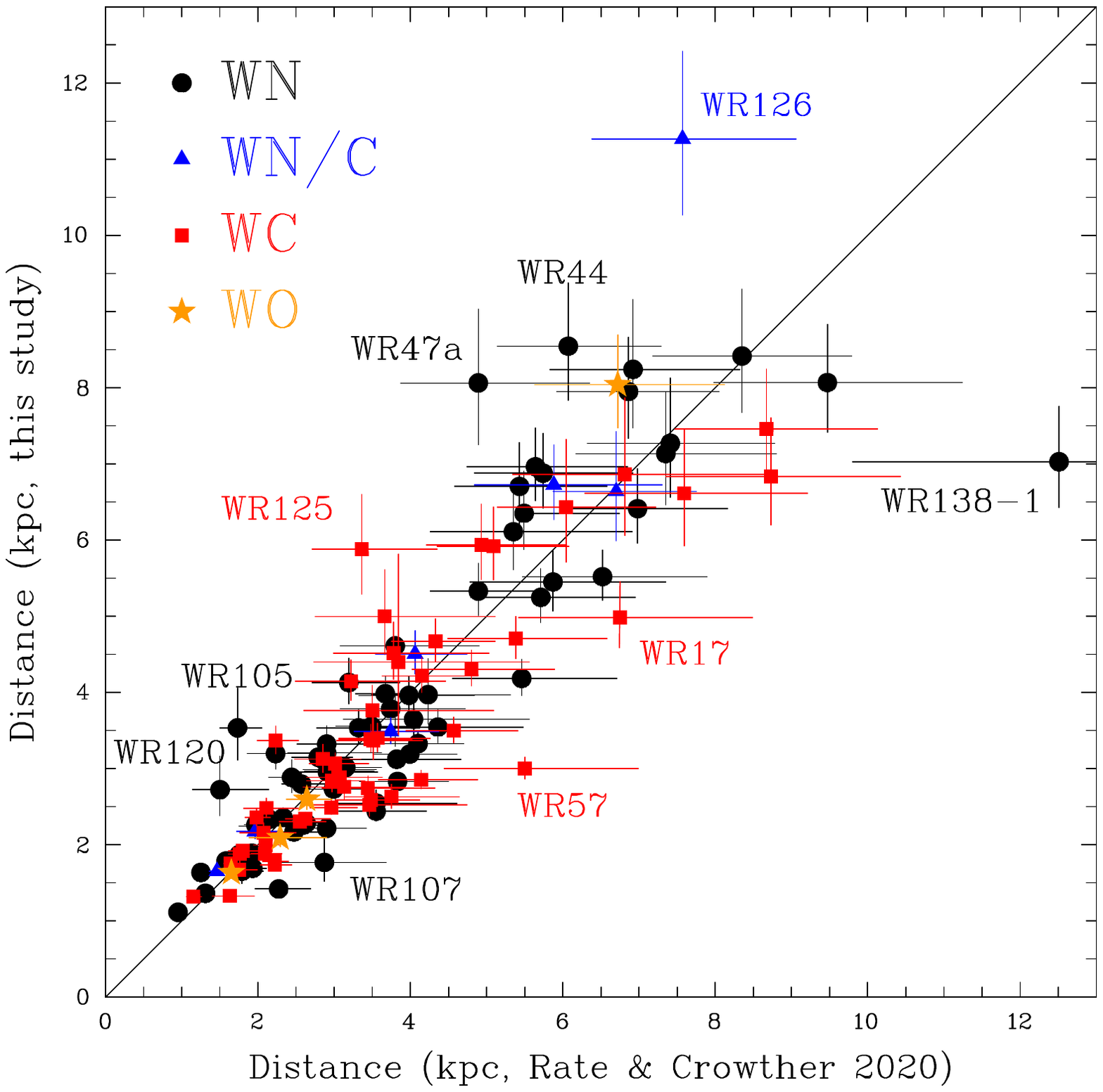}
	\centering
  \caption{Comparison between Galactic WR distances (in kpc) following a common Bayesian methodology of \citet{2020MNRAS.493.1512R} using either {\it Gaia} DR2 parallaxes incorporating global zero point corrections of \citet{2018A&A...616A...2L}, versus DR3 parallaxes incorporating local zero point corrections of \citet{2021A&A...649A...4L} and \citet{2022A&A...657A.130M}. The key distinguishes between WN (black circles), WN/C (blue triangles), WC (red squares) and WO (orange stars) subtypes. In general, DR3 plus revised zero points provides improved uncertainties on distances, with some outliers labelled.} 
	\label{DR2_DR3} 
\end{figure}

\section{Observations}\label{obs}

Our primary objective is to provide line luminosities of prominent optical emission lines in Milky Way WR stars, requiring (i) spectrophotometric datasets; (ii) interstellar extinctions drawn from  spectroscopic studies; (iii) distances from {\it Gaia} DR3 parallaxes.

\subsection{Milky Way WR stars}

A subset of flux calibrated spectroscopic datasets involved observations through large apertures, while others involved relatively flux calibrated datasets adjusted to literature photometry, primarily narrow-band $ubvr$ photometry of \citet{1984ApJ...281..789M} or \citet{1988AJ.....96.1076T}.  In instances of discrepant photometric measurements, inspection of large aperture low resolution (LORES) LWR or LWP (1850--3350\AA) spectrophotometry from IUE\footnote{INES datasets obtained from http://sdc.cab.inta-csic.es/ines/} provided the primary reference. We also utilise IUE SWP (1150--1980\AA) spectrophotometry, where available. 

Our primary optical spectroscopic datasets are summarised in Table~\ref{spectroscopy}, with typical spectral resolutions of 1--5\AA, although some historical datasets were obtained at low spectral resolution. In view of the (typically) broad WR emission lines, this  doesn't adversely impact on line flux measurements, but FWHM of narrow lined stars (typically WN8--11) are impacted, so all quoted FWHM are corrected for instrumental broadening (typical uncertainties are $\pm$100 km\,s$^{-1}$).
All datasets have been previously discussed with the exception of INT IDS spectroscopy from 16-17 Sep 2013 (PI Rosslowe), involving observations of WR138--1\footnote{\citet{2009MNRAS.400..524G} discovered WR138a, which was subsequently reclassified as WN9 by \citet{2014AJ....148...34F} and its WR catalogue number was revised to WR138-1 following \citet{2015MNRAS.447.2322R}.}, WR140 and WR149 with the 235\,mm camera, R300V grating and EEV10 detector. Relative flux calibration was achieved using standard star BD+25$^{\circ}$ 4655. Flux calibrated UVES POP datasets \citep{2003Msngr.114...10B, 2022arXiv221109130B} are included if the key C\,{\sc iv} $\lambda\lambda$5801,12 line, missing due to the UVES detector gap, is available from other sources. In some instances spectroscopy extends to or beyond 1$\mu$m (e.g. ANU 2.3m/DBS, VLT/XShooter) which are supplemented by additional near-IR spectrophotometry \citep[e.g. IRTF/SpeX,][]{2006MNRAS.372.1407C}. For inclusion in the present sample we require He\,{\sc ii} $\lambda$4686 for WN and Of/WN stars, C\,{\sc iv} $\lambda\lambda$5801,12 for WC stars, and O\,{\sc iv} $\lambda\lambda$3811,34 for WO stars, thereby excluding heavily reddened stars, such as the rich WR population in Westerlund~1 \citep{2006MNRAS.372.1407C}. We also omit WR+WR binaries involving differing subtypes \citep[e.g. WR70--16,][]{2020MNRAS.495.3323C}.

\begin{figure}
\centering
	\includegraphics[width=0.525\linewidth,bb=175 50 430 535, angle=-90]{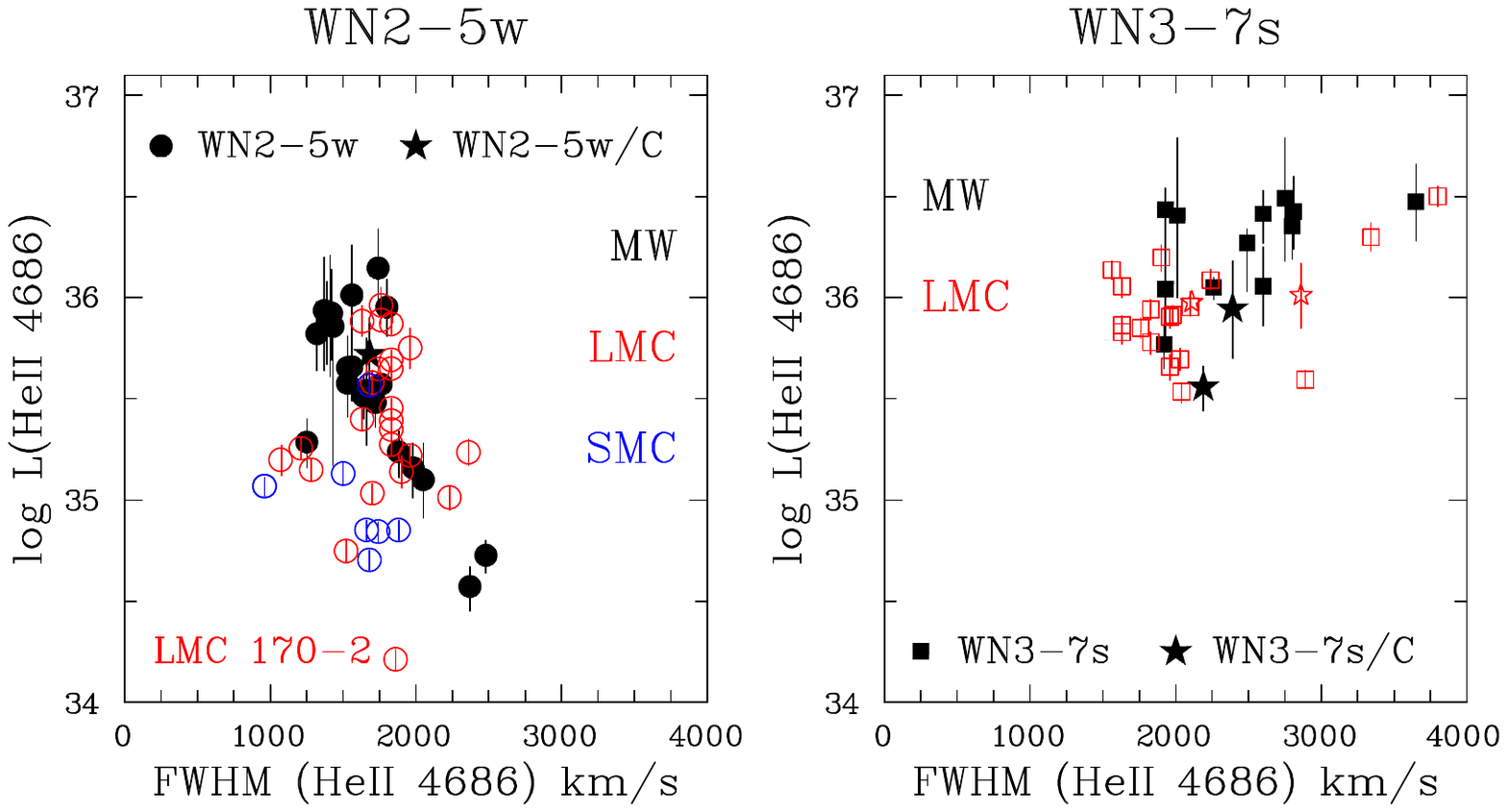}
	\includegraphics[width=0.525\linewidth,bb=175 50 430 535, angle=-90]{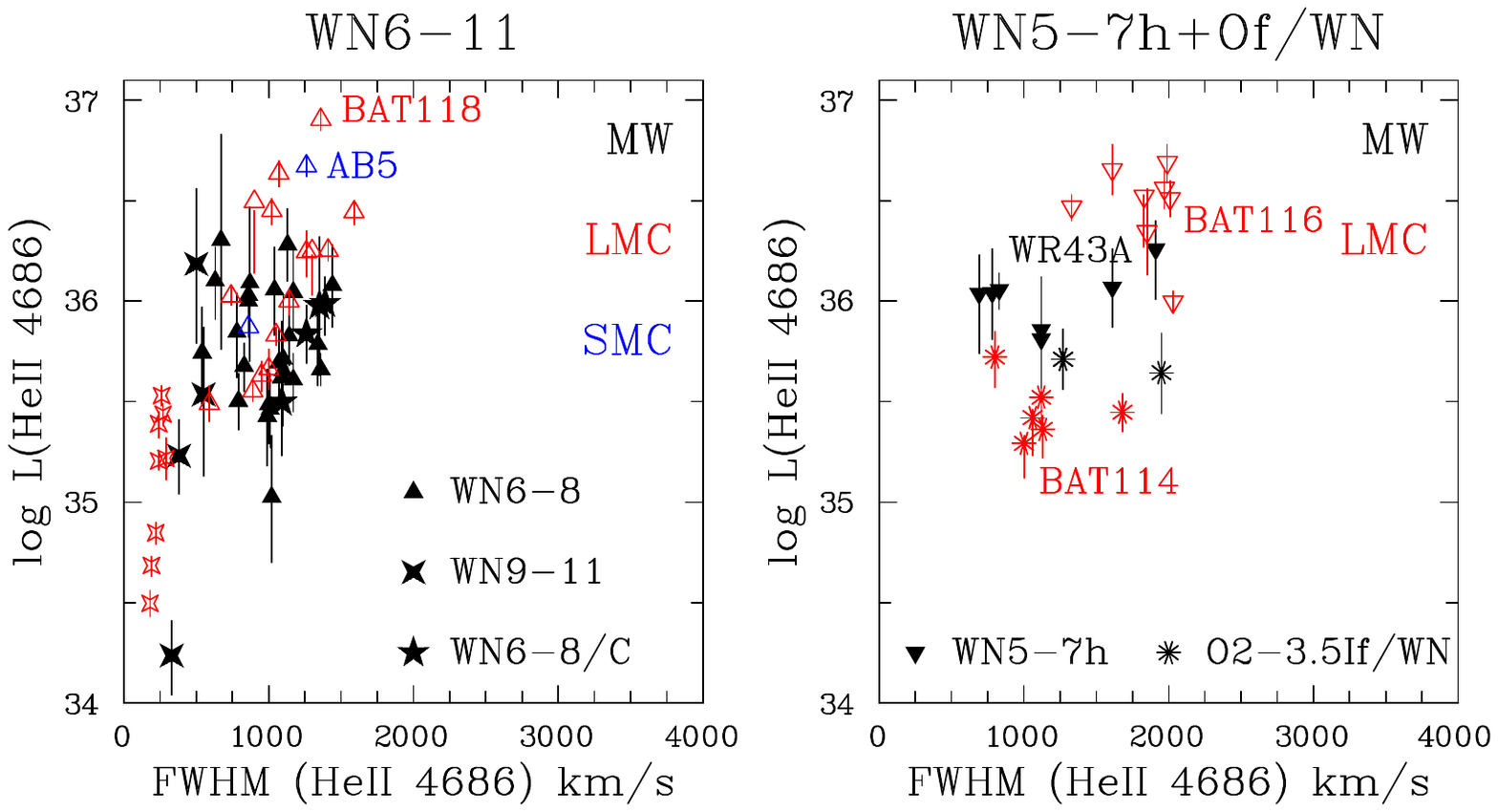}
	\centering
  \caption{He\,{\sc ii} $\lambda$4686 line luminosities of WN2--5w (top left), WN3--7s (top right), WN6--8  (bottom left) and WN5--7h+Of/WN (bottom right) stars in the Milky Way (black, filled), LMC (red, open) and SMC (blue, open) versus FWHM (km\,s$^{-1}$). Weak and strong-lined early-type WN stars are indicated as circles and squares, respectively, WN/C stars as stars, WN6--8 as triangles, WN9--11 as crosses, H-rich main sequence WN5--7h stars as inverted triangles and Of/WN stars as asterisks.}
	\label{4686}
\end{figure}

\begin{figure}
\centering
	\includegraphics[width=0.525\linewidth,bb=175 50 430 535, angle=-90]{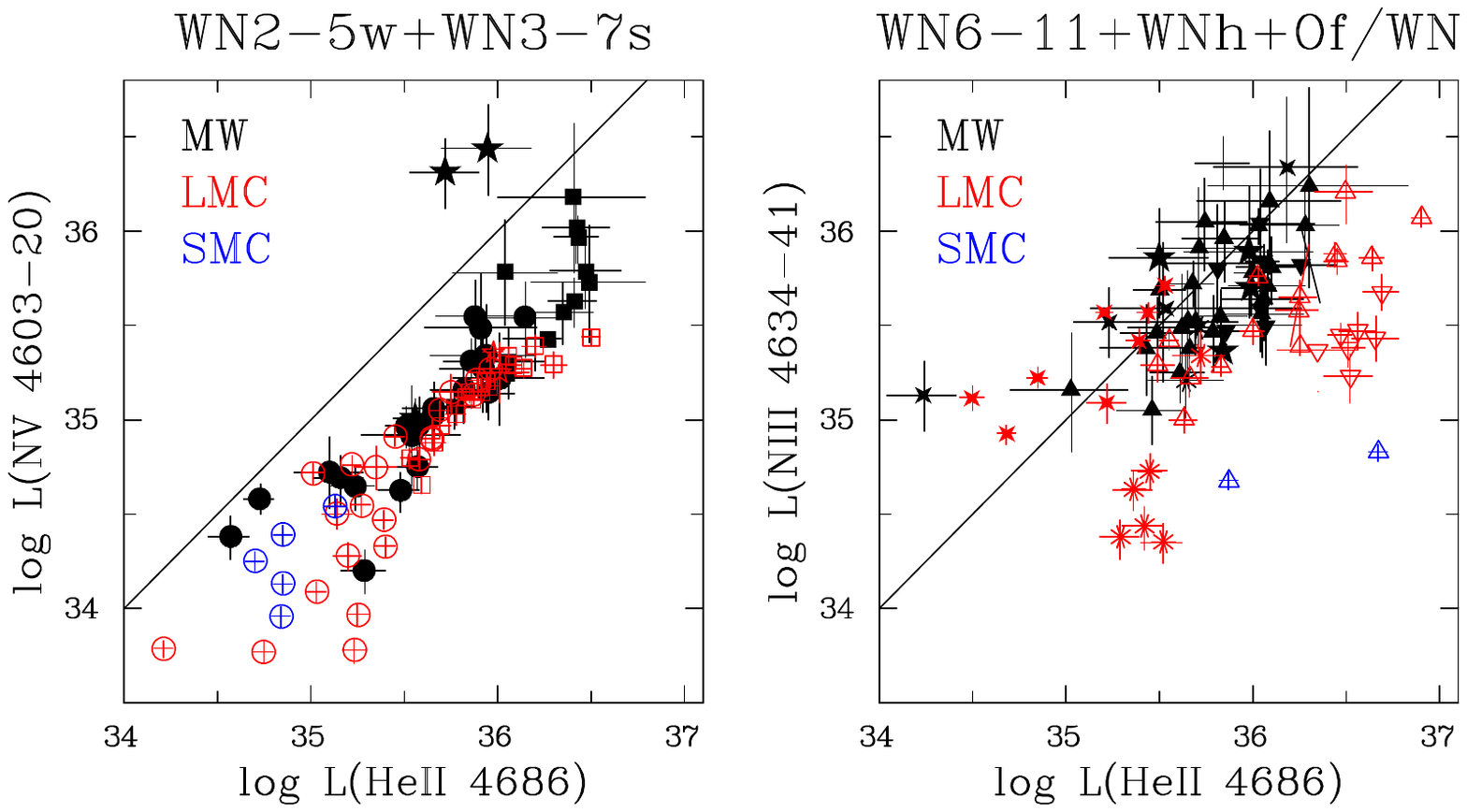}
	\includegraphics[width=0.525\linewidth,bb=175 50 430 535, angle=-90]{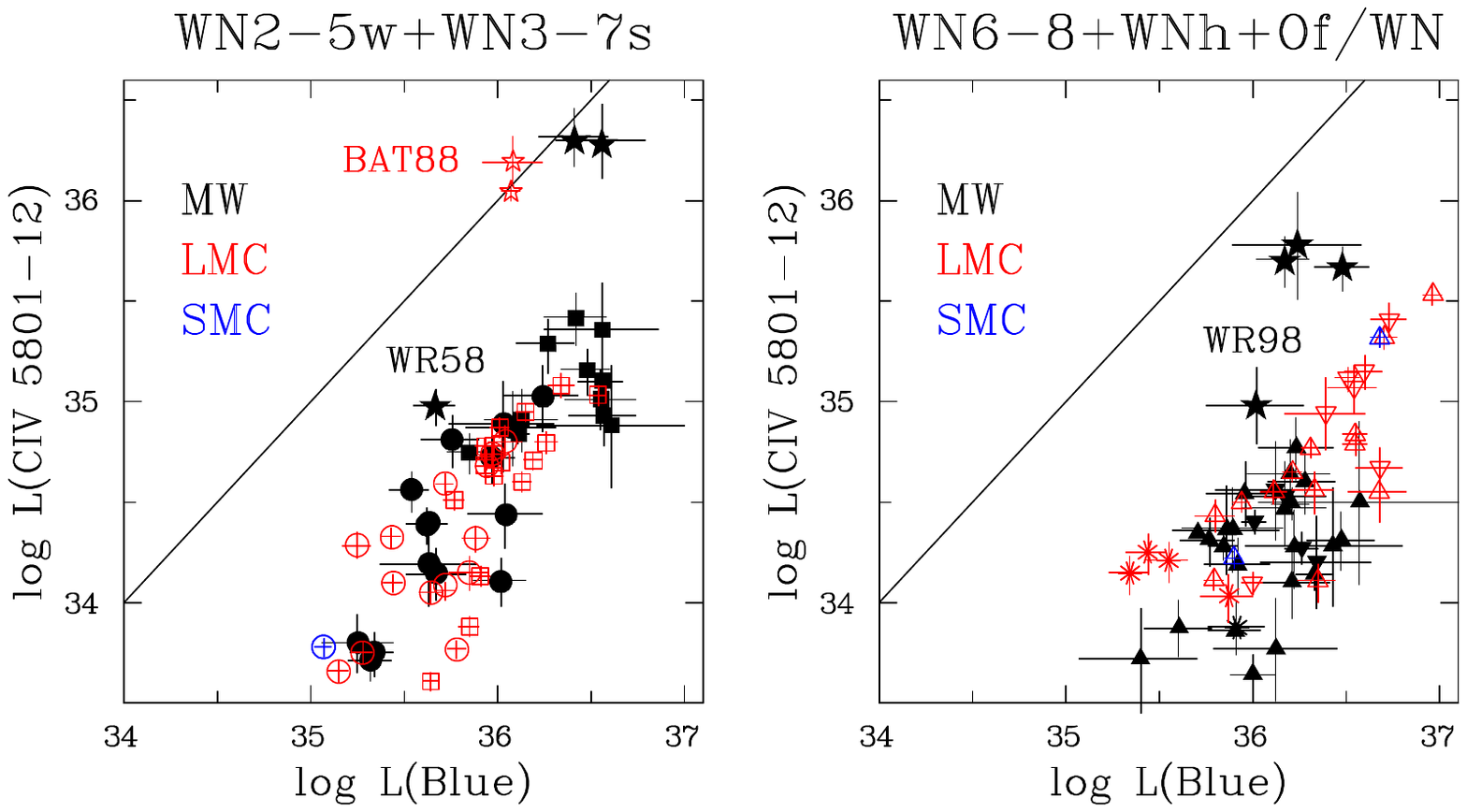}
	\centering
  \caption{(Upper panels): He\,{\sc ii} $\lambda$4686 versus N\,{\sc v} $\lambda\lambda$4603,20+N\,{\sc iii} $\lambda\lambda$4634,41 line luminosities of WN, Of/WN and WN/C stars in the Milky Way (black), LMC (red) and SMC (blue). C\,{\sc iii} $\lambda$4647,51 also contributes for WN/C subtypes, N\,{\sc iii} $\lambda\lambda$4634,41 for WN6--7s subtypes and N\,{\sc ii} $\lambda\lambda$460,43 for WN10--11 subtypes; (Lower panels): Blue (He\,{\sc ii} $\lambda$4686, N\,{\sc v} $\lambda\lambda$4603,20, N\,{\sc iii} $\lambda\lambda$4634,41, C\,{\sc iii} $\lambda$4647,51) versus C\,{\sc iv} $\lambda\lambda$5801,12 line luminosities. C\,{\sc iv} $\lambda\lambda$5801,12 is an order of magnitude stronger in WN/C stars than WN stars. The solid line indicates equal line luminosities. Symbols as in Figure~\ref{4686} (WN9--11 are omitted from lower panels since C\,{\sc iv} $\lambda\lambda$5801,12 is weak/absent).}
	\label{blue_5808_WN}
\end{figure}

Line fluxes were measured using the multiple gaussian fitting routines ELF within the Starlink DIPSO spectroscopic package \citep{dipso}. Spectral coverage varies by instrument/telescope, but usually includes $\lambda$4000--$\lambda$7000, so may exclude some violet (e.g. O\,{\sc iv} $\lambda\lambda$3403,13, N\,{\sc iv} $\lambda\lambda$3478,85) or red (e.g. N\,{\sc iv} $\lambda\lambda$7103,29, C\,{\sc iii} $\lambda\lambda$9701,19 diagnostics). We adopt uncertainties of 10\% for the majority of line flux measurements, 20\% for Mt Stromlo 1.9m/Coud\'{e} and INT/IDS spectroscopy from 1991 and 2013. Comparisons between He\,{\sc ii} $\lambda$4686 line fluxes and literature results for WN stars from \citet{2019AJ....158..192L} reveal $\log F_{\rm HeII 4686} - F_{\rm HeII 4686}^{\rm L19} = 0.00 \pm 0.10$ for 35 stars in common, while comparisons between C\,{\sc iv} $\lambda\lambda$5801,12 line fluxes and literature results for WC stars from \citet{1990ApJ...358..229S} reveal $\log F_{\rm CIV 5801,12} - F_{\rm CIV 5801,12}^{\rm S90b} = 0.00 \pm 0.11$ for 48 stars in common.

In the era of reliable Galactic distances courtesy of {\it Gaia}, interstellar extinctions are of critical importance for line luminosity determinations. Here we adopt E$_{\rm B-V}$ and R$_{\rm V}$ values from contemporary spectroscopic studies \citep{2006A&A...457.1015H, 2012A&A...540A.144S}. For stars (usually WR+O binaries) lacking modern extinction determinations, we follow the approach adopted by \citet{2006A&A...449..711C}  involving the relationship between the He\,{\sc ii} $\lambda$4686 equivalent width ($W_{\rm HeII~4686}$) and  intrinsic colour of WN stars, or adopt $(b-v)_{0}$ = --0.30 mag, plus $E_{\rm B-V} = 1.21 \times E_{b-v}$ and R$_{\rm V}$  = 3.1. 
In order to determine intrinsic fluxes we adopt the Milky Way extinction law of \citet{1979MNRAS.187P..73S}, parameterised by \citet{1983MNRAS.203..301H}, which provided the default law for the majority of Galactic WN and WC stars studied by \citet{2006A&A...457.1015H} and \citet{2012A&A...540A.144S}, respectively\footnote{Contemporary extinction laws are available \citep[e.g.][]{2019ApJ...886..108F}}. For error calculations we assume A$_{\rm V}$ magnitudes are reliable to 10\% throughout. By way of example, we measure $W_{\rm HeII~4686}$ = 19\AA\ for WR138--1, which implies $(b-v)_{0} \sim$  --0.31 mag  \citep[][their fig.~1]{2006A&A...449..711C}, so $E_{b-v}$ = 1.87 mag, based on $b-v$ = 1.56 mag from our INT/IDS spectroscopy, resulting in $A_{\rm V} = 7.0\pm$0.7 mag \citep[versus  $A_{\rm V} = 7.4$~mag from][]{2009MNRAS.400..524G}.


Distances to Galactic WR stars follow from {\it Gaia} DR3 parallaxes \citep{2021A&A...649A...1G}, using star-specific zero point corrections from \citet{2021A&A...649A...4L} updated by \citet{2022A&A...657A.130M}, and the Bayesian methods set out in \citet{2020MNRAS.493.1512R}. Inferred distances are generally in good agreement with \citet{2021AJ....161..147B}. A comparison between distances obtained from  \citet{2020MNRAS.493.1512R} based on DR2 parallaxes and the global zero point correction of \citet{2018A&A...616A...2L} versus DR3 parallaxes and updated zero-point corrections is presented in Fig.~\ref{DR2_DR3}. 
In general, DR3 plus star-specific zero point corrections provides significantly improved uncertainties on WR distances. Sources with DR3 warning flags (e.g. WR2, WR31, WR66, WR104, WR115) are excluded, while we include H-rich WN stars within NGC~3603 (WR43) for which we adopt the distance of 7.27$_{-0.35}^{+0.38}$~kpc determined by \citet{2021MNRAS.508.4952D}. Average distances to Galactic WN and WN/C stars in our sample are 4.0$\pm$2.2~kpc while average distances to WC and WO stars are 3.5$\pm$1.7~kpc. For Galactic stars, {\it Gaia} DR3 distance uncertainties are typically $\leq$10\%, so extinctions usually dominate uncertainties in line luminosities. 


\begin{figure*}
\centering
\begin{minipage}[c]{0.75\linewidth}
	\includegraphics[width=0.35\linewidth,bb=175 50 430 762, angle=-90]{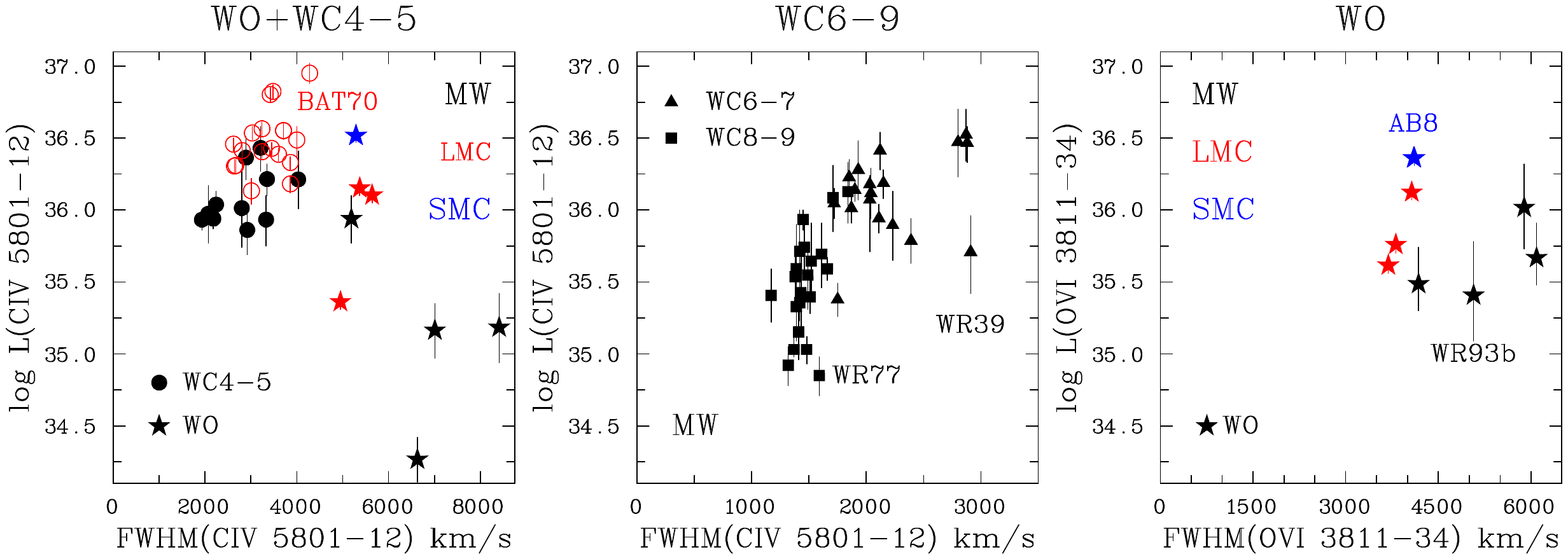}
\end{minipage}\hfill
\begin{minipage}[c]{0.25\linewidth}
	\centering
  \caption{(left): C\,{\sc iv} $\lambda$5808 line luminosities of WO (stars) and WC4--5 (circules) stars in the Milky Way,  LMC and SMC versus FWHM (km\,s$^{-1}$); (centre): C\,{\sc iv} $\lambda\lambda$5801,12 line luminosities of Milky Way WC6--7 (black triangles) and WC8--9 (black squares) versus FWHM (km\,s$^{-1}$); (right): O\,{\sc vi} $\lambda\lambda$3811,34 line luminosities of WO stars in the Milky Way, LMC and SMC versus FWHM (km\,s$^{-1}$).}
	\label{5808}
	\end{minipage}
\end{figure*}

We group WN stars into five categories, strong-lined ("WN3--7s")  weak-lined, early-type ("WN2--5w") stars, weak-lined late-type ("WN6--8") or very late-type ("WN9--11"), and (main sequence) very massive H-rich stars ("WN5--7h") stars, and additionally include transition Of/WN stars ("O2--3.5If/WN"). The division between strong and weak-lined stars was set at $W_{\rm HeII~5412}$ = 40\AA~\citep{1991A&A...248..166K}. We follow this approach with the exception of BAT99-5 which is assigned as a weak-lined star despite $W_{\rm HeII~5412}$ = 41$\pm$1 \AA. Weak/strong lined categories are preferred to narrow/broad-lined  \citep{1996MNRAS.281..163S} whose threshold is FWHM(He\,{\sc ii} $\lambda$4686) = 30\AA\  (corresponding to $\sim$1900 km\,s$^{-1}$), since a subset of broad-lined stars possess a very weak emission line spectrum \citep[e.g. WR46,][]{1995A&A...302..457C}, although the majority of strong-lined stars also possess broad lines, and \citep{1996MNRAS.281..163S} included $W_{\lambda}$(He\,{\sc ii} $\lambda$5412) = 40\AA\ as one of their criteria for broad-lined WN stars. The  majority of strong-lined stars have WN3--4 subtypes, although examples in the Milky Way extend as late as WN6 (WR134, WR136) or WN7 (WR91). LMC 170--2 \citep[WN3/O3,][]{2017ApJ...841...20N} is included as a WN2--5w star, although H$\beta$ absorption would usually favour an O-type classification in preference to either WN or Of/WN  \citep{2011MNRAS.416.1311C}. In order to discriminate between classical WN and main-sequence WN5--7h stars, we consider spectral morphologies as well as association with young star-forming regions (e.g. Carina Nebula, NGC~3603). WC stars are categorised as early-type ("WC4--5"), mid-type ("WC6--7") or late-type ("WC8--9"). 

Line luminosities of individual Galactic WN, WN/C, WC and WO stars are presented in Tables~\ref{WN-all}-\ref{WO-all} in the Appendix. For WN and WN/C stars, aside from the blends at $\lambda$4100 and $\lambda$4630, we have endeavoured to exclude the contribution of He\,{\sc i} $\lambda$7065 and He\,{\sc ii} $\lambda$7177 from the N\,{\sc iv} $\lambda\lambda$7103,29 multiplet. For early-type WC and WO stars, C\,{\sc iv} $\lambda$4658 (6-5), $\lambda$4686 (8-6), $\lambda$4689 (11-7) will also contribute to the C,{\sc iii} $\lambda\lambda$4647,51+He\,{\sc ii} $\lambda$4686 blend. Uncertainties in absolute luminosities involve a combination of absolute flux calibration, distance and extinction, with the latter dominating uncertainties for the majority of Galactic WR stars. Uncertainties on stellar luminosities from literature results account for distance and extinction, but exclude systematic uncertainties resulting from individual studies which varies between studies (typically 0.1--0.2 dex).

In general, WR stars within unresolved star-forming regions are detected via blue ($\lambda$4686) or yellow ($\lambda$5808) bumps. Binaries comprise a significant subset of our sample, but WR fluxes/luminosities are not generally impacted with a few exceptions. Strong absorption lines of OB companions impacts measurements of features in the proximity of (usually) upper Balmer lines, while a few systems are host to multiple WR stars in which case calibrations adopt equal contribution from each component, which is a reasonable approximation for systems with mass ratios of order unity \citep[e.g. WR43A,][]{2008MNRAS.389L..38S}. Excess line emission is also observed in colliding wind systems, usually witnessed in C\,{\sc iii} $\lambda$5696 in WC+O binaries \citep{1997PASP..109..504L}. 

\subsection{Magellanic Cloud WR stars}

We supplement Galactic WR datasets with Magellanic Cloud counterparts, updated from \citet{2006A&A...449..711C} to include additional datasets \citep[e.g.][]{2015A&A...581A.110T, 2017ApJ...841...20N, 2022ApJ...931..157A}. Observations of SMC WN2--5w stars suffer from poor S/N, so we also calibrate higher quality datasets from \citet{2003MNRAS.338..360F}. We also make use of archival ultraviolet IUE or HST spectrophotometry, the latter involving FOS \citep{2002A&A...392..653C}, STIS \citep{2016MNRAS.458..624C, 2022ApJ...931..157A} and COS \citep{2022ApJ...931..157A} instruments. VLT/Xshooter and ANU 2.3m/DBS spectrophotometry extend to the near-IR, with Magellan FIRE datasets also utilised for LMC WC and WO stars \citep{2022ApJ...931..157A}. Again, we assume typical uncertainties of 10\% for line flux measurements obtained from spectrophotometric datasets \citep[e.g.][]{2006A&A...449..711C}.  

Comparisons between He\,{\sc ii} $\lambda$4686 line fluxes and literature results for LMC WN stars from \citet{2019AJ....158..192L} reveal $\log F_{\rm HeII 4686} - F_{\rm HeII 4686}^{\rm L19} = +0.02 \pm 0.05$ for 30 stars in common, while comparisons between C\,{\sc iv} $\lambda$5808 line fluxes and literature results for LMC WC stars from \citet{1990ApJ...348..471S} reveal $\log F_{\rm CIV 5801,12} - F_{\rm CIV 5801,12}^{\rm S90a} = -0.07 \pm 0.09$ for 10 stars in common. Several Magellanic Cloud systems are host to multiple WN+WN stars,  including BAT99-116 \citep[Mk~34,][]{2019MNRAS.484.2692T}, and BAT99-118 \citep[R144,][]{2021A&A...650A.147S}, so we adopt equal line contributions from each component since mass ratios are close to unity, which should be a reasonable assumption with the potential exception of AB5 \citep[HD~5980,][]{2014AJ....148...62K}. Since AB5 is known to be spectroscopically variable \citep{2022RMxAA..58..403K}, we have selected the HST/STIS dataset from 2016 (phase 0.36) instead of 2014 (phase 0.0) \citep{2019MNRAS.486..725H}. R140a is omitted since it involves a mix of WN and WC populations which are difficult to deblend \citep{2018A&A...614A.147C}. 

\begin{figure}
\centering
	\includegraphics[width=0.525\linewidth,bb=175 50 430 535, angle=-90]{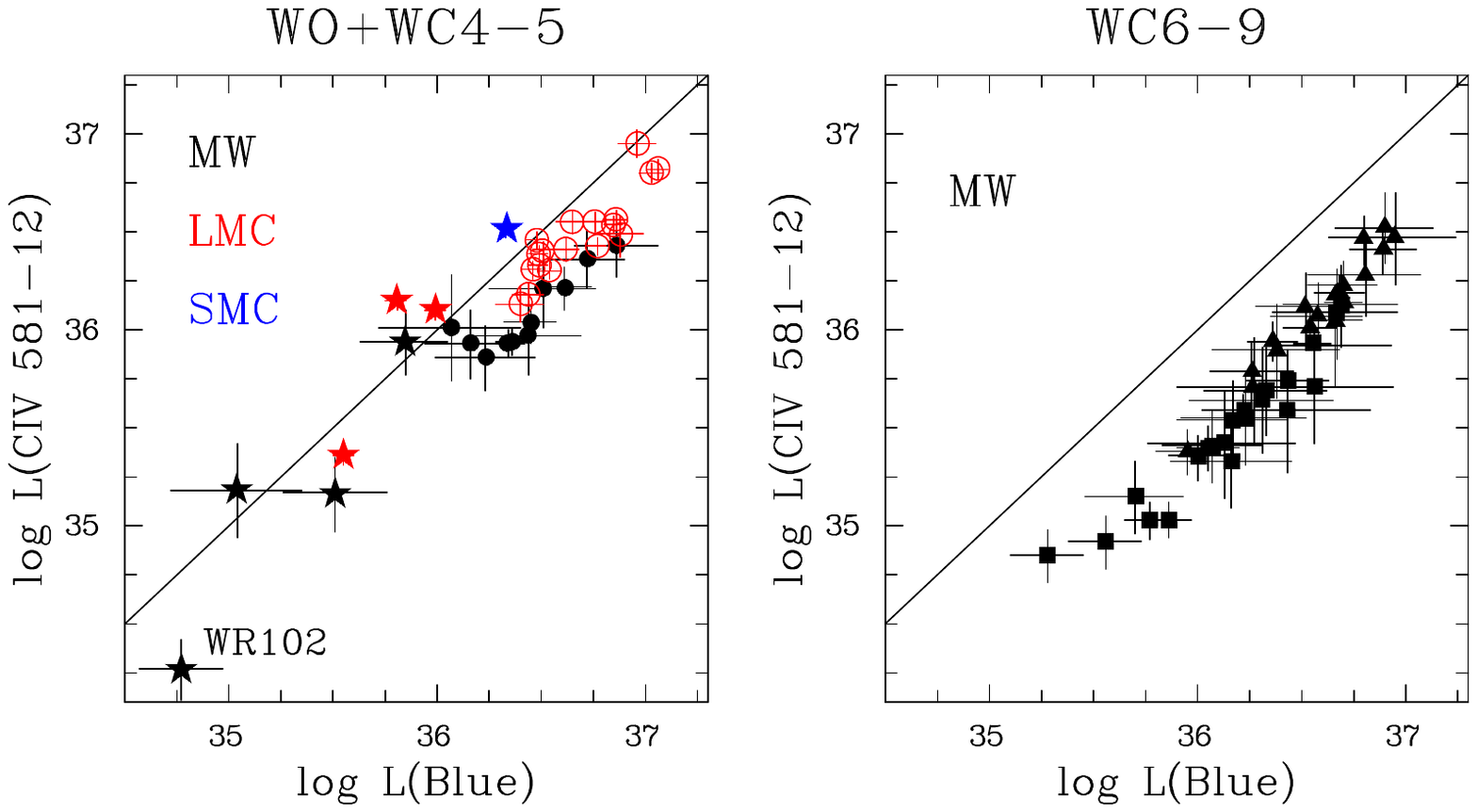}
	\centering
  \caption{Blue (C\,{\sc iii} $\lambda\lambda$4647,51, C\,{\sc iv} $\lambda$4658, He\,{\sc ii} $\lambda$4686) versus yellow (C\,{\sc iv} $\lambda\lambda$5801,12) line luminosities of WC and WO stars in the Milky Way (black), LMC (red) and SMC (blue). Symbols as in Figure~\ref{5808}. The solid line indicates equal blue and yellow luminosities.}
	\label{blue_5808_WC_WO}
\end{figure}

Revised interstellar extinctions are again drawn from spectroscopic studies  where available \citep[e.g.][]{2014A&A...565A..27H}, or  $(b-v)_{0}$ = --0.30 mag otherwise, plus E$_{\rm B-V}$  = 1.21 $E_{b-v}$ and R$_{\rm V}$  = 3.1. Again, we assume that literature $A_{\rm V}$ values are reliable to 10\%. For reference, the average difference between values of $E_{\rm B-V}$ adopted here for LMC WC+O binaries is --0.03$\pm$0.13 mag with respect to \citet{2001MNRAS.324...18B}. The LMC distance of 49.6~kpc adopted from
\citet{2019Natur.567..200P} and a SMC distance of 61.2~kpc obtained from the 0.458 mag difference in distance moduli between the Magellanic Clouds \citep{2014ApJ...780...59G}. We assume an uncertainty of 2\% for individual distances within the Magellanic Clouds. We adopt the LMC extinction law of \citet{1983MNRAS.203..301H} for both Clouds, in common with the studies of Magellanic Cloud WR stars by \citet{2014A&A...565A..27H} and \citet{2015A&A...581A..21H}, noting that alternative prescriptions are available \citep{2003ApJ...594..279G, 2014A&A...564A..63M}. 

Line luminosities of individual Magellanic Cloud WN, Of/WN, WN/WC, WC and WO stars are included in Tables~\ref{WN-all}-\ref{WO-all} in the Appendix. The largest source of uncertainty in line luminosities of Magellanic Cloud stars is usually  flux calibration since distances are well established and extinctions are generally low, with a few exceptions (e.g. VFTS~682, Rubin-Diez et al. in prep). Since Rubin-Diez et al. adopt the 30 Dor extinction law of \citet{2014A&A...564A..63M} for the heavily reddened star VFTS~682 (WN5--7h) we are able to quantify differences in line intensities following \citet{1983MNRAS.203..301H}. We typically obtain line intensities within 10--20\%, higher intensities for violet lines following \citet{1983MNRAS.203..301H}, and higher intensities for red lines following \citet{2014A&A...564A..63M}. Uncertainties on stellar luminosities from literature results account for distance and extinction, but exclude systematic uncertainties resulting from individual studies (variable, albeit typically 0.1--0.2 dex).



\begin{table*}
\caption{WN and Of/WN line luminosity calibrations for Milky Way, LMC and SMC stars, including He\,{\sc ii} $\lambda$4686 FWHM in km\,s$^{-1}$. The complex at $\lambda$4100 involves N\,{\sc iii} $\lambda\lambda$4097,4103, Si\,{\sc iv} $\lambda\lambda$4088,4116, He\,{\sc ii} $\lambda$4100+H$\delta$, while the feature at $\lambda$4630 involves N\,{\sc v} $\lambda\lambda$4603,20, N\,{\sc iii} $\lambda\lambda$4634,41 (or N\,{\sc ii} $\lambda\lambda$4601,43 for very late WN subtypes). 
Line luminosities have been adjusted for systems host to WN+WN binaries (marked with $\diamond$), namely WR43A, BAT99-116, BAT99-118 and AB5.}
\label{WN-calib}
\begin{tabular}{l@{\hspace{1.5mm}}l@{\hspace{1mm}}c@{\hspace{1mm}}l@{\hspace{1mm}}l@{\hspace{1.5mm}}l@{\hspace{1.5mm}}l@{\hspace{1.5mm}}l@{\hspace{1.5mm}}l@{\hspace{1.5mm}}l@{\hspace{1.5mm}}l@{\hspace{1.5mm}}l@{\hspace{1.5mm}}l@{\hspace{1.5mm}}l}
\hline
Category & N & HeII 4686 & \multicolumn{2}{c}{$L_{\rm HeII 4686}$}  & $\underline{L_{\rm NIV~3478,85}}$  & $\underline{L_{\rm NIV~4058}}$ & $\underline{L_{\rm 4100}}$ & $\underline{L_{\rm 4630}}$  & $\underline{L_{\rm HeII~5412}}$ &$\underline{L_{\rm CIV~5801,12}}$ & 
$\underline{L_{\rm HeI~5876} }$ & $\underline{L_{{\rm H}\alpha}}$ & $\underline{L_{\rm NIV~7103,29}}$ \\  
               &    & FWHM  & 10$^{35}$ erg\,s$^{-1}$                   &  $10^{-3} L_{\rm Bol}$                                    & $L_{\rm HeII~4686}$                   &  $L_{\rm HeII~4686}$                     & $L_{\rm HeII~4686}$                & $L_{\rm HeII~4686}$                   & $L_{\rm HeII~4686}$                   &  $L_{\rm HeII~4686}$  &  $L_{\rm HeII~4686}$                   & $L_{\rm HeII~4686}$                 & $L_{\rm HeII~4686}$ \\
\hline
\multicolumn{13}{c}{Milky Way ($Z_{\odot}$)} \\
WN2--5w & 22 &  1680$\pm$310    & \phantom{0}4.9$\pm$\phantom{0}3.7 & 0.37$\pm$0.23 & 0.68$\pm$0.21 &  0.26$\pm$0.14 &   0.27$\pm$0.14      & 0.29$\pm$0.15 &  0.12$\pm$0.03 & 0.06$\pm$0.05 & 0.03$\pm$0.02 & 0.20$\pm$0.08 & 0.21$\pm$0.15 \\
WN3--7s & 12 &   2480$\pm$510   & 20.3$\pm$\phantom{0}8.9 & 1.25$\pm$0.40 & 0.49$\pm$0.14 & \multicolumn{2}{c}{--- 0.40$\pm$0.25 ---} & 0.27$\pm$0.16 &  0.13$\pm$0.02 & 0.07$\pm$0.03 & 0.05$\pm$0.03 & 0.15$\pm$0.04 & 0.16$\pm$0.04 \\
WN6--8  & 25 & \phantom{9}990$\pm$240      & \phantom{0}7.4$\pm$\phantom{0}5.0 & 0.44$\pm$0.20& 0.30$\pm$0.16 & 0.18$\pm$0.09 & 0.65$\pm$0.28 & 0.86$\pm$0.44 &  0.11$\pm$0.04 & 0.03$\pm$0.02 & 0.32$\pm$0.31 & 0.51$\pm$0.43 & 0.12$\pm$0.10 \\
WN9--11  & 4  & \phantom{1}440$\pm$100        & \phantom{0}5.1$\pm$\phantom{0}6.9 & 0.13$\pm$0.11 & $\cdots$ & 0.00$\pm$0.00 &  6.2$\pm$10.2 & 3.1$\pm$3.2 & 0.03$\pm$0.03 & 0.01$\pm$0.01 & 9.7$\pm$18.0 & 45$\pm$87 & 0.01$\pm$0.0  \\
WN5--7h &  7$\diamond$ & 1260$\pm$510    & 11.0$\pm$\phantom{0}3.8 & 0.15$\pm$0.04 & 0.37$\pm$0.39 &  0.18$\pm$0.07 & 0.54$\pm$0.23  &0.58$\pm$0.32  &  0.06$\pm$0.01 & 0.03$\pm$0.02 & 0.05$\pm$0.02 & 0.64$\pm$0.13 & 0.09$\pm$0.01          \\
Of/WN & 2 & 1610$\pm$470    &   \phantom{0}4.8$\pm$0.5 & 0.05$\pm$0.01 & 0.16$\pm$0.08  & 0.25$\pm$0.09 & 0.26$\pm$0.24 & 0.48$\pm$0.16 & 0.06                      & 0.01                &  0.00                  & 0.58                 & 
$\cdots$ \\
\multicolumn{13}{c}{LMC ($0.4 Z_{\odot}$)} \\
WN2--5w & 24 &  1750$\pm$280 & \phantom{0}3.3$\pm$\phantom{0}2.6 & 0.21$\pm$0.14 & 0.79$\pm$0.19 &  0.10$\pm$0.11 & 0.13$\pm$0.12   & 0.19$\pm$0.12 &  0.11$\pm$0.03 & 0.04$\pm$0.03 & 0.00$\pm$0.01 & 0.22$\pm$0.06 & 0.08$\pm$0.04 \\
WN3--7s & 18 & 2170$\pm$600     & 10.1$\pm$\phantom{0}6.9 & 0.70$\pm$0.24 & 0.50$\pm$0.23 & \multicolumn{2}{c}{--- 0.12$\pm$0.04 --- } & 0.16$\pm$0.03 &  0.13$\pm$0.01 & 0.05$\pm$0.03 & 0.01$\pm$0.02 & 0.16$\pm$0.03 & 0.07$\pm$0.03 \\
WN6--8   & 13$\diamond$ & 1080$\pm$270    & 17.8$\pm$14.0 & 0.39$\pm$0.16     & 0.20$\pm$0.08   & 0.17$\pm$0.03    &  0.33$\pm$0.14 & 0.34$\pm$0.19  & 0.09$\pm$0.03 & 0.03$\pm$0.02  & 0.09$\pm$0.06 & 0.41$\pm$0.24 & 0.06$\pm$0.01 \\ 
WN9--11 & 8 & \phantom{1}240$\pm$\phantom{1}40  & \phantom{0}1.6$\pm$1.1 & 0.09$\pm$0.07 & $\cdots$ & 0.01$\pm$0.02 & 1.50$\pm$0.93 & 1.9$\pm$1.1 & 0.05$\pm$0.09 & 0.00$\pm$0.00 & 1.8$\pm$1.5 & 6.5$\pm$5.1 & $\cdots$ \\
WN5--7h & 8$\diamond$ &  1830$\pm$250   & 30.2$\pm$13.6 & 0.24$\pm$0.12 & 0.44$\pm$0.14 &  0.18$\pm$0.05 & 0.18$\pm$0.09  & 0.07$\pm$0.03  &  0.09$\pm$0.02 & 0.03$\pm$0.01 & 0.01$\pm$0.01 & 0.34$\pm$0.08 & 0.11$\pm$0.04 \\
Of/WN & 6 &  1130$\pm$300 & \phantom{0}3.7$\pm$1.9  & 0.05$\pm$0.02 & 0.30 & 0.22 & 0.00 & 0.18$\pm$0.12 & 0.02$\pm$0.02 & 0.04$\pm$0.04 & 0.00$\pm$0.00 & 0.59$\pm$0.13 & 0.11$\pm$0.05 \\
\multicolumn{13}{c}{SMC ($0.2 Z_{\odot}$)} \\
WN2--5w & 9 &  1630$\pm$280 & \phantom{0}1.7$\pm$\phantom{0}1.3 & 0.05$\pm$0.03 & 0.58 & 0.04$\pm$0.08 & 0.00$\pm$0.01  & 0.19$\pm$0.13 &  0.06$\pm$0.02 & 0.01$\pm$0.02 & 0.00$\pm$0.00 & 0.25$\pm$0.14 & 0.22\\
WN6--8    & 2$\diamond$ &   1060$\pm$280  & 15.5$\pm$11.4 & 0.35$\pm$0.06 & 0.23$\pm$0.07 &  0.09$\pm$0.07 & 0.10$\pm$0.03 & 0.04$\pm$0.03 &  0.10$\pm$0.00 & 0.03$\pm$0.02 & 0.03$\pm$0.02 & 0.23$\pm$0.02 & 0.06$\pm$0.01     \\
\hline
All WN2--8       & 140$\diamond$ &  1570$\pm$610   & \phantom{0}9.7$\pm$10.1 & 0.42$\pm$0.37 &  0.48$\pm$0.27 & 0.14$\pm$0.11 & 0.26$\pm$0.25 & 0.34$\pm$0.33 & 0.10$\pm$0.04 & 0.04$\pm$0.03 & 0.08$\pm$0.17 & 0.30$\pm$0.23 & 0.12$\pm$0.14 \\
All WN9--11      & 12                    & \phantom{0}300$\pm$120      & \phantom{0}2.8$\pm$4.1   & 0.10$\pm$0.07  & $\cdots$             & 0.01$\pm$0.01 & 3.1\phantom{0}$\pm$5.9 & 2.3$\pm$1.9 & 0.02$\pm$0.02 & 0.00$\pm$0.01 & 4.4$\pm$10.3 & 20$\pm$52 & 0.01$\pm$0.00 \\
All Of/WN        & 8                        &  1250$\pm$380  & \phantom{0}3.5$\pm$1.3    & 0.05$\pm$0.02 &  0.21$\pm$0.10  & 0.24$\pm$0.07 & 0.18$\pm$0.23 & 0.26$\pm$0.19 & 0.03$\pm$0.02 & 0.03$\pm$0.03 & 0.00$\pm$0.00 & 0.59$\pm$0.12 & 0.11$\pm$0.05 \\
\hline
\end{tabular}
\begin{footnotesize}
\end{footnotesize}
\end{table*}

\begin{table*}
\caption{WN/C line luminosity calibrations for Milky Way and LMC stars, including He\,{\sc ii} $\lambda$4686 FWHM in km\,s$^{-1}$. The complex at $\lambda$4100 involves N\,{\sc iii} $\lambda\lambda$4097,4103, Si\,{\sc iv} $\lambda\lambda$4088,4116, He\,{\sc ii} $\lambda$4100+H$\delta$, while the feature at $\lambda$4603--51 involves N\,{\sc v} $\lambda\lambda$4603,20, N\,{\sc iii} $\lambda\lambda$4634,41 and C\,{\sc iii} $\lambda\lambda$4647,51.}
\label{WNC-calib}
\begin{tabular}{l@{\hspace{1.5mm}}r@{\hspace{1mm}}c@{\hspace{1mm}}l@{\hspace{1mm}}l@{\hspace{1.5mm}}l@{\hspace{1.5mm}}l@{\hspace{1.5mm}}l@{\hspace{1.5mm}}l@{\hspace{1.5mm}}l@{\hspace{1.5mm}}l@{\hspace{1.5mm}}l@{\hspace{1.5mm}}l@{\hspace{1.5mm}}l}
\hline
Category & N & HeII 4686 & \multicolumn{2}{c}{$L_{\rm HeII 4686}$}  & 
$\underline{L_{\rm NIV~3478-85}}$ & $\underline{L_{\rm NIV~4058}}$ & $\underline{L_{\rm 4100}}$ & $\underline{L_{\rm 4603,51}}$ & $\underline{L_{\rm HeII~5412}}$ & $\underline{L_{\rm CIII~5696}}$ & $\underline{L_{\rm CIV~5801,12}}$  & $\underline{L_{{\rm H}\alpha}}$ & $\underline{L_{\rm NIV~7103,29}}$ \\  
               &    & FWHM  & 10$^{35}$ erg\,s$^{-1}$                   &  $10^{-3} L_{\rm Bol}$                                    & $L_{\rm HeII~4686}$                   &  $L_{\rm HeII~4686}$                     & $L_{\rm HeII~4686}$                & $L_{\rm HeII~4686}$                   & $L_{\rm HeII~4686}$             &  $L_{\rm HeII~4686}$       &  $L_{\rm HeII~4686}$                   & $L_{\rm HeII~4686}$                 & $L_{\rm HeII~4686}$ \\
\hline
\multicolumn{13}{c}{3 Milky Way ($Z_{\odot}$) and 2 LMC ($0.4 Z_{\odot}$)} \\
WNE/C &  5  &  2240$\pm$430 & \phantom{0}7.5$\pm$2.9 & 0.47$\pm$0.30
& 0.64$\pm$0.04 & 0.14$\pm$0.12 & 0.26$\pm$0.19 & 1.54$\pm$1.82 & 0.18$\pm$0.08 & 0.01$\pm$0.02 & 1.38$\pm$1.39 & 0.21$\pm$0.08 & 0.19$\pm$0.20 \\
\multicolumn{13}{c}{Milky Way ($Z_{\odot}$)} \\
WNL/C & 4    &  1270$\pm$130 & \phantom{0}7.3$\pm$3.1 & 0.36$\pm$0.17 
& 0.82 & 0.91 & 0.87                 &  1.75$\pm$1.32 & 0.15$\pm$0.05 & 0.06$\pm$0.08 & 0.53$\pm$0.17 & 0.18$\pm$0.09                   & 0.20                 \\
\hline
All  WN/C    & 9     &  1810$\pm$600 & \phantom{0}7.4$\pm$2.8 &  0.43$\pm$0.25 
& 0.68$\pm$0.10 & 0.27$\pm$0.33 & 0.36$\pm$0.30 & 1.63$\pm$1.53 & 0.17$\pm$0.06 & 0.03$\pm$0.06 & 1.26$\pm$1.21  & 0.20$\pm$0.08 & 0.19$\pm$0.14 \\
\hline
\end{tabular}
\begin{footnotesize}
\end{footnotesize}
\end{table*}

\begin{table*}
\caption{WC line luminosity calibrations for Milky Way and LMC stars, including C\,{\sc iv} $\lambda\lambda$5801,12 FWHM in km\,s$^{-1}$.  The blue feature involves C\,{\sc iii} $\lambda\lambda$4647,51, C\,{\sc iv} $\lambda$4658 and He\,{\sc ii} $\lambda$4686, while the feature at $\lambda$6559,81 involves He\,{\sc ii} $\lambda$6560 and C\,{\sc ii} $\lambda\lambda$6559,81.}
\label{WC-calib}
\begin{tabular}{l@{\hspace{1.5mm}}r@{\hspace{1.5mm}}c@{\hspace{1.5mm}}l@{\hspace{1.5mm}}l@{\hspace{1.5mm}}l@{\hspace{1.5mm}}l@{\hspace{1.5mm}}l@{\hspace{1.5mm}}l@{\hspace{1.5mm}}l@{\hspace{1.5mm}}l@{\hspace{1.5mm}}l@{\hspace{1.5mm}}l}
\hline
Category & N & CIV 5801,12 & \multicolumn{2}{c}{$L_{\rm CIV 5801.12}$} & $\underline{L_{\rm OIV~3403,13}}$ & $\underline{L_{\rm Blue}}$  & $\underline{L_{\rm CIII~5696}}$ &$\underline{L_{\rm HeI~5876}}$ & 
$\underline{L_{\rm 6559,81} }$ & $\underline{L_{\rm CIII~6727,73}}$ & $\underline{L_{\rm CIV~7725}}$  & $\underline{L_{\rm CIII~9701,19}}$\\  
               &    & FWHM & 10$^{35}$ erg\,s$^{-1}$                   &  $10^{-3} L_{\rm Bol}$                                    & $L_{\rm CIV~5801,12}$                   &  $L_{\rm CIV~5801,12}$                     & $L_{\rm CIV~5801,12}$                & $L_{\rm CIV~5801,12}$                   & $L_{\rm CIV~5801,12}$                   &  $L_{\rm CIV~5801,12}$                   & $L_{\rm CIV~5801,12}$                  & $L_{\rm CIV~5801,12}$ \\
\hline
\multicolumn{13}{c}{Milky Way ($Z_{\odot}$)} \\
WC4--5 & 11 &  2790$\pm$630  & 13.3$\pm$\phantom{0}6.6 & 1.43$\pm$0.44 
& 0.55$\pm$0.25 & 2.30$\pm$0.51 & 0.02$\pm$0.03 & 0.08$\pm$0.01 & 0.07$\pm$0.03 & 0.11$\pm$0.03 & 0.09$\pm$0.02 & 0.13$\pm$0.07 \\
WC6--7 & 18 &  2180$\pm$400    & 15.3$\pm$\phantom{0}9.0 & 0.89$\pm$0.43 
& 0.77$\pm$0.24 & 3.11$\pm$0.50    & 0.35$\pm$0.24 & 0.16$\pm$0.11 & 0.16$\pm$0.06 & 0.17$\pm$0.03 & 0.11$\pm$0.01             & 0.28$\pm$0.06 \\
WC8--9 & 21 &  1480$\pm$140    & \phantom{0}4.1$\pm$\phantom{0}3.5 & 0.41$\pm$0.20 
& 1.32$\pm$0.92 & 4.82$\pm$1.19    & 3.21$\pm$1.01 & 0.56$\pm$0.26 & 1.12$\pm$0.48 & 0.60$\pm$0.27 &0.11$\pm$0.04 & 0.99$\pm$0.28\\
\multicolumn{13}{c}{LMC ($0.4 Z_{\odot}$)} \\
WC4--5 & 18 &  3370$\pm$490  & 34.1$\pm$19.8 & 2.19$\pm$0.27 
& 0.55$\pm$0.26 & 1.63$\pm$0.40    & 0.02$\pm$0.03 & 0.02$\pm$0.03 & 0.05$\pm$0.01 & 0.06$\pm$0.02 & 0.06$\pm$0.01 & 0.06$\pm$0.03 \\
\hline
All  WC      & 68 &  2380$\pm$850   & 16.4$\pm$16.2 & 0.98$\pm$0.68 
&  0.79$\pm$0.53 & 3.12$\pm$1.48    & 1.09$\pm$1.54 & 0.26$\pm$0.28 & 0.45$\pm$0.56 & 0.26$\pm$0.27 & 0.08$\pm$0.03   & 0.34$\pm$0.38  \\
\hline
\end{tabular}
\begin{footnotesize}
\end{footnotesize}
\end{table*}

\begin{table*}
\caption{WO line luminosity calibrations for Milky Way, LMC and SMC stars, including C\,{\sc iv} $\lambda\lambda$5808 FWHM in km\,s$^{-1}$. The blue feature involves C\,{\sc iv} $\lambda$4658 and He\,{\sc ii} $\lambda$4686.} 
\label{WO-calib}
\begin{tabular}{l@{\hspace{2mm}}r@{\hspace{1mm}}c@{\hspace{1mm}}l@{\hspace{1mm}}l@{\hspace{5mm}}l@{\hspace{2mm}}l@{\hspace{2mm}}l@{\hspace{2mm}}l@{\hspace{2mm}}l@{\hspace{2mm}}l}
\hline
Subtype & N & CIV~5801,12 & \multicolumn{2}{c}{$L_{\rm CIV~5801,12}$}  & $\underline{L_{\rm OIV~3403,13}}$ & $\underline{L_{\rm OVI~3811,34}}$ & $\underline{L_{\rm Blue}}$ & $\underline{L_{\rm OV~5572,607}}$  & $\underline{L_{\rm HeII~6560}}$ & $\underline{L_{\rm CIV~7725} }$  \\  
               &    & FWHM  & 10$^{35}$ erg\,s$^{-1}$                   &  $10^{-3} L_{\rm Bol}$                                    & $L_{\rm CIV~5801,12}$                   &  $L_{\rm CIV~5801,12}$                     & $L_{\rm CIV~5801,12}$                & $L_{\rm CIV~5801,12}$                   & $L_{\rm CIV~5801,12}$                   &  $L_{\rm CIV~5801,12}$       \\
\hline
\multicolumn{11}{c}{Milky Way ($Z_{\odot}$)} \\
WO2--4 & 4 &  6800$\pm$1300  & \phantom{0}3.0$\pm$3.9 & 0.25$\pm$0.20 & 1.8$\pm$0.8 & 8.6$\pm$12.2 & 1.7$\pm$1.2 & 0.5$\pm$0.6 & 0.22$\pm$0.16 & 0.45$\pm$0.47  \\
\multicolumn{11}{c}{LMC ($0.4 Z_{\odot}$)} \\
WO3--4 & 3 &   5300$\pm$300  & \phantom{0}9.7$\pm$6.5 & 1.18$\pm$0.84 & 1.9$\pm$2.0 & 2.2$\pm$3.1 & 0.9$\pm$0.6 &0.16$\pm$0.14 & 0.08$\pm$0.08 & 0.14$\pm$0.13 \\
\multicolumn{11}{c}{SMC ($0.2 Z_{\odot}$)} \\
WO4 & 1 &   5300 & 32.9                & 0.59                  & 1.0                 & 0.7                 &  0.65                & 0.16                   & 0.04          & 0.19          \\
\hline
All  WO & 8 &   6100$\pm$1200   & \phantom{0}9$\pm$11 & 0.70$\pm$0.69 &  1.7$\pm$1.3 & 5.2$\pm$8.5 & 1.3$\pm$1.0 & 0.3$\pm$0.4 & 0.13$\pm$0.13 & 0.30$\pm$0.36 \\
\hline
\end{tabular}
\begin{footnotesize}
\end{footnotesize}
\end{table*}

\section{Results}\label{results}

\subsection{WN, Of/WN and WN/C stars}

In Fig.~\ref{4686} we compare He\,{\sc ii} $\lambda$4686 FWHM and line luminosities for Milky Way and Magellanic Cloud WN and OIf/WN stars. Galactic strong-lined WN stars exhibit the highest line luminosities and largest FWHM $\lambda$4686. LMC strong-lined WN stars overlap with Galactic counterparts, but  typically have lower  luminosities and FWHM. Weak-lined WN2--5 stars in the Milky Way span a broad range of line luminosities, ranging from $\log (L_{\rm 4686}$/erg\,s$^{-1}$) = 36.1$\pm$0.2 (WR141) to 
34.6$\pm$0.1 (WR3), with similar results obtained for LMC stars, except that LMC~170--2 \citep{2017ApJ...841...20N} has a very low luminosity of $\log (L_{\rm 4686}$/erg\,s$^{-1}$) = 34.2, and is morphologically similar to WR3 and WR46 in the Milky Way \citep{1995A&A...302..457C}. SMC WN2--5w stars possess uniformly low $\lambda$4686 line luminosities,  spanning $\log (L_{\rm 4686}$/erg\,s$^{-1}$) = 35.6 (AB3) to 34.7 (AB11), which has previously been highlighted by \citet{2006A&A...449..711C}. \citet{2020MNRAS.491.4406S} have investigated wind driving of WR stars, and establish a dependence on metallicity (Fe-group elements) and Eddington parameter $\Gamma_{e}$ ($\propto L/M$) so strong-lined WN stars likely differ in their wind properties from weak-lined WN stars owing to higher luminosity-to-mass ratios. He\,{\sc ii} $\lambda$4686 properties of Galactic and LMC WN/C stars are similar to normal WN stars in these galaxies.

Galactic weak-lined WN6--8 stars possess relatively low FWHM $\lambda$4686, albeit a wide range of line luminosities, spanning 
$\log (L_{\rm 4686}$/erg\,s$^{-1}$) = 36.3$\pm$0.5 (WR147) to  35.0$\pm$0.3 (WR107).  LMC weak-lined WN6--8 stars extend the Galactic sequence to higher luminosities, including R144 (BAT99--118), which is host to a multiple WN+WN system \citep{2021A&A...650A.147S} and R145 (BAT99--119) which is also a binary system \citep{2019A&A...627A.151S}. There are too few SMC late-type WN stars to draw robust conclusions, although  the high luminosity SMC system AB~5 (HD~5980) is also multiple \citep{2014AJ....148...62K, 2022RMxAA..58..403K}. Milky Way and LMC WN9--11 stars possess very low FWHM and line luminosities, with the notable exception of WR105 (WN9) whose $\lambda$4686 line luminosity is comparable to WN6--8 stars. H-rich main sequence WN6--7h stars in the Milky Way form a relatively homogenous group, overlapping with weak-lined classical WN stars in He\,{\sc ii} $\lambda$4686 properties, and includes the WN+WN binary system WR43A. LMC counterparts extend to higher luminosity, broader lines (WN5h) stars, including the WN5h+WN5h binary Mk~34 \citep[BAT99-116,][]{2019MNRAS.484.2692T} with Of/WN stars, such as Mk~35  \citep[BAT99-114,][]{2011MNRAS.416.1311C}, possessing lower line luminosities. 

Fig.~\ref{blue_5808_WN} (upper panels) compares He\,{\sc ii} $\lambda$4686 and N\,{\sc v} $\lambda\lambda$4603,20+N\,{\sc iii} $\lambda$4634,41 line luminosities of WN, Of/WN and WN/C stars. For early subtypes N\,{\sc v} $\lambda\lambda$4603,20/He\,{\sc ii} $\lambda$4686 $\sim$ 0.2--0.3 in all environments, with the exception of Galactic WN/C stars which also exhibit a contribution from C\,{\sc iii} $\lambda\lambda$4647,51. In contrast, for late subtypes N\,{\sc iii} $\lambda\lambda$4634,41/He\,{\sc ii} $\lambda$4686 $\sim$ 0.8 in the Milky Way, $\sim$0.2 in the LMC and $\leq$0.1 in the SMC, with the exception of WN9--11 stars for which N\,{\sc ii} $\lambda\lambda$4601,43+N\,{\sc iii} $\lambda\lambda$4634,41/He\,{\sc ii} $\lambda$4686 $\geq$ 1. Fig.~\ref{blue_5808_WN} (lower panels) compares the blue (He\,{\sc ii} $\lambda$4686, N\,{\sc v} $\lambda\lambda$4603,20, N\,{\sc iii} $\lambda$4634,41, C\,{\sc iii} $\lambda\lambda$4647,51) and C\,{\sc iv} $\lambda\lambda$5801,12 line luminosities of WN, Of/WN and WN/C stars. For Galactic early-type WN stars  the line luminosity of C\,{\sc iv} $\lambda\lambda$5801,12 is an order of magnitude weaker than the blue feature, with late-type WN stars somewhat weaker still (and usually resolved into individual components). A broadly similar dependence is obtained for Magellanic Cloud counterparts, with C\,{\sc iv} $\lambda\lambda$5801,12 undetected in a subset of WN stars from all environments. In contrast, C\,{\sc iv} $\lambda\lambda$5801,12 luminosities of WN/C stars exceed normal WN stars by an order of magnitude, and are more typical of WC stars. The least extreme examples of WN/C stars in our sample are WR58 and WR98 \citep{1989ApJ...337..251C}.

Tables~\ref{WN-calib}-\ref{WNC-calib} provide calibrations of He\,{\sc ii} $\lambda$4686  line luminosities for WN2--8 (N=140), WN9--11 (N=12), Of/WN (N=8) and WN/C (N=9) stars, respectively, together with other prominent optical emission lines. This highlights  features that may be detectable in high S/N observations of WR galaxies, notably the $\lambda$4100 complex in WN stars which is largely unaffected by nebular emission, in contrast with H$\alpha$, H$\beta$, He\,{\sc i} $\lambda$5876. Although H$\beta$ is omitted here,  $L_{{\rm H} \alpha}/L_{{\rm H} \beta} = 1.9 \pm$0.8 for  141 Milky Way and Magellanic Cloud WN  and WN/C stars.

\subsection{WC and WO stars}

In Fig.~\ref{5808} (left and centre panels) we compare the FWHM and line luminosities of C\,{\sc iv} $\lambda\lambda$5801,12 for Milky Way and Magellanic Cloud WO and WC stars. WO stars possess exceptionally high FWHM though span a wide range of C\,{\sc iv} $\lambda\lambda$5801,12 line luminosities, ranging from $\log (L_{\rm CIV~5801,12}$/erg\,s$^{-1})$ = 36.52$\pm$0.05 (AB8, SMC) to 34.25$\pm$0.15 (WR102, Milky Way). In contrast, LMC WC4--5 stars possess uniformly  high luminosities, with the WC4+O binary BAT99-70 possessing the highest luminosity C\,{\sc iv} $\lambda\lambda$5801,12. Galactic counterparts overlap with LMC stars, though extend to lower luminosities. Galactic WC6--7 stars overlap with WC4--5 subtypes, although these extend to lower luminosities, unusually so for WR39. Galactic WC8--9 stars exhibit a narrow range of FWHM but a broad range of line luminosities, ranging from  $\log (L_{\rm CIV~5801,12}$/erg\,s$^{-1})$ = 36.1$\pm$0.2 (WR60) to 34.85$\pm$0.15 (WR77). \citet{1990ApJ...358..229S} have previously studied Galactic WC stars in clusters and associations to suggest uniform C\,{\sc iv} $\lambda$5801--12 fluxes for Galactic WC5--7 stars, with lower line fluxes for WC8--9 stars. Fig.~\ref{5808} (right panel) compares O\,{\sc vi} $\lambda\lambda$3811,34 FWHM and line luminosities of Milky Way and Magellanic Cloud WO stars, revealing uniformly high FWHM, though an order of magnitude spread in line luminosity from $\log (L_{\rm OVI~3811,34}$/erg\,s$^{-1})$ = 36.35$\pm$0.05 (AB8) to 35.4$\pm$0.3 (WR93b). 

Fig.~\ref{blue_5808_WC_WO} compares blue (C\,{\sc iii} $\lambda\lambda$4647,51, C\,{\sc iv} $\lambda$4658, He\,{\sc ii} $\lambda$4686) to yellow (C\,{\sc iv} $\lambda\lambda$5801,12) line luminosities of WC and WO stars. For Galactic WC4--5 stars the blue and yellow features are well correlated, with the former  exceeding C\,{\sc iv} $\lambda\lambda$5801,12 by a factor of  2--3, while these features are more comparable in strength in LMC WC4--5 and WO stars. Blue and yellow features are also tightly correlated for WC6--9 stars, with the former exceeding C\,{\sc iv} $\lambda\lambda$5801,12 by a factor of 3 (WC6--7) to 5 (WC8--9).

Tables~\ref{WC-calib}-\ref{WO-calib} provide calibrations of  C\,{\sc iv} $\lambda\lambda$5801,12 line luminosities for WC (N=68) and WO (N=8) stars, respectively. Other strong WC features include C\,{\sc iii} $\lambda\lambda$4647,51, $\lambda$5696, the latter having occasionally been reported in WR galaxies \citep[e.g.][]{1999A&A...341..399S}, with O\,{\sc iv} $\lambda\lambda$3403,13 also prominent in WC and WO stars.

\begin{table*}
\begin{center}
\caption{Comparison to literature calibrations of violet (O\,{\sc iv} $\lambda\lambda$3403,11, N\,{\sc iv} $\lambda\lambda$3478,85, O\,{\sc vi} $\lambda\lambda$3811,34), 
blue (He\,{\sc ii} $\lambda$4686, C\,{\sc iii} $\lambda\lambda$4647,51, N\,{\sc iii--v} $\lambda\lambda$4603,41),  yellow (C\,{\sc iii} $\lambda$5696, C\,{\sc iv} $\lambda\lambda$5801,12) and red (N\,{\sc iv} $\lambda\lambda$7103,29, C\,{\sc iii} $\lambda\lambda$9701,19) WR line luminosities. Wind features associated with strong nebular emission (H$\alpha$, $\beta$) are excluded.}
\label{literature}
\begin{tabular}{
c@{\hspace{1mm}}l@{\hspace{1mm}}
c@{\hspace{1mm}}l@{\hspace{1mm}}
c@{\hspace{1mm}}l@{\hspace{1mm}}
c@{\hspace{1mm}}l@{\hspace{1mm}}
c@{\hspace{1mm}}l@{\hspace{1mm}}
c@{\hspace{1mm}}l@{\hspace{1mm}}
c@{\hspace{1mm}}l@{\hspace{1mm}}
c@{\hspace{1mm}}l@{\hspace{3mm}}
l@{\hspace{1mm}}l}
\hline
$L_{\rm OIV~3403,13}$ & N &
$L_{\rm NIV~3478,85}$  & N &
$L_{\rm OVI~3811,34}$   & N &
$L_{\rm Blue}$           &  N & 
$L_{\rm CIII~5696}$ & N & 
$L_{\rm CIV~5801,12}$ & N &
$L_{\rm NIV~7103,29}$ &  N &
$L_{\rm CIII~9701,19}$ & N &  Sample & Ref\\
      10$^{35}$ erg\,s$^{-1}$ & &
      10$^{35}$ erg\,s$^{-1}$ & &
      10$^{35}$ erg\,s$^{-1}$ & &
      10$^{35}$ erg\,s$^{-1}$  &  &
      10$^{35}$ erg\,s$^{-1}$ &  &
      10$^{35}$ erg\,s$^{-1}$ & &
      10$^{35}$ erg\,s$^{-1}$ & &
      10$^{35}$ erg\,s$^{-1}$ &
      &              &     \\
\hline
\multicolumn{18}{c}{WN3--7s} \\
   $\cdots$ &  & \phantom{1}9.5$\pm$5.7 & 10 &  $\cdots$ &  & 26$\pm$12 & 12     & $\cdots$ &     & \phantom{0}1.3$\pm$0.7  & 11 & 3.3$\pm$1.7 & \phantom{1}8   & $\cdots$  & & MW       & 3\\
    $\cdots$ &  & \phantom{1}5.4$\pm$3.7 & 15 & $\cdots$ &   & 12$\pm$\phantom{1}7 & 18      & $\cdots$ &     & \phantom{0}0.5$\pm$0.4 & 18  & 0.7$\pm$0.5 &  16  & $\cdots$ & & LMC       & 3 \\
 \hline
\multicolumn{18}{c}{WN2--5w} \\
    $\cdots$ &    & $\cdots$ &  & $\cdots$ & & \phantom{1}5$\pm$3  & 26   &  $\cdots$  &  &  \phantom{0}0.4$\pm$0.1 & \phantom{1}5   &$\cdots$ &   & $\cdots$   &  & MW+LMC$\dag$ & 1 \\ 
    $\cdots$ &  & \phantom{1}3.8$\pm$2.7 & 13 & 0.1$\sharp$ & 1 & \phantom{1}6$\pm$5 & 23  & $\cdots$ &     & \phantom{0}0.3$\pm$0.3  & 19 & 1.0$\pm$0.7 & 11   & $\cdots$  & & MW       & 3\\
    $\cdots$ &  & \phantom{1}3.0$\pm$1.8 & 12 & $\cdots$ &   & \phantom{1}4$\pm$3 & 24  &  $\cdots$ &    & \phantom{0}0.2$\pm$0.2 & 23 & 0.3$\pm$0.3 &  20  & $\cdots$ & & LMC       & 3 \\
    $\cdots$ &  & \phantom{1}2.1 & \phantom{1}1   & $\cdots$ &   & \phantom{1}2$\pm$1 & \phantom{1}9   &  $\cdots$  &     & \phantom{0}0.01$\pm$0.02 & \phantom{1}7 & 0.3 &  \phantom{1}1  &$\cdots$ & & SMC      & 3 \\
\hline
\multicolumn{18}{c}{WN6--8} \\
   $\cdots$ &          & $\cdots$        &     & $\cdots$ &  & 21$\pm$18  & 19                  & $\cdots$ &   & \phantom{0}1.0$\pm$0.6  & 10 & $\cdots$ &  & $\cdots$ &  & MW+LMC$\ddag$ & 1\\
   $\cdots$ &          & \phantom{1}2.2$\pm$1.4 &  12  & $\cdots$ &  & 13$\pm$\phantom{1}8  & 25                & $\cdots$ &   & \phantom{0}0.2$\pm$0.1   & 24 & 0.9$\pm$1.0 & \phantom{1}6   & $\cdots$ &  & MW         & 3\\
   $\cdots$ &          & \phantom{1}2.8$\pm$2.6 &   \phantom{1}8  & $\cdots$ &  & 26$\pm$25 & 15
                    & $\cdots$  &    &  \phantom{0}0.6$\pm$0.6 & 13 & 1.3$\pm$0.1 & \phantom{1}4   & $\cdots$ & & LMC         & 3 \\
  $\cdots$  &          & \phantom{1}3.2$\pm$1.6 & \phantom{1}2      & $\cdots$ &  & 10$\pm$\phantom{1}3 & \phantom{1}2                  & $\cdots$  &    & \phantom{0}0.6$\pm$0.6 & \phantom{1}2  & 0.9$\pm$0.6 & \phantom{1}2 & $\cdots$  & & SMC     & 3\\
\hline
\multicolumn{18}{c}{WN9--11} \\
 $\cdots$ &        & $\cdots$    &      & $\cdots$   &       &  \phantom{0}7$\pm$10 & 12 &  $\cdots$ &  & \phantom{0}0.01$\pm$0.01 & \phantom{0}4 & 0.03$\pm$0.02 & \phantom{0}2 & $\cdots$ &  & MW+LMC & 3 \\
\hline
\multicolumn{18}{c}{WN5--7h+Of/WN} \\
  $\cdots$ &           & \phantom{1}3.1$\pm$3.8   & \phantom{1}8       & $\cdots$  & & 14$\pm$\phantom{1}7 & \phantom{1}9                 & $\cdots$ &  & \phantom{0}0.2$\pm$0.1  & \phantom{1}6 & $\cdots$ &  & $\cdots$ & & MW & 3 \\
  $\cdots$ &           & 11.8$\pm$7.1 & \phantom{1}8       & $\cdots$ &  &  20$\pm$18 &                14                 & $\cdots$ &  & \phantom{0}0.7$\pm$0.7 &                   13 & 1.9$\pm$1.6 & \phantom{1}8 & $\cdots$ & & LMC & 3 \\
\hline
\multicolumn{18}{c}{WN/C} \\
  $\cdots$ &       & \phantom{1}4.9$\pm$1.6 &   \phantom{1}4    & $\cdots$ &   & 18$\pm$10 & \phantom{1}9 & \phantom{1}0.1$\pm$0.2 & \phantom{1}3 & \phantom{0}9.5$\pm$7.8  & \phantom{1}9 & 1.4$\pm$0.8 & \phantom{1}3 & $\cdots$ & & MW+LMC & 3 \\
\hline
\multicolumn{18}{c}{WC4--5} \\
    $\cdots$ &       &     $\cdots$   &        &  $\cdots$ &   & 50\phantom{000}                & \phantom{1}5      & $\cdots$ &   & 32\phantom{000}  & \phantom{1}5  &$\cdots$ &    & $\cdots$    &            & LMC           & 2 \\
    $\cdots$ &        &    $\cdots$    &       & $\cdots$  &   & 49$\pm$17 & 32 & \phantom{1}0.7$\pm$0.4& 20 & 28$\pm$12  & 20 &$\cdots$ &   & $\cdots$               &    & MW+LMC   & 1 \\
\phantom{0}6$\pm$\phantom{0}4 &  \phantom{1}8   &  $\cdots$  & & 0.9$\pm$0.5  & \phantom{1}9  & 31$\pm$18  & 11 & \phantom{1}0.2$\pm$0.3  & 11         &13$\pm$\phantom{0}7  &  11   & $\cdots$ &  & 1.5$\pm$1.1  & 4 & MW               & 3 \\
 16$\pm$\phantom{0}7                  & \phantom{1}7   & $\cdots$ &  & 2.0$\pm$1.6  & 15 &  53$\pm$29  & 17  & \phantom{1}0.5$\pm$0.8  & 17        & 34$\pm$20 &  17 & $\cdots$ & & 2.1$\pm$1.2 & 7 & LMC                & 3 \\
 \hline
 \multicolumn{18}{c}{WC6--7} \\
    $\cdots$      &      &   $\cdots$ &     & $\cdots$       &  &   37$\pm$11 & 25 & \phantom{1}5.5$\pm$4.1 & 25 & 12$\pm$5 & \phantom{1}7 & $\cdots$ &   & $\cdots$ & & MW & 1 \\
 \phantom{1}9$\pm$\phantom{0}5 &  12   & $\cdots$  &  & 1.1$\pm$0.7 & 15 & 45$\pm$23  &18 & \phantom{1}5.6$\pm$6.9  &  18 &      15$\pm$9   & 18 & $\cdots$ &  & 4.7$\pm$2.1 & 7 & MW & 3 \\
  \hline
 \multicolumn{18}{c}{WC8--9} \\
     $\cdots$ &      & $\cdots$ &      & $\cdots$ &   &   10$\pm$\phantom{1}2     & 24 & \phantom{1}7.3$\pm$1.4     &   25 &  \phantom{0}2.9$\pm$2.5   & \phantom{1}5 & $\cdots$ &   & $\cdots$  &  & MW & 1 \\
    \phantom{0}9$\pm$\phantom{0}7 &  \phantom{1}7    & $\cdots$ &      & $\cdots$ &   & 19$\pm$14   & 21 & 11$\pm$8     & 21 & \phantom{0}4.1$\pm$3.5 & 21 & $\cdots$  & & 2.7$\pm$1.6 & 5 & MW & 3 \\
 \hline
 \multicolumn{18}{c}{WO} \\
   19$\pm$14   & \phantom{1}3   & $\cdots$ &  & 15$\pm$9 &  \phantom{1}3 &  \phantom{0}8$\pm$3   & \phantom{1}3  & $\cdots$   &            & 11$\pm$\phantom{1}2                   &  \phantom{1}3  & $\cdots$ &  & $\cdots$  & &  MC+IC1613 & 1 \\
    \phantom{1}9$\pm$10    & \phantom{1}8    & $\cdots$ &  & \phantom{1}8$\pm$7  & \phantom{1}8  & \phantom{0}7$\pm$7   & \phantom{1}8   & $\cdots$  &             & \phantom{0}9$\pm$11 & \phantom{1}8 & $\cdots$ &  & $\cdots$ & & MW+MC & 3 \\
\hline
\end{tabular}
\end{center}
\begin{footnotesize}
1: \citet{1998ApJ...497..618S};
2: \citet{1990ApJ...348..471S};
3: This work\\
$\sharp$: O\,{\sc vi} is observed in WR46 \citep{1995A&A...302..457C} with $L_{\rm OVI~3811,34}/L_{\rm HeII~4686}$ = 0.17;
$\dag$: WN2--4 weak and strong lined;
$\ddag$: WN6--9 weak and strong lined.
\end{footnotesize}
\end{table*}

\subsection{Calibrations}

In order to analyse distant, unresolved star-forming regions host to WR stars we shall primarily focus  on the broad blue bump, involving a blend of N\,{\sc v} $\lambda\lambda$4603,20, N\,{\sc iii} $\lambda\lambda$4634,41, C\,{\sc iii} $\lambda\lambda$4647,51. He\,{\sc ii} $\lambda$4686, plus the yellow C\,{\sc iv} $\lambda\lambda$5801,12 bump. Although C\,{\sc iv} $\lambda\lambda$5801,12 is present in the majority of WN stars, it is typically much weaker than the blue bump (Fig.~\ref{blue_5808_WN}), with upper limits measured for many weak-lined WN stars. WN/C stars possess high C\,{\sc iv} $\lambda\lambda$5801,12 line luminosities, with an average value of $\log (L_{\rm CIV~5801,12}$/erg\,s$^{-1}$) = 36.0, although one would not expect prominent C\,{\sc iv} emission from such stars in an unresolved stellar population since they only comprise a small fraction of the overall WR population. The strength of the bumps are tightly correlated for WC stars, confirming results from \citet{1990ApJ...358..229S},  with $L_{\rm Blue}/L_{\rm CIV~5801,12}$ = 1.9$\pm$0.6 for WC4--5 stars, 3.0$\pm$0.4 for WC6--7 stars and 4.8$\pm$1.2 for WC8--9 stars. In contrast, these features are  poorly correlated in WO stars.

Previous optical calibrations have been obtained by \citet{1990ApJ...348..471S} for LMC WC stars, \citet{1998ApJ...497..618S} for Milky Way and LMC WR stars, and \citet{2006A&A...449..711C} for Magellanic Cloud WN stars, which have also been adapted for other metallicities \citep{2010A&A...516A.104L}.  In Table~\ref{literature} we compare our current results with previous calibrations. Overall we reinforce the decrease in line luminosity of early-type WN stars with metallicity established by \citet{2006A&A...449..711C} by the addition of Milky Way stars, plus higher line luminosities of WC4--5 stars in the LMC with respect to the Milky Way. However, we find no clear metallicity dependence for late-type WN or WO stars (hindered by low number of SMC stars), with significantly improved statistics for Galactic WC6--7 and WC8--9 stars with respect to \citet{1990ApJ...358..229S} and \citet{1998ApJ...497..618S}. The lack of strong metallicity dependence for some subtypes is somewhat surprising, recalling \citet{2020MNRAS.491.4406S}. However, the lower (luminosity) threshold to the formation of WR stars increases at lower metallicity \citep[][their figure~3]{2020A&A...634A..79S} so reduced wind strengths at low metallicity are countered by a shift to higher stellar luminosities on average. 

\citet{1998ApJ...497..618S} have previously adopted $L_{\rm HeII~4686} = 6.5 \times 10^{35}$ erg\,s$^{-1}$ and 2.5$\times 10^{35}$ erg\,s$^{-1}$ for (LMC) Of/WN and Of stars, respectively. We obtain $L_{\rm HeII~4686} = (3.5 \pm 1.3) \times 10^{35}$ erg\,s$^{-1}$ for Of/WN stars (Table~\ref{WN-calib}), whereas $L_{\rm HeII~4686} \leq 10^{35}$ erg\,s$^{-1}$ is more typical of Magellanic Cloud Of supergiants. For example, $L_{\rm HeII~4686}  = 8 \times 10^{34}$ erg\,s$^{-1}$ for the O2 supergiant Mk~42 (BAT99-105) based on our MUSE dataset and the interstellar extinction from \citet{2014A&A...565A..27H}. Definitive results await analysis of VLT/Xshooter flux calibrated spectroscopy of Magellanic Cloud OB stars obtained via the X-Shooting ULLYSES initiative (Vink et al., submitted).

Beyond the usual blue and yellow WR bumps, we include line luminosities for prominent violet features, O\,{\sc iv} $\lambda\lambda$3403,13, N\,{\sc iv} $\lambda\lambda$3478,85 and O\,{\sc iv} $\lambda\lambda$3811,34, noting that the latter is challenging to measure in WC stars due to line blends with other features, primarily the $\lambda$3700 complex of O\,{\sc iii-iv} and C\,{\sc iv} on its blue wing. Although these are weaker than the standard WR diagnostics, they may be detectable in suitable host galaxies, especially those below the Balmer jump.  The absence of O\,{\sc vi} $\lambda\lambda$3811,34 from WO stars in integrated populations is not wholly unexpected in view of the low line luminosity with respect to C\,{\sc iv} $\lambda\lambda$5801,12 in WC4--5 stars (Table~\ref{literature}) and location adjacent to stellar absorption lines \citep[e.g. H9 at $\lambda$3835,][]{2006MNRAS.370..799S}. Longward of the visible range, N\,{\sc iv} $\lambda\lambda$7103,29 and He\,{\sc ii} $\lambda$10124 are the strongest emission lines in WNE and WNE/C stars, with $L_{\rm HeII~10124}/L_{\rm HeII~4686}$ =  0.13$\pm$0.02 while He\,{\sc i} $\lambda$10830 is the strongest feature in WN6--8 (WN9--11) stars with $L_{\rm HeI~10830}/L_{\rm HeII~4686}$ =  0.7$\pm0.6$ (0.9$\pm$0.8). C\,{\sc iv} $\lambda$7725 and C\,{\sc iii} $\lambda\lambda$9701,19 are the strongest red features in WC4--5 stars, albeit an order of magnitude weaker than C\,{\sc iv} $\lambda\lambda$5801,12, the former also prominent in WO stars. C\,{\sc iii} $\lambda\lambda$9701,19 is prominent in late WC  subtypes, with $L_{\rm CIII~9701-19}/L_{\rm CIV~5801-12} = 0.3\pm0.1$ and $1.0\pm0.3$ for WC6--7 and WC8--9 stars, respectively. C\,{\sc ii} $\lambda\lambda$7231,37 is prominent in WC8--9 subtypes, with He\,{\sc i} $\lambda$10830  also very strong with $L_{\rm HeI~10830}/L_{\rm CIV~5801-12} $= 0.5$\pm$0.4. 

In the vacuum ultraviolet attention has usually
been focused on He\,{\sc ii} $\lambda$1640 ($n$=3--2) since it is prominent in most WR subtypes, and its line luminosity  exceeds He\,{\sc ii} $\lambda$4686 ($n$=4-3) by an order of magnitude \citep[][their fig.~3]{2006A&A...449..711C}. \citet{1998ApJ...497..618S} obtained $L_{\rm 1640}/L_{\rm 4686} = 7.8$ for Milky Way and LMC WN stars, supported by the  comprehensive study of  \citet{2019AJ....158..192L}, while \citet{2006A&A...449..711C} obtained $L_{\rm 4686}/L_{\rm 1640} = 10$ from LMC WN stars and theoretical models. Based on our adopted extinctions and $\lambda$1640 line fluxes measured from low resolution IUE/SWP spectroscopy we obtain $L_{\rm 4686}/L_{\rm 1640} = 12\pm4$ for Galactic, and Magellanic Cloud WN and WN/C stars. N\,{\sc iv} $\lambda\lambda$1238,42 is also strong in early WN stars, although this line and C\,{\sc iv} $\lambda\lambda$1548,51 are present in early O stars \citep{2016MNRAS.458..624C}. N\,{\sc iv}] $\lambda$1486 and N\,{\sc iv} $\lambda$1718 are also prominent in early WN stars, with emission line luminosities 20--30\% of He\,{\sc ii} $\lambda$1640. In late WN stars He\,{\sc ii} $\lambda$1640 is less dominant, in part due to the forest of iron lines in its vicinity \citep[][their fig.~2]{1995A&A...293..427C}, especially Fe\,{\sc iv} $\lambda$1633, with C\,{\sc  iv} $\lambda\lambda$1548,51, Si\,{\sc iv} $\lambda\lambda$1393,1402 and N\,{\sc iv} $\lambda$1718 P Cygni profiles also prominent. In the near-UV, the strongest WN emission line is He\,{\sc ii} $\lambda$3203 ($n$=5--3) with $L_{\rm 3203}/L_{\rm 4686} = 0.7\pm 0.2$ for Milky Way and Magellanic Cloud WN stars.


The strongest far-ultraviolet feature in WC4--7 stars is C\,{\sc iv} $\lambda\lambda$1548,51 \citep[][their fig.~5]{2006A&A...449..711C}, with He\,{\sc ii} $\lambda$1640, C\,{\sc iii}] $\lambda$1909, O\,{\sc iv} $\lambda\lambda$1338,43, O\,{\sc iv}] $\lambda\lambda$1397,1407 and C\,{\sc iii} $\lambda$1247 also prominent. For LMC and Milky Way WC4--5 stars, $L_{\rm CIV~1548,51}/L_{\rm CIV~5801,12} =8\pm$2 and $L_{\rm HeII~1640}/L_{\rm CIV~5801,12} = 5\pm2$, somewhat higher than the ratios of 6.0 and 2.6 obtained by \citet{2006A&A...449..711C} for single LMC WC4--5 stars indicating the sensitivity to extinction determinations and sample size.  
C\,{\sc iv} $\lambda\lambda$1548,51 remains the strongest far-UV line in WC6--7 stars, with C\,{\sc iii}] $\lambda$1909 the strongest far-UV emission line in WC8--9 stars. We obtain  $L_{\rm CIII]~1909}/L_{\rm CIV~5801,12} = 4\pm$2, 7$\pm$4, 23$\pm$8 for Galactic WC4--5, WC6--7 and WC8--9 stars, respectively. In the near-UV, the strongest WC emission line is C\,{\sc iii} $\lambda$2297, followed by C\,{\sc iv} $\lambda$2530 in early WC stars. $L_{\rm CIII~2297}/L_{\rm CIV~5801,12} =5\pm$2, 6$\pm$4, 23$\pm$10 for Galactic WC4--5, WC6--7 and WC8--9 stars, respectively.

\begin{figure}
\centering
	\includegraphics[width=0.525\linewidth,bb=175 50 430 535, angle=-90]{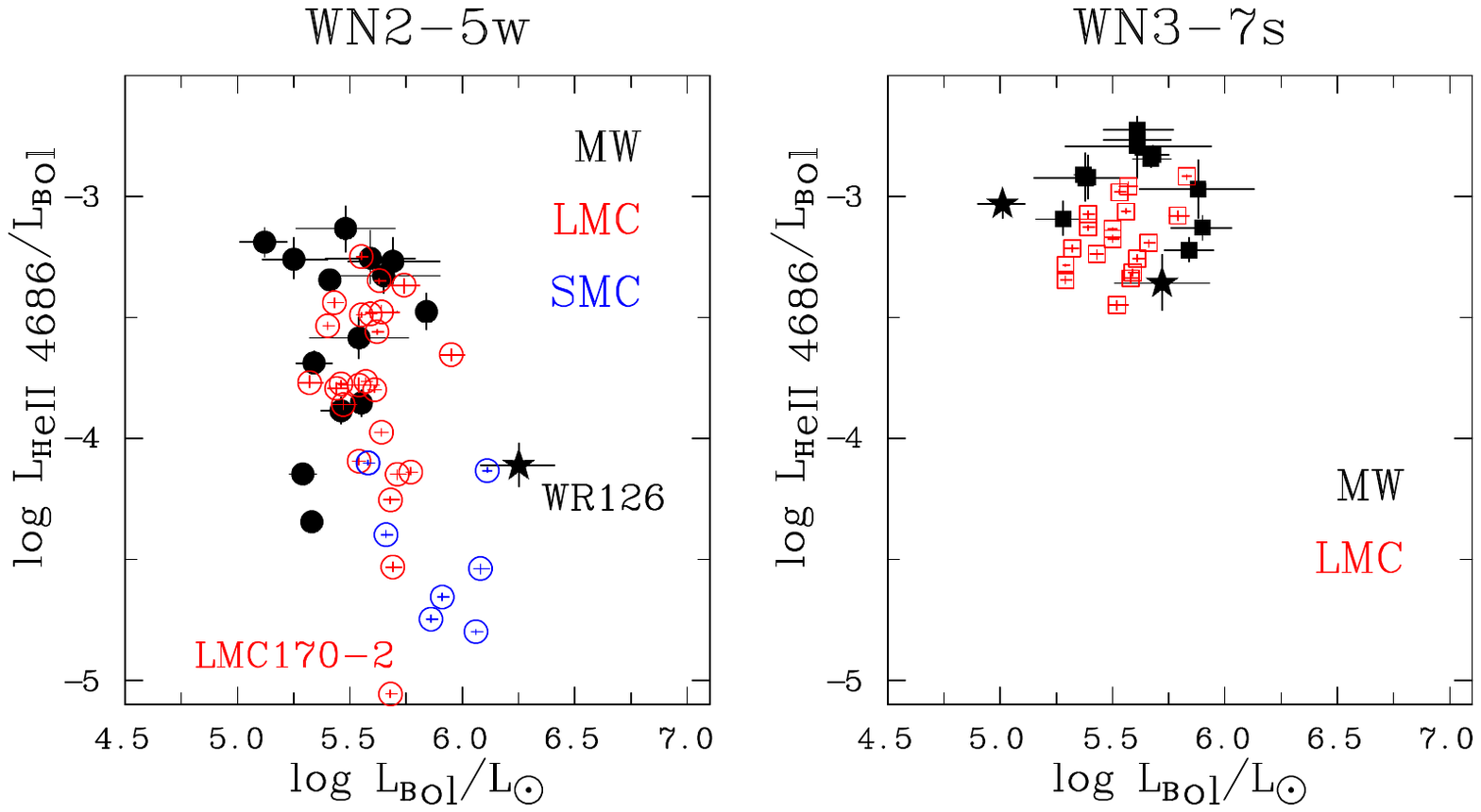}
	\includegraphics[width=0.525\linewidth,bb=175 50 430 535, angle=-90]{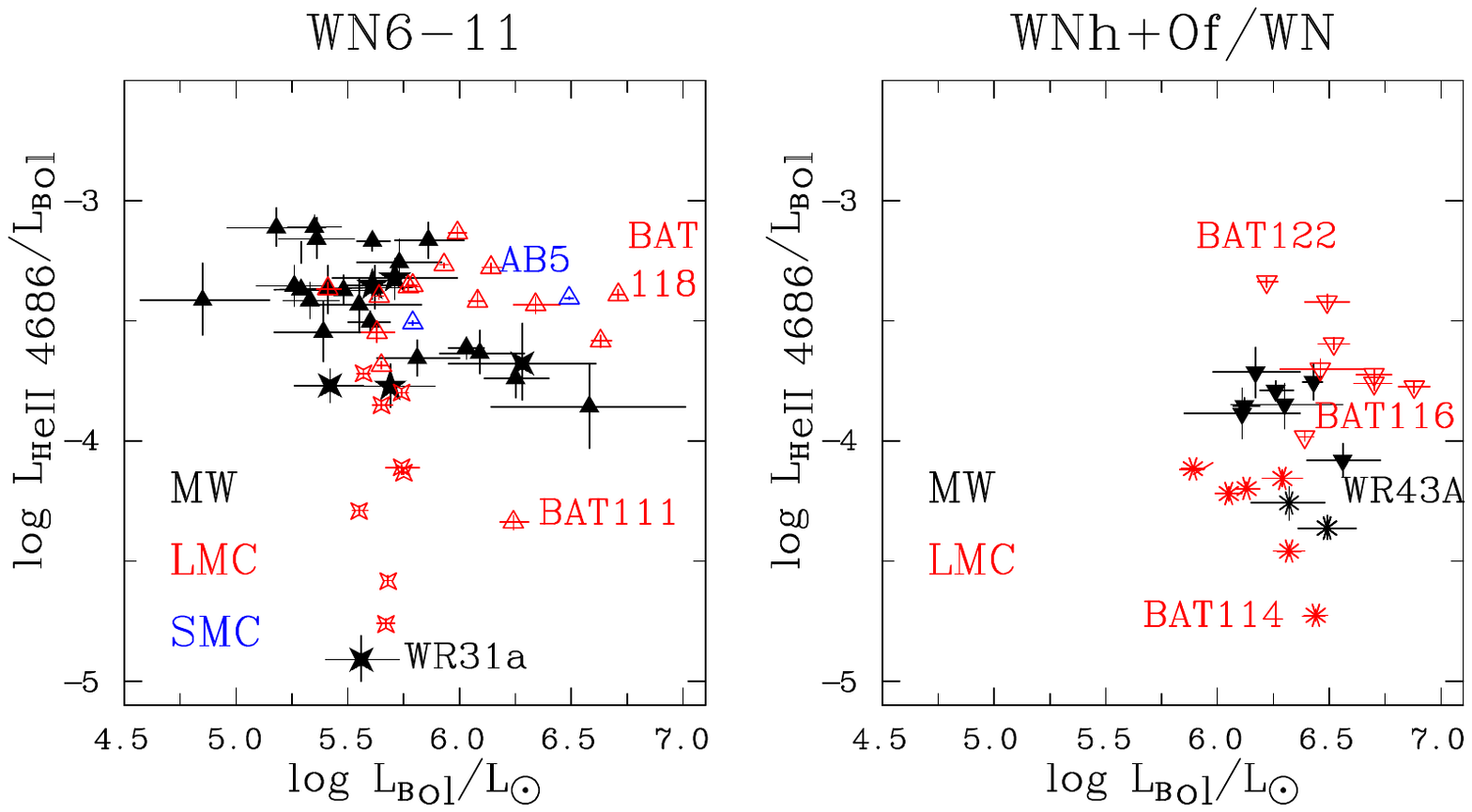}
	\centering
  \caption{Bolometric luminosity versus ratio of He\,{\sc ii} $\lambda$4686 line to bolometric luminosity for weak-lined WN2--5 (top left), strong lined WN3--7 (top right), weak lined WN6--11 (bottom left) and WNh+Of/WN (bottom right) stars in the Milky Way (black), LMC (red) and SMC (blue). Symbols as in Figure~\ref{4686}.}
	\label{logL_4686}
\end{figure}

\subsection{Line to bolometric luminosity ratio}

In view of the range of properties of WR stars within different environments, one can compare the ratios of line to bolometric luminosities for WR stars whose physical properties have been determined from spectroscopic analysis. Results are primarily drawn from \citet{2019A&A...625A..57H} and \citet{2019A&A...621A..92S} for Galactic WR stars, \citet{2014A&A...565A..27H}, \citet{2019A&A...627A.151S}, \citet{2022ApJ...924...44A}, and  \citet{2022ApJ...931..157A} for LMC WR stars, plus \citet{2015A&A...581A..21H} and \citet{2016A&A...591A..22S} for SMC WR stars. These are supplemented by studies of WO stars \citep{2015A&A...581A.110T} and very massive WNh and Of/WN stars in dense clusters \citep{2010MNRAS.408..731C, 2022A&A...663A..36B}. 

In Figure~\ref{logL_4686} we compare bolometric luminosities to the ratio of He\,{\sc ii} $\lambda$4686 line to bolometric luminosity for Milky Way, LMC and SMC WN, WN/C and Of/WN stars, with average values provided in Table~\ref{WN-calib}. This reinforces the difference in line strength of He\,{\sc ii} $\lambda$4686 between strong and weak-lined WN stars in view of their comparable bolometric luminosities, since  $\log L_{\rm HeII~4686}/L_{\rm Bol}$ = --3.0$_{-0.3}^{+0.1}$ for strong-lined WN stars, whereas $\log L_{\rm HeII~4686}/L_{\rm Bol}$ = --3.6$_{-0.9}^{+0.2}$ for weak-lined WN2--5 stars, with the lowest ratio for LMC~170--2 \citep{2017ApJ...841...20N}. The metallicity dependence for weak-lined early-type WN stars is also apparent (compare Milky Way and SMC stars). The situation is less clear for weak-lined WN6--8 stars, since Milky Way and Magellanic Cloud stars are broadly similar in their ratios of line to bolometric luminosities, $\log L_{\rm HeII~4686}/L_{\rm Bol}$ = --3.4$_{-0.2}^{+0.2}$. WN9--11 stars universally exhibit very low ratios \citep[e.g. WR31a,][]{1994A&A...281..833S} with $\log L_{\rm HeII~4686}/L_{\rm Bol}$ = --4.0$_{-0.5}^{+0.3}$. Figure~\ref{logL_4686} also reveals a low line to bolometric luminosity ratio for BAT99-111 (R136b) which has previously been classified O4\,If/WN8 \citep{2016MNRAS.458..624C} although WN8 is preferred on the basis of its H$\beta$ morphology from MUSE/NFM observations \citep{2011MNRAS.416.1311C, 2021Msngr.182...50C}. Bolometric luminosities of WN5--7h stars are high in both galaxies, with $\log L/L_{\rm Bol}$ = 6.3$\pm $0.2 and 6.5$\pm$0.2 for the Milky Way and LMC respectively, with typical ratios of  $\log L_{\rm HeII~4686}/L_{\rm Bol}$ = --3.7$_{-0.3}^{+0.2}$, and ratios of line to bolometric luminosities for Of/WN stars somewhat lower still,  $\log L_{\rm HeII~4686}/L_{\rm Bol}$ =  --4.3$_{-0.2}^{+0.2}$.

\begin{figure*}
\centering
\begin{minipage}[c]{0.75\linewidth}
	\includegraphics[width=0.35\linewidth,bb=175 50 430 762, angle=-90]{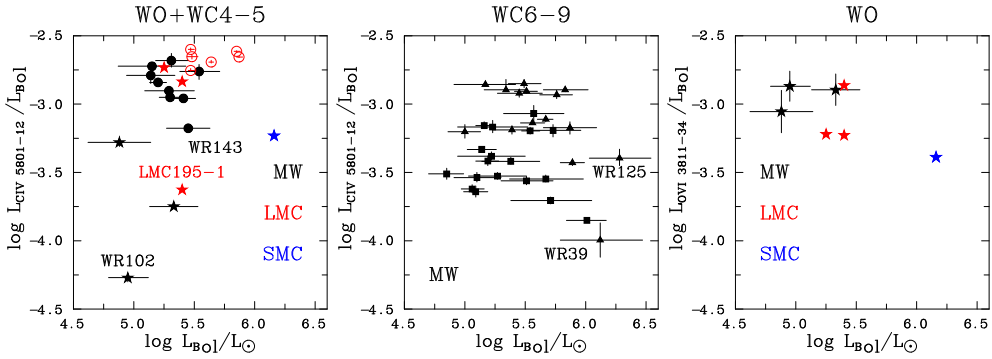} 
\end{minipage}\hfill
\begin{minipage}[c]{0.25\linewidth}
	\centering
  \caption{(left) Bolometric luminosity versus ratio of C\,{\sc iv} $\lambda\lambda$5801,12 line to bolometric luminosity for WO and WC4--5 stars; (centre) Bolometric luminosity versus ratio of C\,{\sc iv} $\lambda\lambda$5801,12 line to bolometric luminosity for WC6--7 (triangles) and WC8--9 (squares) stars; (right) Bolometric luminosity versus ratio of O\,{\sc vi} $\lambda\lambda$3811,34 line to bolometric luminosity for WO stars. Symbols as in Figure~\ref{5808}}
	\label{logL_5808}
	\end{minipage}
\end{figure*}

Figure~\ref{logL_5808} compares bolometric luminosities to the ratio of C\,{\sc iv} $\lambda\lambda$5801,12 line to bolometric luminosity for Milky Way and Magellanic Cloud WC and WO stars, and bolometric luminosities to the ratio of O\,{\sc vi} $\lambda$3811,34 line to bolometric luminosity for WO stars. Galactic WC4--5 stars reveal significantly higher ratios than WC8--9 stars, with  intermediate ratios for most WC6--7 stars (WR39 is an notable outlier),
and very high ratios for LMC WC4--5 stars. WO stars show a broad range of C\,{\sc iv} $\lambda\lambda$5801,12 line ratios, some in common with WC4--5 stars (BAT99-123 and LH\,41-1042) with others revealing far lower ratios (e.g. WR102 in Milky Way, LMC195--1 in LMC). Line ratios of O\,{\sc vi} $\lambda\lambda$3811,34 to bolometric luminosity for WO stars suggest a weak metallicity dependence, with higher ratios for Milky Way stars than those in Magellanic Clouds. 

\subsection{Templates}\label{templates}

In addition to calibrations, we have produced WR emission line templates for each environment, although this is hindered by the variation in spectral coverage of individual datasets (Table~\ref{spectroscopy}), so stars with limited spectral coverage are generally excluded. We have degraded composite spectra to a resolution of 10\AA, and produced continuum-subtracted templates based on single WR stars and all (single and binary) WR stars, since the latter are often contaminated by (Balmer) absorption lines from companion OB stars. Templates incorporate velocity corrections of 284 km\,s$^{-1}$ and 162 km\,s$^{-1}$ for the LMC and SMC, respectively \citep{2016AJ....152...50T}. Templates generally correct for atmospheric telluric effects, but no adjustment for interstellar atomic features (Ca\,{\sc ii} H\&K, Na\,{\sc i} D) lines or diffuse interstellar bands (e.g. $\lambda$4430) is made. 

Fig.~\ref{WNE-lum} presents templates for Galactic, LMC and SMC early-type WN stars from single and combined (single and binary) samples, separated into weak and strong lined subsets. Aside from He\,{\sc ii} $\lambda$4686, the strongest feature in metal-rich early-type WN stars is N\,{\sc iv} $\lambda\lambda$3478,85 followed by He\,{\sc ii} $\lambda$5412, $\lambda$6560 and the complex involving N\,{\sc iv} $\lambda$4058, He\,{\sc ii} $\lambda$4100, Si\,{\sc iv} $\lambda\lambda$4088,4116 and N\,{\sc iii} $\lambda\lambda$4097,4103. C\,{\sc iv} $\lambda\lambda$5801,12 is present, though relatively weak, in Galactic and LMC early-type WN stars.

Fig.~\ref{WNL-lum} presents templates for Galactic, LMC and SMC late-type WN stars, separated into classical (WN6--8) and main-sequence H-rich (WN5--7h) subsets.  The scarcity of LMC main-sequence WN stars required the use of 30 Doradus MUSE datasets \citep{2018A&A...614A.147C} which exclude the spectral region shortward of $\lambda$4600. He\,{\sc ii} $\lambda$4686 aside, the strongest stellar features in WN6--8 stars include N\,{\sc iv} $\lambda\lambda$3478,85, the $\lambda$4100 complex, N\,{\sc iii} $\lambda\lambda$4634,41, plus H$\alpha$, H$\beta$ and He\,{\sc i} $\lambda$5876 in the Milky Way. The change in N\,{\sc iii} $\lambda\lambda$4634,41/He\,{\sc ii} $\lambda$4686 ratio from high to low metallicity is apparent. H-rich main sequence WN5--7h stars resemble WN6--8 stars aside from the relative strength of N\,{\sc iv} $\lambda$4058, weakness of N\,{\sc iii} $\lambda$4634,41, plus He\,{\sc i} lines and higher $\lambda$4686 FWHM (Fig.~\ref{4686}). 

Fig.~\ref{OfWN-lum} presents templates for Of/WN and WN9--11 stars in the Milky Way and LMC. Of/WN stars reveal weak, narrow He\,{\sc ii} $\lambda$4686 emission, plus N\,{\sc iv} $\lambda$4058 and H$\alpha$, with the upper Balmer series in absorption. WN9--11 stars also exhibit weak, narrow weak He\,{\sc ii} $\lambda$4686 emission, with either the N\,{\sc iii} $\lambda\lambda$4634,41 triplet or N\,{\sc ii} $\lambda\lambda$4601,43 multiplet prominent and strong Balmer and He\,{\sc i} emission lines. \citet{2000ApJ...531..776G} have highlighted N\,{\sc iii} $\lambda\lambda$4512,48 and Si\,{\sc iii} $\lambda\lambda$4554,76 from late WN stars in WR galaxies. Such features are present in late WN stars, albeit with the $\lambda$4100 complex and N\,{\sc iii} $\lambda\lambda$4634,41/N\,{\sc ii} $\lambda\lambda$4601,43 more prominent. 

Fig.~\ref{WNC-lum} presents templates for Galactic and LMC WN/C stars, separated into early and late-type categories. Early subtypes are dominated by C\,{\sc iv} $\lambda\lambda$5801,12 plus the blend of C\,{\sc iii} $\lambda\lambda$4647,51 and He\,{\sc ii} $\lambda$4686, while the latter dominates late WN/C subtypes. C\,{\sc iii} $\lambda$5696 is present, albeit weakly in late WN/C templates. We combine Galactic and LMC WN/C stars to construct templates, noting that some exhibit strong C\,{\sc iii} $\lambda\lambda$4647,51 (e.g. WR26, WR126), while this feature is absent in others (e.g. WR58, BAT99-36, BAT99-88) such that spectroscopy of C\,{\sc iv} $\lambda\lambda$5801,12 is essential in discriminating between WN and WN/C stars. 

Fig.~\ref{WC-gal-lum} presents templates for Galactic WC4--5, WC6--7 and WC8--9 stars from single and combined (single plus binary) samples. C\,{\sc iii} $\lambda\lambda$4647,51 (plus He\,{\sc ii} $\lambda$4686) dominates the blue WR bump for all subtypes, while the transition from C\,{\sc iv} $\lambda\lambda$5801,12 to C\,{\sc iii} $\lambda$5696 in the yellow from early to late subtypes is apparent, with O\,{\sc iv} $\lambda\lambda$3403,13 prominent in WC4--7 stars. At late subtypes, C\,{\sc ii} $\lambda\lambda$6559,81, $\lambda\lambda$7231,7 are particularly strong in the red. The forest of blue features in WC8--9 stars primarily involve C\,{\sc ii-iii} \citep{2006ApJ...636.1033C}. 

The upper panel of Fig.~\ref{WC4-WO-lum} presents templates of LMC WC4--5 emission lines. Aside from C\,{\sc iii} $\lambda$4647--51+He\,{\sc ii} $\lambda$4686 and C\,{\sc iv} $\lambda\lambda$5801,12, O\,{\sc iv} $\lambda\lambda$3403,13 is also prominent, in common with Milky Way counterparts. The lower panel presents templates of our combined WO sample, incorporating Milky Way, LMC and SMC stars, 
highlighting the defining O\,{\sc vi} $\lambda\lambda$3811,34 feature, together with prominent O\,{\sc iv} $\lambda\lambda$3403,13, C\,{\sc iv} $\lambda$4658+He\,{\sc ii} $\lambda$4686 and C\,{\sc iv} $\lambda\lambda$5801,12. Comparing WO and LMC WC4--5 stars, the average O\,{\sc vi} $\lambda\lambda$3811,34 line luminosity is factor of $\sim$4 stronger in the former group, while C\,{\sc iii} $\lambda\lambda$4647,51+C\,{\sc iv} $\lambda$4658+He\,{\sc ii} $\lambda$4686 and C\,{\sc iv} $\lambda$5801,12 are 4--8 times weaker (Table~\ref{literature}). 

\subsection{Cumulative WR line luminosities}

Armed with line luminosities for a large subset of (optically detected) WR stars in the Milky Way and Magellanic Clouds we can also consider their cumulative blue and yellow line luminosities, and contributions from different WR populations. Our sample includes all known WR stars in the SMC, 66\% of the WR population in the LMC \citep{2018ApJ...863..181N}\footnote{The global WR population of the LMC is 153, following \citet{2018ApJ...863..181N} after including the WN8 star BAT99-111 (R136b), excluding two Of stars (Mk~42, LH~99-3), and noting that their catalogue entry 125 is Mk~37 rather than Mk~30 (also listed as entry 129).}, though only $\sim$7\% of the estimated Milky Way population \citep[N=1900$\pm$250,][]{2015MNRAS.447.2322R}.  18\% of the LMC population are carbon or oxygen sequence WR stars, whereas these comprise 21\% of our sample. In the Milky Way carbon and oxygen sequence stars  represent 42\% of the known population, in close agreement with our dataset\footnote{275 WC or WO stars from a total of 667 stars according to v1.25 of the Galactic WR catalogue https://pacrowther.staff.shef.ac.uk/WRcat }.

\begin{table*}
\begin{center}
\caption{Cumulative blue and yellow line luminosities for our sample of WR systems in the Milky Way, LMC and SMC, with completeness fractions of $\sim$7\%, 66\% and 100\% \citep{2015MNRAS.447.2322R, 2018ApJ...863..181N}. Systems involved WN+WN binaries (WR43A, BAT99-116, BAT199-118, AB5) are reflected in + statistics. WC and WO stars dominate the C\,{\sc iv} $\lambda\lambda$5801,12 feature, and also contribute the majority of the integrated blue feature in Milky Way and LMC. AB5 (HD~5980) alone contributes half of the integrated blue WR bump of the SMC. We estimate total LMC (Milky Way) blue and yellow line luminosities of 2.6$\times 10^{38}$ erg\,s$^{-1}$ (3.6$\times 10^{39}$ erg\,s$^{-1}$) and 8.8$\times 10^{37}$ erg\,s$^{-1}$ (7.7$\times 10^{38}$ erg\,s$^{-1}$), respectively.}
\label{WR-bumps}
\begin{tabular}{l@{\hspace{-0.5mm}}r@{\hspace{1mm}}c@{\hspace{1mm}}l@{\hspace{2mm}}r@{\hspace{1mm}}c@{\hspace{1mm}}l@{\hspace{2mm}}r@{\hspace{1mm}}c@{\hspace{1mm}}l
                         @{\hspace{2mm}}r@{\hspace{1mm}}c@{\hspace{1mm}}l@{\hspace{2mm}}r@{\hspace{1mm}}c@{\hspace{1mm}}l@{\hspace{2mm}}r@{\hspace{1mm}}c@{\hspace{1mm}}l}
\hline
Galaxy &  \multicolumn{3}{c}{Milky Way} & \multicolumn{3}{c}{LMC} & \multicolumn{3}{c}{SMC} & \multicolumn{3}{c}{Milky Way} & \multicolumn{3}{c}{LMC} & \multicolumn{2}{c}{SMC} \\
           & N & $\Sigma L$  erg\,s$^{-1}$ & \% & N & $\Sigma L$  erg\,s$^{-1}$ &   \%     & N & $\Sigma L$ erg\,s$^{-1}$ &   \% & N & $\Sigma L$ erg\,s$^{-1}$ & \% & N & $\Sigma L$ erg\,s$^{-1}$ &  \% & N & $\Sigma L$ erg\,s$^{-1}$ & \%  \\
\hline
 Subtype & \multicolumn{9}{c}{------ Blue bump (N\,{\sc iii-v} $\lambda\lambda$4603,41, C\,{\sc iii} $\lambda\lambda$4647,51, He\,{\sc ii} $\lambda$4686) ------} & \multicolumn{9}{c}{------ Yellow bump (C\,{\sc iv} $\lambda\lambda$5801,12) ------} \\
WN3--7s  &  12 & \phantom{0}$3.1\times 10^{37}$  & \phantom{1}12                    & 18  &  \phantom{0}$2.1\times 10^{37}$ &                     \phantom{1}10 & $\cdot$ & $\cdots$ & $\cdots$   
           &  12 & \phantom{0}$1.4\times 10^{36}$ &    \phantom{10}2 & 18  &  \phantom{0}$0.9\times 10^{36}$ &  \phantom{10}1 & $\cdot$ &  $\cdots$ &  $\cdots$ \\
WN2--5w & 22 & \phantom{0}$1.4\times 10^{37}$  & \phantom{10}5                    & 24  &  \phantom{0}$0.9\times 10^{37}$ &  \phantom{10}5 & 9 & $1.7\times 10^{36}$ & \phantom{1}18   
           &  19 & \phantom{0}$0.6\times 10^{36}$ &    \phantom{10}1 & 22  &  \phantom{0}$0.3\times 10^{36}$ &  \phantom{10}0 & 7 &  \phantom{0}0.1$\times 10^{35}$ &  \phantom{10}0\\
WN6--11      & 29 &  \phantom{0}$3.9\times 10^{37}$  & \phantom{1}15 & 21+1 &  \phantom{0}$4.2\times 10^{37}$                    & \phantom{1}21 & 2+1  & $5.6\times 10^{36}$                     & \phantom{1}59   &     
                 25 &   \phantom{0}$0.5\times 10^{36}$ &    \phantom{10}1  & 15+1 &  \phantom{0}$1.0\times 10^{36}$  &  \phantom{10}1 & 2+1 &  \phantom{0}2.2$\times 10^{35}$ &  \phantom{10}6 \\
WN5-7h$\ast$  & 9+1 & \phantom{0}$1.3\times 10^{37}$ & \phantom{10}5                         & 14+1 & \phantom{0}$3.0\times 10^{37}$ & \phantom{1}15    & $\cdot$ & $\cdots$ & $\cdots$ &
                6+1 & \phantom{0}$0.1\times 10^{36}$ & \phantom{10}0                          & 13+1 & \phantom{0}$1.0\times 10^{36}$ & \phantom{10}1    & $\cdot$ & $\cdots$ & $\cdots$ \\            
WN/C                 &   7 &  \phantom{0}$1.4\times 10^{37}$    &   \phantom{10}5  & 2   &  \phantom{0}$0.2\times 10^{37}$ &    \phantom{10}1  & $\cdot$ & $\cdots$ & $\cdots$ &    7   &   \phantom{0}$5.9\times 10^{36}$ & \phantom{1}10 &  2  &  \phantom{0}$2.7\times 10^{36}$ &  \phantom{10}4 & $\cdot$ & $\cdots$ & $\cdots$ \\
WC4--5              &  11 &  \phantom{0}$3.4\times 10^{37}$    &\phantom{1}13 & 18  &  \phantom{0}$9.7\times 10^{37}$ & \phantom{1}48 & $\cdot$ & $\cdots$ & $\cdots$  & 11    & $14.7\times 10^{36}$ & \phantom{1}24 & 17 & $61.5\times 10^{36}$ & \phantom{1}88 &  $\cdot$ & $\cdots$ & $\cdots$ \\
WC6--9              & 39 & $12.0\times 10^{37}$ & \phantom{1}45 & $\cdot$ & $\cdots$ & $\cdots$  & $\cdot$ & $\cdots$ & $\cdots$  & 39 & $36.1\times 10^{36}$ & \phantom{1}59 & $\cdot$ & $\cdots$ & $\cdots$ & $\cdot$ & $\cdots$ & $\cdots$ \\
WO                    & 4   &  \phantom{0}$0.1\times 10^{37}$    &  \phantom{10}0  & 3  &  \phantom{0}$0.2\times 10^{37}$  &  \phantom{10}1  & 1  & $2.2\times 10^{36}$ & \phantom{1}23    & 4     &  \phantom{0}$1.2\times 10^{36}$ &  \phantom{10}2 & 3 &  \phantom{0}$2.9\times 10^{36}$ &  \phantom{10}4 & 1 & $32.9\times 10^{35}$ & \phantom{1}93 \\
\hline
Total                 & 133+1 & $26.7\times 10^{37}$ & 100 & 100+2 & 20.4$\times 10^{37}$ & 100 & 12+1 & $9.5\times 10^{36}$ & 100 &  123+1 & $60.4\times 10^{36}$ & 100 & 91+2 & $70.2\times 10^{36}$ & 100 & 10+1 & $35.2\times 10^{35}$ & 100\\ 
\hline
\end{tabular}
\begin{footnotesize}
$\ast$ incl. Of/WN stars.
\end{footnotesize}
\end{center}
\end{table*}

Table~\ref{WR-bumps} summarises cumulative blue and yellow WR line luminosities from our current dataset in the Milky Way, LMC and SMC. As expected, WC and WO stars dominate the integrated C\,{\sc iv} $\lambda\lambda$5801,12 line luminosity in all environments, but also contribute significantly to the blue feature: 58\% in the Galaxy, 49\% in the LMC and 23\% in the SMC, for the samples included. AB5 (HD~5980) alone contributes half of the integrated blue WR bump of the SMC. If we utilise our LMC calibrations for the remainder of the known  WR population, we estimate total blue and yellow line luminosities of 2.6$\times 10^{38}$ erg\,s$^{-1}$ (51\% from WN, Of/WN and WN/C stars, 49\% from WC and WO stars) and 8.8$\times 10^{37}$ erg\,s$^{-1}$ (8\% from WN, Of/WN and WN/C stars, 92\% from WC and WO stars), respectively. The fractional contributions from WN and WC stars are not affected because the overwhelming majority of omitted nitrogen-sequence stars possess weak lines (weak-lined WNE, WN10--11, Of/WN and LMC~170--2 type). 

Accounting for line luminosities of all known WR stars in the Milky Way, we obtain cumulative blue and C\,{\sc iv} $\lambda\lambda$5801,12 line luminosities of 1.2$\times 10^{39}$ erg\,s$^{-1}$ and 2.7$\times 10^{38}$ erg\,s$^{-1}$, respectively. If these are representative of the total ($\sim$1900) the blue (3.6$\times 10^{39}$ erg\,s$^{-1}$) and yellow (7.7$\times 10^{38}$ erg\,s$^{-1}$) sum is an order of magnitude higher than the LMC. This is reasonable, given their relative star-formation rates of 1.9 $M_{\odot}$\,yr$^{-1}$ \citep{2011AJ....142..197C} and 0.25 $M_{\odot}$\,yr$^{-1}$ \citep{2008ApJS..178..247K}, respectively, and environmental differences in WR line luminosities. 

\begin{figure}
\centering
	\includegraphics[width=0.65\linewidth,bb=25 50 550 775,angle=-90]{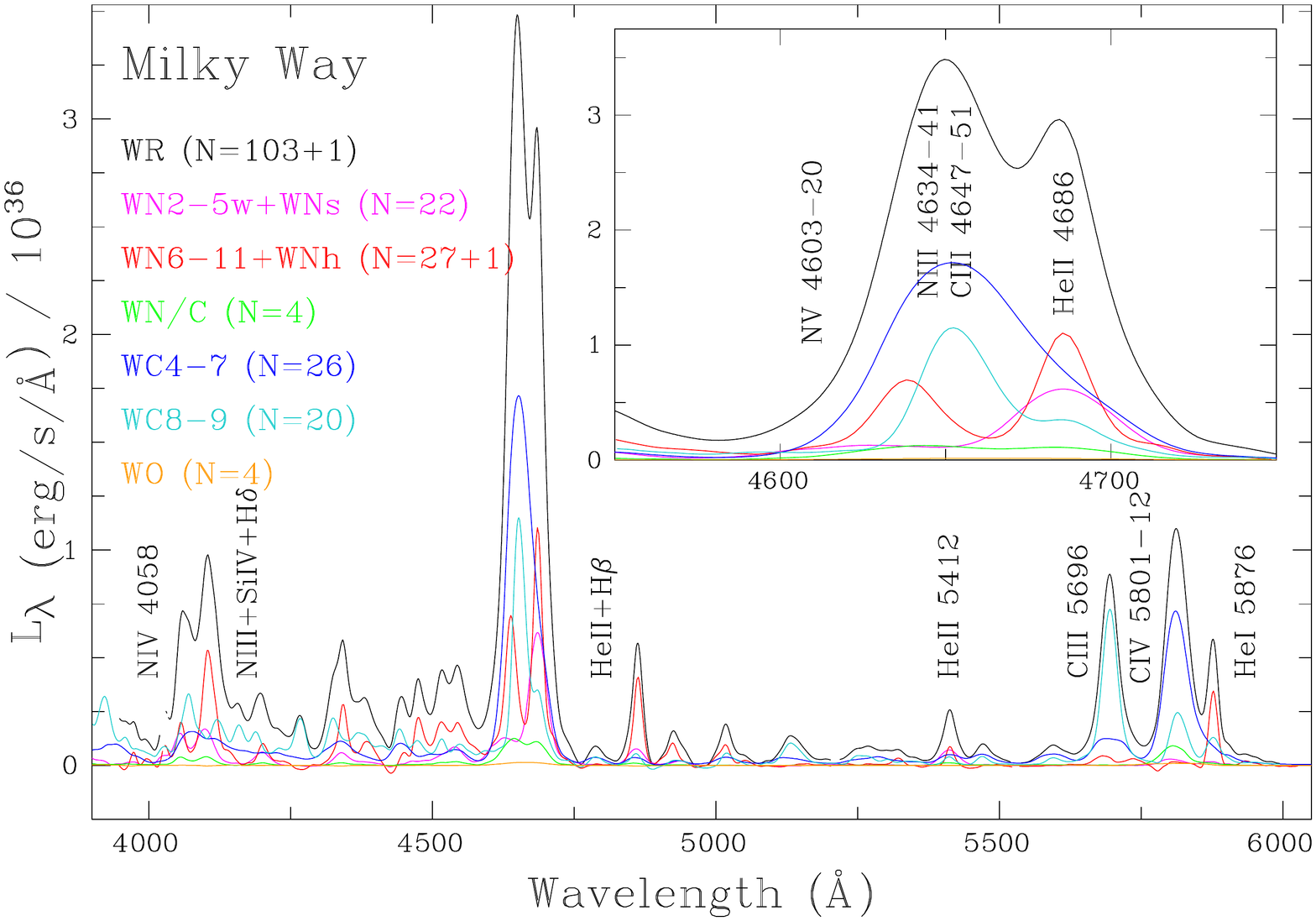}
         \includegraphics[width=0.65\linewidth,bb=25 50 550 775,angle=-90]{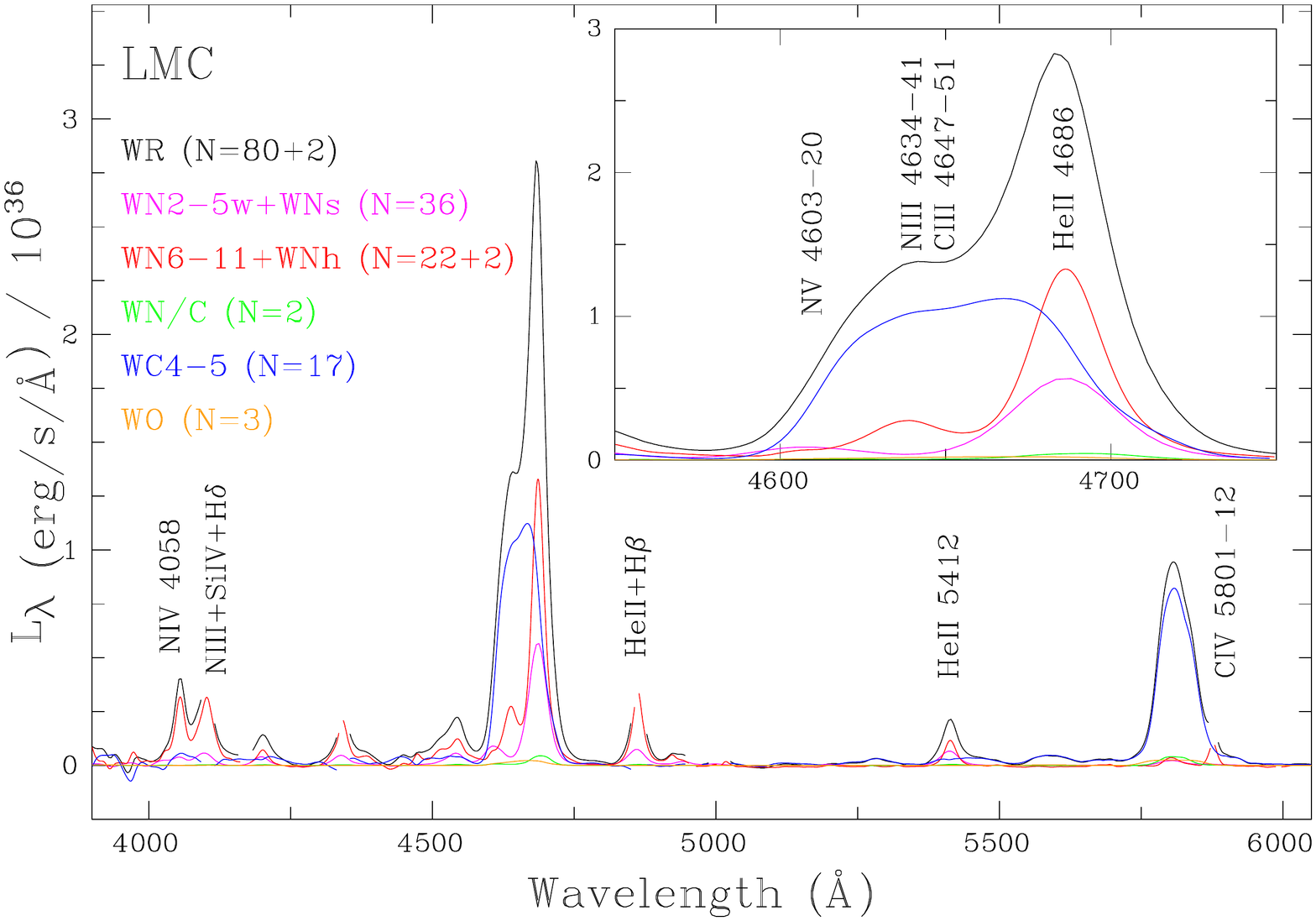}
         \includegraphics[width=0.65\linewidth,bb=25 50 550 775,angle=-90]{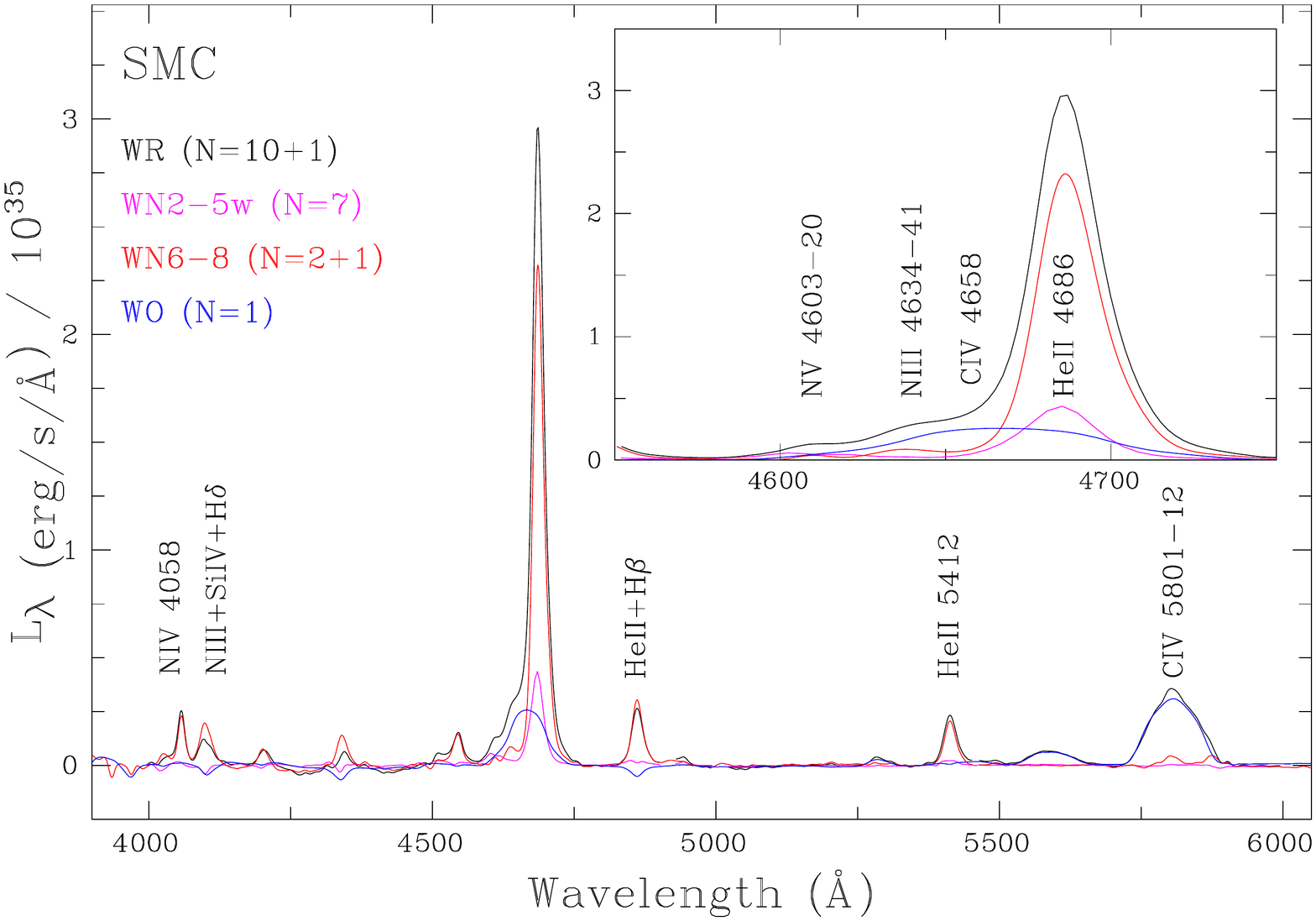}
	\centering
  \caption{(Top): Cumulative $\lambda\lambda$3950--6050 emission line spectrum of 103+1 Milky Way WR stars (black) together, plus inset for the blue WR bump, including contributions from WN2--5w+WN3--7s (N=22), WN6--11+WNh (N=27+1), WN/C (N=4), WC4--7 (N=26), WC8--9 (N=20) and WO (N=4) stars, indicating the primary contributors to the blue and yellow WR bumps. The global Milky Way WR line luminosity is estimated to be a factor of $\sim$13 times higher than presented here based on the known WR subtype distribution; (Middle): As above, for 80+2 LMC WR stars (black), highlighting contributions from WN2--5w+WN3--7s (N=36), WNL+WNh (N=22+2), WN/C (N=2), WC4--5 (N=17) and WO (N=3) stars. The global LMC WR line luminosity is estimated to be $\sim$25\%  higher than presented here, accounting for incompleteness; (bottom): As above for 10+1 SMC WR stars (black), highlighting contributions from WNE (N=7), WNL (N=2+1) and WO (N=1) stars. The global SMC WR line luminosity is estimated to be a few percent higher than presented here, owing to the omission of two WN2--5w stars.} 
	\label{gal-lum} 
\end{figure}

The fractional contribution to the cumulative WR bumps are further highlighted in Fig.~\ref{gal-lum} which provides the integrated line spectrum of 102 Galactic stars (upper panel), 80 LMC stars (middle panel) and 10 SMC stars (lower panel). The integrated Milky Way template highlights that strong lined and late-type WN stars are the primary contributors to He\,{\sc ii} $\lambda$4686, WC4--7, WC8--9 and late-type WN subtypes dominating the $\lambda$4650 feature, and WC4--7 and WC8--9 subtypes dominating C\,{\sc iv} $\lambda\lambda$5801,12 and C\,{\sc iii} $\lambda$5696 respectively. Similar results are obtained for the LMC, aside from the lack of WC6--9 stars, while the cumulative SMC bumps are dominated by late-type WN stars (primarily AB5) and the WO star AB8. Fig.~\ref{gal-lum} is reminiscent of WR bumps observed in star-forming galaxies \citep{1992ApJ...401..543V, 1999A&A...341..399S, 2008A&A...485..657B, 2021MNRAS.503.6112S}, but serves as a caution to approximating Wolf-Rayet populations by restricted WC and WN flavours, which additional (violet or red) WR diagnostics would help discriminate.

\section{Applications}\label{WR-gal}

We now undertake analysis of two WR galaxies with our revised line luminosity calibration, selected at low and high metallicities. Historically, two methods have been employed to quantify the number of WR stars from optical emission line bumps, gaussian fits to emission lines calibrated with reference line luminosities or template fitting. The former approach is more straightforward, but discards line profile information and may result in unphysical solutions through independent solutions to blue and red bumps. The latter approach incorporates line profile information but may involve templates ill suited to the host galaxy metallicity. Here we apply both approaches to our selected WR galaxies.

\subsection{NGC~3049 (Mrk~710)}

\begin{figure}
\centering
	\includegraphics[width=0.65\linewidth,bb=25 40 525 795,angle=-90]{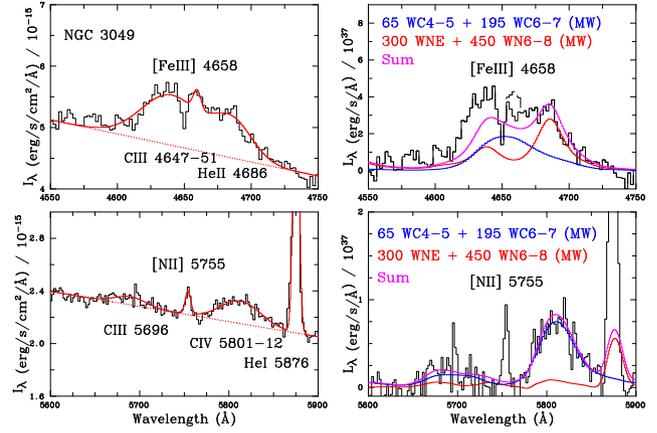}
	\centering
  \caption{Left panels: De-reddened, velocity corrected, blue and yellow WR bumps in NGC~3049 from \citet{1999A&A...341..399S}, including gaussian fits (red) to broad (C\,{\sc iii} $\lambda\lambda$4647,51, He\,{\sc ii} $\lambda$4686, C\,{\sc iii} $\lambda$5696, C\,{\sc iv} $\lambda\lambda$5801,12) and nebular ([Fe\,{\sc iii}] $\lambda$4658, [N\,{\sc ii}] $\lambda$5755, He\,{\sc i} $\lambda$5876) lines, with the continuum indicated as dotted lines. $\sim$1300 WN6-8 and $\sim$380 WC6--7 stars are estimated from Galactic calibrations, adopting a distance of 19.3~Mpc. Right panels: Continuum subtracted luminosities of blue and yellow WR bumps in NGC~3049 (black,
  nebular [Fe\,{\sc iii}] component indicated in dashed lines) plus composite template fits adopting 750 Milky Way WN stars (red, distributed 1:1:3 between WN2--5w:WN3-7s:WN6--8 stars as observed in the Milky Way), 260 Milky Way WC4--7 stars (blue, distributed between WC4--5 and WC6--7 stars as observed in the Milky Way) plus their sum (pink). Nebular  [N\,{\sc ii}] $\lambda$5755 and He\,{\sc i} $\lambda$5876 are not reproduced by the stellar WR templates.The poor match to the $\lambda$4640--4650 blend may arise from the use of solar metallicity WN6---8 stars (NGC~3049 is considered to be supersolar).}
	\label{n3049-WR} 
\end{figure}

Our high metallicity target, NGC~3049 is a barred spiral at a distance of 19.3~Mpc \citep{2013AJ....146...86T} which was identified as a WR galaxy by \citet{1986A&A...169...71K}. WR populations of NGC~3049 have been studied using ESO 2.2m/EFOSC2 \citep{1999A&A...341..399S} and HST/STIS \citep{2002ApJ...580..824G}, and exhibits broad emission at N\,{\sc iii} $\lambda\lambda$4634,41/C\,{\sc iii} $\lambda\lambda$4647,51, He\,{\sc ii} $\lambda$4686, C\,{\sc iii} $\lambda$5696 and C\,{\sc iv} $\lambda\lambda$5801,12. NGC~3049 is unusual in the $\lambda$4634,41 feature being comparable in strength to He\,{\sc ii} $\lambda$4686 \citep[see][]{1999NewA....4..197S}.
Fig.~\ref{n3049-WR} (left panels) presents de-reddened, velocity corrected WR bumps including gaussian fits to stellar and nebular emission lines. Table~\ref{ngc3049} provides a summary of nebular and stellar line luminosities in NGC~3049, remeasured from EFOSC observations which provides a spectral coverage of 4370--6550\AA\ at a resolving power of $R\sim$750 \citep{1999A&A...341..399S}. The spectroscopic L(H$\beta$) corresponds to an ionizing output of $Q_{\rm H} = 1.6 \times 10^{52}$ ph\,s$^{-1}$, comparable to 30 Doradus in the LMC, although the integrated output is likely to be considerably higher.

\begin{table}
\begin{center}
\caption{Line luminosities of selected nebular and stellar emission lines in NGC~3049 \citep{1999A&A...341..399S} based on $c$(H$\beta$)=0.22, a distance of 19.3~Mpc and Milky Way calibrations from this study. Results in parenthesis are from Galactic template fits.}
\label{ngc3049}
\begin{tabular}{l@{\hspace{2mm}}l@{\hspace{2mm}}l@{\hspace{2mm}}l@{\hspace{2mm}}l@{\hspace{-4mm}}l}
\hline
Line  &  FWHM & $F_{\lambda}$  & $I_{\lambda}$ & $L_{\lambda}$ & N(WR) \\
         & km\,s$^{-1}$    & \multicolumn{2}{c}{$10^{-15}$ erg\,s$^{-1}$\,cm$^{-2}$} & 10$^{37}$ erg\,s$^{-1}$ & \\
\hline
C\,{\sc iii} $\lambda\lambda$4647,51 & 3300$\pm$800 &  26.2$\pm$5.8    &  \phantom{0}45.2$\pm$9.9 &  202$\pm$44 & \\

[Fe\,{\sc iii}] $\lambda$4658 &  $\cdots$    &   \phantom{0}1.3$\pm$1.1        & \phantom{00}2.3$\pm$2.1 & \phantom{0}10$\pm$9 & \\

He\,{\sc ii} $\lambda$4686 & 2200$\pm$600  & 13.0$\pm$4.9   &  \phantom{0}21.9$\pm$8.2  & \phantom{0}98$\pm$37  &  1300 WN6--8 \\

H$\beta$                            & $\cdots$     & 98.9$\pm$1.2   &  169\phantom{.0}$\pm$2                   &   755$\pm$9                & (750 WN2--8) \\  

[O\,{\sc iii}] $\lambda$4959 & $\cdots$     &   11.8$\pm$0.5     &   \phantom{0}19.8$\pm$0.9      &  \phantom{0}88$\pm$4                & \\

[O\,{\sc iii}] $\lambda$5007  & $\cdots$     & 36.2$\pm$0.7    & \phantom{0}60.6$\pm$1.1      &  271$\pm$5                       &                   \\

C\,{\sc iii} $\lambda$5696   & 2100$\pm$500 & \phantom{0}1.9$\pm$0.5 & \phantom{00}2.7$\pm$0.7    & \phantom{0}12$\pm$3        & \\

[N\,{\sc ii}] $\lambda$5755     & $\cdots$ &  \phantom{0}0.9$\pm$0.2    & \phantom{00}1.3$\pm$0.3    & \phantom{00}6$\pm$1  & \\

C\,{\sc iv} $\lambda\lambda$5801,12 &  2700$\pm$300 &  \phantom{0}6.5$\pm$0.7  &  \phantom{00}9.6$\pm$1.0   & \phantom{0}43$\pm$5   &  380 WC6--7 \\

He\,{\sc i} $\lambda$5876 & $\cdots$      & 11.9$\pm$0.2    & \phantom{0}18.1$\pm$0.3 & \phantom{0}81$\pm$1 &  (260 WC4--7) \\
\hline
\end{tabular}
\end{center}
\begin{footnotesize}
\end{footnotesize}
\end{table}

Strong line calibrations indicate NGC~3049 is supersolar \citep[$\log$ O/H+12 $\geq$ 8.9,][]{2000ApJ...531..776G}  so we apply Galactic calibrations to estimate its WR populations. The C\,{\sc iv} $\lambda\lambda$5801,12 to C\,{\sc iii} $\lambda$5696 ratio suggest a dominant WC6--7 subclass, requiring 380 WC stars. The contribution from WC6--7 stars to the He\,{\sc ii} $\lambda$4686 is considered to be negligible, so we adopt a WN6--8 population of $\sim$1300,

Fig.~\ref{n3049-WR} (right panels) present continuum subtracted line luminosities of blue and yellow WR bumps in NGC~3049, together with emission line templates for Milky Way WN (red) and WC (blue) stars, plus their sum (pink). For the yellow bump we prefer a reduced WC4--7 content of 260 stars in order to reproduce the C\,{\sc iv} $\lambda\lambda$5801,12 and C\,{\sc iii} $\lambda$5696 emission line, whose FWHM is well reproduced (we adopt a similar mix of WC4--5 and WC6--7 stars to the Milky Way sample i.e. 1:3). For the blue bump, 
N\,{\sc v} $\lambda$4603--20 is not apparent, but we adhere to a mix of WNEw:WN3--7s:WN6--8 stars corresponding to the Milky Way (1:1:3), requiring $\sim$750 stars to reproduce the He\,{\sc ii} $\lambda$4686 line, accounting for the contribution of WC stars to this region. The combination of N\,{\sc iii} $\lambda\lambda$4634,41 (from WN6--8 stars) and C\,{\sc iii} $\lambda\lambda$4647,51 (from WC4--7 stars) reproduce the majority of the blue bump reasonably well. The residual mismatch likely arises from stronger N\,{\sc iii} $\lambda\lambda$4634,41 emission than our Galactic templates since N\,{\sc iii}/He\,{\sc ii} is observed to scale with metallicity (recall Fig.~\ref{blue_5808_WN}) and NGC~3049 is likely supersolar. C\,{\sc iii} $\lambda$5696 is reasonably well matched, and includes modest contributions from WN6--8 stars via N\,{\sc iv} $\lambda$5737/Si\,{\sc iiii} $\lambda$5740 and the N\,{\sc ii} $\lambda$5680 multiplet (note also nebular [N\,{\sc ii}] $\lambda$5755). The use of empirical templates requires fewer WR stars, owing to the contribution of WN stars to the yellow bump, and WC stars to the blue bump, which are usually not accounted for, and suggests a WN to WC ratio of $\sim$3. 

\subsection{IC~4870 (ESO~105--IG~011)}

Our low metallicity target IC~4870 is a nearby barred irregular galaxy at a distance of 8.5~Mpc \citep{2013AJ....146...86T} which was identified as a WR galaxy by \citet{2001A&A...380...19J}. Spectroscopy reported  by \citet{2001A&A...380...19J} focus on the blue WR bump, so we have utilised archival NTT/EMMI spectroscopy from May 1999 (PA = 0, $T_{\rm exp}$ = 600 s). Grism \#5 and a 1 arcsec slit provided a spectral coverage of 3975--6685\AA\ at a resolving power of $R$ = 1000--1300. The brightest knot reveals a strong emission line spectrum, including broad C\,{\sc iii} $\lambda$4647--51/C\,{\sc iv} $\lambda$4658, He\,{\sc ii} $\lambda$4686 and C\,{\sc iv} $\lambda\lambda$5801,12, plus extended nebular emission. After standard reduction techniques, including flux calibration, we obtained $c$(H$\beta$)=0.19 from the observed flux ratio $F$(H$\alpha$)/$F$(H$\beta$) = 3.37. Since the spectral coverage does not extend to [O\,{\sc ii}] $\lambda$3727, we are unable to determine a direct oxygen abundance, but strong line calibrations \citep[N2, O3N2,][]{2004MNRAS.348L..59P} indicate IC~4870 is metal-poor ($\log$ O/H+12 $\sim$7.9, or 1/6 $Z_{\odot}$), which is supported by the high electron temperature of $T_{e}$ = 13.7kK obtained from [O\,{\sc iii}] $\lambda$4363/5007.

\begin{table}
\begin{center}
\caption{Line luminosities of selected emission lines in IC~4870 from gaussian fits, based on $c$(H$\beta$)=0.18, a distance of 8.5~Mpc and calibrations from this study. Separate nebular (n) and broad (b) entries are provided for He\,{\sc ii} $\lambda$4686. Calibrations are from SMC (WN6--8) and LMC (WC4--5) samples, while results shown in parenthesis arise from LMC (WC) and SMC (WN) template fits.}
\label{ic4870}
\begin{tabular}{l@{\hspace{2mm}}l@{\hspace{-1mm}}l@{\hspace{2mm}}l@{\hspace{2mm}}l@{\hspace{0mm}}l}
\hline
Line  &  FWHM & $F_{\lambda}$  & $I_{\lambda}$ & $L_{\lambda}$ & N(WR) \\
         & km\,s$^{-1}$      & \multicolumn{2}{c}{$10^{-15}$ erg\,s$^{-1}$\,cm$^{-2}$} & 10$^{37}$ erg\,s$^{-1}$ & \\
\hline
H$\delta$                            & $\cdots$    & \phantom{0}13.5$\pm$0.2  & \phantom{0}22.5$\pm$0.4   & \phantom{0}19.5$\pm$0.3         & \\

H$\gamma$                        & $\cdots$    & \phantom{0}23.2$\pm$0.2 &  \phantom{0}37.8$\pm$0.4 &  \phantom{0}32.8$\pm$0.3           & \\

[O\,{\sc iii}] $\lambda$4363 & $\cdots$    & \phantom{00}4.5$\pm$0.2 & \phantom{00}7.2$\pm$0.3 &    \phantom{00}6.3$\pm$0.3                  & \\

He\,{\sc i} $\lambda$4471   & $\cdots$    & \phantom{00}2.1$\pm$0.2 & \phantom{00}3.4$\pm$0.3 &   \phantom{00}2.9$\pm$0.3                  & \\

C\,{\sc iii} $\lambda\lambda$4647,51  & 5700$\pm$2200 & \phantom{00}7.0$\pm$3.2 &  \phantom{0}10.8$\pm$4.8 &  \phantom{00}9.0$\pm$4.0 & \\

He\,{\sc ii} $\lambda$4686 (b) & 1300$\pm$300 &  \phantom{00}5.4$\pm$0.9 &  \phantom{00}8.5$\pm$1.4   & \phantom{00}7.3$\pm$1.2  &  47 WN6--8 \\

He\,{\sc ii} $\lambda$4686 (n) & $\cdots$  &  \phantom{00}0.8$\pm$0.4 &  \phantom{00}1.3$\pm$0.6   & \phantom{00}1.1$\pm$0.5  &  (120 WN2--8) \\

[Ar\,{\sc iv}] $\lambda$4711 & $\cdots$      & \phantom{00}1.0$\pm$0.3 & \phantom{00}1.5$\pm$0.5     & \phantom{00}1.4$\pm$0.3 & \\

H$\beta$                            & $\cdots$     & \phantom{0}51.0$\pm$0.3                   &  \phantom{0}78.6$\pm$0.5                    &   \phantom{0}68.1$\pm$0.4                \\  

[O\,{\sc iii}] $\lambda$4959 & $\cdots$     & 114\phantom{.0}$\pm$2                  &   174\phantom{.0}$\pm$4                   &  151\phantom{.0}$\pm$3                & \\

[O\,{\sc iii}] $\lambda$5007  & $\cdots$     & 339\phantom{.0}$\pm$3                    & 515\phantom{.0}$\pm$4                          &  446\phantom{.0}$\pm$3                  & \\

C\,{\sc iv} $\lambda\lambda$5801,12 &  4300$\pm$200   &  \phantom{00}7.3$\pm$0.5 &  \phantom{0}10.3$\pm$0.7   & \phantom{00}8.9$\pm$0.6   &  26 WC4--5 \\

He\,{\sc i} $\lambda$5876 & $\cdots$      & \phantom{00}6.6$\pm$0.1 & \phantom{00}9.3$\pm$0.1 & \phantom{00}8.1$\pm$0.1 & (22 WC4--5) \\

H$\alpha$                          & $\cdots$      & 172\phantom{.0}$\pm$1                  & 231\phantom{.0}$\pm$2                       & 200\phantom{.0}$\pm$2 \\

[N\,{\sc ii}]  $\lambda$6584  & $\cdots$      & \phantom{00}2.7$\pm$1.1 & \phantom{00}3.6$\pm$1.5 & \phantom{00}3.1$\pm$1.3 \\
\hline
\end{tabular}
\end{center}
\begin{footnotesize}
\end{footnotesize}
\end{table}

\begin{figure}
\centering
	\includegraphics[width=0.65\linewidth,bb=25 40 525 795,angle=-90]{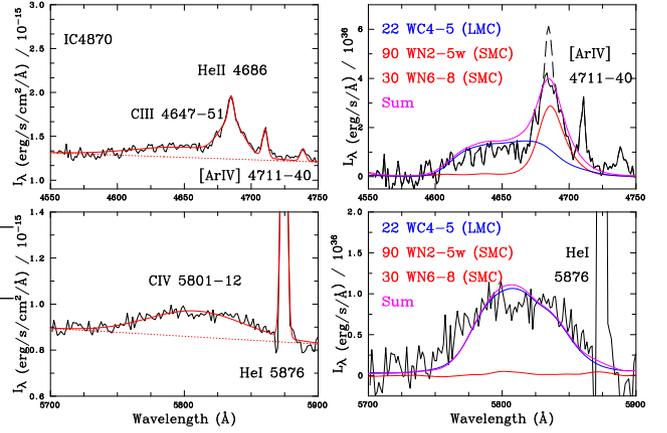}
	\centering
  \caption{Left panels: De-reddened, velocity corrected, blue and yellow WR bumps in IC~4870 (black), including gaussian fits (red) to broad (C\,{\sc iii} $\lambda\lambda$4647,51, He\,{\sc ii} $\lambda$4686, C\,{\sc iv} $\lambda\lambda$5801,12) and nebular ([Ar\,{\sc iv}] $\lambda\lambda$4711,40, He\,{\sc i} $\lambda$5876) lines, with the continuum indicated as dotted lines. $\sim$47 SMC WN6--8 and 26 LMC WC4--5 stars are estimated from calibrations, adopting a distance of 8.5~Mpc. Right panels: Continuum subtracted luminosities of blue and yellow WR bumps in IC~4870 (black, nebular He\,{\sc ii} $\lambda$4686 component indicated in dashed lines) plus template fits adopting 120 SMC WN stars (red), 22 LMC WC4--5 stars (blue) plus their sum (pink).}
	\label{ic4870-WR} 
\end{figure}

Fig.~\ref{ic4870-WR} (left panels) presents de-reddened, velocity corrected WR bumps including gaussian fits to stellar and nebular emission lines. Fluxes of selected nebular and stellar emission lines in IC~4870 measured from  fits are presented in Table~\ref{ic4870}. He\,{\sc ii} $\lambda$4686 comprises both stellar (86\%) and nebular (14\%) components, with a nebular He\,{\sc ii} $\lambda$4686)/H$\beta$ intensity ratio of 2\%. Indeed, \citet{2001A&A...380...19J} attribute 16\% of the He\,{\sc ii} $\lambda$4686 to nebular origin in their higher resolution 1.5m ESO Boller \& Chivens spectroscopy of IC~4870 ($R$ = 2100 at $\lambda$4686). The spectroscopic H$\alpha$ luminosity corresponds to an ionizing output of $Q_{\rm H} = 1.5 \times 10^{51}$ ph\,s$^{-1}$, typical of nearby giant H\,{\sc ii} regions (e.g. NGC~595 in M33), although the integrated ionizing output is likely to be significantly larger. The yellow C\,{\sc iv} $\lambda\lambda$5801,12 feature is very broad, exceeding the average value of LMC WC4--5 stars (Table~\ref{WC-calib}).

Ideally one would apply SMC-metallicity calibrations to IC~4870, but since WC stars are not observed in the SMC we revert to LMC calibrations, allowing us to estimate a population of $\sim$26 WC4--5 stars in IC~4870 from the yellow bump. FWHM(He\,{\sc ii} $\lambda$4686) is intermediate between early and late WN subtypes (Table~\ref{WN-calib}), so we apply a SMC WN6--8 (WN2--5w) calibration to estimate a population of 47 (440) stars, excluding the nebular component. At SMC metallicity, N\,{\sc iii} $\lambda\lambda$4634,41/He\,{\sc ii} $\lambda$4686 $\sim$ 0.04, so accounting for this modest contribution to the broad C\,{\sc iii} $\lambda\lambda$4647,51 feature, we can estimate the number of WC4--5 stars from this diagnostic, again based on LMC calibrations. We estimate a slightly lower population of 19 WC4--5 stars from the blue bump. 

Fig.~\ref{ic4870-WR} (right panels) present continuum subtracted line luminosities of blue and yellow WR bumps in IC~4870, together with emission line templates for WN and WC stars, plus their sums (pink).  For the yellow WR bump, 22 (LMC) WC4--5 stars provide a satisfactory match to the broad C\,{\sc iv} $\lambda\lambda$5801,12 emission line, with negligible contribution from WN stars. For the blue bump, a mix of early and late WN stars are more plausible than a purely early or late WN population, recalling the WNE:WNL ratio is 2 (LMC) to 3 (SMC) in local metal-poor galaxies. Consequently we adopt a ratio of 3:1 for the WN2--5w:WN6--8 population in IC~4870, with both comprising SMC templates.  We require a total of 120 WN stars to reproduce the non-nebular He\,{\sc ii} $\lambda$4686 feature, accounting for the contribution of 22 WC4--5 stars, which also provide a satisfactory fit to C\,{\sc iii} $\lambda\lambda$4647,51+C\,{\sc iv} $\lambda$4658.

Although there is a no direct evidence supporting the presence of WN2--5w stars in IC~4870, they are capable of producing the observed nebular He\,{\sc ii} $\lambda$4686 emission, in contrast to WN6--8 or WC4--5 stars \citep{2006A&A...449..711C}. The measured nebular He\,{\sc ii} $\lambda$4686 luminosity (Table~\ref{ic4870}) corresponds to an ionizing budget of Q(He$^{+}$) = $1.3\times10^{49}$ photon\,s$^{-1}$, an order of magnitude higher than individual nebular He\,{\sc iii} $\lambda$4686 producing H\,{\sc ii} regions in Local Group galaxies \citep{1991ApJ...373..458G}. Adopting  Q(He$^{+}$) ionizing photons of SMC WN2--5w stars from \citet{2015A&A...581A..21H}, only $\sim$10 stars are required to produce sufficient ionizing photons in order to reproduce the observed nebular emission. Consequently only $\sim$10\% of the inferred population would need to be located within H\,{\sc ii} regions to match observations. This more realistic WN population inferred for this metal poor galaxy produces a WN:WC ratio of $\sim$5.5:1, which is more plausible than 2:1 if solely WN6--8 and WC4--5 calibrations are adopted (Table~\ref{ic4870}). The modest He\,{\sc ii} $\lambda$4686  line luminosity of metal-poor WN2--5w stars would severely hinder their direct detection, so the spectroscopic absence of WR bumps in nebular $\lambda$4686 emitting galaxies does not necessarily exclude them from being responsible for the observed nebular emission \citep{1996ApJ...467L..17S, 2012MNRAS.421.1043S}, although several alternative candidates exist \citep{2019A&A...629A.134G, 2019A&A...622L..10S, 2022A&A...661A..67O}.

\section{Outlook and conclusions}\label{summary}

We report calibrations of line luminosities of optical emission lines of Galactic WR stars, greatly expanding on previous studies owing to the availability of {\it Gaia} DR3 parallaxes, and provide revised line luminosities for Magellanic Cloud counterparts, updated from previous calibrations \citep{1990ApJ...348..471S, 1998ApJ...497..618S, 2006A&A...449..711C}. 

Overall we confirm previous results for a reduction in He\,{\sc ii} $\lambda$4686 line luminosity for early-type WN stars at low metallicity, providing separate calibrations for strong-lined and weak-lined stars. \citet{2020MNRAS.491.4406S} have investigated wind driving of WR stars, and establish a dependence on metallicity (Fe-group elements) and Eddington parameter $\Gamma_{e}$ ($\propto L/M$) so strong-lined stars likely differ from weak-lined stars in their wind properties owing to their higher luminosity-to-mass ratios. 

The situation is less clear for late-type WN stars since average line luminosities of LMC stars are  higher than Galactic counterparts and one of the two late WN stars in the SMC is the complex multiple system HD~5980, whose primary potentially dominates the line emission \citep{2019MNRAS.486..725H}. This does not necessarily argue against a metallicity dependence of wind strengths since the lower (luminosity) threshold to the formation of WR stars increases at lower metallicity \citep{2020A&A...634A..79S}, so reduced wind strengths at low metallicity are countered by a shift to higher stellar luminosities on average. We find very massive main-sequence stars in the Milky Way and LMC possess similar He\,{\sc ii} $\lambda$4686 line luminosities to classical late WN stars, albeit with broader lines plus weaker N\,{\sc iii} $\lambda\lambda$4634,41 and He\,{\sc i} $\lambda$5876 lines. C\,{\sc iv} $\lambda\lambda$5801,12 is an order of magnitude weaker than the blue bump in early WN stars, and is weaker still in late subtypes, though has comparable strength in WN/C subtypes.

LMC WC4--5 stars possess higher C\,{\sc iv} $\lambda\lambda$5801,12 line luminosities than Galactic counterparts, while we provide the first calibration of line luminosities for WC6--7 and WC8--9 stars based on large samples.  C\,{\sc iii} $\lambda\lambda$4647,51+He\,{\sc ii} $\lambda$4686 is strong for all WC subtypes, with C\,{\sc iii} $\lambda$5696 is the dominant emission line in the yellow region for WC9 stars. The blue WC bump is well correlated with 
C\,{\sc iv}$ \lambda\lambda$5801,12 in all subtypes.  We also provide line luminosities of O\,{\sc vi} $\lambda\lambda$3811,34 in WO stars, plus other prominent lines (O\,{\sc iv} $\lambda\lambda$3403,13, C\,{\sc iv} $\lambda$4658+He\,{\sc ii} $\lambda$4686, C\,{\sc iv} $\lambda\lambda$5801,12). We also utilise spectroscopic studies to quantify the line  to bolometric luminosity ratios for each spectral type, which ranges from $L_{\rm HeII~4686}/L_{\rm Bol} = 3 \times 10^{-5}$ for SMC WNEw stars, to $L_{\rm CIV~5801,12}/L_{\rm Bol} = 2 \times 10^{-3}$ for LMC WC4--5 stars. 

Figures~\ref{4686} and \ref{5808} emphasise the diversity in line luminosities within our various WR categories, so average properties should be used with caution. For example, \citet{2012AJ....143..136L} have presented optical line fluxes of WR142--1 (WN6w) from Hobby--Eberly Telescope (HET) observations. Applying their derived extinction ($A_{\rm V} = 8.1_{-0.7}^{+0.6}$ mag) and the average He\,{\sc ii} $\lambda$4686 line luminosity of Galactic WN6--8 stars  (Table~\ref{WN-calib}) implies a distance of 2.1$_{-0.9}^{+0.6}$ kpc, whereas the {\it Gaia} DR3 distance to WR142-1 is 1.58$\pm$0.05 kpc.

Flux calibrated spectroscopy also permits us to  produce templates for single and composite (single and binary) subtypes at each metallicity. We construct cumulative emission line luminosities for the Milky Way ($\sim$7\% completeness), LMC (66\% complete) and SMC (100\% complete), with global blue WR line luminosities of $\approx$3.7$\times 10^{39}$ erg\,s$^{-1}$, 2.6$\times 10^{38}$ erg\,s$^{-1}$ and 9.5$\times 10^{37}$ erg\,s$^{-1}$, respectively. We also apply our templates to representative metal-poor (IC~4870) and metal-rich (NGC~3049) galaxies that exhibit strong WR bumps at optical wavelengths. Reasonable consistency is achieved, although templates are preferred since they exploit line profile information and incorporate contributions from WN stars to the yellow WR bump, and contributions from WC stars to the blue WR bump.

The prospect for additional empirical WR line luminosities is promising at high metallicity, thanks to large numbers of detections in M31 \citep{2019Galax...7...74N}, while the situation is less satisfactory at low metallicity. IC10 has a relatively high star-formation rate, although it suffers from high foreground extinction and its metallicity is close to that of the LMC \citep{2017MNRAS.472.4618T}. Other nearby dwarf galaxies suffer from very low star formation rates \citep[][their fig.11]{2019Galax...7...88C}, so only a handful of WR stars are known in NGC~6822 and IC~1613  \citep{1991AJ....102..927A}. Consequently there is little alternative to the use of theoretical spectra below SMC metallicity, providing these are carefully calibrated against empirical results. 

\section*{Acknowledgements}

PAC and JMB are supported by the Science and Technology Facilities Council research grant ST/V000853/1 (PI. V. Dhillon). Thanks to Bill Vacca for sharing his IRTF/SpeX dataset, Frank Tramper for providing his VLT/Xshooter spectra of WO stars, Nidia Morrell for sharing her Magellan/MagE spectrum of LMC 170--2, Paco Najarro for providing his VLT/Xshooter spectrum of VFTS~682, Gloria Koenigsberger for providing her HST/STIS spectra of HD~5980, and Daniel Schaerer for providing his flux calibrated ESO 2.2m EFOSC2 spectrum of NGC~3049.

This research has made extensive use of NASA's Astrophysics Data System Bibliographic Services, and the SIMBAD database, operated at CDS, Strasbourg, France, plus the European Space Agency (ESA) {\it Gaia} archive whose website is https://archives.esac.esa.int/gaia and the ESO Science Archive Facility (IC~4870 was observed under programme 59.A-9004). The INES System has been developed by the ESA IUE Project at VILSPA. Data and access software are distributed and maintained by INTA through the INES Principal Centre at LAEFF. The Starlink software \citep{2014ASPC..485..391C} is currently supported by the East Asian Observatory. 

For the purpose of open access, the author has applied a Creative Commons Attribution (CC BY) license to any Author Accepted Manuscript version arising.


\section*{Data Availability} 


Template continuum subtracted emission line spectra of Milky Way and Magellanic Cloud WR stars will be publicly available in a Zenodo collection upon publication (ascii format, wavelength units of \AA, monochromatic luminosity units of erg\,s$^{-1}$\,\AA$^{-1}$) at \url{10.5281/zenodo.7573775}, separately for single and single+binary WN2--5w, WN3--7s, WN6--8, WN9--11, WN5--7h+Of/WN (main sequence), WN/C, WC4--5, WC6--7, WC8--9, WO stars.

\bibliographystyle{mnras}








\appendix

\section{Line luminosities of Galactic and Magellanic Cloud WR stars}

Line luminosities of individual Galactic and Magellanic Cloud WN+Of/WN, WN/WC, WC and WO stars are provided in Tables~\ref{WN-all}-\ref{WO-all}, sorted by catalogue number (WR for Milky Way, BAT for LMC, AB for SMC). Optical datasets refer to Table~\ref{spectroscopy}, Galactic distances are from {\it Gaia} DR3 parallaxes \citep{2021A&A...649A...1G}, using zero point corrections from \citet{2021A&A...649A...4L} updated by \citet{2022A&A...657A.130M}, and the Bayesian methods set out in \citet{2020MNRAS.493.1512R}, while Magellanic Cloud distances are from \citet{2019Natur.567..200P} and \citet{2014ApJ...780...59G}. References are provided for A$_{\rm V}$ (mag) and stellar luminosities ($L_{\odot}$), with the latter adjusted to our adopted distances. Observed line fluxes, line luminosities and FWHM of He\,{\sc ii} $\lambda$4686 (for WN, WN/C, Of/WN stars) or C\,{\sc iv} $\lambda\lambda$5801,12 are provided in erg\,s$^{-1}$\,cm$^{-2}$, erg\,s$^{-1}$ and km\,s$^{-1}$, the latter corrected for instrumental broadening. The He\,{\sc ii} $\lambda$4686 or C\,{\sc iv} $\lambda\lambda$5801,12 line luminosity (in 10$^{-3} L_{\rm Bol}$) is restricted to stars with a measured stellar luminosity, with other optical line luminosities provided with respect to He\,{\sc ii} $\lambda$4686 or C\,{\sc iv} $\lambda\lambda$5801,12 ($\cdots$ refers to a lack of data). For WN and WN/C stars, aside from the blends at $\lambda$4100 and $\lambda$4630, we have endeavoured to exclude the contribution of He\,{\sc i} $\lambda$7065 and He\,{\sc ii} $\lambda$7177 to the N\,{\sc iv} $\lambda\lambda$7103,29 multiplet. For early-type WC and WO stars, C\,{\sc iv} $\lambda$4658 (6-5), $\lambda$4686 (8-6), $\lambda$4689 (11-7) will also contribute to the C,{\sc iii} $\lambda\lambda$4647,51+He\,{\sc ii} $\lambda$4686 blend.

\begin{landscape}
\begin{table}
\caption{
Luminosities of prominent optical emission lines of Galactic (WR), LMC (BAT99) and SMC (AB) WN stars, categorised as  weak-lined early-type (WN2--5w), strong-lined (WN3--7s), late-type (WN6--8), very late-type (WN9--11) or (main sequence) very massive H-rich (WN5--7h) stars. He\,{\sc ii} $\lambda$4686 FWHM are also provided in km\,s$^{-1}$ ($\pm$100 km\,s$^{-1}$). The feature at $\lambda$4100 involves N\,{\sc iii} $\lambda\lambda$4097,4103, Si\,{\sc iv} $\lambda\lambda$4088,4116, He\,{\sc ii} $\lambda$4100+H$\delta$, while the feature at $\lambda$4630 involves N\,{\sc v} $\lambda\lambda$4603,20, N\,{\sc iii} $\lambda\lambda$4634,41. 
Stars included in spectral templates are indicated in bold (the remainder are excluded owing to limited spectral coverage).}
\label{WN-all}
\begin{center}
\begin{small}
 \begin{tabular}{l@{\hspace{1mm}}l@{\hspace{1.5mm}}l@{\hspace{1.5mm}}l@{\hspace{1.5mm}}l@{\hspace{1.5mm}}l@{\hspace{1mm}}c@{\hspace{1mm}}r@{\hspace{1mm}}c@{\hspace{0mm}}c@{\hspace{1mm}}c
 @{\hspace{1.5mm}}l@{\hspace{1mm}}l@{\hspace{1mm}}l@{\hspace{1mm}}l@{\hspace{1mm}}l@{\hspace{1mm}}l@{\hspace{1mm}}l@{\hspace{1mm}}l@{\hspace{1mm}}l}   %
\hline
Star & Category & Data & $d$ & $A_{\rm V}$ & $\log$  & Ref & $\log F_{\rm HeII~4686}$ & HeII 4686 & \multicolumn{2}{c}{$L_{\rm HeII~4686}$} 
& $\underline{L_{\rm NIV~3478,85}}$ 
& $\underline{L_{\rm NIV~4058}}$ 
& $\underline{L_{4100}}$ 
& $\underline{L_{4630}}$ 
&$\underline{L_{\rm HeII~5412}}$ 
& $\underline{L_{\rm CIV~5801,12}}$ 
&$\underline{L_{\rm HeI~5876}}$ 
& $\underline{L_{{\rm H}\alpha}}$ 
& $\underline{L_{\rm NIV~7103,29}}$ \\  
        &    &   ID    &   kpc     & mag              & $L_{\rm Bol}/L_{\odot}$        &        & erg\,s$^{-1}$\,cm$^{-2}$ & FWHM & 10$^{35}$ erg\,s$^{-1}$ &  $10^{-3} L_{\rm Bol}$   & $L_{\rm HeII~4686}$ &  $L_{\rm HeII~4686}$  & $L_{\rm HeII~4686}$ & $L_{\rm HeII~4686}$  & $L_{\rm HeII~4686}$ &  $L_{\rm HeII~4686}$ &   $L_{\rm HeII~4686}$ &
 $L_{\rm HeII~4686}$  & $L_{\rm HeII~4686}$ \\
\hline
{\bf WR1}  &  WN3--7s  & II91        & \phantom{0}3.0$\pm$0.1 & 2.5$\pm$0.3         & 5.84$_{-0.11}^{+0.11}$ & a& --10.06$^{+0.08}_{-0.10}$ & 2500                    &   15.9$_{-5.1}^{+6.1}$ & 0.60$\pm$0.07 & 0.28$_{-0.02}^{+0.03}$ & \multicolumn{2}{c}{--- 0.21$_{-0.00}^{+0.01}$ ---} & 0.17 & 0.16$_{-0.01}^{+0.01}$ & 0.12$_{-0.01}^{+0.01}$ & 0.04 & 0.21$_{-0.02}^{+0.02}$ & 0.21$_{-0.03}^{+0.02}$ \\ [2pt]
{\bf WR3}  & WN2--5w & II91         & \phantom{0}2.2$\pm$0.1 & 1.4$_{-0.1}^{+0.2}$  & 5.33$_{-0.07}^{+0.07}$ & a & --10.89$_{-0.10}^{+0.08}$ & 2400 & \phantom{0}0.4$_{-0.1}^{+0.1}$ & 0.05$_{-0.01}^{+0.00}$ & $\cdots$    & $\cdots$    & $\cdots$    & 0.65 & 0.16$_{-0.00}^{+0.01}$ & 0.02 & 0.00 & 0.24$_{-0.01}^{+0.01}$ & $\cdots$\\   [2pt]
{\bf WR6}  & WN3--7s  & VU         & \phantom{0}1.4$\pm$0.1 & 0.4$_{-0.0}^{+0.1}$         & 5.37$_{-0.04}^{+0.05}$ & a &   --8.55$_{-0.05}^{+0.04}$ & 2300                &  11.2$_{-1.4}^{+1.4}$ & 1.23$_{-0.05}^{+0.26}$ & 0.60$_{-0.01}^{+0.01}$  & \multicolumn{2}{c}{--- 0.28 ---} & 0.16 & 0.13 & \#0.06 & 0.03 & 0.14 & 0.15$_{-0.01}^{+0.00}$ \\ [2pt]
WR7 & WN3--7s   & AR     & \phantom{0}4.0$\pm$0.4 & 2.0$\pm$0.2         & 5.28$_{-0.12}^{+0.12}$ & a & --10.48$_{-0.04}^{+0.04}$ & 1900             & \phantom{0}5.9$_{-1.4}^{+1.7}$ & 0.81$_{-0.12}^{+0.14}$ & 0.50$_{-0.03}^{+0.03}$ & 0.02      & 0.07$_{-0.00}^{+0.01}$              & 0.20 & 0.15$_{-0.00}^{+0.01}$ & 0.09$_{-0.00}^{+0.01}$ & 0.02 & 0.18$_{-0.02}^{+0.01}$ & \ddag0.07$_{-0.00}^{+0.01}$ \\ [2pt]
{\bf WR10} & WN2--5w & CS         & \phantom{0}4.2$\pm$0.2 & 2.2$\pm$0.2         & 5.55$_{-0.10}^{+0.10}$ & a& --11.10$_{-0.04}^{+0.04}$ & 1300       & \phantom{0}1.9$_{-0.5}^{+0.6}$ & 0.14$_{-0.02}^{+0.02}$ & 0.63$_{-0.03}^{+0.05}$  & 0.62$_{-0.02}^{+0.03}$  &     0.38$_{-0.01}^{+0.01}$                       & 0.08 & 0.08$_{-0.00}^{+0.01}$ & 0.03 & 0.00 & 0.27$_{-0.02}^{+0.02}$ &$\cdots$\\ [2pt]
{\bf WR12} & WN6--8   & CS         & \phantom{0}5.2$_{-0.3}^{+0.4}$ & 3.6$_{-0.4}^{+0.3}$ & 5.86$_{-0.15}^{+0.16}$ & a & --10.94$_{-0.04}^{+0.05}$ & 1100                &        19.1$_{-6.6}^{+9.4}$ & 0.68$_{-0.10}^{+0.13}$ & 0.18$_{-0.01}^{+0.02}$ & 0.22$_{-0.01}^{+0.01}$   & 0.58$_{-0.03}^{+0.03}$                           & 0.56$_{-0.00}^{+0.01}$ & 0.09$_{-0.00}^{+0.01}$ & 0.01 & 0.18$_{-0.01}^{+0.02}$ & 0.50$_{-0.05}^{+0.06}$ &$\cdots$\\ [2pt]
{\bf WR16} & WN6--8     & CS        & \phantom{0}2.3$\pm$0.1 & 2.3$_{-0.3}^{+0.2}$ & 5.60$_{-0.10}^{+0.09}$ & a  & --10.21$_{-0.04}^{+0.05}$ & \phantom{0}600 & \phantom{0}4.7$_{-1.2}^{+1.4}$ & 0.31$_{-0.03}^{+0.04}$ & $\cdots$ & 0.10 & 0.83$_{-0.02}^{+0.03}$        & 1.11$_{-0.00}^{+0.01}$ & 0.08$_{-0.01}^{+0.00}$ & 0.01 & 0.33$_{-0.02}^{+0.02}$ & 0.65$_{-0.05}^{+0.05}$ &$\cdots$\\ [2pt]
WR18 & WN3--7s  & AR      & \phantom{0}3.1$\pm$0.1 & 3.3$_{-0.4}^{+0.3}$ & 5.90$_{-0.14}^{+0.13}$ & a & --10.27$_{-0.04}^{+0.04}$ & 2800               &       22.6$_{-7.2}^{+10.0}$ & 0.74$_{-0.08}^{+0.09}$ & $\cdots$ & \multicolumn{2}{c}{--- 0.28$_{-0.01}^{+0.01}$ ---}  & 0.16$_{-0.00}^{+0.01}$  & 0.12$_{-0.00}^{+0.01}$ & 0.12$_{-0.01}^{+0.00}$ & 0.02 & 0.15$_{-0.01}^{+0.02}$ & 0.17$_{-0.02}^{+0.03}$ \\   [2pt]
{\bf WR20} & WN2--5w  & CS     & \phantom{0}6.4$_{-0.4}^{+0.5}$ & 4.8$\pm$0.5        & 5.69$_{-0.20}^{+0.21}$ & a & --12.03$_{-0.04}^{+0.04}$ & 1600               &         10.3$_{-4.5}^{+7.5}$ & 0.54$_{-0.11}^{+0.14}$ &  0.64$_{-0.08}^{+0.10}$ & 0.24$_{-0.02}^{+0.02}$   & 0.29$_{-0.02}^{+0.02}$                          & 0.16 & 0.12$_{-0.01}^{+0.01}$ & 0.07$_{-0.01}^{+0.01}$ & 0.02 & 0.10$_{-0.02}^{+0.02}$ & 0.15$_{-0.03}^{+0.04}$ \\ [2pt]
{\bf WR21} & WN2--5w+O& CS   & \phantom{0}3.2$\pm$0.1  & 2.1$\pm$0.2  & $\cdots$ & b &--10.56$_{-0.04}^{+0.05}$ & 1800  & \phantom{0}3.7$_{-0.9}^{+1.1}$ & $\cdots$ & 0.61$_{-0.04}^{+0.03}$  & 0.09$_{-0.00}^{+0.01}$ & 0.12$_{-0.00}^{+0.01}$                  & 0.15 & 0.08$_{-0.01}^{+0.00}$ & 0.07$_{-0.01}^{+0.00}$ & 0.00 & 0.15$_{-0.01}^{+0.02}$ & 0.28$_{-0.03}^{+0.03}$ \\ [2pt]
{\bf WR22} & WN5--7h+O & VU & \phantom{0}2.4$_{-0.2}^{+0.1}$  & 1.6$\pm$0.2 & 6.26$_{-0.07}^{+0.08}$ & a & --9.52$_{-0.05}^{+0.04}$ & \phantom{0}800                  &       11.4$_{-2.2}^{+2.5}$ & 0.16$_{-0.01}^{+0.02}$  & 0.03$_{-0.00}^{+0.01}$ & 0.13 & 0.55$_{-0.01}^{+0.01}$                          & 0.61 & 0.05$_{-0.00}^{+0.01}$ & \#0.02 & 0.05 & 0.59$_{-0.03}^{+0.03}$  & 0.09$_{-0.01}^{+0.00}$ \\ [2pt]
{\bf WR24} & WN5--7h      & VU  & \phantom{0}2.4$_{-0.1}^{+0.2}$ & 0.9$\pm$0.1       &6.12$_{-0.05}^{+0.07}$  & a  & --9.44$_{-0.04}^{+0.05}$ & 1100      & \phantom{0}7.2$_{-1.1}^{+1.1}$ & 0.14$_{-0.01}^{+0.01}$ & 0.09  & 0.20 & 0.34                         & 0.41 & 0.06$_{-0.00}^{+0.01}$ & \#0.04$_{-0.01}^{+0.00}$ & 0.00 & 0.49$_{-0.01}^{+0.02}$ & 0.10 \\ [2pt]
{\bf WR25} & Of/WN+O & {\bf CS}   & \phantom{0}2.3$\pm$0.1 & 3.0$\pm$0.3 & 6.49$_{-0.13}^{+0.13}$ & a,c & --10.47$_{-0.05}^{+0.04}$ & 1300    & \phantom{0}5.1$_{-1.5}^{+2.0}$ & 0.04$_{-0.00}^{+0.01}$ & 0.10$_{-0.01}^{+0.01}$ & 0.31$_{-0.01}^{+0.01}$ & 0.43$_{-0.01}^{+0.01}$                        & 0.60 & 0.06$_{-0.00}^{+0.01}$ & 0.01 & 0.00 & 0.58$_{-0.04}^{+0.05}$ & $\cdots$ \\  [2pt]
{\bf WR28}  & WN6--8+O   & CS   & \phantom{0}6.9$\pm$0.5   & 4.5$_{-0.4}^{+0.5}$       & 6.09$_{-0.18}^{+0.20}$ & a & --11.91$_{-0.04}^{+0.04}$ & 1200               &           11.1$_{-4.6}^{+7.4}$ & 0.23$_{-0.04}^{+0.06}$ & 0.29$_{-0.04}^{+0.04}$   & 0.21$_{-0.01}^{+0.02}$ & 0.39$_{-0.02}^{+0.03}$                        & 0.33 & 0.09$_{-0.01}^{+0.01}$ & 0.03$_{-0.01}^{+0.00}$ & 0.02 & 0.26$_{-0.05}^{+0.05}$ & $\cdots$ \\ [2pt]
{\bf WR29} & WN6--8+O    & CS  & \phantom{0}7.0$\pm$0.5 & 3.6$\pm$0.4 & $\cdots$ & b & --12.06$_{-0.04}^{+0.04}$ & 1000 & \phantom{0}2.9$_{-1.0}^{+1.4}$ & $\cdots$ & 0.15$_{-0.02}^{+0.01}$  & 0.19$_{-0.01}^{+0.01}$ & 0.37$_{-0.02}^{+0.02}$               & 0.39 & 0.08$_{-0.01}^{+0.00}$ & 0.03$_{-0.01}^{+0.00}$ & 0.04 & 0.27$_{-0.04}^{+0.04}$ & $\cdots$ \\ [2pt]
{\bf WR31a} & WN9--11 & AR & \phantom{0}7.1$_{-0.6}^{+0.8}$ & 3.5$_{-0.3}^{+0.4}$ & 5.56$_{-0.16}^{+0.17}$ & d & --13.28$_{-0.04}^{+0.05}$ &  300 & \phantom{0}0.2$_{-0.1}^{+0.1}$ & 0.01$_{-0.00}^{+0.01}$ & $\cdots$  & 0.00 & 21$_{-1}^{+2}$               & 7.8 & 0.00 & 0.00 & 37$_{-4}^{+4}$ & 175$_{-23}^{+26}$ & $\cdots$ \\ [2pt]
WR34  & WN2-5w      & CS & \phantom{0}7.3$_{-0.7}^{+0.8}$  & 4.4$_{-0.4}^{+0.5}$        & 5.59$_{-0.20}^{+0.20}$ & a    & --12.04$_{-0.05}^{+0.04}$ & 1400           & \phantom{0}8.4$_{-3.5}^{+5.6}$  & 0.55$_{-0.12}^{+0.17}$ & 0.64$_{-0.08}^{+0.09}$ & 0.24$_{-0.01}^{+0.02}$ & 0.26$_{-0.02}^{+0.02}$ & 0.22$_{-0.00}^{+0.01}$ &$\cdots$&$\cdots$&$\cdots$&$\cdots$&$\cdots$\\ [2pt]
{\bf WR35}& WN6--8       & CS    & \phantom{0}7.9$_{-0.6}^{+0.8}$  & 4.3$\pm$0.4        & 5.73$_{-0.19}^{+0.19}$ & a    & --11.93$_{-0.04}^{+0.04}$ & 1000    &    11.4$_{-4.6}^{+7.3}$ & 0.55$_{-0.11}^{+0.14}$ & 0.40$_{-0.05}^{+0.05}$ &0.30$_{-0.02}^{+0.02}$  & 0.46$_{-0.03}^{+0.03}$                           & 0.38$_{-0.00}^{+0.01}$ & 0.07$_{-0.00}^{+0.01}$ & 0.04$_{-0.01}^{+0.00}$ & 0.02$_{-0.00}^{+0.01}$ & 0.20$_{-0.03}^{+0.04}$ & $\cdots$ \\ [2pt]
{\bf WR36} & WN3--7s    & CS    & \phantom{0}6.7$_{-0.5}^{+0.6}$ & 3.8$_{-0.4}^{+0.3}$        & 5.39$_{-0.16}^{+0.17}$ & a    & --11.51$_{-0.04}^{+0.05}$  & 2600   & 11.4$_{-4.2}^{+6.2}$ & 1.20$_{-0.21}^{+0.27}$ & 0.58$_{-0.06}^{+0.07}$ & 0.10$_{-0.01}^{+0.01}$  & 0.19$_{-0.01}^{+0.01}$                           & 0.18 & 0.11$_{-0.01}^{+0.01}$ & 0.07$_{-0.01}^{+0.01}$ & 0.07$_{-0.01}^{+0.00}$ & 0.12$_{-0.02}^{+0.01}$  & 0.18$_{-0.03}^{+0.03}$ \\ [2pt]
{\bf WR37} & WN3--7s    & CS    & \phantom{0}6.8$_{-0.6}^{+0.7}$   & 6.1$\pm$0.6        & 5.88$_{-0.26}^{+0.25}$ & a    & --12.24$_{-0.04}^{+0.05}$  & 2800   &     31.0$_{-16.0}^{+31.0}$ & 1.07$_{-0.26}^{+0.37}$ & 0.36$_{-0.06}^{+0.07}$ & 0.03 & 0.15$_{-0.02}^{+0.01}$                          & 0.17 & 0.09$_{-0.01}^{+0.01}$ & 0.07$_{-0.01}^{+0.02}$ & 0.02 & 0.11$_{-0.03}^{+0.03}$   & $\cdots$ \\ [2pt]
{\bf WR40} & WN6--8       & CS    & \phantom{0}2.8$_{-0.1}^{+0.2}$   & 1.6$_{-0.1}^{+0.2}$ & 5.61$_{-0.07}^{+0.08}$ & a  & --9.74$_{-0.05}^{+0.04}$  & \phantom{0}900   &  10.7$_{-2.1}^{+2.5}$ & 0.68$_{-0.06}^{+0.07}$  & 0.33$_{-0.02}^{+0.01}$ & 0.11$_{-0.01}^{+0.00}$ & 0.76$_{-0.02}^{+0.02}$                       & 0.99$_{-0.00}^{+0.01}$  & 0.08$_{-0.00}^{+0.01}$ & 0.01 & 0.48$_{-0.02}^{+0.02}$ & 0.71$_{-0.04}^{+0.05}$   & 0.28$_{-0.02}^{+0.02}$ \\ [2pt]
WR43A & 2$\times$WN5--7h & HF & \phantom{0}7.3$\pm$0.4         & 3.9$\pm$0.4         & 6.56$_{-0.16}^{+0.17}$ & e   & --11.64$_{-0.04}^{+0.04}$  & 1600     & 11.7$_{-4.3}^{+6.6}$ & 0.08$_{-0.01}^{+0.02}$    & 0.19$_{-0.02}^{+0.03}$ & 0.31$_{-0.02}^{+0.02}$ & 0.47$_{-0.02}^{+0.03}$           & 0.27 &$\cdots$&$\cdots$&$\cdots$&$\cdots$&$\cdots$\\ [2pt]
WR43B & WN5--7h   & HF           & \phantom{0}7.3$\pm$0.4        & 3.9$\pm$0.4         & 6.43$_{-0.16}^{+0.16}$ & e    & --11.44$_{-0.04}^{+0.05}$ & 1900        &     18.1$_{-6.7}^{+10.0}$ & 0.18$_{-0.03}^{+0.03}$    & 0.26$_{-0.03}^{+0.03}$ & 0.25$_{-0.02}^{+0.01}$ & 0.46$_{-0.02}^{+0.03}$           & 0.37$_{-0.01}^{+0.00}$ &$\cdots$&$\cdots$&$\cdots$&$\cdots$&$\cdots$\\ [2pt]
{\bf WR43C} & Of/WN  & HF           & \phantom{0}7.3$\pm$0.4        & 3.9$\pm$0.4         & 6.32$_{-0.17}^{+0.16}$ & e     & --12.07$_{-0.05}^{+0.04}$ & 1900      &  \phantom{0}4.4$_{-1.6}^{+2.5}$   & 0.06$_{-0.01}^{+0.01}$    &0.21$_{-0.02}^{+0.03}$ & 0.18$_{-0.01}^{+0.01}$ & 0.10$_{-0.01}^{+0.00}$          & 0.37$_{-0.01}^{+0.00}$ &$\cdots$&$\cdots$&$\cdots$&$\cdots$&$\cdots$\\ [2pt]
WR44  & WN2--5w+O & CS & \phantom{0}8.5$_{-0.7}^{+0.9}$ & 2.7$\pm$0.3 & 5.84$_{-0.13}^{+0.14}$ & a    & --11.27$_{-0.05}^{+0.04}$ & 1800  & \phantom{0}9.0$_{-2.6}^{+3.3}$   & 0.33$_{-0.05}^{+0.07}$  & 0.83$_{-0.06}^{+0.06}$ & 0.11$_{-0.01}^{+0.00}$ & 0.15$_{-0.00}^{+0.01}$          & 0.16$_{-0.01}^{+0.00}$ &$\cdots$&$\cdots$&$\cdots$&$\cdots$&$\cdots$\\ [2pt]
{\bf WR46}  & WN2--5w    & CS    & \phantom{0}2.3$_{-0.1}^{+0.0}$ & 1.3$\pm$0.1 & 5.29$_{-0.06}^{+0.06}$ & a   & --10.68$_{-0.05}^{+0.04}$ & 2500  & \phantom{0}0.5$_{-0.1}^{+0.1}$   & 0.07$_{-0.00}^{+0.01}$ & $\cdots$ &$\cdots$           &  $\cdots$ & 0.72$_{-0.01}^{+0.00}$ & 0.11 & 0.01 & 0.00 & 0.16$_{-0.00}^{+0.01}$ & $\cdots$ \\ [2pt]
{\bf WR47} & WN6--8+O   & CS    & \phantom{0}3.6$\pm$0.2 & 4.0$_{-0.4}^{+0.3}$ &  $\cdots$ & b & --11.03$_{-0.05}^{+0.04}$ & 1400      &    12.0$_{-4.5}^{+6.9}$ & $\cdots$ & $\cdots$ & 0.24$_{-0.01}^{+0.02}$ & 0.41$_{-0.02}^{+0.03}$               & 0.43$_{-0.01}^{+0.00}$ & 0.12$_{-0.01}^{+0.00}$ & 0.05$_{-0.01}^{+0.00}$ & 0.06$_{-0.00}^{+0.01}$ & 0.13$_{-0.02}^{+0.02}$ & $\cdots$ \\ [2pt]
{\bf WR47a} & WN6--8      & AD   & \phantom{0}8.1$_{-0.8}^{+1.0}$ & 7.6$_{-0.7}^{+0.8}$ & $\cdots$                   & b & --13.53$_{-0.04}^{+0.04}$ & \phantom{0}900 & 12.3$_{-7.3}^{+16.9}$ & $\cdots$ & $\cdots$ & 0.10$_{-0.01}^{+0.01}$ & 0.72$_{-0.08}^{+0.09}$ & 1.17$_{-0.02}^{+0.02}$ & 0.07$_{-0.01}^{+0.01}$ & 0.02$_{-0.01}^{+0.00}$ & 0.61$_{-0.12}^{+0.14}$ & 1.00$_{-0.26}^{+0.35}$ & 0.03$_{-0.01}^{+0.02}$ \\ [2pt]
{\bf WR49} & WN2--5w     & CS    & \phantom{0}8.4$_{-0.7}^{+0.9}$ & 3.0$\pm$0.3       & 5.25$_{-0.14}^{+0.15}$  & a   & --11.82$_{-0.04}^{+0.04}$  & 1500      & \phantom{0}3.8$_{1.2}^{+1.6}$ & 0.55$_{-0.10}^{+0.12}$   & 0.66$_{-0.05}^{+0.07}$ & 0.29$_{-0.02}^{+0.01}$ & 0.27$_{-0.01}^{+0.01}$                    & 0.25  & 0.11$_{-0.01}^{+0.01}$ & 0.04$_{-0.01}^{+0.00}$ & 0.01 & 0.22$_{-0.02}^{+0.03}$ & $\cdots$ \\ [2pt]
{\bf WR51} & WN2--5w    & CS    & \phantom{0}4.0$\pm$0.2   & 5.3$_{-0.6}^{+0.5}$      & 5.54$_{-0.22}^{+0.22}$  & a    & --12.30$_{-0.05}^{+0.04}$ & 1700      & \phantom{0}3.5$_{-1.6}^{+2.8}$ & 0.26$_{-0.05}^{+0.06}$   & 0.71$_{-0.11}^{+0.11}$   & 0.20$_{-0.02}^{+0.02}$ & 0.27$_{-0.02}^{+0.02}$                  & 0.24  & 0.11$_{-0.01}^{+0.01}$  & 0.04$_{-0.00}^{+0.01}$ & 0.01 & 0.20$_{-0.04}^{+0.04}$ & $\cdots$ \\ [2pt]
WR54   & WN2--5w & CS      & \phantom{0}5.5$_{-0.3}^{+0.4}$ & 3.1$\pm$0.3        & 5.41$_{-0.13}^{+0.14}$   & a  & --11.41$_{-0.05}^{+0.04}$ & 1500       & \phantom{0}4.5$_{-1.4}^{+1.9}$  & 0.45$_{-0.06}^{+0.08}$   & 1.01$_{-0.08}^{+0.10}$ & 0.28$_{-0.01}^{+0.02}$ & 0.32$_{-0.02}^{+0.01}$      & 0.23 &$\cdots$&$\cdots$&$\cdots$&$\cdots$&$\cdots$\\ [2pt]
WR55   & WN6--8    & CS     & \phantom{0}3.0$\pm$0.3  & 2.8$\pm$0.1 & 5.35$_{-0.12}^{+0.12}$ & a  & --10.55$_{-0.05}^{+0.04}$ & 1100             & \phantom{0}6.7$_{-1.9}^{+2.5}$ & 0.78$_{-0.10}^{+0.10}$    & 0.12$_{-0.01}^{+0.01}$ & 0.24$_{-0.01}^{+0.01}$ & 0.49$_{-0.02}^{+0.01}$     & 0.53 &$\cdots$&$\cdots$&$\cdots$&$\cdots$&$\cdots$\\ [2pt]
\hline
\end{tabular}
\end{small}
\end{center}
\end{table}
\end{landscape}

\addtocounter{table}{-1}

\begin{landscape}
\begin{table}
\caption{continued}
\begin{center}
\begin{footnotesize}
 \begin{tabular}{l@{\hspace{1mm}}l@{\hspace{1.5mm}}l@{\hspace{1.5mm}}l@{\hspace{1.5mm}}l@{\hspace{1.5mm}}l@{\hspace{1mm}}c@{\hspace{1mm}}r@{\hspace{1mm}}c@{\hspace{0mm}}c@{\hspace{1mm}}c
 @{\hspace{1.5mm}}l@{\hspace{1mm}}l@{\hspace{1mm}}l@{\hspace{1mm}}l@{\hspace{1mm}}l@{\hspace{1mm}}l@{\hspace{1mm}}l@{\hspace{1mm}}l@{\hspace{1mm}}l}   %
\hline
Star & Category & Data & $d$ & $A_{\rm V}$ & $\log$  & Ref & $\log F_{\rm HeII~4686}$ & HeII 4686 & \multicolumn{2}{c}{$L_{\rm HeII~4686}$} 
& $\underline{L_{\rm NIV~3478,85}}$ 
& $\underline{L_{\rm NIV~4058}}$ 
& $\underline{L_{4100}}$ 
& $\underline{L_{4630}}$ 
&$\underline{L_{\rm HeII~5412}}$ 
& $\underline{L_{\rm CIV~5801,12}}$ 
&$\underline{L_{\rm HeI~5876}}$ 
& $\underline{L_{{\rm H}\alpha}}$ 
& $\underline{L_{\rm NIV~7103,29}}$ \\  
        &    &      ID     &   kpc     & mag              & $L_{\rm Bol}/L_{\odot}$        &        & erg\,s$^{-1}$\,cm$^{-2}$ & FWHM & 10$^{35}$ erg\,s$^{-1}$ &  $10^{-3} L_{\rm Bol}$   & $L_{\rm HeII~4686}$ &  $L_{\rm HeII~4686}$  & $L_{\rm HeII~4686}$ & $L_{\rm HeII~4686}$  & $L_{\rm HeII~4686}$ &  $L_{\rm HeII~4686}$ &   $L_{\rm HeII~4686}$ &
 $L_{\rm HeII~4686}$  & $L_{\rm HeII~4686}$ \\
\hline
{\bf WR61}   & WN2--5w  & CS     & \phantom{0}6.4$\pm$0.5  & 1.9$\pm$0.2 & 5.12$_{-0.11}^{+0.10}$ & a  & --11.13$_{-0.04}^{+0.05}$ & 1600       & \phantom{0}3.2$_{-0.8}^{+0.9}$ & 0.65$_{-0.09}^{+0.10}$   & 0.53$_{-0.03}^{+0.03}$ & 0.25$_{-0.01}^{+0.01}$ & 0.32$_{-0.01}^{+0.01}$     & 0.28$_{-0.00}^{+0.01}$ & 0.12$_{-0.00}^{+0.01}$ & 0.07$_{-0.00}^{+0.01}$ & 0.02$_{-0.00}^{+0.01}$ & 0.13$_{-0.01}^{+0.01}$ & $\cdots$ \\ [2pt]
{\bf WR67} & WN6--8      & CS     & \phantom{0}3.2$\pm$0.2 & 3.9$\pm$0.4          & 5.36$_{-0.17}^{+0.17}$ & a   & --11.23$_{-0.04}^{+0.05}$ & 1300     & \phantom{0}6.1$_{-2.3}^{+3.5}$ & 0.69$_{-0.12}^{+0.15}$    & 0.40$_{-0.05}^{+0.05}$ & 0.30$_{-0.02}^{+0.02}$ & 0.46$_{-0.02}^{+0.03}$     & 0.48  & 0.11$_{-0.00}^{+0.01}$  & 0.06$_{-0.01}^{+0.00}$ & 0.04 & 0.14$_{-0.02}^{+0.02}$ & $\cdots$ \\ [2pt]
{\bf WR71} & WN6--8      & CS    & \phantom{0}4.1$\pm$0.3  & 1.1$_{-0.1}^{+0.2}$ & 5.29$_{-0.07}^{+0.08}$ & a   & --10.23$_{-0.05}^{+0.04}$  & 1400     & \phantom{0}4.6$_{-0.8}^{+0.9}$ & 0.61$_{-0.05}^{+0.06}$    & 0.62$_{-0.02}^{+0.03}$ & 0.32$_{-0.01}^{+0.01}$ & 0.55$_{-0.01}^{+0.01}$     & 0.52  & 0.12$_{-0.00}^{+0.01}$ & 0.04 & 0.04  & 0.13$_{-0.01}^{+0.00}$ & $\cdots$ \\ [2pt]
{\bf WR74}  & WN6--8     & CS     & \phantom{0}4.0$_{-0.3}^{+0.2}$   & 5.6$_{-0.5}^{+0.6}$        & 5.41$_{-0.24}^{+0.23}$  & a   & --12.40$_{-0.05}^{+0.04}$ & 1100     & \phantom{0}4.2$_{-2.0}^{+3.7}$ & 0.43$_{-0.09}^{+0.10}$ & $\cdots$ & 0.16$_{-0.02}^{+0.01}$ & 0.29$_{-0.03}^{+0.02}$  & 0.74$_{-0.01}^{+0.01}$  & 0.16$_{-0.02}^{+0.01}$ & 0.06$_{-0.01}^{+0.00}$ & 0.16$_{-0.02}^{+0.03}$ & 0.15$_{-0.03}^{+0.04}$ & 0.20$_{-0.04}^{+0.07}$ \\ [2pt]
WR75  & WN3--7s  &  CS     & \phantom{0}3.5$_{-0.2}^{+0.3}$   & 3.5$_{-0.4}^{+0.3}$        & 5.61$_{-0.15}^{+0.15}$ & a    & --10.45$_{-0.05}^{+0.04}$ & 2800     &                26.6$_{-9.2}^{+13.2}$  & 1.71$_{-0.27}^{+0.32}$  & 0.72$_{-0.07}^{+0.08}$ & \multicolumn{2}{c}{--- 0.77$_{-0.04}^{+0.03}$ --- } & 0.39 & 0.12$_{-0.01}^{+0.01}$ & 0.03$_{-0.00}^{+0.01}$ & 0.08$_{-0.00}^{+0.01}$  & 0.11$_{-0.01}^{+0.02}$ &$\cdots$ \\  [2pt]
{\bf WR78} & WN6--8 & VU     & \phantom{0}1.6$\pm$0.1    & 1.8$\pm$0.2       & 6.03$_{-0.08}^{+0.08}$  & a    & --9.37$_{-0.04}^{+0.04}$   & \phantom{0}900    &  10.0$_{-2.1}^{+2.5}$ & 0.24$_{-0.02}^{+0.03}$    & 0.05$_{-0.00}^{+0.01}$ & 0.10 & 0.47$_{-0.02}^{+0.01}$         & 0.61$_{-0.01}^{+0.00}$ & 0.11$_{-0.01}^{+0.00}$ & \#0.03 & 0.12$_{-0.00}^{+0.01}$ & 0.37$_{-0.02}^{+0.03}$ & 0.09$_{-0.01}^{+0.01}$  \\ [2pt]
{\bf WR82}   & WN6--8    & CS    & \phantom{0}3.8$\pm$0.3  & 3.8$_{-0.4}^{+0.3}$         & 5.26$_{-0.17}^{+0.16}$ & a    & --11.58$_{-0.05}^{+0.04}$  & 1000     & \phantom{0}3.1$_{-1.1}^{+1.7}$ & 0.44$_{-0.08}^{+0.10}$ & 0.31$_{-0.04}^{+0.03}$ & 0.23$_{-0.01}^{+0.02}$ & 0.73$_{-0.04}^{+0.05}$        & 0.94$_{-0.00}^{+0.01}$    & 0.11$_{-0.00}^{+0.01}$  & 0.01 & 0.17$_{-0.02}^{+0.02}$ & 0.40$_{-0.05}^{+0.07}$ & $\cdots$ \\ [2pt]
{\bf WR83}   & WN2--5w & CS     & \phantom{0}4.6$_{-0.3}^{+0.4}$  & 4.0$\pm$0.4  &  $\cdots$ & b &  --11.46$_{-0.05}^{+0.04}$ & 1400 & \phantom{0}7.6$_{-2.9}^{+4.4}$ & $\cdots$ & 0.75$_{-0.08}^{+0.09}$ & 0.37$_{-0.02}^{+0.03}$       & 0.50$_{-0.03}^{+0.03}$    & 0.47$_{-0.01}^{+0.00}$  & 0.10$_{-0.01}^{+0.00}$  & 0.04$_{-0.01}^{+0.00}$ & 0.01 & 0.19$_{-0.03}^{+0.03}$ & $\cdots$ \\  [2pt]
{\bf WR84}   & WN6--8   & CS      & \phantom{0}2.7$\pm$0.1 & 5.4$_{-0.5}^{+0.6}$         & 5.18$_{-0.22}^{+0.22}$  & a   & --11.94$_{-0.05}^{+0.04}$  & 1100     & \phantom{0}4.5$_{-2.1}^{+3.8}$ & 0.77$_{-0.11}^{+0.16}$ & 0.31$_{-0.04}^{+0.06}$ & 0.27$_{-0.03}^{+0.02}$ & 0.65$_{-0.05}^{+0.05}$       & 0.74$_{-0.00}^{+0.01}$     & 0.14$_{-0.01}^{+0.01}$ & 0.05$_{-0.01}^{+0.01}$ & 0.09$_{-0.01}^{+0.02}$ & 0.12$_{-0.02}^{+0.03}$  & $\cdots$ \\ [2pt]
{\bf WR85}  & WN6--8      &  CS    & \phantom{0}2.2$_{-0.0}^{+0.1}$  & 3.5$_{-0.4}^{+0.3}$ & 5.48$_{-0.14}^{+0.15}$ & a& --10.75$_{-0.04}^{+0.04}$ & 1100       & \phantom{0}4.9$_{-1.7}^{+2.4}$ & 0.42$_{-0.05}^{+0.07}$   & 0.45$_{-0.04}^{+0.05}$ & 0.34$_{-0.02}^{+0.02}$ & 0.62$_{-0.03}^{+0.03}$      & 0.67$_{-0.00}^{+0.01}$     & 0.11$_{-0.00}^{+0.01}$ & 0.03 & 0.04$_{-0.01}^{+0.00}$ & 0.32$_{-0.04}^{+0.04}$   & $\cdots$ \\ [2pt]
{\bf WR87} & WN5--7h    & CS     & \phantom{0}3.0$_{-0.2}^{+0.1}$ & 6.4$_{-0.7}^{+0.6}$        & 6.11$_{-0.26}^{+0.26}$   & a & --12.32$_{-0.05}^{+0.04}$ & 1100     & \phantom{0}6.5$_{-3.4}^{+6.8}$ & 0.13$_{-0.03}^{+0.03}$ & 0.57$_{-0.10}^{+0.12}$ & 0.14$_{-0.01}^{+0.01}$  & 0.70$_{-0.06}^{+0.07}$ &   1.04$_{-0.01}^{+0.01}$     & 0.06$_{-0.01}^{+0.01}$ & 0.06$_{-0.01}^{+0.01}$ & 0.06$_{-0.01}^{+0.02}$ & 0.77$_{-0.17}^{+0.22}$   & $\cdots$ \\ [2pt]
{\bf WR89} & WN5--7h    & CS     & \phantom{0}3.2$\pm$0.2   & 5.9$\pm$0.6       & 6.30$_{-0.24}^{+0.26}$    & a  & --11.95$_{-0.04}^{+0.04}$  & \phantom{0}700    &                   10.9$_{-5.5}^{+10.4}$ & 0.14$_{-0.03}^{+0.03}$  &1.07$_{-0.17}^{+0.21}$ & 0.16$_{-0.02}^{+0.01}$  & 0.95$_{-0.08}^{+0.09}$  & 1.01$_{-0.01}^{+0.01}$      & 0.03$_{-0.00}^{+0.01}$ & 0.01$_{-0.00}^{+0.01}$ & 0.07$_{-0.01}^{+0.01}$ & 0.73$_{-0.15}^{+0.19}$    & $\cdots$ \\ [2pt]
WR91  & WN3--7s    & CS    & \phantom{0}3.6$_{-0.2}^{+0.3}$   & 8.0$_{-0.8}^{+0.7}$      & 5.61$_{-0.32}^{+0.33}$     & a  & --12.68$_{-0.05}^{+0.04}$ & 2000    &            25.4$_{-15.3}^{+36.9}$ & 1.61$_{-0.40}^{+0.54}$ & $\cdots$ & $\cdots$ & $\cdots$ & 0.60$_{-0.01}^{+0.00}$    & 0.12$_{-0.02}^{+0.02}$ & 0.03$_{-0.01}^{+0.01}$ & 0.11$_{-0.02}^{+0.03}$ & 0.14$_{-0.04}^{+0.05}$     & $\cdots$ \\ [2pt]
WR94  & WN2--5w & CS    & \phantom{0}1.1$\pm$0.0    & 6.1$\pm$0.6 &5.65$_{-0.24}^{+0.25}$   & a & --11.21$_{-0.04}^{+0.04}$   & 1400             & \phantom{0}8.1$_{-4.1}^{+7.9}$ & 0.47$_{-0.07}^{+0.08}$  & 1.04$_{-0.16}^{+0.19}$ & 0.41$_{-0.04}^{+0.04}$ & 0.54$_{-0.04}^{+0.05}$   & 0.38$_{-0.01}^{+0.00}$ &$\cdots$&$\cdots$&$\cdots$&$\cdots$&$\cdots$\\ [2pt]
{\bf WR97}  & WN3--7s+O& CS  & \phantom{0}2.3$_{-0.0}^{+0.1}$     & 3.7$\pm$0.4     &  $\cdots$ & b & --11.53$_{-0.04}^{+0.05}$ & 2000 & \phantom{0}1.3$_{-0.5}^{+0.7}$ & $\cdots$ & 0.54$_{-0.06}^{+0.06}$ & 0.12$_{-0.01}^{+0.01}$ & 0.07  & 0.42    & 0.09$_{-0.01}^{+0.01}$ & 0.05$_{-0.00}^{+0.01}$ & 0.00 & 0.26$_{-0.04}^{+0.04}$ & $\cdots$ \\ [2pt]
WR100  &  WN3--7s  & CS         & \phantom{0}2.5$_{-0.1}^{+0.2}$ & 5.6$_{-0.5}^{+0.6}$         & 5.38$_{-0.23}^{+0.23}$ & a & --11.60$_{-0.04}^{+0.05}$ & 1900               &   11.0$_{-5.3}^{+9.8}$ & 1.19$_{-0.24}^{+0.31}$   & 0.37$_{-0.06}^{+0.07}$ & 0.26$_{-0.02}^{+0.02}$  & 0.59$_{-0.05}^{+0.05}$ & 0.55$_{-0.00}^{+0.01}$ &$\cdots$&$\cdots$&$\cdots$&$\cdots$&$\cdots$\\ [2pt]
{\bf WR105} & WN9--11     & II96        & \phantom{0}3.5$_{-0.4}^{+0.6}$  & 7.7$_{-0.8}^{+0.7}$       & 6.28$_{-0.33}^{+0.33}$  & f & --12.72$_{-0.05}^{+0.04}$ & \phantom{0}500&        15.3$_{-9.1}^{+21.0}$ & 0.21$_{-0.06}^{+0.10}$ & $\cdots$ & 0.00 & 0.72$_{-0.08}^{0.08}$ & 1.42$_{-0.02}^{+0.02}$ & 0.02: & 0.00 & 0.82$_{-0.15}^{+0.19}$ & 1.6$_{-0.4}^{+0.5}$ & $\cdots$ \\ [2pt]
{\bf WR107} & WN6--8     & II96        & \phantom{0}1.8$\pm$0.3  & 6.3$\pm$0.6 & 4.85$_{-0.28}^{+0.30}$ & a  &--12.54$_{-0.05}^{+0.04}$ & 1000  &     \phantom{0}1.1$_{-0.6}^{+1.0}$ & 0.38$_{-0.10}^{+0.17}$ & $\cdots$ & 0.22$_{-0.02}^{+0.02}$ & 0.87$_{-0.07}^{+0.07}$ & 1.36$_{-0.01}^{+0.01}$ & 0.17$_{-0.01}^{+0.02}$ & 0.05$_{-0.01}^{+0.01}$ & 0.49$_{-0.07}^{+0.07}$ & 0.17$_{-0.03}^{+0.04}$ & $\cdots$ \\ [2pt]
{\bf WR108} & WN9--11    & II91        & \phantom{0}3.1$_{-0.1}^{+0.2}$  & 3.7$_{-0.4}^{+0.3}$         & 5.42$_{-0.16}^{+0.15}$ & f  & --11.63$_{-0.05}^{+0.04}$ & \phantom{0}400 &\phantom{0}1.7$_{-0.6}^{+0.9}$ &0.17$_{-0.02}^{+0.03}$ &$\cdots$ &0.00 & 1.20$_{-0.06}^{+0.07}$ & 1.93$_{-0.01}^{+0.01}$ & 0.02 &0.02 & 0.25$_{-0.02}^{+0.03}$ & 1.13$_{-0.14}^{+0.17}$ & 0.01 \\   [2pt]
{\bf WR110} & WN3--7s   & AR       & \phantom{0}1.8$\pm$0.1   & 3.8$\pm$0.4 & 5.61$_{-0.15}^{+0.16}$ & a & --9.93$_{-0.05}^{+0.04}$    & 3600    &  29.8$_{-10.7}^{+15.8}$ & 1.88$_{-0.23}^{+0.27}$ & 0.57$_{-0.05}^{+0.07}$ & \multicolumn{2}{c}{--- 0.47$_{-0.03}^{+0.02}$ ---} & 0.20$_{-0.00}^{+0.01}$ & 0.11$_{-0.01}^{+0.01}$ & 0.03$_{-0.00}^{+0.01}$ & 0.07$_{-0.00}^{+0.01}$ & 0.13$_{-0.01}^{+0.02}$ & 0.17$_{-0.03}^{+0.02}$ \\ [2pt] 
{\bf WR116} & WN6--8     & II96    & \phantom{0}2.9$\pm$0.2    & 6.6$_{-0.7}^{+0.6}$         & 5.56$_{-0.27}^{+0.27}$ & a & --12.49$_{-0.05}^{+0.04}$ & 1100            & \phantom{0}5.2$_{-2.8}^{+5.7}$  & 0.37$_{-0.09}^{+0.11}$ & $\cdots$ & 0.11$_{-0.01}^{+0.01}$ & 1.25$_{-0.12}^{+0.12}$  & 1.58$_{-0.02}^{+0.02}$ & 0.07$_{-0.00}^{+0.01}$ & 0.01 & 0.69$_{-0.11}^{+0.14}$ & 1.04$_{-0.24}^{+0.30}$ & $\cdots$ \\  [2pt]
WR120 & WN6--8     & WI94      & \phantom{0}2.7$_{-0.4}^{+0.5}$ & 4.7$\pm$0.5           & 5.39$_{-0.22}^{+0.23}$ & a  & --11.81$_{-0.05}^{+0.04}$ & 1000           & \phantom{0}2.7$_{-1.2}^{+1.9}$ & 0.28$_{-0.07}^{+0.10}$ & $\cdots$ & $\cdots$ & $\cdots$ & 0.89$_{-0.01}^{+0.01}$ & 0.24$_{-0.02}^{+0.02}$ & 0.08$_{-0.01}^{+0.01}$ & 0.41$_{-0.05}^{+0.06}$ & $\cdots$ & $\cdots$ \\ [2pt]
{\bf WR123}  & WN6--8    &  AR+II91      & \phantom{0}6.1$_{-0.5}^{+0.6}$ & 2.5$_{-0.2}^{+0.3}$  & 5.33$_{-0.12}^{+0.13}$ & a & --11.41$_{-0.05}^{+0.04}$   & \phantom{0}800          & \phantom{0}3.2$_{-0.9}^{+1.2}$ & 0.38$_{-0.06}^{+0.08}$ & $\cdots$ & 0.12$_{-0.01}^{+0.00}$ & 1.15$_{-0.05}^{+0.05}$ & 1.54$_{-0.01}^{+0.00}$ & 0.17$_{-0.01}^{+0.01}$ & 0.02 & 0.68$_{-0.05}^{+0.06}$ & 0.15$_{-0.01}^{+0.02}$ & \ddag0.08$_{-0.01}^{+0.01}$ \\   [2pt]
WR124 & WN6--8     & II91    & \phantom{0}5.4$_{-0.3}^{+0.5}$  & 3.8$\pm$0.4  & 5.62$_{-0.16}^{+0.17}$ & a & --11.58$_{-0.10}^{+0.08}$ & \phantom{0}800          & \phantom{0}7.0$_{-2.9}^{+4.1}$ & 0.43$_{-0.07}^{+0.10}$ & $\cdots$ & $\cdots$ & $\cdots$ & 1.31$_{-0.01}^{+0.01}$ & 0.08$_{-0.01}^{+0.00}$ & 0.02 & 0.92$_{-0.09}^{+0.11}$ & 1.29$_{-0.19}^{+0.23}$ & $\cdots$ \\ [2pt]
WR128 & WN2--5w  & II91     & \phantom{0}3.3$_{-0.2}^{+0.3}$ & 1.4$\pm$0.1 & 5.34$_{-0.08}^{+0.08}$ & a   & --10.55$_{-0.09}^{+0.08}$ & 1900         & \phantom{0}1.7$_{-0.4}^{+0.5}$ & 0.20$_{-0.02}^{+0.02}$ & $\cdots$ & $\cdots$ & $\cdots$ & 0.26 & 0.11$_{-0.01}^{+0.00}$ & 0.03 & 0.00 & 0.23$_{-0.01}^{+0.02}$ & 0.13$_{-0.01}^{+0.01}$ \\ [2pt]
{\bf WR131} & WN5--7h   & II96+WI94   & \phantom{0}8.2$_{-0.7}^{+1.0}$ & 4.3$\pm$0.4           & 6.17$_{-0.19}^{+0.20}$  & a & --11.97$_{-0.05}^{+0.04}$ & \phantom{0}800            &   11.0$_{-4.5}^{+7.1}$  & 0.19$_{-0.04}^{+0.05}$ & $\cdots$ & 0.10$_{-0.00}^{+0.01}$ & 0.29$_{-0.02}^{+0.02}$ & 0.35 & 0.06$_{-0.00}^{+0.01}$ & 0.03 & 0.05$_{-0.01}^{+0.00}$ & $\cdots$ &$\cdots$\\ [2pt]
WR133 & WN2--5w+O & II91 & \phantom{0}1.8$_{-0.1}^{+0.0}$ & 1.2$\pm$0.1   & $\cdots$ & b & --9.66$_{-0.09}^{+0.08}$ & 1700              & \phantom{0}3.0$_{-0.7}^{+0.8}$ & $\cdots$&$\cdots$ & $\cdots$ & $\cdots$ & 0.14 &0.10$_{-0.00}^{+0.01}$ & 0.12 & 0.00 & 0.20$_{-0.01}^{+0.01}$ & 0.24$_{-0.02}^{+0.01}$ \\ [2pt]
{\bf WR134} & WN3--7s  & II91     & \phantom{0}1.9$_{-0.1}^{+0.0}$  & 1.9$\pm$0.2 & 5.67$_{-0.08}^{+0.09}$ & a   & --9.14$_{-0.09}^{+0.08}$ & 2600            &     25.9$_{-7.2}^{+8.1}$ & 1.43$_{-0.12}^{+0.14}$   & 0.55$_{-0.03}^{+0.03}$ & \multicolumn{2}{c}{ --- 0.43$_{-0.01}^{+0.02}$ ---}  & 0.16 & 0.13$_{-0.00}^{+0.01}$  & 0.06$_{-0.01}^{+0.00}$ & 0.04 & 0.16$_{-0.02}^{+0.01}$ & 0.20$_{-0.01}^{+0.02}$ \\ [2pt] 
{\bf WR136}& WN3--7s & II91     & \phantom{0}1.7$_{-0.1}^{+0.0}$   & 1.7$\pm$0.2         & 5.68$_{-0.08}^{+0.07}$  & a   & --8.92$_{-0.10}^{+0.07}$ & 1900            &    27.2$_{-7.2}^{+7.9}$ & 1.49$_{-0.11}^{+0.12}$   & 0.33$_{-0.02}^{+0.02}$  &\multicolumn{2}{c}{--- 0.56$_{-0.01}^{+0.02}$  ---}   & 0.34 & 0.15$_{-0.01}^{+0.00}$  & 0.05$_{-0.01}^{+0.00}$ & 0.06 & 0.24$_{-0.02}^{+0.01}$ & 0.18$_{-0.02}^{+0.01}$ \\ [2pt]
{\bf WR138} & WN2--5w+O & II91 & \phantom{0}2.2$\pm$0.1 & 1.9$\pm$0.2   & $\cdots$ & b & --9.84$_{-0.09}^{+0.08}$  & 1400   & \phantom{0}7.2$_{-2.0}^{+2.3}$ & $\cdots$ & $\cdots$ & 0.19$_{-0.00}^{+0.01}$ & 0.21$_{-0.00}^{+0.01}$ & 0.28$_{-0.00}^{+0.01}$ & 0.15$_{-0.00}^{+0.01}$ & 0.07$_{-0.00}^{+0.01}$ & 0.05 & 0.29$_{-0.02}^{+0.02}$ & 0.28$_{-0.03}^{+0.02}$ \\  [2pt]
{\bf WR138-1} & WN9--11 & II13 & \phantom{0}7.0$_{-0.6}^{+0.8}$ & 7.0$_{-0.7}^{+0.7}$   & $\cdots$ & b & --13.68$_{-0.10}^{+0.08}$  & \phantom{0}500   & \phantom{0}3.3$_{-1.9}^{+4.0}$ & $\cdots$ & $\cdots$ & 0.00 & 1.25$_{-0.12}^{+0.14}$ & 1.17$_{-0.01}^{+0.02}$ & 0.07$_{-0.01}^{+0.01}$ & 0.00 & 1.04$_{-0.19}^{+0.22}$ & 1.59$_{-0.28}^{+0.40}$ & 0.01$_{-0.00}^{+0.01}$ \\  [2pt]
WR139 & WN2--5w+O & II91 & \phantom{0}1.4$_{-0.1}^{+0.0}$ & 2.5$_{-0.2}^{+0.3}$ & $\cdots$ & b & --9.93$_{-0.10}^{+0.08}$ & 1600        & \phantom{0}4.5$_{-1.4}^{+1.8}$ & $\cdots$ & $\cdots$ & $\cdots$ & $\cdots$ & 0.25& 0.14$_{-0.01}^{+0.01}$ & 0.14$_{-0.01}^{+0.01}$ & 0.05 & 0.21$_{-0.02}^{+0.02}$ & 0.28$_{-0.03}^{+0.04}$ \\ [2pt]
WR141 & WN2--5w+O & II91 & \phantom{0}1.9$_{-0.1}^{+0.0}$ & 3.7$_{-0.4}^{+0.3}$  & $\cdots$ & b &--10.27$_{-0.09}^{+0.08}$ & 1700    &  14.0$_{-5.5}^{+7.6}$ & $\cdots$& $\cdots$ & $\cdots$ & $\cdots$ & 0.25& 0.12$_{-0.00}^{+0.01}$ & 0.08$_{-0.01}^{+0.00}$ & 0.04$_{-0.00}^{+0.01}$ & 0.13$_{-0.02}^{+0.01}$ & 0.13$_{-0.02}^{+0.03}$ \\ [2pt]
{\bf WR147} & WN6--8+B & II96   & \phantom{0}1.7$_{-0.1}^{+0.1}$ & 10.7$\pm$1.1  &    6.58$_{-0.44}^{+0.43}$     & a    & --13.43$_{-0.05}^{+0.04}$ & \phantom{0}700  &   20.1$_{-14.4}^{+47.0}$ & 0.14$_{-0.05}^{+0.07}$  & $\cdots$ & 0.06$_{-0.01}^{+0.01}$ & 0.54$_{-0.08}^{+0.09}$ & 0.86$_{-0.02}^{+0.02}$ & 0.11$_{-0.02}^{+0.03}$ & 0.02$_{-0.01}^{+0.00}$ & 0.81$_{-0.21}^{+0.27}$ & 0.60$_{-0.20}^{+0.32}$ & $\cdots$ \\  [2pt]
{\bf WR148} & WN6--8    & II91     & \phantom{0}8.1$\pm$0.7 & 3.0$\pm$0.3 &6.25$_{-0.14}^{+0.15}$ & a      & --11.27$_{-0.10}^{+0.08}$ & \phantom{0}600  & 12.6$_{-4.5}^{+5.8}$ & 0.18$_{-0.03}^{+0.04}$  & $\cdots$ & 0.09$_{-0.01}^{+0.00}$ & 0.38$_{-0.02}^{+0.01}$  & 0.51 & 0.07$_{-0.00}^{+0.01}$ & 0.03 & 0.16$_{-0.01}^{+0.01}$ & 0.38$_{-0.04}^{+0.07}$ & $\cdots$ \\   [2pt]
\hline
\end{tabular}
\end{footnotesize}
\end{center}
\end{table}
\end{landscape}

\addtocounter{table}{-1}

\begin{landscape}
\begin{table}
\caption{continued}
\begin{center}
\begin{footnotesize}
 \begin{tabular}{l@{\hspace{1mm}}l@{\hspace{1.5mm}}l@{\hspace{1.5mm}}l@{\hspace{1.5mm}}l@{\hspace{1.5mm}}l@{\hspace{1mm}}c@{\hspace{1mm}}r@{\hspace{1mm}}c@{\hspace{0mm}}c@{\hspace{1mm}}c
 @{\hspace{1.5mm}}l@{\hspace{1mm}}l@{\hspace{1mm}}l@{\hspace{1mm}}l@{\hspace{1mm}}l@{\hspace{1mm}}l@{\hspace{1mm}}l@{\hspace{1mm}}l@{\hspace{1mm}}l}   %
\hline
Star & Category & Data & $d$ & $A_{\rm V}$  & $\log$  & Ref & $\log F_{\rm HeII~4686}$ & HeII 4686 & \multicolumn{2}{c}{$L_{\rm HeII~4686}$} 
& $\underline{L_{\rm NIV~3478,85}}$ 
& $\underline{L_{\rm NIV~4058}}$ 
& $\underline{L_{4100}}$ 
& $\underline{L_{4630}}$ 
&$\underline{L_{\rm HeII~5412}}$ 
& $\underline{L_{\rm CIV~5801,12}}$ 
&$\underline{L_{\rm HeI~5876}}$ 
& $\underline{L_{{\rm H}\alpha}}$ 
& $\underline{L_{\rm NIV~7103,29}}$ \\  
        &    &    ID          &   kpc     & mag              & $L_{\rm Bol}/L_{\odot}$        &        & erg\,s$^{-1}$\,cm$^{-2}$ & FWHM  & 10$^{35}$ erg\,s$^{-1}$ &  $10^{-3} L_{\rm Bol}$   & $L_{\rm HeII~4686}$ &  $L_{\rm HeII~4686}$  & $L_{\rm HeII~4686}$ & $L_{\rm HeII~4686}$  & $L_{\rm HeII~4686}$ &  $L_{\rm HeII~4686}$ &   $L_{\rm HeII~4686}$ &
 $L_{\rm HeII~4686}$  & $L_{\rm HeII~4686}$ \\
\hline
WR149 & WN2--5w  & II13    & \phantom{0}5.3$_{-0.3}^{+0.4}$ & 5.3$_{-0.5}^{+0.6}$        & 5.48$_{-0.22}^{+0.22}$ & a     & --12.20$_{-0.09}^{+0.08}$ & 1300              & \phantom{0}8.5$_{-4.2}^{+7.2}$ & 0.73$_{-0.15}^{+0.18}$  & $\cdots$ & 0.20$_{-0.02}^{+0.01}$ & 0.23$_{-0.02}^{+0.02}$  & 0.21$_{-0.00}^{+0.00}$ & 0.13$_{-0.02}^{+0.01}$ & 0.09$_{-0.01}^{+0.01}$ & 0.04$_{-0.00}^{+0.01}$ & 0.13$_{-0.03}^{+0.02}$ & 0.16$_{-0.04}^{+0.04}$ \\ [2pt]
WR152 & WN2--5w  & II91     &\phantom{0}3.5$_{-0.2}^{+0.3}$  & 1.9$\pm$0.2 & 5.46$_{-0.09}^{+0.09}$ & a & --10.96$_{-0.10}^{+0.08}$ & 2000  & \phantom{0}1.4$_{-0.4}^{+0.5}$ & 0.13$_{-0.01}^{+0.02}$    & $\cdots$& 0.00 & 0.04  &  0.34 & 0.09 & 0.01 & 0.00 & 0.18$_{-0.01}^{+0.01}$ & 0.04$_{-0.01}^{+0.00}$ \\   [2pt]
WR155 & WN6--8+O & II91   &\phantom{0}2.7$\pm$0.1     & 2.2$\pm$0.2  & $\cdots$ & b &--10.42$_{-0.10}^{+0.08}$ & 1200 & \phantom{0}4.1$_{-1.2}^{+1.4}$ &$\cdots$& $\cdots$ & $\cdots$ & $\cdots$ & 0.44$_{-0.01}^{+0.00}$ & 0.09$_{-0.01}^{+0.00}$ & 0.05 & 0.06$_{-0.00}^{+0.01}$ & 0.30$_{-0.02}^{+0.03}$ & 0.13$_{-0.01}^{+0.01}$ \\  [2pt]
WR156 & WN6--8     & II91    &\phantom{0}3.3$_{-0.1}^{+0.2}$     & 4.6$_{-0.5}^{+0.4}$   & 5.81$_{-0.18}^{+0.19}$ & a        & --11.61$_{-0.10}^{+0.07}$ & \phantom{0}500 &\phantom{0}5.5$_{-2.5}^{+3.9}$ & 0.22$_{-0.03}^{+0.04}$ &$\cdots$ & $\cdots$ & $\cdots$ & 2.02$_{-0.02}^{+0.02}$  & 0.05$_{-0.00}^{+0.01}$ & 0.03$_{-0.00}^{+0.01}$ & 0.45$_{-0.05}^{+0.06}$ & 1.27$_{-0.21}^{+0.25}$ & $\cdots$ \\ [2pt]
WR157& WN2--5w  & II91    &\phantom{0}2.8$_{-0.1}^{+0.2}$     & 2.8$\pm$0.3  & $\cdots$ & b &--10.53$_{-0.10}^{+0.08}$ & 1300           & \phantom{0}6.6$_{-2.3}^{+2.8}$ & $\cdots$ & $\cdots$ & $\cdots$ & $\cdots$ & 0.22 & 0.16$_{-0.01}^{+0.00}$ & $\cdots$&$\cdots$&$\cdots$&$\cdots$\\ [2pt]
{\bf BAT99-1} & WN3--7s & NE & 49.6$\pm$1.0                 & 0.5$_{-0.0}^{+0.1}$ & 5.29$_{-0.02}^{+0.03}$ & g  & --12.19$_{-0.04}^{+0.04}$ & 2000         & \phantom{0}3.4$_{-0.4}^{+0.4}$ & 0.45$_{-0.01}^{+0.02}$ & \ddag0.13   & 0.00 & 0.05 & 0.18 & 0.14 & 0.01 & 0.00 & 0.19 & 0.00 \\ [2pt]
{\bf BAT99-3} & WN3--7s &  NE  & 49.6$\pm$1.0                  & 0.5$_{-0.1}^{+0.0}$ & 5.50$_{-0.02}^{+0.03}$ & g & --11.77$_{-0.05}^{+0.04}$ & 2000          & \phantom{0}8.2$_{-0.9}^{+0.9}$ & 0.67$_{-0.02}^{+0.02}$ & \ddag0.66$_{-0.02}^{+0.01}$  & 0.03 & 0.08 & 0.17 & 0.14 & 0.05 & 0.02 & 0.16$_{-0.01}^{+0.00}$ & 0.08$_{-0.01}^{+0.00}$ \\ [2pt]
{\bf BAT99-5} & WN2--5w & NE & 49.6$\pm$1.0               & 1.0$\pm$0.1 & 5.44$_{-0.04}^{+0.05}$ & g & --12.72$_{-0.05}^{+0.04}$ & 2400     & \phantom{0}1.7$_{-0.2}^{+0.3}$ & 0.16$_{-0.01}^{+0.01}$ & $\cdots$          & 0.00 & 0.04$_{-0.00}^{+0.01}$ & 0.04 & 0.15$_{-0.01}^{+0.00}$ & 0.00 & 0.00 & 0.17$_{-0.00}^{+0.01}$ & 0.00 \\ [2pt]
{\bf BAT99-7} & WN3--7s  & NE  & 49.6$\pm$1.0                & 0.3$\pm$0.0 & 5.83$_{-0.02}^{+0.02}$ & g & --11.11$_{-0.05}^{+0.04}$ & 3800    &     31.7$_{-3.4}^{+3.4}$ & 1.21$_{-0.03}^{+0.03}$ & \ddag0.30$_{-0.01}^{+0.00}$   & 0.00 & 0.05$_{-0.01}^{+0.00}$ & 0.09 & 0.11 & 0.03 & 0.00 & 0.15$_{-0.01}^{+0.00}$ & 0.02 \\ [2pt]
{\bf BAT99-13} & WN9--11 & AR+SC & 49.6$\pm$1.0   & 0.8$_{-0.1}^{+0.0}$ & 5.55$_{-0.03}^{+0.04}$ & g & --12.98$_{-0.05}^{+0.04}$ & \phantom{0}200   &   0.7$_{-0.1}^{+0.1}$ & 0.05$_{-0.00}^{+0.00}$ & $\cdots$ & 0.00 & 3.7$_{-0.1}^{+0.0}$ & 2.4 & 0.01 & 0.00 & 3.3$_{-0.0}^{+0.0}$ & 11.1$_{-0.3}^{+0.3}$  & $\cdots$ \\ [2pt]
{\bf BAT99-15} & WN3--7s & NE  & 49.6$\pm$1.0              & 0.3$\pm$0.0 & 5.56$_{-0.02}^{+0.02}$ & g & --11.53$_{-0.04}^{+0.04}$ & 2200          &    12.2$_{-1.3}^{+1.3}$ & 0.87$_{-0.02}^{+0.02}$ & \ddag0.55$_{-0.01}^{+0.01}$ & 0.04 & 0.12 & 0.16 & 0.13 & 0.07 & 0.03 & 0.16$_{-0.01}^{+0.00}$ & 0.07 \\ [2pt]
{\bf BAT99-16} & WN6--8    & AR+SC & 49.6$\pm$1.0   & 0.3$_{-0.0}^{+0.1}$ & 5.79$_{-0.02}^{+0.02}$ & g & --11.61$_{-0.04}^{+0.05}$ & \phantom{0}700   &   10.6$_{-1.1}^{+1.2}$ & 0.44$_{-0.01}^{+0.02}$ & \ddag0.16$_{-0.00}^{+0.01}$ & 0.23 & 0.50$_{-0.00}^{+0.01}$ & 0.54 & 0.10 & 0.04 & 0.19$_{-0.01}^{+0.00}$ & 0.51$_{-0.00}^{+0.01}$  & $\cdots$ \\ [2pt]
{\bf BAT99-17} & WN2--5w+B  & NE  & 49.6$\pm$1.0         & 0.4$_{-0.1}^{+0.0}$ & 5.63$_{-0.02}^{+0.03}$ & h & --11.78$_{-0.04}^{+0.04}$ & 1800         & \phantom{0}7.4$_{-0.8}^{+0.8}$  & 0.45$_{-0.01}^{+0.01}$ & \ddag0.75$_{-0.01}^{+0.02}$  & 0.08 & 0.10 & 0.18 & 0.13 & 0.07 & 0.01 & 0.15$_{-0.00}^{+0.01}$ & 0.10 \\ [2pt]
{\bf BAT99-18} & WN2--5w & NE & 49.6$\pm$1.0               & 0.4$_{-0.1}^{+0.0}$ & 5.62$_{-0.02}^{+0.03}$ & g & --12.00$_{-0.05}^{+0.04}$ & 1800          & \phantom{0}4.5$_{-0.5}^{+0.5}$ & 0.28$_{-0.01}^{+0.00}$ & \ddag0.80$_{-0.02}^{+0.01}$   & 0.09 & 0.11 & 0.18 & 0.10$_{-0.00}^{+0.01}$ & 0.03 & 0.00 & 0.20 & 0.07 \\ [2pt]
{\bf BAT99-22} & WN9--11   &  AR+SC & 49.6$\pm$1.0   & 0.5$_{-0.1}^{+0.0}$ & 5.74$_{-0.02}^{+0.03}$ & g & --12.17$_{-0.05}^{+0.04}$ & \phantom{0}300            & \phantom{0}3.4$_{-0.4}^{+0.4}$ & 0.16$_{-0.01}^{+0.01}$ & \ddag0.00   & 0.00 &  1.18$_{-0.01}^{+0.01}$ & 1.52 & 0.00 & 0.03 & 0.77$_{-0.01}^{+0.01}$ & 3.27$_{-0.06}^{+0.06}$ & $\cdots$ \\ [2pt] 
{\bf BAT99-23} & WN2--5w & NE  & 49.6$\pm$1.0              & 2.3$_{-0.3}^{+0.2}$ & 5.54$_{-0.09}^{+0.09}$ & g & --13.20$_{-0.05}^{+0.04}$ & 1800          & \phantom{0}2.2$_{-0.5}^{+0.7}$ & 0.17$_{-0.02}^{+0.01}$ & $\cdots$         & 0.08$_{-0.01}^{+0.00}$ & 0.19$_{-0.01}^{+0.01}$ & 0.25 & 0.10 & 0.00 & 0.00 & 0.26$_{-0.02}^{+0.02}$ & 0.04$_{-0.01}^{+0.00}$ \\ [2pt]
{\bf BAT99-24} & WN3--7s & NE  & 49.6$\pm$1.0               & 0.4$_{-0.1}^{+0.0}$ & 5.53$_{-0.02}^{+0.03}$ & g & --11.51$_{-0.05}^{+0.04}$ & 2600           &    13.7$_{-1.5}^{+1.5}$ & 1.04$_{-0.03}^{+0.03}$ & \ddag0.54$_{-0.01}^{+0.01}$  & 0.04 & 0.11 & 0.14 & 0.13 & 0.04 & 0.02 & 0.15$_{-0.00}^{+0.01}$ & 0.09 \\ [2pt]
{\bf BAT99-25} & WN2--5w & NE  & 49.6$\pm$1.0              & 0.6$_{-0.1}^{+0.0}$ & 5.54$_{-0.03}^{+0.03}$ & g & --12.71$_{-0.04}^{+0.05}$ & 1700           & \phantom{0}1.1$_{-0.1}^{+0.1}$ & 0.08$_{-0.00}^{+0.00}$ & $\cdots$       &  0.07 & 0.00 & 0.11 & 0.08 & 0.01 & 0.00 & 0.30$_{-0.01}^{+0.00}$ & 0.06 \\ [2pt]
{\bf BAT99-26} & WN3--7s & NE  & 49.6$\pm$1.0               & 0.5$_{-0.0}^{+0.1}$ & 5.61$_{-0.02}^{+0.03}$ & g & --11.78$_{-0.05}^{+0.04}$ & 1800           & \phantom{0}8.7$_{-1.0}^{+1.0}$ & 0.55$_{-0.02}^{+0.02}$ & \ddag0.83$_{-0.02}^{+0.02}$  & 0.07  & 0.11 & 0.18 & 0.13 & 0.09$_{-0.01}^{+0.00}$ & 0.01 & 0.13 & 0.10$_{-0.00}^{+0.01}$ \\ [2pt]
{\bf BAT99-30} &  WN6--8 & AR    & 49.6$\pm$1.0              & 0.3$_{-0.1}^{+0.0}$ & 5.64$_{-0.02}^{+0.02}$ & g & --11.76$_{-0.05}^{+0.04}$ & 1000            & \phantom{0}6.8$_{-0.7}^{+0.7}$ & 0.40$_{-0.01}^{+0.01}$ & \ddag0.29$_{-0.01}^{+0.00}$  & 0.20  & 0.39  & 0.28 & 0.10 & 0.05 & 0.05 & \ddag0.50$_{-0.00}^{+0.01}$ & $\cdots$ \\ [2pt]
{\bf BAT99-31} & WN3--7s & NE  & 49.6$\pm$1.0              & 0.6$_{-0.0}^{+0.1}$ & 5.32$_{-0.03}^{+0.03}$ &  g & --12.08$_{-0.05}^{+0.04}$ & 2000           & \phantom{0}4.9$_{-0.6}^{+0.6}$ & 0.61$_{-0.02}^{+0.02}$ & \ddag0.57$_{-0.01}^{+0.02}$ & 0.05 & 0.10 & 0.19 & 0.13$_{-0.01}^{+0.00}$ & 0.07$_{-0.01}^{+0.00}$ & 0.01 & 0.27$_{-0.01}^{+0.00}$ & 0.08 \\ [2pt]
{\bf BAT99-32} & WN6--8+O & AR+SC & 49.6$\pm$1.0  & 0.3$\pm$0.0 & 5.93$_{-0.02}^{+0.02}$ & g & --11.36$_{-0.05}^{+0.04}$ & 1400         &    17.9$_{-1.9}^{+1.9}$ & 0.54$_{-0.01}^{+0.01}$ & \ddag0.36 & 0.19 & 0.26  & 0.14 & 0.12 & 0.03 & 0.04 & 0.24$_{-0.00}^{+0.01}$ & $\cdots$ \\ [2pt]
 {\bf BAT99-35} & WN2--5w & NE     & 49.6$\pm$1.0         & 0.4$_{-0.0}^{+0.1}$ & 5.59$_{-0.02}^{+0.03}$ & g  & --11.97$_{-0.05}^{+0.04}$ & 1800            & \phantom{0}4.9$_{-0.5}^{+0.6}$ & 0.33$_{-0.01}^{+0.01}$ & \ddag0.80$_{-0.01}^{+0.02}$ & 0.03 & 0.07 & 0.23 & 0.13 & 0.01 & 0.00 & 0.19 & 0.06 \\ [2pt]
{\bf BAT99-37} & WN2--5w & NE    & 49.6$\pm$1.0         & 1.9$\pm$0.2 & 5.64$_{-0.07}^{+0.08}$ & g & --12.62$_{-0.05}^{+0.04}$ & 2000           & \phantom{0}5.6$_{-1.2}^{+1.4}$ & 0.33$_{-0.02}^{+0.03}$ & $\cdots$         & 0.04$_{-0.01}^{+0.00}$ & 0.09$_{-0.00}^{+0.01}$ & 0.25 & 0.12$_{-0.01}^{+0.00}$ & 0.02 & 0.00 & 0.14$_{-0.00}^{+0.02}$ & 0.03 \\ [2pt]
{\bf BAT99-40} & WN2--5w+O &  NE     & 49.6$\pm$1.0         & 0.6$_{-0.1}^{+0.0}$ & 5.61$_{-0.03}^{+0.03}$ & g & --12.34$_{-0.05}^{+0.04}$ & 1600          & \phantom{0}2.5$_{-0.3}^{+0.3}$ & 0.16$_{-0.01}^{+0.01}$ & \ddag0.63$_{-0.02}^{+0.02}$ & 0.06 & 0.12 & 0.08$_{-0.00}^{+0.01}$ & 0.08 & 0.08$_{-0.00}^{+0.01}$ & 0.00 & 0.23$_{-0.01}^{+0.00}$ & 0.07 \\ [2pt]
{\bf BAT99-41} & WN3--7s & NE      & 49.6$\pm$1.0         & 0.5$_{-0.1}^{+0.0}$ & 5.59$_{-0.02}^{+0.03}$ & g & --11.82$_{-0.05}^{+0.04}$ & 1600              & \phantom{0}7.3$_{-0.8}^{+0.8}$ & 0.48$_{-0.01}^{+0.02}$ & \ddag0.07 & 0.04 & 0.10$_{-0.00}^{+0.01}$  & 0.18 & 0.13$_{-0.00}^{+0.01}$ & 0.08 & 0.01 & 0.14 & 0.10 \\ [2pt]
 {\bf BAT99-44} & WN6--8 & AR       & 49.6$\pm$1.0          & 0.5$_{-0.1}^{+0.0}$ & 5.65$_{-0.02}^{+0.03}$ & g & --12.13$_{-0.05}^{+0.04}$ & \phantom{0}900             & \phantom{0}3.6$_{-0.4}^{+0.4}$ & 0.21$_{-0.01}^{+0.00}$ & \ddag0.15$_{-0.00}^{+0.01}$ & 0.18 & 0.63$_{-0.01}^{+0.01}$  & 0.74 & 0.08 & 0.04 & 0.19 & \ddag1.01$_{-0.02}^{+0.02}$ & $\cdots$ \\ [2pt]
 {\bf BAT99-45} & WN9--11 & AR+SC & 49.6$\pm$1.0   & 0.3$_{-0.0}^{+0.0}$ & 5.75$_{-0.02}^{+0.02}$ & l & --12.40$_{-0.05}^{+0.05}$ & \phantom{0}200   &   1.6$_{-0.2}^{+0.2}$ & 0.07$_{-0.00}^{+0.01}$ & $\cdots$ & 0.03 & 2.2$_{-0.0}^{+0.0}$ & 2.3 & 0.05 & 0.00 & 2.8 & 6.8$_{-0.1}^{+0.0}$  & $\cdots$ \\ [2pt] 
 {\bf BAT99-46} & WN2--5w & NE    & 49.6$\pm$1.0       & 0.8$\pm$0.1 & 5.43$_{-0.03}^{+0.04}$ & g & --12.27$_{-0.04}^{+0.04}$ & 1700  & \phantom{0}3.8$_{-0.5}^{+0.5}$ & 0.36$_{-0.01}^{+0.02}$ & \ddag0.78$_{-0.03}^{+0.03}$ & 0.05 & 0.11$_{-0.01}^{+0.00}$ & 0.16 & 0.12$_{-0.00}^{+0.01}$ & 0.03 & 0.01 & 0.17$_{-0.01}^{+0.00}$ & 0.10 \\  [2pt]
BAT99-47 & WN3--7s & NE  & 49.6$\pm$1.0           & 0.8$_{-0.1}^{+0.0}$ & 5.58$_{-0.03}^{+0.04}$ & g & --12.00$_{-0.05}^{+0.04}$ & 1600     & \phantom{0}6.8$_{-0.9}^{+0.9}$ & 0.46$_{-0.02}^{+0.02}$ & $\cdots$     & 0.01 & 0.07 & 0.19$_{-0.00}^{+0.01}$ & 0.13 & 0.02 & 0.00 & 0.13$_{-0.00}^{+0.01}$ & 0.02 \\ [2pt]
BAT99-48 & WN3--7s & NE  & 49.6$\pm$1.0          & 0.4$_{-0.1}^{+0.0}$ & 5.39$_{-0.02}^{+0.03}$ & g & --11.74$_{-0.05}^{+0.05}$ & 2000     & \phantom{0}8.0$_{-0.9}^{+0.9}$ & 0.84$_{-0.02}^{+0.03}$ & \ddag0.70$_{-0.01}^{+0.01}$ & 0.03 & 0.09 & 0.19 & 0.13 & 0.08$_{-0.01}^{+0.00}$ & 0.02 & 0.14$_{-0.01}^{+0.00}$ & 0.08 \\ [2pt]
BAT99-50 & WN2--5w & NE  & 49.6$\pm$1.0          & 0.7$_{-0.1}^{+0.0}$ & 5.64$_{-0.03}^{+0.03}$ & g & --12.54$_{-0.04}^{+0.04}$ & 1200             & \phantom{0}1.8$_{-0.2}^{+0.2}$ & 0.11$_{-0.01}^{+0.00}$ & $\cdots$      & 0.27 & 0.20  & 0.05 & 0.07 & 0.03 & 0.00 & 0.27$_{-0.01}^{+0.01}$ & 0.15$_{-0.01}^{+0.00}$ \\ [2pt]
{\bf BAT99-51} & WN3--7s & NE  & 49.6$\pm$1.0           & 0.1$_{-0.0}^{+0.0}$ & 5.29$_{-0.01}^{+0.02}$ & g & --11.91$_{-0.05}^{+0.04}$ & 2900             & \phantom{0}3.9$_{-0.4}^{+0.4}$ & 0.52$_{-0.00}^{+0.01}$ & \ddag0.48$_{-0.00}^{+0.01}$    & 0.03 & 0.06 & 0.11 & 0.15 & 0.01 & 0.00 & 0.19 & 0.05 \\ [2pt]
       BAT99-54 & WN9--11 & AR      & 49.6$\pm$1.0           & 1.9$\pm$0.2 & 5.74$_{-0.07}^{+0.08}$ & g & --13.15$_{-0.05}^{+0.04}$ & \phantom{0}300            & \phantom{0}1.7$_{-0.4}^{+0.4}$ & 0.08$_{-0.01}^{+0.00}$ & $\cdots$       & 0.03 & 0.17$_{-0.01}^{+0.00}$  & 0.74 & 0.04 & 0.01 & 0.17$_{-0.01}^{+0.00}$ & $\cdots$ & $\cdots$ \\ [2pt]
 {\bf BAT99-56} & WN3--7s & NE  & 49.6$\pm$1.0          & 0.5$_{-0.1}^{+0.0}$ & 5.55$_{-0.02}^{+0.03}$ & g & --11.80$_{-0.04}^{+0.05}$ & 1800            & \phantom{0}7.8$_{-0.9}^{+0.9}$ & 0.56$_{-0.01}^{+0.02}$ & \ddag0.85$_{-0.02}^{+0.01}$   & 0.04 & 0.10 & 0.21 & 0.13 & 0.07 & 0.02 & 0.14 & 0.08 \\ [2pt]
  \hline
\end{tabular}
\end{footnotesize}
\end{center}
\end{table}
\end{landscape}

\addtocounter{table}{-1}

\begin{landscape}
\begin{table}
\caption{continued}
\begin{center}
\begin{footnotesize}
 \begin{tabular}{l@{\hspace{1mm}}l@{\hspace{1.5mm}}l@{\hspace{1.5mm}}l@{\hspace{1.5mm}}l@{\hspace{1.5mm}}l@{\hspace{1mm}}c@{\hspace{1mm}}r@{\hspace{1mm}}c@{\hspace{0mm}}c@{\hspace{1mm}}c
 @{\hspace{1.5mm}}l@{\hspace{1mm}}l@{\hspace{1mm}}l@{\hspace{1mm}}l@{\hspace{1mm}}l@{\hspace{1mm}}l@{\hspace{1mm}}l@{\hspace{1mm}}l@{\hspace{1mm}}l}   %
\hline
Star & Category & Data & $d$ & $A_{\rm V}$  & $\log$  & Ref & $\log F_{\rm HeII~4686}$ & HeII 4686 & \multicolumn{2}{c}{$L_{\rm HeII~4686}$} 
& $\underline{L_{\rm NIV~3478,85}}$ 
& $\underline{L_{\rm NIV~4058}}$ 
& $\underline{L_{4100}}$ 
& $\underline{L_{4630}}$ 
&$\underline{L_{\rm HeII~5412}}$ 
& $\underline{L_{\rm CIV~5801,12}}$ 
&$\underline{L_{\rm HeI~5876}}$ 
& $\underline{L_{{\rm H}\alpha}}$ 
& $\underline{L_{\rm NIV~7103,29}}$ \\  
        &    &   ID          &   kpc     & mag              & $L_{\rm Bol}/L_{\odot}$        &        & erg\,s$^{-1}$\,cm$^{-2}$ & FWHM & 10$^{35}$ erg\,s$^{-1}$ &  $10^{-3} L_{\rm Bol}$   & $L_{\rm HeII~4686}$ &  $L_{\rm HeII~4686}$  & $L_{\rm HeII~4686}$ & $L_{\rm HeII~4686}$  & $L_{\rm HeII~4686}$ &  $L_{\rm HeII~4686}$ &   $L_{\rm HeII~4686}$ &
 $L_{\rm HeII~4686}$  & $L_{\rm HeII~4686}$ \\
\hline
 {\bf BAT99-57} & WN3--7s & NE  & 49.6$\pm$1.0          & 0.4$_{-0.1}^{+0.0}$ & 5.39$_{-0.02}^{+0.03}$ & g & --11.80$_{-0.04}^{+0.04}$ & 1800           & \phantom{0}7.1$_{-0.8}^{+0.8}$ & 0.75$_{-0.02}^{+0.02}$ & \ddag0.66$_{-0.02}^{+0.01}$   & 0.02 & 0.10$_{-0.01}^{+0.00}$ & 0.20 & 0.13 & 0.07 & 0.01 & 0.15 & 0.07$_{-0.01}^{+0.00}$ \\ [2pt]
 {\bf BAT99-58} & WN6--8  & AR   & 49.6$\pm$1.0          & 1.9$\pm$0.2 & 5.63$_{-0.07}^{+0.08}$ & g & --12.71$_{-0.04}^{+0.04}$ & 1000            & \phantom{0}4.6$_{-1.0}^{+1.2}$ & 0.28$_{-0.02}^{+0.02}$ & \ddag0.21$_{-0.02}^{+0.01}$    & 0.19$_{-0.01}^{+0.01}$ & 0.41$_{-0.02}^{+0.01}$ & 0.36 & 0.10 & 0.06 & 0.06$_{-0.01}^{+0.00}$ & \ddag0.40$_{-0.02}^{+0.03}$ & $\cdots$ \\ [2pt]
 {\bf BAT99-60} & WN2--5w+O & NE & 49.6$\pm$1.0   & 0.4$_{-0.1}^{+0.0}$ & 5.46$_{-0.02}^{+0.03}$ & h & --12.37$_{-0.05}^{+0.04}$ & 1800           & \phantom{0}1.9$_{-0.2}^{+0.2}$ & 0.17$_{-0.01}^{+0.00}$ & \ddag0.75$_{-0.01}^{+0.02}$    & 0.15$_{-0.01}^{+0.00}$ & 0.00 & 0.19 & 0.10 & 0.02 & 0.00 & 0.22$_{-0.00}^{+0.01}$ & 0.03 \\ [2pt]
 {\bf BAT99-62} & WN2--5w      & NE  & 49.6$\pm$1.0   & 0.5$_{-0.1}^{+0.0}$ & 5.40$_{-0.02}^{+0.03}$ & g & --12.23$_{-0.05}^{+0.04}$ & 1800           & \phantom{0}2.8$_{-0.3}^{+0.4}$ & 0.29$_{-0.01}^{+0.01}$ & \ddag1.02$_{-0.02}^{+0.02}$   & 0.01 & 0.09$_{-0.01}^{+0.00}$ & 0.29 & 0.13 & 0.01 & 0.00 & 0.20 & 0.05 \\ [2pt]
 {\bf BAT99-63} & WN2--5w      & NE  & 49.6$\pm$1.0 & 0.4$_{-0.1}^{+0.0}$ & 5.57$_{-0.02}^{+0.03}$ & g & --12.26$_{-0.04}^{+0.04}$ & 1800            & \phantom{0}2.5$_{-0.3}^{+0.3}$ & 0.17$_{-0.00}^{+0.01}$ & \ddag0.70$_{-0.01}^{+0.02}$   & 0.13$_{-0.00}^{+0.01}$ & 0.15$_{-0.01}^{+0.00}$ & 0.12 & 0.09$_{-0.01}^{+0.00}$ & 0.05 & 0.00 & 0.25$_{-0.00}^{+0.01}$ & 0.08$_{-0.01}^{+0.00}$ \\ [2pt]
 {\bf BAT99-65} & WN2--5w      & NE  & 49.6$\pm$1.0  & 1.74$\pm$0.2 & 5.74$_{-0.07}^{+0.07}$ & g & --12.32$_{-0.04}^{+0.04}$ & 1800            & \phantom{0}9.2$_{-1.8}^{+2.1}$ & 0.43$_{-0.03}^{+0.03}$ & $\cdots$      & 0.05 & 0.10$_{-0.00}^{+0.01}$ & 0.20 & 0.12 & 0.07 & 0.01 & 0.12$_{-0.00}^{+0.01}$ & 0.08$_{-0.01}^{+0.00}$ \\ [2pt]
 {\bf BAT99-66} & WN2--5w      & NE  &  49.6$\pm$1.0  & 0.5$_{-0.1}^{+0.0}$ & 5.77$_{-0.02}^{+0.03}$ &  g & --12.48$_{-0.05}^{+0.04}$ & 2000          & \phantom{0}1.7$_{-0.2}^{+0.2}$ & 0.07$_{-0.00}^{+0.01}$ &$\cdots$      & 0.00 &  0.08 & 0.35 & 0.12 & 0.01 & 0.00 & 0.24$_{-0.01}^{+0.00}$ & 0.02 \\ [2pt]
{\bf BAT99-67} & WN2--5w      & NE  & 49.6$\pm$1.0  & 1.2$_{-0.1}^{+0.2}$ & 5.95$_{-0.05}^{+0.06}$ & g & --12.18$_{-0.05}^{+0.04}$ & 1600           & \phantom{0}7.7$_{-1.3}^{+1.3}$ & 0.22$_{-0.01}^{+0.01}$ & \ddag0.55$_{-0.03}^{+0.03}$ & 0.27$_{-0.01}^{+0.01}$ & 0.52$_{-0.01}^{+0.01}$ & 0.00  & 0.13$_{-0.00}^{+0.01}$ & 0.03 & 0.00 & 0.30$_{-0.02}^{+0.01}$ & 0.15$_{-0.00}^{+0.01}$ \\ [2pt]
 {\bf BAT99-73} & WN2--5w     & NE  & 49.6$\pm$1.0  & 0.8$_{-0.1}^{+0.0}$ & 5.71$_{-0.03}^{+0.04}$ & g & --12.68$_{-0.05}^{+0.04}$ & 1300             & \phantom{0}1.4$_{-0.2}^{+0.2}$ & 0.07$_{-0.00}^{+0.00}$ & $\cdots$    & 0.22 & 0.13  & 0.00 & 0.08 & 0.03 & 0.00 & 0.32$_{-0.01}^{+0.01}$ & 0.15$_{-0.01}^{+0.00}$ \\ [2pt]
 {\bf BAT99-74} & WN2--5w     & NE  & 49.6$\pm$1.0  & 0.8$_{-0.1}^{+0.0}$ & 5.68$_{-0.03}^{+0.04}$ & g & --12.82$_{-0.04}^{+0.05}$ & 2200            & \phantom{0}1.0$_{-0.1}^{+0.1}$ & 0.06$_{-0.01}^{+0.00}$ & $\cdots$    & 0.00 & 0.00 & 0.50 & 0.09 & 0.01 & 0.00 & 0.21$_{-0.00}^{+0.01}$ & 0.00 \\ [2pt]
 {\bf BAT99-75} & WN2--5w     & NE  & 49.6$\pm$1.0  & 0.3$_{-0.1}^{+0.0}$ & 5.55$_{-0.02}^{+0.02}$ & g & --11.95$_{-0.04}^{+0.04}$ & 1700            & \phantom{0}4.4$_{-0.4}^{+0.5}$ & 0.32$_{-0.00}^{+0.01}$ & $\cdots$   & 0.11 & 0.17 & 0.18  & 0.14$_{-0.01}^{+0.00}$ & 0.09 & 0.02 & 0.17 & 0.11$_{-0.00}^{+0.01}$ \\ [2pt]
 {\bf BAT99-76} & WN9--11 & AR+SC & 49.6$\pm$1.0  & 1.0$\pm$0.1 & 5.65$_{-0.04}^{+0.05}$ & g & --12.55$_{-0.05}^{+0.04}$ & \phantom{0}200         & \phantom{0}2.4$_{-0.3}^{+0.4}$ & 0.14$_{-0.01}^{+0.01}$ & \ddag0.00 & 0.00 & 0.88$_{-0.02}^{+0.01}$ & 1.07 & 0.01 & 0.01 & 0.57$_{-0.02}^{+0.01}$ & 1.66$_{-0.06}^{+0.06}$ & $\cdots$ \\ [2pt]
 BAT99-78        & WN2--5w & HF     & 49.6$\pm$1.0  & 0.8$_{-0.1}^{+0.0}$ & 5.69$_{-0.03}^{+0.04}$ & g & --13.08$_{-0.05}^{+0.04}$  & 1500             & \phantom{0}0.6$_{-0.1}^{+0.0}$ & 0.03$_{-0.00}^{+0.00}$ & 0.56$_{-0.01}^{+0.02}$           & 0.11 & 0.15$_{-0.01}^{+0.00}$ & 0.11 & $\cdots$ & $\cdots$ & $\cdots$ & $\cdots$ & $\cdots$ \\ [2pt]
 {\bf BAT99-81} & WN2--5w & NE    & 49.6$\pm$1.0    & 1.2$_{-0.1}^{+0.2}$ & 5.47$_{-0.05}^{+0.06}$ & g & --12.87$_{-0.04}^{+0.04}$ & 1100           & \phantom{0}1.6$_{-0.3}^{+0.3}$ & 0.14$_{-0.01}^{+0.01}$ & $\cdots$    & 0.43$_{-0.01}^{+0.01}$ & 0.42$_{-0.01}^{+0.01}$ & 0.12 & 0.08 & 0.12$_{-0.00}^{+0.01}$ & 0.00 & 0.35$_{-0.01}^{+0.02}$ & 0.11$_{-0.01}^{+0.00}$ \\ [2pt]
 {\bf BAT99-82} & WN3--7s & NE     & 49.6$\pm$1.0  & 1.0$\pm$0.1 & 5.52$_{-0.04}^{+0.05}$ & g & --12.30$_{-0.04}^{+0.04}$ & 2000             & \phantom{0}4.6$_{-0.7}^{+0.7}$ & 0.36$_{-0.02}^{+0.01}$ & \ddag0.13$_{-0.00}^{+0.01}$ & 0.00 & 0.06 & 0.17 & 0.13 & 0.00 & 0.00 & 0.15$_{-0.01}^{+0.01}$ & $\cdots$ \\ [2pt]
  {\bf BAT99-86} & WN2--5w & NE  & 49.6$\pm$1.0     & 1.4$_{-0.2}^{+0.1}$ & 5.32$_{-0.05}^{+0.06}$ & g & --12.98$_{-0.04}^{+0.04}$ & 1900           & \phantom{0}1.4$_{-0.3}^{+0.2}$ & 0.17$_{-0.01}^{+0.01}$ & \ddag1.25$_{-0.07}^{+0.08}$ & 0.02 & 0.10 & 0.23 & 0.12$_{-0.00}^{+0.01}$ & 0.02 & 0.01 & 0.28$_{-0.02}^{+0.01}$ & 0.02 \\ [2pt]
 BAT99-89        & WN6--8    & AR    & 49.6$\pm$1.0     & 1.1$_{-0.2}^{+0.1}$ & 5.77$_{-0.04}^{+0.05}$ & g & --11.97$_{-0.05}^{+0.04}$ & 1100           &   10.0$_{-1.5}^{+1.6}$ & 0.44$_{-0.02}^{+0.02}$ & $\cdots$     & 0.17$_{-0.01}^{+0.00}$ & 0.32$_{-0.00}^{+0.01}$ & 0.30 & 0.12 & 0.04$_{-0.01}^{+0.00}$ & 0.10$_{-0.00}^{+0.01}$ & $\cdots$ & $\cdots$ \\ [2pt]
 BAT99-91       & WN6--8 & HF     & 49.6$\pm$1.0  & 1.2$_{-0.2}^{+0.1}$ & 5.41$_{-0.05}^{+0.06}$ & g & --12.43$_{-0.05}^{+0.04}$ & \phantom{0}900    & \phantom{0}4.3$_{-0.7}^{+0.7}$ & 0.43$_{-0.02}^{+0.02}$ & 0.17$_{-0.01}^{+0.01}$          & 0.16$_{-0.01}^{+0.00}$ & 0.28$_{-0.01}^{+0.00}$ & 0.24$_{-0.01}^{+0.00}$ & $\cdots$ & $\cdots$ & $\cdots$ & $\cdots$ & $\cdots$ \\ [2pt]
 {\bf BAT99-94} & WN3--7s  & NE & 49.6$\pm$1.0     & 1.1$\pm$0.1 & 5.79$_{-0.04}^{+0.05}$ & g & --11.69$_{-0.05}^{+0.04}$ & 3300          &   20.0$_{-3.0}^{+3.3}$ & 0.83$_{-0.04}^{+0.05}$ & \ddag0.65$_{-0.03}^{+0.03}$           & 0.00 & 0.10 & 0.10 & 0.13 & 0.06 & 0.00 & 0.15$_{-0.00}^{+0.01}$ & 0.09$_{-0.00}^{+0.01}$ \\ [2pt]
 {\bf BAT99-95} & WN6--8 & AR+SC & 49.6$\pm$1.0  & 0.9$\pm$0.1 & 5.99$_{-0.04}^{+0.04}$ & g & --11.48$_{-0.04}^{+0.05}$ & 1600      &     27.8$_{-3.9}^{+4.2}$ & 0.73$_{-0.03}^{+0.04}$ & $\cdots$     & 0.14 & 0.32$_{-0.00}^{+0.01}$ & 0.27 & 0.11$_{-0.01}^{+0.00}$ & 0.02 & 0.13$_{-0.01}^{+0.00}$ & 0.28$_{-0.01}^{+0.01}$ & $\cdots$ \\ [2pt]
 BAT99-96         & WN6--8 & VM &49.6$\pm$1.0  & 2.6$_{-0.2}^{+0.3}$ & 6.34$_{-0.10}^{+0.11}$ &  g & --12.24$_{-0.09}^{+0.08}$ & \phantom{0}900     &    31.3$_{-10.1}^{+12.3}$ & 0.37$_{-0.03}^{+0.03}$ & $\cdots$      & $\cdots$ & $\cdots$ & 0.51$_{-0.00}^{+0.01}$ & 0.10$_{-0.01}^{+0.00}$ & 0.01 & 0.15$_{-0.01}^{+0.01}$ & 0.21$_{-0.02}^{+0.02}$ & 0.05$_{-0.00}^{+0.01}$ \\ [2pt]
 BAT99-97        & Of/WN & VM &  49.6$\pm$1.0 & 2.3$_{-0.3}^{+0.2}$ & 6.29$_{-0.09}^{+0.09}$ & g & --12.83$_{-0.10}^{+0.08}$ & \phantom{0}800 & 5.3$_{-1.6}^{+1.8}$ & 0.07$_{-0.01}^{+0.01}$ & $\cdots$ & $\cdots$ & $\cdots$ & 0.41$_{-0.00}^{+0.01}$ & 0.01 & 0.02 & 0.00 & 0.58$_{-0.4}^{+0.05}$ & $\cdots$ \\ [2pt]
BAT99-98         & WN6--8 & VM            & 49.6$\pm$1.0 & 1.5$_{-0.2}^{+0.1}$ & $\cdots$ & b & --11.93$_{-0.10}^{+0.08}$ & 1260 &  17.6$_{-4.4}^{+4.7}$ & $\cdots$ & $\cdots$ & $\cdots$ & $\cdots$ & 0.21 & 0.08 & 0.02 & 0.02 & 0.35$_{-0.02}^{+0.02}$ & 0.07$_{-0.01}^{+0.00}$ \\ [2pt]
 BAT99-99       &Of/WN+O & VM       & 49.6$\pm$1.0  & 1.1$_{-0.1}^{+0.1}$ & 5.89$_{-0.05}^{+0.05}$ & g & --12.65$_{-0.09}^{+0.08}$ & 1100    & \phantom{0}2.3$_{-0.5}^{+0.6}$ & 0.08$_{-0.01}^{+0.00}$ & $\cdots$  & $\cdots$ & $\cdots$ & 0.19 & 0.01 & 0.08 & 0.00 & 0.82$_{-0.04}^{+0.03}$ & 0.15$_{-0.01}^{+0.01}$ \\ [2pt]
{\bf BAT99-100} & WN6--8         & HF+VM & 49.6$\pm$1.0  & 1.1$_{-0.2}^{+0.1}$ & 6.14$_{-0.04}^{+0.05}$ & g & --11.52$_{-0.05}^{+0.04}$ & 1000 &    28.2$_{-4.2}^{+4.5}$  & 0.53$_{-0.03}^{+0.02}$ & 0.10$_{-0.00}^{+0.01}$       & 0.12 & 0.25$_{-0.00}^{+0.01}$  & 0.25 & 0.10$_{-0.01}^{+0.00}$ & 0.02 & 0.12$_{-0.01}^{+0.00}$ & 0.24$_{-0.01}^{+0.01}$ &0.04 \\ [2pt]
 BAT99-103  & WN6--8+O        & VM          & 49.6$\pm$1.0  & 0.9$\pm$0.1 & 6.08$_{-0.04}^{+0.03}$ & h & --11.67$_{-0.09}^{+0.08}$ & 1300  &  17.8$_{-4.0}^{+4.1}$ & 0.38$_{-0.02}^{+0.03}$  & $\cdots$ & $\cdots$ & $\cdots$ & 0.25 & 0.10$_{-0.00}^{+0.01}$ & 0.01 & 0.05 & 0.21$_{-0.01}^{+0.01}$ & 0.07$_{-0.00}^{+0.01}$ \\ [2pt]
 {\bf BAT99-104} & Of/WN & VM        & 49.6$\pm$1.0  & 1.4$_{-0.1}^{+0.2}$ & 6.05$_{-0.06}^{+0.06}$ & g & --12.74$_{-0.09}^{+0.08}$ & 1100  & \phantom{0}2.6$_{-0.6}^{+0.7}$ & 0.06$_{-0.00}^{+0.00}$ & $\cdots$  & $\cdots$ & $\cdots$ &  0.10 & 0.04 & 0.01 & 0.00 & 0.46$_{-0.02}^{+0.02}$ & 0.06$_{-0.01}^{+0.00}$ \\ [2pt]
 {\bf BAT99-106} & WN5--7h & HF   & 49.6$\pm$1.0  & 1.7$\pm$0.2 & 6.52$_{-0.07}^{+0.07}$ & e & --11.74$_{-0.05}^{+0.04}$ & 2000 &                  32.4$_{-6.3}^{+7.2}$ & 0.25$_{-0.01}^{+0.02}$ & 0.43$_{-0.02}^{+0.03}$          & 0.22$_{-0.00}^{+0.01}$ & 0.21$_{-0.01}^{+0.00}$ & 0.07 & 0.10$_{-0.01}^{+0.00}$ & 0.04 & 0.00 & 0.40$_{-0.02}^{+0.02}$ & $\cdots$ \\ [2pt]
 {\bf BAT99-108} & WN5--7h & HF   & 49.6$\pm$1.0  & 1.8$\pm$0.2 & 6.88$_{-0.07}^{+0.07}$ & e & --11.61$_{-0.05}^{+0.04}$ & 2000 &                  49.1$_{-9.9}^{+11.5}$ & 0.17$_{-0.01}^{+0.01}$ & 0.48$_{-0.03}^{+0.03}$          & 0.19$_{-0.01}^{+0.00}$ & 0.18$_{-0.01}^{+0.00}$ & 0.10 & 0.12$_{-0.01}^{+0.00}$ & 0.05 & 0.0 & 0.36$_{-0.02}^{+0.02}$  & $\cdots$ \\ [2pt]
 {\bf BAT99-109} & WN5--7h & HF   & 49.6$\pm$1.0  & 1.9$\pm$0.2 & 6.70$_{-0.08}^{+0.08}$ & e & --11.79$_{-0.04}^{+0.05}$ & 2000 &                 36.5$_{-7.7}^{+8.9}$ & 0.19$_{-0.01}^{+0.01}$ & 0.45$_{-0.03}^{+0.03}$          & 0.19$_{-0.01}^{+0.00}$ & 0.17$_{-0.01}^{+0.00}$ & 0.08 & 0.11 & 0.04 & 0.00 & 0.33$_{-0.01}^{+0.02}$ & $\cdots$ \\ [2pt]
 {\bf BAT99-110} & Of/WN & HF   & 49.6$\pm$1.0  & 1.7$_{-0.1}^{+0.2}$ & 6.32$_{-0.07}^{+0.07}$ & e,j & --12.83$_{-0.04}^{+0.04}$ & 1700 &                  2.8$_{-0.6}^{+0.6}$ & 0.03$_{-0.00}^{+0.01}$ & 0.30$_{-0.01}^{+0.02}$          & 0.22$_{-0.01}^{+0.01}$ & 0.00 & 0.19 & 0.00 & 0.00 & 0.00 & 0.55$_{-0.03}^{+0.03}$ & $\cdots$ \\ [2pt]
 {\bf BAT99-111} & WN6--8$\dag$    & HF   & 49.6$\pm$1.0  & 1.6$_{-0.1}^{+0.2}$         & 6.24$_{-0.06}^{+0.07}$ & g & --12.75$_{-0.05}^{+0.04}$ & \phantom{0}600   & \phantom{0}3.1$_{-0.6}^{+0.7}$ & 0.05$_{-0.01}^{+0.00}$ & 0.00       & 0.13$_{-0.01}^{+0.00}$ & 0.11 & 0.63  &0.00 & 0.00 & 0.04 & 0.69$_{-0.04}^{+0.05}$ & $\cdots$ \\ [2pt]
 BAT99-112      & WN5--7h   &  HF+VM   & 49.6$\pm$1.0 & 2.5$_{-0.2}^{+0.3}$ & 6.49$_{-0.10}^{+0.10}$ & e & --11.97$_{-0.05}^{+0.04}$ & 1600  &    45.2$_{-11.5}^{+14.6}$  & 0.38$_{-0.03}^{+0.03}$ & 0.33$_{-0.03}^{+0.03}$       & 0.15$_{-0.00}^{+0.01}$ & 0.16$_{-0.00}^{+0.01}$ & 0.06 & 0.06$_{-0.00}^{+0.01}$ & 0.01 & 0.00 & 0.24$_{-0.02}^{+0.02}$ & 0.06$_{-0.00}^{+0.01}$ \\ [2pt]
BAT99-113   & Of/WN+O & VM & 49.6$\pm$1.0  & 1.2$_{-0.1}^{+0.2}$          & 6.13$_{-0.05}^{+0.06}$ & h & --12.54$_{-0.10}^{+0.07}$ & 1100   & \phantom{0}2.9$_{-0.7}^{+0.7}$ & 0.06$_{-0.01}^{+0.00}$ & $\cdots$    & $\cdots$ & $\cdots$ & 0.09 & 0.03 & 0.04 & 0.00 & 0.50$_{-0.02}^{+0.02}$ & $\cdots$ \\  [2pt]
 BAT99-114 & Of/WN        & VM & 49.6$\pm$1.0  & 1.2$_{-0.1}^{+0.2}$         & 6.43$_{-0.05}^{+0.05}$ & g & --12.74$_{-0.09}^{+0.08}$ & 1000    & \phantom{0}2.0$_{-0.5}^{+0.4}$ & 0.02$_{-0.00}^{+0.00}$ & $\cdots$    & $\cdots$ & $\cdots$ & 0.12 & 0.03 & 0.07 & 0.00 & 0.64$_{-0.03}^{+0.03}$ & 0.13$_{-0.00}^{+0.01}$ \\ [2pt]
  \hline
\end{tabular}
\end{footnotesize}
\end{center}
\end{table}
\end{landscape}

\addtocounter{table}{-1}

\begin{landscape}
\begin{table}
\caption{continued}
\begin{center}
\begin{footnotesize}
 \begin{tabular}{l@{\hspace{1mm}}l@{\hspace{1.5mm}}l@{\hspace{1.5mm}}l@{\hspace{1.5mm}}l@{\hspace{2mm}}l@{\hspace{1mm}}c@{\hspace{1mm}}r@{\hspace{1mm}}c@{\hspace{0mm}}c@{\hspace{1mm}}c
 @{\hspace{1.5mm}}l@{\hspace{1mm}}l@{\hspace{1mm}}l@{\hspace{1mm}}l@{\hspace{1mm}}l@{\hspace{1mm}}l@{\hspace{1mm}}l@{\hspace{1mm}}l@{\hspace{1mm}}l}   %
\hline
Star & Category & Data & $d$ & $A_{\rm V}$  & $\log$  & Ref & $\log F_{\rm HeII~4686}$ & HeII 4686 & \multicolumn{2}{c}{$L_{\rm HeII~4686}$} 
& $\underline{L_{\rm NIV~3478,85}}$ 
& $\underline{L_{\rm NIV~4058}}$ 
& $\underline{L_{4100}}$ 
& $\underline{L_{4630}}$ 
&$\underline{L_{\rm HeII~5412}}$ 
& $\underline{L_{\rm CIV~5801,12}}$ 
&$\underline{L_{\rm HeI~5876}}$ 
& $\underline{L_{{\rm H}\alpha}}$ 
& $\underline{L_{\rm NIV~7103,29}}$ \\  
        &    &   ID          &   kpc     & mag              & $L_{\rm Bol}/L_{\odot}$        &        & erg\,s$^{-1}$\,cm$^{-2}$ & FWHM  & 10$^{35}$ erg\,s$^{-1}$ &  $10^{-3} L_{\rm Bol}$   & $L_{\rm HeII~4686}$ &  $L_{\rm HeII~4686}$  & $L_{\rm HeII~4686}$ & $L_{\rm HeII~4686}$  & $L_{\rm HeII~4686}$ &  $L_{\rm HeII~4686}$ &   $L_{\rm HeII~4686}$ &
 $L_{\rm HeII~4686}$  & $L_{\rm HeII~4686}$ \\
\hline
 {\bf BAT99-116} & 2$\times$WN5--7h& VM & 49.6$\pm$1.0  & 2.0$\pm$0.2 & 6.70$_{-0.09}^{+0.08}$ & k & --11.87$_{-0.10}^{+0.08}$ & 1800 &      33.3$_{-9.3}^{+10.4}$ & 0.17$_{-0.01}^{+0.01}$ & $\cdots$    &  $\cdots$ & $\cdots$ & 0.05 & 0.07 & 0.04$_{-0.01}^{+0.00}$ & 0.03$_{-0.00}^{+0.01}$ & 0.39$_{-0.02}^{+0.03}$ & 0.11$_{-0.01}^{+0.01}$ \\ [2pt]
 {\bf BAT99-117} & WN5--7h  & NE  & 49.6$\pm$1.0  & 0.7$\pm$0.1       & 6.39$_{-0.03}^{+0.04}$ & g & --11.82$_{-0.04}^{+0.05}$ & 2000    & \phantom{0}9.9$_{-1.12}^{+1.3}$ & 0.10$_{-0.00}^{+0.01}$ & \ddag0.68$_{-0.02}^{+0.02}$ & 0.09$_{-0.00}^{+0.01}$        & 0.03        & 0.00 & 0.07$_{-0.00}^{+0.01}$ & 0.01 & 0.00 & 0.24$_{-0.01}^{+0.01}$ & 0.11 \\ [2pt]
 {\bf BAT99-118} & 2$\times$WN6--8 & AR+SC & 49.6$\pm$1.0  & 0.6$_{-0.0}^{+0.1}$ & 6.71$_{-0.03}^{+0.03}$ & l & --10.86$_{-0.05}^{+0.04}$ & 1400 &     80.1$_{-9.7}^{+10.0}$ & 0.41$_{-0.02}^{+0.01}$ & $\cdots$    & 0.15         & 0.26$_{-0.01}^{+0.00}$       & 0.15 & 0.09 & 0.04 & 0.04 & 0.41$_{-0.01}^{+0.00}$ & $\cdots$ \\ [2pt]
 {\bf BAT99-119} & WN6--8+O   & AR+SC & 49.6$\pm$1.0  & 1.1$_{-0.2}^{+0.1}$ & 6.63$_{-0.04}^{+0.05}$ & h & --11.34$_{-0.04}^{+0.04}$ & 1100 &  43.3$_{-6.5}^{+7.0}$ & 0.26$_{-0.01}^{+0.01}$ & 0.17$_{-0.01}^{+0.01}$          & 0.17         & 0.23$_{-0.00}^{+0.01}$       & 0.17 & 0.10$_{-0.00}^{+0.01}$ & 0.05 & 0.03 & 0.32$_{-0.01}^{+0.02}$ & $\cdots$ \\ [2pt]
 {\bf BAT99-120} & WN9--11 & AR+SC & 49.6$\pm$1.0  & 0.6$_{-0.1}^{+0.0}$ & 5.57$_{-0.03}^{+0.03}$ & g & --12.30$_{-0.05}^{+0.04}$ & \phantom{0}300  & \phantom{0}2.7$_{-0.3}^{+0.4}$ &  0.19$_{-0.01}^{+0.01}$ &$\cdots$    &  0.03        & 0.86$_{-0.01}^{+0.01}$      & 1.34$_{-0.00}^{+0.01}$  & 0.02 & 0.01 & 0.45$_{-0.01}^{+0.00}$ & 1.54$_{-0.03}^{+0.03}$ & $\cdots$ \\ [2pt]
 {\bf BAT99-122} & WN5--7h  & NE & 49.6$\pm$1.0  & 1.1$_{-0.2}^{+0.1}$         & 6.22$_{-0.04}^{+0.05}$ & g & --11.50$_{-0.05}^{+0.04}$  & 1300  &   29.5$_{-4.4}^{+4.7}$  & 0.46$_{-0.02}^{+0.02}$ & \ddag0.23$_{-0.01}^{+0.01}$ & 0.25$_{-0.01}^{+0.00}$        & 0.35$_{-0.01}^{+0.01}$      & 0.10$_{-0.01}^{+0.00}$ & 0.11 & 0.04 & 0.02 & 0.31$_{-0.02}^{+0.01}$ & 0.11$_{-0.00}^{+0.01}$ \\ [2pt]
 BAT99-128 & WN3--7s  & NE & 49.6$\pm$1.0 & 0.6$_{-0.0}^{+0.1}$         & 5.43$_{-0.03}^{+0.03}$ & g & --12.00$_{-0.04}^{+0.05}$  & 1800   & \phantom{0}6.0$_{-0.7}^{+0.8}$  & 0.58$_{-0.02}^{+0.02}$ & $\cdots$  & 0.01       & 0.06       & 0.18 & 0.15$_{-0.01}^{+0.00}$ & 0.01 & 0.08 & 0.17$_{-0.00}^{+0.01}$ & 0.01 \\ [2pt]
 {\bf BAT99-130} & WN9--11 & AR+SC & 49.6$\pm$1.0  & 0.9$_{-0.1}^{+0.1}$ & 5.67$_{-0.04}^{+0.04}$ & g & --13.42$_{-0.05}^{+0.04}$ & \phantom{0}200  & \phantom{0}0.3$_{-0.0}^{+0.1}$ &  0.02$_{-0.00}^{+0.00}$ &$\cdots$    &  0.00        & 2.2$_{-0.0}^{+0.1}$      & 4.2$_{-0.0}^{+0.0}$  & 0.00 & 0.00 & 3.3$_{-0.1}^{+0.1}$ & 12$_{-0}^{+0}$ & $\cdots$ \\ [2pt]
BAT99-131 & WN3--7s  & NE & 49.6$\pm$1.0  & 0.5$_{-0.1}^{+0.0}$         & 5.66$_{-0.02}^{+0.03}$ & g & --11.65$_{-0.04}^{+0.04}$ & 1600     &    11.4$_{-1.3}^{+1.3}$ & 0.64$_{-0.02}^{+0.02}$  & $\cdots$ & 0.06      & 0.10$_{-0.01}^{+0.00}$        & 0.19 & 0.14 & 0.04 & 0.01 & 0.14$_{-0.00}^{+0.01}$ & 0.09$_{-0.00}^{+0.01}$ \\ [2pt]
 {\bf BAT99-132} & WN3--7s & NE & 49.6$\pm$1.0  & 0.9$_{-0.1}^{+0.0}$        & 5.57$_{-0.04}^{+0.04}$ & g & --11.69$_{-0.04}^{+0.05}$  & 1900    &    15.8$_{-2.2}^{+2.2}$ & 1.10$_{-0.06}^{+0.05}$ & \ddag0.64$_{-0.02}^{+0.03}$ & 0.06    & 0.11$_{-0.00}^{+0.01}$        & 0.16 & 0.14 & 0.04 & 0.01 & 0.18$_{-0.01}^{+0.01}$ & 0.11$_{-0.00}^{+0.01}$ \\ [2pt]
 BAT99-133 & WN9--11 & AR+SC & 49.6$\pm$1.0  & 0.4$_{-0.0}^{+0.1}$ & 5.68$_{-0.02}^{+0.02}$ & g & --12.98$_{-0.05}^{+0.04}$ & \phantom{0}200  & \phantom{0}0.5$_{-0.1}^{+0.0}$ &  0.03$_{-0.01}^{+0.00}$ &$\cdots$    &  0.00        & 1.5$_{-0.1}^{+0.0}$      & 1.7$_{-0.0}^{+0.0}$  & 0.00 & 0.00 & 2.2$_{-0.1}^{+0.0}$ & 5.4$_{-0.1}^{+0.1}$ & $\cdots$\\ [2pt]
 {\bf BAT99-134} & WN3--7s & NE & 49.6$\pm$1.0  & 0.2$\pm$0.0         & 5.50$_{-0.02}^{+0.02}$ & g & --11.62$_{-0.05}^{+0.04}$ & 2100      & \phantom{0}9.0$_{-0.9}^{+0.9}$ & 0.73$_{-0.01}^{+0.02}$ & \ddag0.57$_{-0.01}^{+0.01}$ & 0.04   & 0.08       & 0.17 & 0.12$_{-0.00}^{+0.01}$ & 0.06 & 0.02 & 0.15$_{-0.00}^{+0.01}$ & 0.05$_{-0.00}^{+0.01}$ \\ [2pt]
  {\bf LMC 170--2}   & WN2--5w & MM & 49.6$\pm$1.0  & 0.4$\pm$0.0         & 5.68$_{-0.02}^{+0.03}$ & n & --13.45$_{-0.05}^{+0.04}$ & 1900         & \phantom{0}0.16$_{-0.02}^{+0.02}$ & 0.01$_{-0.00}^{+0.00}$ & 0.00          & 0.01    & 0.00       & 0.38 & 0.04 & 0.01 & 0.00 & 0.19 & 0.00 \\ [2pt]
{\bf VFTS 682} & WN5--7h & VX & 49.6$\pm$1.0  & 4.6$\pm$0.5 &  6.46$_{-0.18}^{+0.19}$ & m & --13.21$_{-0.05}^{+0.04}$ & 1800 &              22.2$_{-8.8}^{+14.0}$ & 0.20$_{-0.02}^{+0.02}$ & 0.50$_{-0.06}^{+0.07}$    & 0.18$_{-0.02}^{+0.01}$ & 0.17$_{-0.01}^{+0.01}$        & 0.10 & 0.08 & 0.04 & 0.00 & 0.46$_{-0.05}^{+0.06}$ & 0.17$_{-0.02}^{+0.02}$ \\        [2pt]
 {\bf AB1}           & WN2--5w  & NE & 61.2$\pm$1.2  & 0.8$\pm$0.1        & 6.08$_{-0.03}^{+0.04}$     & o & --12.92$_{-0.04}^{+0.04}$ & 1500 & \phantom{0}1.4$_{-0.2}^{+0.2}$ & 0.03$_{-0.00}^{+0.00}$ & $\cdots$   & 0.06 & 0.00         & 0.26 & 0.10  & 0.00 & 0.00 & 0.38$_{-0.01}^{+0.01}$ & 0.00 \\ [2pt]
 {\bf AB2}          & WN2--5w   & NE & 61.2$\pm$1.2  & 0.4$_{-0.1}^{+0.0}$        & 5.58$_{-0.02}^{+0.03}$     & o & --12.76$_{-0.05}^{+0.04}$ & 1000 & \phantom{0}1.2$_{-0.2}^{+0.1}$ & 0.08$_{-0.00}^{+0.00}$ & $\cdots$   & 0.25 & 0.00         & 0.00 & 0.05  & 0.05 & 0.00 & 0.46$_{-0.00}^{+0.01}$ & 0.22 \\ [2pt]
 AB3        & WN2--5w+O & CS      & 61.2$\pm$1.2 & 0.6$_{-0.1}^{+0.0}$        & 5.94$_{-0.02}^{+0.03}$     & p & --12.36$_{-0.04}^{+0.04}$ & 1600 & \phantom{0}3.6$_{-0.4}^{+0.5}$ & 0.11$_{-0.01}^{+0.00}$ & 0.58$_{-0.01}^{+0.02}$         & 0.05 & 0.03         & 0.23 & $\cdots$ & $\cdots$ & $\cdots$ &$\cdots$ &$\cdots$\\ [2pt]
 {\bf AB4}        & WN6--8        & AR  & 61.2$\pm$1.2  & 0.3$_{-0.0}^{+0.1}$       & 5.79$_{-0.02}^{+0.03}$      & o & --11.94$_{-0.05}^{+0.04}$   & \phantom{0}900  & \phantom{0}7.4$_{-0.8}^{+0.8}$ & 0.31$_{-0.01}^{+0.01}$ & \ddag0.28$_{-0.01}^{+0.00}$ & 0.13 & 0.12 & 0.06 & 0.10 & 0.02 & 0.00 & 0.24 & 0.07 \\ [2pt]
 {\bf AB5}        & 2$\times$WN6--8 & HS & 61.2$\pm$1.2 & 0.2$_{-0.0}^{+0.1}$      & 6.49$_{-0.02}^{+0.02}$      & q,r & --11.09$_{-0.05}^{+0.04}$ & 1300     &  47.0$_{-4.9}^{+5.0}$ & 0.39$_{-0.01}^{+0.01}$ & 0.18$_{-0.00}^{+0.01}$           & 0.04  & 0.08 & 0.01 & 0.10 & 0.04 & 0.04 & 0.22$_{-0.01}^{+0.00}$ & 0.06 \\ [2pt]
AB6        & WN2--5w+O   & AR+CS  & 61.2$\pm$1.2  & 0.2$\pm$0.0       & 6.29$_{-0.02}^{+0.02}$      & p & --12.35$_{-0.05}^{+0.04}$ & 1900 & \phantom{0}2.5$_{-0.3}^{+0.3}$ & 0.03$_{-0.00}^{+0.00}$ & $\cdots$ & 0.00  & 0.00 & 0.21 & $\cdots$ & $\cdots$ & $\cdots$ &$\cdots$&$\cdots$ \\ [2pt]
 {\bf AB7}        &WN2--5w+O   & AR & 61.2$\pm$1.2  &  0.2$_{-0.0}^{+0.1}$      & 6.11$_{-0.02}^{+0.02}$      & p & --12.20$_{-0.05}^{+0.04}$ & 1700  & \phantom{0}3.7$_{-0.4}^{+0.4}$ & 0.07$_{-0.00}^{+0.01}$ & $\cdots$ & 0.00 & 0.00 & 0.00 & 0.05 & 0.00 & 0.00 & 0.13$_{-0.00}^{+0.01}$ & 0.00 \\ [2pt]
 {\bf AB9}        & WN2--5w      & NE & 61.2$\pm$1.2  & 0.3$_{-0.0}^{+0.1}$     & 6.06$_{-0.02}^{+0.03}$       & o & --12.96$_{-0.05}^{+0.04}$ & 1700   & \phantom{0}0.7$_{-0.1}^{+0.1}$ & 0.02$_{-0.00}^{+0.00}$ & $\cdots$ & 0.00  & 0.00 & 0.19 & 0.07 & 0.00 & 0.00 & 0.18 & 0.00\\ [2pt]
 {\bf AB10}      & WN2--5w      & NE & 61.2$\pm$1.2  & 0.5$_{-0.1}^{+0.0}$      & 5.66$_{-0.02}^{+0.03}$      & o & --13.02$_{-0.04}^{+0.04}$ & 1900   & \phantom{0}0.7$_{-0.1}^{+0.1}$ & 0.04$_{-0.00}^{+0.00}$ & $\cdots$ & 0.02 & 0.00  & 0.35  & 0.06 & 0.00  & 0.00  & $\diamond$0.33 & 0.00 \\ [2pt]
 {\bf AB11}       & WN2--5w     & NE & 61.2$\pm$1.2  & 0.4$_{-0.0}^{+0.1}$     & 5.86$_{-0.02}^{+0.03}$       & o & --13.15$_{-0.04}^{+0.04}$ & 1700   & \phantom{0}0.5$_{-0.1}^{+0.1}$ & 0.02$_{-0.00}^{+0.00}$ & $\cdots$ & 0.00 & 0.00 & 0.36 & 0.02 & 0.03 & 0.00 & 0.20$_{-0.00}^{+0.01}$ & 0.00 \\ [2pt]
 {\bf AB12}      & WN2--5w      & NE & 61.2$\pm$1.2 & 0.5$_{-0.0}^{+0.1}$     & 5.91$_{-0.02}^{+0.03}$       & o & --13.06$_{-0.05}^{+0.04}$ & 1700  &\phantom{0}0.7$_{-0.1}^{+0.1}$ & 0.02$_{-0.00}^{+0.00}$ & $\cdots$ & 0.00 & 0.00 & 0.13 & 0.05  & 0.00 & 0.00 & 0.14$_{-0.01}^{+0.00}$ &0.00 \\ [2pt]
\hline
\end{tabular}
\end{footnotesize}
\end{center}
\footnotesize{
a: \citet{2006A&A...457.1015H}; 
b: \citet{2006A&A...449..711C}; 
c: \citet{1995A&A...293..427C}; 
d: \citet{1994A&A...281..833S}; 
e: \citet{2010MNRAS.408..731C}, 
f: \citet{1999ApJ...511..374B} 
g: \citet{2014A&A...565A..27H}; 
h: \citet{2019A&A...627A.151S}; 
i: \citet{1997A&A...320..500C}; 
j: \citet{2022A&A...663A..36B}; 
k:  \citet{2019MNRAS.484.2692T}; 
l: \citet{2021A&A...650A.147S}
m: Rubio-Diez et al. (in prep); 
n: \citet{2017ApJ...841...20N}; 
o:  \citet{2015A&A...581A..21H}; 
p: \citet{2016A&A...591A..22S}; 
q: \citet{2014AJ....148...62K}; 
r: \citet{2019MNRAS.486..725H} \\
Note: \# CASPEC observations \citep{1997A&A...321..268S}; \ddag: SIT observations \citep{1988AJ.....96.1076T}; $\dag$: MUSE/NFM observations \citep{2021Msngr.182...50C}; $\diamond$: CTIO/R-C spectrograph observations \citep{2003MNRAS.338..360F}
}
\end{table}
\end{landscape}

\begin{landscape}
\begin{table}
\caption{Luminosities of prominent optical emission lines of Galactic(WR) and LMC (BAT99) WN/C, categorised as categorised as either stong-lined (WN3--7s/C) or early/late weak-lined (WN2--5w/C or WN6--8/C) for consistency with WN stars. He\,{\sc ii} $\lambda$4686 FWHM are also provided in km\,s$^{-1}$ ($\pm$100 km\,s$^{-1}$). The feature at $\lambda$4100 involves N\,{\sc iii} $\lambda\lambda$4097,4103, Si\,{\sc iv} $\lambda\lambda$4088,4116, He\,{\sc ii} $\lambda$4100+H$\delta$, while the feature at $\lambda\lambda$4630 involves N\,{\sc v} $\lambda\lambda$4603,20, N\,{\sc iii} $\lambda\lambda$4634,41 and C\,{\sc iii} $\lambda\lambda$4647,51. Stars included in spectral templates are indicated in bold.}
\label{WNC-all}
\begin{center}
\begin{footnotesize}
 \begin{tabular}{l@{\hspace{1mm}}l@{\hspace{1.5mm}}l@{\hspace{1.5mm}}l@{\hspace{1.5mm}}l@{\hspace{1.5mm}}l@{\hspace{0mm}}c@{\hspace{1mm}}r@{\hspace{1mm}}c@{\hspace{0mm}}c@{\hspace{1mm}}c
 @{\hspace{1.5mm}}l@{\hspace{1mm}}l@{\hspace{1mm}}l@{\hspace{1mm}}l@{\hspace{1mm}}l@{\hspace{1mm}}l@{\hspace{1mm}}l@{\hspace{1mm}}l@{\hspace{1mm}}l}   %
\hline
Star & Category & Data & $d$ & $A_{\rm V}$  & $\log$  & Ref & $\log F_{\rm HeII~4686}$ & HeII 4686 & \multicolumn{2}{c}{$L_{\rm HeII~4686}$} 
& $\underline{L_{\rm NIV~3478,85}}$ 
& $\underline{L_{\rm NIV~4058}}$ 
& $\underline{L_{4100}}$ 
& $\underline{L_{4630}}$ 
& $\underline{L_{\rm HeII~5412}}$ 
& $\underline{L_{\rm CIII~5696}}$ 
&$\underline{L_{\rm CIV~5801,12}}$ 
& $\underline{L_{\rm HeII~6560}}$ 
& $\underline{L_{\rm NIV~7103,29}}$ \\  
        &    &     ID         &   kpc     & mag              & $L_{\rm Bol}/L_{\odot}$        &        & erg\,s$^{-1}$\,cm$^{-2}$ & FWHM  & 10$^{35}$ erg\,s$^{-1}$ &  $10^{-3} L_{\rm Bol}$   & $L_{\rm HeII~4686}$ &  $L_{\rm HeII~4686}$  & $L_{\rm HeII~4686}$ & $L_{\rm HeII~4686}$  & $L_{\rm HeII~4686}$ &  $L_{\rm HeII~4686}$ &  $L_{\rm HeII~4686}$ &
 $L_{\rm HeII~4686}$  & $L_{\rm HeII~4686}$ \\
\hline
{\bf WR8} &  WN6--8/C  & CS         & \phantom{0}3.5$\pm$0.2 & 2.7$\pm$0.3 & 5.61$_{-0.12}^{+0.11}$        & a              & --10.64$_{-0.05}^{+0.04}$ & 1300 & \phantom{1}6.9$_{-2.0}^{+2.6}$ & 0.45$_{-0.06}^{+0.06}$      & 0.82$_{-0.06}^{+0.07}$ & 0.91$_{-0.04}^{+0.04}$ & 0.87$_{-0.03}^{+0.04}$                               & 3.36$_{-0.01}^{+0.01}$ & 0.13$_{-0.00}^{+0.01}$ & 0.03$_{-0.00}^{+0.01}$ & 0.67$_{-0.04}^{+0.06}$ & 0.15$_{-0.01}^{+0.02}$ & $\cdots$ \\ [2pt]
{\bf WR26}& WN3--7s/C   & CS         & \phantom{0}6.6$_{-0.6}^{+0.8}$ & 4.7$\pm$0.5  & 5.72$_{-0.21}^{+0.21}$     & a              & --12.07$_{-0.04}^{+0.04}$ & 2400 & \phantom{1}8.8$_{-3.8}^{+6.3}$ & 0.44$_{-0.10}^{+0.13}$       & 0.64$_{-0.08}^{+0.09}$  & \multicolumn{2}{c}{--- 0.68$_{-0.05}^{+0.04}$ ---} & 3.09$_{-0.01}^{+0.01}$ & 0.21$_{-0.02}^{+0.02}$ & 0.00 & 2.28$_{-0.27}^{+0.32}$ & 0.17$_{-0.03}^{+0.04}$ & $\cdots$ \\ [2pt]
{\bf WR58} & WN3--7s/C  & CS         & \phantom{0}6.7$_{-0.4}^{+0.5}$ & 1.9$\pm$0.2        & 5.01$_{-0.11}^{+0.10}$     & a        & --11.13$_{-0.04}^{+0.04}$ & 2200 & \phantom{1}3.6$_{--0.8}^{+1.0}$ & 0.93$_{-0.12}^{+0.14}$        & 0.67$_{-0.04}^{+0.04}$      & 0.17$_{-0.00}^{+0.01}$      & 0.21$_{-0.01}^{+0.01}$                             & 0.28 & 0.15$_{-0.00}^{+0.01}$ & 0.00 & 0.26$_{-0.01}^{+0.02}$ & 0.21$_{-0.02}^{+0.02}$ & $\cdots$ \\ [2pt]
{\bf WR98} & WN6--8/C & CS          & \phantom{0}2.2$\pm$0.1 & 5.0$\pm$0.5 & 5.69$_{-0.21}^{+0.20}$     & a        & --11.80$_{-0.04}^{+0.04}$ & 1100 & \phantom{1}3.1$_{-1.4}^{+2.5}$ & 0.17$_{-0.03}^{+0.04}$        & $\cdots$& $\cdots$ & $\cdots$                      & 2.29$_{-0.01}^{+0.01}$ & 0.12$_{-0.01}^{+0.01}$ & 0.18$_{-0.03}^{+0.03}$ & 0.30$_{-0.04}^{+0.06}$ & 0.11$_{-0.02}^{+0.04}$ & $\cdots$ \\ [2pt]
{\bf WR126} & WN2--5w/C & KI         & 11.3$_{-1.0}^{+1.1}$           & 3.6$_{-0.4}^{+0.3}$       & 6.25$_{-0.17}^{+0.16}$     & a               & --12.20$_{-0.05}^{+0.04}$ & 1700 & \phantom{1}5.2$_{-1.9}^{+2.7}$ & 0.08$_{-0.02}^{+0.02}$  & $\cdots$ &  \multicolumn{2}{c}{--- 0.73$_{-0.04}^{+0.04}$  ---} & 3.93$_{-0.02}^{+-0.01}$ & 0.30$_{-0.02}^{+0.02}$ & 0.00 & 3.97$_{-0.38}^{+0.41}$ & 0.33$_{-0.05}^{+0.05}$ & 0.34$_{-0.06}^{+0.06}$ \\ [2pt]
{\bf WR145} &WN6--8/C+O& WI94        & \phantom{0}1.6$\pm$0.0 & 7.0$\pm$0.7        & 5.71$_{-0.28}^{+0.28}$     & a    & --11.94$_{-0.05}^{+0.04}$ & 1300 & \phantom{1}9.5$_{-5.3}^{+11.4}$ & 0.48$_{-0.09}^{+0.10}$   & $\cdots$& $\cdots$ & $\cdots$                  & 0.82$_{-0.00}^{+0.01}$ & 0.13$_{-0.02}^{+0.02}$ & 0.01 & 0.63$_{-0.11}^{+0.14}$ & $\cdots$ & $\cdots$ \\ [2pt]
{\bf WR153} &WN6--8/C+O& II91         & \phantom{0}4.5$\pm$0.3 & 2.1$_{-0.2}^{+0.3}$ & $\cdots$ & b & --10.45$_{-0.09}^{+0.08}$ & 1400 & \phantom{1}9.7$_{-2.9}^{+3.4}$ & $\cdots$ & $\cdots$& $\cdots$ & $\cdots$        & 0.52 & 0.22$_{-0.01}^{+0.01}$ & 0.00 & 0.53$_{-0.03}^{+0.03}$ & 0.28$_{-0.02}^{+0.02}$ & 0.20$_{-0.02}^{+0.02}$ \\ [2pt]
{\bf BAT99-36} & WN3--7s/C & AR      & 49.6$\pm$1.0                    & 0.5$_{-0.1}^{+0.0}$        & 5.70$_{-0.02}^{+0.03}$     & c                  & --11.73$_{-0.04}^{+0.05}$ & 2100 & \phantom{1}9.5$_{-1.1}^{+1.1}$ & 0.49$_{-0.02}^{+0.02}$     & \ddag0.60$_{-0.01}^{+0.01}$ & 0.03 & 0.08                       & 0.24 & 0.12 & 0.00 & 1.17$_{-0.01}^{+0.02}$ & $\cdots$ & $\cdots$ \\ [2pt]
{\bf BAT99-88} & WN3--7s/C & NE     & 49.6$\pm$1.0                    & 3.2$_{-0.4}^{+0.3}$       & 5.79$_{-0.12}^{+0.13}$     & c                  & --12.97$_{-0.05}^{+0.04}$ & 2900 & 10.3$_{-3.2}^{+4.5}$ & 0.43$_{-0.04}^{+0.05}$   & $\cdots$    & 0.00 & 0.09$_{-0.01}^{+0.00}$                       & 0.17 & 0.11$_{-0.01}^{+0.01}$ & 0.00 & 1.52$_{-0.12}^{+0.12}$ & 0.15$_{-0.02}^{+0.02}$ & 0.05$_{-0.01}^{+0.01}$ \\ [2pt]
\hline
\end{tabular}
\end{footnotesize}
\end{center}
\footnotesize{a: \citet{2012A&A...540A.144S}, b: \citet{2006A&A...449..711C}, c: \citet{2014A&A...565A..27H} } \\
Note: \ddag: SIT observations \citep{1988AJ.....96.1076T}

\end{table}
\end{landscape}

\begin{landscape}
\begin{table}
\caption{Luminosities of prominent optical emission lines of Galactic (WR) and LMC (BAT99) WC stars, categorised as early-type (WC4--5), mid-type (WC6--7) or late-type (WC8--9) stars. C\,{\sc iv} $\lambda\lambda$5801,12 FWHM are also provided in km\,s$^{-1}$ ($\pm$100 km\,s$^{-1}$). The feature at 6559--81 involves He\,{\sc ii} 6560 and C\,{\sc ii} $\lambda\lambda$6559,81. Stars included in spectral templates are indicated in bold (the remainder are excluded owing to limited spectral coverage)}
\label{WC-all}
\begin{center}
\begin{footnotesize}
  \begin{tabular}{l@{\hspace{1mm}}l@{\hspace{1.5mm}}l@{\hspace{1.5mm}}l@{\hspace{1.5mm}}l@{\hspace{2mm}}l@{\hspace{-1mm}}r@{\hspace{1mm}}r@{\hspace{1mm}}c@{\hspace{-1mm}}c@{\hspace{1mm}}c
 @{\hspace{1.5mm}}l@{\hspace{1mm}}l@{\hspace{1mm}}l@{\hspace{1mm}}l@{\hspace{1mm}}l@{\hspace{1mm}}l@{\hspace{1mm}}l@{\hspace{1mm}}l@{\hspace{1mm}}l}   %
\hline
WR & Category & Data & $d$ & $A_{\rm V}$  & $\log$  & Ref & $\log F_{\rm CIV~5801,12}$ & CIV 5801,12 & \multicolumn{2}{c}{$L_{\rm CIV~5801,12}$} 
& $\underline{L_{\rm OIV~3403,13}}$ 
& $\underline{L_{\rm CIII~4647,51}}$ 
& $\underline{L_{\rm HeII~4686}}$ 
& $\underline{L_{\rm CIII~5696}}$ 
&$\underline{L_{\rm HeI~5876}}$ 
& $\underline{L_{\rm 6559-81}}$ 
& $\underline{L_{\rm CIII~6727,73}}$ 
& $\underline{L_{\rm CIV~7725}}$ 
& $\underline{L_{\rm CIII~9701,19}}$  \\  
        &    &    {\bf ID}          &   kpc     & mag              & $L_{\rm Bol}/L_{\odot}$        &        & erg\,s$^{-1}$\,cm$^{-2}$ & FWHM & 10$^{35}$ erg\,s$^{-1}$ &  $10^{-3} L_{\rm Bol}$   & $L_{\rm 5801,12}$ &  $L_{\rm 5801,12}$  & $L_{\rm 5801,12}$ & $L_{\rm 5801,12}$  & $L_{\rm 5801-12}$ &  $L_{\rm 5801,12}$ & $L_{\rm 5801,12}$ &
 $L_{\rm 5801,12}$  & $L_{\rm 5801,12}$ \\
\hline
{\bf WR4}  & WC4--5  & WI02       & \phantom{0}2.6$_{-0.1}^{+0.2}$ & 2.3$_{-0.3}^{+0.2}$         & 5.41$_{-0.10}^{+0.10}$    & a      & --9.72$_{-0.05}^{+0.04}$ &  2200 &           10.9$_{-2.3}^{+2.6}$ & 1.10$_{-0.08}^{+0.08}$     & 1.01$_{-0.13}^{+0.14}$      & \multicolumn{2}{c}{--- 2.60$_{-0.16}^{+0.16}$ ---}    & 0.07 & 0.06 & 0.06 & 0.11$_{-0.00}^{+0.01}$ & 0.10$_{-0.01}^{+0.00}$ & $\cdots$ \\ [2pt]
{\bf WR5}  & WC6--7  & WI02        & \phantom{0}2.8$_{-0.1}^{+0.2}$ & 3.2$\pm$0.3         & 5.49$_{-0.13}^{+0.14}$    & a      & --9.95$_{-0.04}^{+0.04}$ & 1900 &           16.9$_{-4.5}^{+5.7}$ & 1.41$_{-0.09}^{+0.11}$     & 0.87$_{-0.15}^{+0.18}$      & 2.58$_{-0.22}^{+0.23}$    & 0.38$_{-0.03}^{+0.04}$                           & 0.09$_{-0.00}^{+0.01}$ & 0.08 & 0.08 & 0.13$_{-0.01}^{+0.01}$ & 0.08$_{-0.00}^{+0.01}$ & $\dagger$0.17$_{-0.02}^{+0.03}$  \\ [2pt]
{\bf WR9}  & WC4--5+O& CS          & \phantom{0}3.5$\pm$0.2 & 3.9$_{-0.4}^{+0.4}$ & $\cdots$& b &--10.19$_{-0.05}^{+0.04}$ & 3200 &        27.0$_{-8.2}^{+11.2}$ &$\cdots$& $\cdots$ &\multicolumn{2}{c}{--- 2.70$_{-0.27}^{+0.31}$ ---} & 0.00 & 0.00 & 0.10$_{-0.00}^{+0.01}$ & 0.15$_{-0.01}^{+0.00}$ & $\cdots$ & $\cdots$ \\ [2pt]
{\bf WR13} & WC6--7  & CS          & \phantom{0}4.3$_{-0.2}^{+0.3}$ & 4.5$_{-0.4}^{+0.5}$         & 5.45$_{-0.18}^{+0.19}$    & a      &--10.92$_{-0.05}^{+0.04}$ & 2000 &           13.1$_{-4.5}^{+6.4}$ & 1.20$_{-0.07}^{+0.09}$   & $\cdots$ & \multicolumn{2}{c}{--- 2.49$_{-0.29}^{+0.34}$ ---}  & 0.08 & 0.08 & 0.10$_{-0.01}^{+0.00}$ & 0.15$_{-0.01}^{+0.01}$ & $\cdots$ & $\cdots$ \\ [2pt]
{\bf WR14} & WC6--7  & AD          & \phantom{0}1.7$_{-0.0}^{+0.1}$ & 2.4$_{-0.2}^{+0.3}$         & 5.56$_{-0.10}^{+0.10}$    & a      & --9.46$_{-0.04}^{+0.05}$ & 1900 &           10.3$_{-2.2}^{+2.6}$ & 0.73$_{-0.03}^{+0.03}$   & 0.75$_{-0.10}^{+0.12}$     & 2.73$_{-0.18}^{+0.19}$    & 0.66$_{-0.05}^{+0.04}$                          & 0.32 & 0.12 & 0.16$_{-0.01}^{+0.00}$ & 0.19$_{-0.01}^{+0.00}$ & 0.12$_{-0.01}^{+0.00}$ & 0.33$_{-0.03}^{+0.04}$ \\ [2pt]
{\bf WR15} & WC6--7  & CS          & \phantom{0}2.5$\pm$0.1 & 4.6$_{-0.4}^{+0.5}$         & 5.83$_{-0.18}^{+0.19}$    & a      &--10.07$_{-0.04}^{+0.04}$ & 2900 &           33.4$_{-11.5}^{+16.6}$ & 1.27$_{-0.05}^{+0.05}$   & $\cdots$ & \multicolumn{2}{c}{--- 2.37$_{-0.27}^{+0.33}$ ---}     & 0.17 & 0.18 & 0.10$_{-0.00}^{+0.01}$ & 0.12$_{-0.01}^{+0.01}$ & $\cdots$ & $\cdots$ \\ [2pt]
{\bf WR17} & WC4--5  & AD          & \phantom{0}5.0$_{-0.4}^{+0.5}$ & 1.2$_{-0.2}^{+0.1}$         & 5.30$_{-0.09}^{+0.09}$    & a      & --9.97$_{-0.05}^{+0.04}$ & 1900 & \phantom{0}8.6$_{-1.4}^{+1.5}$ & 1.12$_{-0.08}^{+0.08}$ & 0.36$_{-0.02}^{+0.03}$  & 2.03$_{-0.07}^{+0.06}$    & 0.51$_{-0.01}^{+0.02}$                    & 0.03 & 0.07 & 0.08$_{-0.01}^{+0.00}$ & 0.11 & 0.08 & 0.13$_{-0.01}^{+0.00}$ \\ [2pt]
{\bf WR19} & WC4--5+O& AD          & \phantom{0}4.7$\pm$0.3 & 5.1$_{-0.5}^{+0.5}$        & $\cdots$& b &--11.11$_{-0.04}^{+0.04}$ & 4000 & 16.3$_{-6.1}^{+9.2}$ &$\cdots$& 0.74$_{-0.19}^{+0.26}$     & \multicolumn{2}{c}{--- 1.98$_{-0.26}^{+0.30}$  ---} & 0.00 & 0.00 & 0.03 & 0.08$_{-0.01}^{+0.01}$ & 0.08$_{-0.01}^{+0.01}$ & 0.18$_{-0.04}^{+0.05}$ \\    [2pt]
{\bf WR23} & WC6--7  & AD          & \phantom{0}2.3$\pm$0.1 & 2.3$_{-0.2}^{+0.3}$         & 5.51$_{-0.10}^{+0.09}$    & a      & --9.49$_{-0.05}^{+0.04}$ & 2100 &           15.4$_{-3.2}^{+3.8}$ & 1.24$_{-0.05}^{+0.06}$      & 0.79$_{-0.09}^{+0.11}$     & 2.50$_{-0.14}^{+0.15}$    & 0.65$_{-0.03}^{+0.04}$                   & 0.23 & 0.13 & 0.13$_{-0.01}^{+0.00}$ & 0.17$_{-0.00}^{+0.01}$ & 0.12$_{-0.01}^{+0.01}$ & 0.29$_{-0.03}^{+0.03}$ \\ [2pt]
{\bf WR27} & WC6--7  & CS           & \phantom{0}2.3$_{-0.0}^{+0.1}$ & 6.3$_{-0.7}^{+0.6}$  & 5.17$_{-0.26}^{+0.25}$     & a      & --11.29$_{-0.04}^{+0.04}$ & 2200 & \phantom{0}7.8$_{-3.4}^{+5.8}$ & 1.39$_{-0.05}^{+0.05}$ & 0.71$_{-0.11}^{+0.26}$ &2.63$_{-0.36}^{+0.40}$ & 0.45$_{-0.06}^{+0.06}$                     & 0.14$_{-0.01}^{+0.00}$ & 0.07 & 0.11$_{-0.00}^{+0.01}$ & 0.19$_{-0.01}^{+0.02}$ & $\cdots$ & $\cdots$ \\ [2pt]
{\bf WR30} & WC6--7+O& CS         & \phantom{0}5.9$\pm$0.5 & 2.1$_{-0.2}^{+0.3}$ & $\cdots$ & b & --10.48$_{-0.04}^{+0.04}$ & 2100 & \phantom{0}8.8$_{-1.9}^{+2.1}$ & $\cdots$ & 0.28$_{-0.03}^{+0.04}$ & 1.84$_{-0.10}^{+0.12}$   & 0.78$_{-0.04}^{+0.04}$          & 0.07 & 0.08 & 0.13$_{-0.01}^{+0.00}$ & 0.13$_{-0.01}^{+0.00}$ & $\cdots$ & $\cdots$ \\ [2pt]
{\bf WR33} & WC4--5    & CS         & \phantom{0}6.6$_{-0.7}^{+0.9}$ & 2.3$_{-0.3}^{+0.2}$       & 5.31$_{-0.13}^{+0.14}$ & a            & --10.34$_{-0.05}^{+0.04}$ & 3400 &                   16.4$_{-3.7}^{+4.4}$ & 2.08$_{-0.22}^{+0.25}$ & $\cdots$& \multicolumn{2}{c}{--- 2.50$_{-0.15}^{+0.16}$ ---}  & 0.00 & 0.00 &  0.10 & 0.13 & $\cdots$ & $\cdots$ \\ [2pt]
{\bf WR38} & WC4--5    & AD       & \phantom{0}6.3$_{-0.7}^{+0.9}$ &  4.2$_{-0.5}^{+0.4}$       & 5.14$_{-0.20}^{+0.20}$ & a            & --11.32$_{-0.04}^{+0.05}$ & 3300   &\phantom{0}8.6$_{-2.9}^{+4.0}$ & 1.62$_{-0.22}^{+0.27}$ & 0.38$_{-0.08}^{+0.11}$ & \multicolumn{2}{c}{--- 1.70$_{-0.19}^{+0.20}$ --- } & 0.00 & 0.00 & 0.02 & 0.05 & 0.06$_{-0.01}^{+0.01}$ & 0.04$_{-0.00}^{+0.01}$ \\  [2pt]
WR39 & WC6--7+O & CS       & \phantom{0}4.4$_{-1.0}^{+1.4}$ & 6.4$_{-0.7}^{+0.6}$      & 6.12$_{-0.33}^{+0.35}$ & a & --12.05$_{-0.05}^{+0.04}$ & 2900 & \phantom{0}5.1$_{-2.5}^{+4.1}$ & 0.10$_{-0.02}^{+0.04}$ & $\cdots$ & \multicolumn{2}{c}{--- 3.61$_{-0.58}^{+0.69}$ --- }  & 0.42$_{-0.01}^{+0.00}$ & 0.25 & 0.32$_{-0.03}^{+0.02}$ & 0.21$_{-0.02}^{+0.02}$ & $\cdots$ & $\cdots$ \\ [2pt]
{\bf WR50} & WC6--7+O  & CS       & \phantom{0}3.4$\pm$0.2 & 3.9$_{-0.4}^{+0.4}$     & 5.39$_{-0.16}^{+0.16}$ & a & --10.82$_{-0.04}^{+0.04}$ & 2400 & \phantom{0}6.1$_{-1.9}^{+2.6}$ & 0.65$_{-0.05}^{+0.05}$ & 0.91$_{-0.18}^{+0.21}$ & \multicolumn{2}{c}{--- 2.98$_{-0.28}^{+0.32}$ ---}  & 0.43 & 0.09 & 0.19$_{-0.00}^{+0.01}$ & 0.16$_{-0.00}^{+0.01}$ & $\cdots$ & $\cdots$ \\ [2pt]
{\bf WR53} & WC8--9   & AD         & \phantom{0}2.9$\pm$0.1 & 2.8$_{-0.3}^{+0.3}$      & 5.14$_{-0.12}^{+0.12}$  & a           & --10.64$_{-0.05}^{+0.04}$  & 1500         & \phantom{0}2.5$_{-0.6}^{+0.7}$ & 0.47$_{-0.03}^{+0.03}$       & 0.84$_{-0.12}^{+0.15}$  & 3.43$_{-0.26}^{+0.27}$ & 1.10$_{-0.08}^{+0.09}$ & 2.60$_{-0.02}^{+0.01}$ & 0.29 & 0.62$_{-0.02}^{+0.02}$ & 0.44$_{-0.02}^{+0.02}$ & 0.14$_{-0.01}^{+0.02}$ & 0.88$_{-0.11}^{+0.12}$ \\   [2pt]
{\bf WR56} & WC6--7   & CS           & \phantom{0}7.5$\pm$0.7 & 2.6$_{-0.2}^{+0.3}$      & 5.00 $_{-0.14}^{+0.13}$ &  a         & --11.42$_{-0.05}^{+0.04}$   & 1700         & \phantom{0}2.4$_{-0.8}^{+1.0}$ & 0.63$_{-0.07}^{+0.07}$      & 1.08$_{-0.16}^{+0.18}$  & 2.95$_{-0.21}^{+0.22}$ & 0.81$_{-0.06}^{+0.06}$ & 0.56 & 0.15 & 0.20$_{-0.01}^{+0.00}$ & 0.18$_{-0.01}^{+0.01}$ & $\cdots$  & $\cdots$ \\ [2pt]
{\bf WR57} & WC8--9   & CS          & \phantom{0}3.0$_{-0.1}^{+0.2}$   & 1.7$_{-0.2}^{+0.1}$     & 5.16$_{-0.07}^{+0.08}$   & a         & --10.06$_{-0.05}^{+0.04}$  & 1700          & \phantom{0}3.9$_{-0.7}^{+0.7}$ & 0.69$_{-0.04}^{+0.05}$     & 3.21$_{-0.26}^{+0.28}$  & 3.37$_{-0.13}^{+0.13}$ & 0.93$_{-0.03}^{+0.04}$ & 1.14$_{-0.01}^{+0.00}$ & 0.18 & 0.32$_{-0.00}^{+0.01}$ & 0.25$_{-0.01}^{+0.00}$ & $\cdots$ & $\cdots$ \\ [2pt]
WR59 & WC8--9   &  CS          & \phantom{0}3.4$\pm$0.2   & 7.6$_{-0.7}^{+0.8}$     & 5.67$_{-0.30}^{+0.31}$   & a         & --12.28$_{-0.04}^{+0.04}$  & 1400          & \phantom{0}5.1$_{-2.5}^{+4.9}$ & 0.28$_{-0.01}^{+0.02}$     & $\cdots$ & 5.55$_{1.04}^{+1.27}$ & 1.54$_{-0.28}^{+0.34}$ & 3.68$_{-0.06}^{+0.05}$ & 0.57$_{-0.00}^{+0.01}$ & 1.21$_{-0.11}^{+0.12}$ & 1.01$_{-0.11}^{+0.12}$ & $\cdots$ & $\cdots$ \\ [2pt]
WR60 & WC8--9  & CS           & \phantom{0}3.4$\pm$0.3   & 5.4$_{-0.5}^{+0.6}$      & 5.73$_{-0.22}^{+0.23}$  & a         & --11.03$_{-0.05}^{+0.04}$  & 1800        &                    13.4$_{-5.3}^{+8.2}$ & 0.64$_{-0.06}^{+0.08}$     & 1.57$_{-0.43}^{+0.59}$ & 2.40$_{-0.34}^{+0.38}$ & 1.26$_{-0.17}^{+0.20}$ & 1.87$_{-0.02}^{+0.02}$ & 0.39 & 0.42$_{-0.03}^{+0.02}$ & 0.28$_{-0.02}^{+0.02}$ & $\cdots$  & $\cdots$ \\ [2pt]
{\bf WR65} & WC8--9+B& CS        & \phantom{0}2.8$\pm$0.2    & 7.5$_{-0.7}^{+0.8}$      & 5.71$_{-0.33}^{+0.34}$  & a         & --12.46$_{-0.04}^{+0.04}$ & 1400          & \phantom{0}2.0$_{-1.0}^{+1.8}$ & 0.10$_{-0.00}^{+0.01}$ & $\cdots$ & 5.56$_{-1.04}^{+1.28}$ & 1.32$_{-0.24}^{+0.29}$ & 4.14$_{-0.07}^{+0.06}$ & 0.86$_{-0.01}^{+0.01}$ & 1.51$_{-0.14}^{+0.15}$ & 0.86$_{-0.09}^{+0.11}$ & $\cdots$ & $\cdots$ \\  [2pt]
{\bf WR68} & WC6--7    & CS        & \phantom{0}5.9$_{-0.5}^{+0.6}$     & 5.3$_{-0.6}^{+0.5}$      & 5.87$_{-0.22}^{+0.22}$ & a         & --11.31$_{-0.04}^{+0.05}$  & 1900          &    19.0$_{-7.3}^{+11.1}$ & 0.67$_{-0.07}^{+0.08}$     & $\cdots$ & 2.55$_{-0.35}^{+0.39}$ & 0.81$_{-0.11}^{+0.12}$ & 0.51$_{-0.00}^{+0.01}$ & 0.26 & 0.20$_{-0.01}^{+0.01}$ & 0.17$_{-0.01}^{+0.02}$ & $\cdots$ & $\cdots$ \\ [2pt]
{\bf WR69} & WC8--9    & CS       & \phantom{0}2.6$\pm$0.1      & 2.0$\pm$0.2      & 5.06$_{-0.09}^{+0.10}$  & a        & --10.63$_{-0.05}^{+0.04}$ & 1500           & \phantom{0}1.1$_{-0.2}^{+0.2}$ & 0.24$_{-0.02}^{+0.02}$     & $\cdots$ & 4.33$_{-0.24}^{+0.26}$ & 1.18$_{-0.06}^{+0.07}$ & 4.48$_{-0.02}^{+0.01}$ & 0.70 & 1.70$_{-0.05}^{+0.04}$ & $\cdots$ & $\cdots$ & $\cdots$ \\ [2pt]
{\bf WR70} & WC8--9+B& CS      & \phantom{0}3.1$_{-0.2}^{+0.1}$       & 4.7$_{-0.4}^{+0.5}$    & $\cdots$ & b & --11.41$_{-0.04}^{+0.04}$ & 1200  & \phantom{0}2.6$_{-0.9}^{+1.3}$ & $\cdots$ & $\cdots$ &3.44$_{-0.42}^{+0.48}$ & 1.19$_{-0.14}^{+0.16}$ & 3.52$_{-0.04}^{+0.03}$ & 0.35$_{-0.01}^{+0.00}$ & 1.65$_{-0.09}^{+0.10}$ & $\cdots$ & $\cdots$ & $\cdots$ \\ [2pt]
{\bf WR73} & WC8--9 & CS         & \phantom{0}6.9$_{-0.8}^{+1.0}$        & 5.7$_{-0.5}^{+0.6}$    & $\cdots$ & b & --12.20$_{-0.05}^{+0.04}$ & 1600 & \phantom{0}4.9$_{-2.0}^{+3.3}$ & $\cdots$ & $\cdots$ & 3.36$_{-0.49}^{+0.57}$ & 0.99$_{-0.14}^{+0.16}$ & 3.20$_{-0.03}^{+0.04}$ & 0.36$_{-0.01}^{+0.00}$ & 1.10$_{-0.07}^{+0.09}$ & 0.49$_{-0.04}^{+0.04}$ & $\cdots$ & $\cdots$ \\ [2pt]
{\bf WR77} & WC8--9+O & CS    & \phantom{0}3.1$\pm$0.2       & 3.4$_{-0.3}^{+0.4}$ & $\cdots$ & b & --11.49$_{-0.05}^{+0.04}$ & 1600      & \phantom{0}0.7$_{-0.2}^{+0.3}$ & $\cdots$ & $\cdots$ & 1.81$_{-0.17}^{+0.17}$ & 0.87$_{-0.07}^{+0.09}$ & 2.15$_{-0.01}^{+0.01}$ & 0.25 & 0.60$_{-0.02}^{+0.03}$ & 0.47$_{-0.02}^{+0.02}$ & $\cdots$ & $\cdots$ \\ [2pt]
{\bf WR80} & WC8--9+O  & CS    & \phantom{0}3.8$\pm$0.3        & 7.0$_{-0.7}^{+0.7}$   & 5.23$_{-0.29}^{+0.29}$ & a           & --12.19$_{-0.05}^{+0.04}$ & 1500           & \phantom{0}4.4$_{-2.1}^{+3.7}$ & 0.68$_{-0.06}^{+0.07}$     & $\cdots$ & 3.58$_{-0.61}^{+0.73}$ & 1.06$_{-0.18}^{+0.20}$ & 3.77$_{-0.05}^{+0.06}$ & 0.49$_{-0.01}^{+0.00}$ & 1.31$_{-0.11}^{+0.12}$ & 1.00$_{-0.09}^{+0.11}$ & $\cdots$ & $\cdots$ \\ [2pt]
{\bf WR81} & WC8--9     & CS   & \phantom{0}2.5$\pm$0.1        & 6.0$_{-0.6}^{+0.6}$ & 5.38$_{-0.24}^{+0.24}$     & a          & --11.56$_{-0.05}^{+0.04}$  & 1500          & \phantom{0}3.5$_{-1.5}^{+2.4}$ & 0.38$_{-0.02}^{+0.02}$     & $\cdots$ & 3.48$_{-0.50}^{+0.59}$ & 1.30$_{-0.18}^{+0.21}$ & 3.90$_{-0.04}^{+0.05}$ & 0.91$_{-0.00}^{+0.01}$ & 1.37$_{-0.10}^{+0.10}$ & 0.51$_{-0.04}^{+0.05}$ & $\cdots$ & $\cdots$ \\ [2pt]
{\bf WR88} & WC8--9     & CS  & \phantom{0}2.7$_{-0.1}^{+0.2}$  & 5.9$_{-0.6}^{+0.6}$ & 5.27$_{-0.24}^{+0.24}$     & a          & --11.86$_{-0.04}^{+0.04}$ & 1400            &  2.1$_{-0.9}^{+1.5}$ & 0.30$_{-0.02}^{+0.02}$     & $\cdots$ & 5.32$_{-0.72}^{+0.72}$ & 1.54$_{-0.20}^{+0.23}$ & 3.23$_{-0.03}^{+0.04}$ & 0.86$_{-0.00}^{+0.01}$ & 0.85$_{-0.06}^{+0.06}$ & 0.35$_{-0.03}^{+0.03}$ & $\cdots$ & $\cdots$ \\ [2pt]
{\bf WR90} & WC6--7     & VU  & \phantom{0}1.3$\pm$0.0      & 1.5$_{-0.1}^{+0.2}$ & 5.67$_{-0.07}^{+0.06}$ &   a          & --8.74$_{-0.04}^{+0.04}$ & 1900   &   13.8$_{-2.2}^{+2.4}$ & 0.77$_{-0.03}^{+0.04}$      & 0.95$_{-0.08}^{+0.08}$ & 2.73$_{-0.11}^{+0.11}$ & 0.94$_{-0.04}^{+0.03}$ & 0.40 & 0.14 & 0.17$_{-0.01}^{+0.00}$ & 0.16$_{-0.00}^{+0.01}$ & 0.11$_{-0.01}^{+0.00}$ & 0.25$_{-0.01}^{+0.02}$ \\   [2pt]
{\bf WR92} & WC8--9    & CS     & \phantom{0}4.5$_{-0.3}^{+0.4}$      & 1.8$_{-0.2}^{+0.2}$ & 5.09$_{-0.10}^{+0.10}$  &  a           & --11.03$_{-0.05}^{+0.04}$ & 1400             & \phantom{0}1.1$_{-0.2}^{+0.2}$ & 0.23$_{-0.02}^{+0.02}$       & $\cdots$ & 5.17$_{-0.27}^{+0.28}$ & 1.62$_{-0.08}^{+0.09}$ & 4.36$_{-0.02}^{+0.01}$ & 0.81$_{-0.00}^{+0.01}$ & 1.86$_{-0.05}^{+0.04}$ & 0.50$_{-0.02}^{+0.01}$ & $\cdots$ & $\cdots$ \\ [2pt]
{\bf WR93} & WC6--7+O & CS   & \phantom{0}1.9$\pm$0.1      & 5.3$_{-0.5}^{+0.6}$ & $\cdots$ & b & --10.43$_{-0.05}^{+0.04}$ & 2000       & \phantom{0}8.4$_{-3.2}^{+4.9}$ & $\cdots$ &$\cdots$ & 2.06$_{-0.28}^{+0.32}$ & 0.99$_{-0.13}^{+0.16}$ & 0.56$_{-0.01}^{+0.00}$ & 0.52$_{-0.01}^{+0.00}$ & 0.20$_{-0.01}^{+0.02}$ & 0.15$_{-0.01}^{+0.01}$ & $\cdots$ & $\cdots$ \\ [2pt]
{\bf WR95} & WC8--9     & CS    & \phantom{0}2.2$_{-0.2}^{+0.1}$       & 6.9$_{-0.6}^{+0.7}$ & 5.22$_{-0.28}^{+0.28}$   &  a         & --11.92$_{-0.05}^{+0.04}$  & 1400              & \phantom{0}2.7$_{-1.3}^{+2.2}$ & 0.41$_{-0.02}^{+0.03}$      & $\cdots$ & 3.98$_{-0.65}^{+0.79}$ & 1.11$_{-0.18}^{+0.20}$ & 3.92$_{-0.05}^{+0.06}$ & 0.54$_{-0.01}^{+0.00}$ & 1.19$_{-0.09}^{+0.10}$ & 1.15$_{-0.11}^{+0.12}$ & $\cdots$ & $\cdots$ \\ [2pt]
{\bf WR103} & WC8--9   & II91   & \phantom{0}2.5$_{-0.1}^{+0.2}$       & 2.0$_{-0.2}^{+0.1}$ & 5.19$_{-0.10}^{+0.09}$    & a          & --10.25$_{-0.10}^{+0.08}$ & 1400            & \phantom{0}2.3$_{-0.6}^{+0.6}$    & 0.38$_{-0.02}^{+0.03}$     & $\cdots$ & 3.43$_{-0.18}^{+0.19}$ & 1.00$_{-0.05}^{+0.06}$ & 3.78$_{-0.01}^{+0.02}$ & 0.83$_{-0.00}^{+0.01}$ & 1.16$_{-0.03}^{+0.03}$ & 0.54$_{-0.02}^{+0.01}$ & $\cdots$ & $\cdots$ \\ [2pt]
{\bf WR106} & WC8--9   & WI02  & \phantom{0}2.9$\pm$0.2       & 4.6$_{-0.4}^{+0.5}$ & 5.10$_{-0.19}^{+0.20}$    & a           & --11.59$_{-0.04}^{+0.05}$ & 1400           & \phantom{0}1.4$_{-0.4}^{+0.6}$    & 0.29$_{-0.02}^{+0.02}$    & 0.43$_{-0.10}^{+0.13}$ & 2.79$_{-0.33}^{+0.39}$ & 0.78$_{-0.09}^{+0.10}$ & 3.66$_{-0.04}^{+0.03}$ & 0.84$_{-0.00}^{+0.01}$ & 1.77$_{-0.10}^{+0.10}$ & 0.95$_{-0.06}^{+0.07}$ & 0.09$_{-0.01}^{+0.01}$ & $\dagger$1.25$_{-0.23}^{+0.29}$ \\  [2pt]
\hline
\end{tabular}
\end{footnotesize}
\end{center}
\end{table}
\end{landscape}

\addtocounter{table}{-1}

\begin{landscape}
\begin{table}
\caption{continued}
\begin{center}
\begin{footnotesize}
 \begin{tabular}{l@{\hspace{1mm}}l@{\hspace{1.5mm}}l@{\hspace{1.5mm}}l@{\hspace{1.5mm}}l@{\hspace{1.5mm}}l@{\hspace{-1mm}}r@{\hspace{1mm}}r@{\hspace{1mm}}c@{\hspace{-1mm}}c@{\hspace{1mm}}c
 @{\hspace{1.5mm}}l@{\hspace{1mm}}l@{\hspace{1mm}}l@{\hspace{1mm}}l@{\hspace{1mm}}l@{\hspace{1mm}}l@{\hspace{1mm}}l@{\hspace{1mm}}l@{\hspace{1mm}}l}   %
\hline
WR & Category & Data & $d$ & $A_{\rm V}$ & $\log$  & Ref & $\log F_{\rm CIV~5801,12}$ & CIV 5801,12 & \multicolumn{2}{c}{$L_{\rm CIV~5801,2}$} 
& $\underline{L_{\rm OIV~3403,13}}$ 
& $\underline{L_{\rm CIII~4647,51}}$ 
& $\underline{L_{\rm HeII~4686}}$ 
& $\underline{L_{\rm CIII~5696}}$ 
& $\underline{L_{\rm HeI~5876}}$ 
& $\underline{L_{\rm 6559-81}}$ 
& $\underline{L_{\rm CIII~6727,73}}$ 
& $\underline{L_{\rm CIV~7725}}$ 
& $\underline{L_{\rm CIII~9701,19}}$  \\  
        &    &       ID       &   kpc     & mag              & $L_{\rm Bol}/L_{\odot}$        &        & erg\,s$^{-1}$\,cm$^{-2}$ & FWHM & 10$^{35}$ erg\,s$^{-1}$ &  $10^{-3} L_{\rm Bol}$   & $L_{\rm 5801,12}$ &  $L_{\rm 5801,12}$  & $L_{\rm 5801,12}$ & $L_{\rm 5801,12}$  & $L_{\rm 5801,12}$ &  $L_{\rm 5801,12}$ & $L_{\rm 5801,12}$ &
 $L_{\rm 5801,12}$  & $L_{\rm 5801,12}$ \\
\hline
{\bf WR111} & WC4--5   & WI02  & \phantom{0}1.3$_{-0.0}^{+0.1}$       & 1.3$_{-0.2}^{+0.1}$ & 5.20$_{-0.07}^{+0.07}$     & a          & --8.86$_{-0.05}^{+0.04}$ & 2200             & \phantom{0}8.7$_{-1.3}^{+1.4}$     & 1.44$_{-0.08}^{+0.09}$   & 0.69$_{-0.05}^{+0.05}$ & \multicolumn{2}{c}{--- 2.64$_{-0.08}^{+0.10}$ ---} & 0.06 & 0.08 & 0.08 & 0.14$_{-0.01}^{+0.00}$ & 0.10$_{-0.01}^{+0.00}$ & $\dagger$0.18$_{-0.01}^{+0.01}$ \\ [2pt]
{\bf WR113} & WC8--9+O& KI  & \phantom{0}1.9$_{-0.0}^{+0.1}$      & 4.1$_{-0.4}^{+0.5}$ & 6.01$_{-0.17}^{+0.16}$     &  a        &--10.47$_{-0.04}^{+0.04}$ & 1500       & \phantom{0}5.5$_{-1.8}^{+2.5}$    & 0.14$_{-0.00}^{+0.01}$ & 1.11$_{-0.20}^{+0.24}$     & 4.05$_{-0.36}^{+0.39}$ & 0.87$_{-0.07}^{+0.08}$ & 2.20$_{-0.01}^{+0.02}$ & 0.29 & 0.75$_{-0.03}^{+0.04}$ & 0.60$_{-0.03}^{+0.03}$ & $\cdots$ & $\cdots$ \\ [2pt]
{\bf WR114} & WC4--5 & KI     & \phantom{0}1.9$\pm$0.1       & 5.1$_{-0.5}^{+0.5}$ & 5.29$_{-0.20}^{+0.21}$    & a          & --10.54$_{-0.05}^{+0.04}$ & 2100         & \phantom{0}9.4$_{-3.5}^{+5.2}$       & 1.25$_{-0.07}^{+0.07}$   & 0.59$_{-0.15}^{+0.20}$          & 2.41$_{-0.32}^{+0.36}$ & 0.49$_{-0.06}^{+0.07}$ & 0.07 & 0.09 & 0.08$_{-0.01}^{+0.00}$ & 0.14$_{-0.01}^{+0.01}$ & $\cdots$ & $\cdots$ \\ [2pt]
{\bf WR117} & WC8--9   & WI02  & \phantom{0}5.0$_{-0.5}^{+0.6}$      & 5.9$_{-0.6}^{+0.5}$ & 5.57$_{-0.25}^{+0.25}$     & a         & --11.57$_{-0.05}^{+0.04}$  & 1700        &    12.2$_{-5.1}^{+8.2}$     & 0.85$_{-0.10}^{+0.14}$   & 0.72$_{-0.21}^{+0.29}$ & 2.95$_{-0.44}^{+0.51}$   & 0.87$_{-0.13}^{+0.14}$  & 2.90$_{-0.04}^{+0.03}$ & 0.38$_{-0.01}^{+0.00}$ & 0.99$_{-0.07}^{+0.07}$ & 0.48$_{-0.04}^{+0.05}$ & 0.14$_{-0.02}^{+0.03}$ & $\cdots$ \\ [2pt]
{\bf WR119} & WC8--9 & KI    & \phantom{0}4.1$_{-0.2}^{+0.3}$       & 3.4$_{-0.4}^{+0.3}$ & 4.85$_{-0.15}^{+0.14}$     & a         & --11.65$_{-0.05}^{+0.04}$  & 1300         &\phantom{0}0.8$_{-0.2}^{+0.3}$        & 0.31$_{-0.03}^{+0.02}$ & $\cdots$   & 3.31$_{-0.30}^{+0.32}$ & 1.03$_{-0.09}^{+0.09}$ & 4.14$_{-0.02}^{+0.03}$   & 0.73$_{-0.01}^{+0.00}$ & 1.56$_{-0.07}^{+0.07}$ & 0.64$_{-0.03}^{+0.04}$ & $\cdots$ & $\dagger$1.26$_{-0.19}^{+0.21}$ \\     [2pt]
{\bf WR121} & WC8--9 & WI02    & \phantom{0}3.4$\pm$0.2     & 5.3$_{-0.6}^{+0.5}$ & 5.51$_{-0.21}^{+0.22}$     & a         & --11.55$_{-0.05}^{+0.04}$ & 1400        & \phantom{0}3.4$_{-1.1}^{+2.0}$        & 0.27$_{-0.01}^{+0.02}$  & $\cdots$ & 3.29$_{-0.45}^{+0.51}$ & 0.99$_{-0.13}^{+0.15}$ & 3.64$_{-0.04}^{+0.03}$ & 0.91$_{-0.00}^{+0.01}$ & 1.24$_{-0.07}^{+0.09}$ & 0.65$_{-0.05}^{+0.06}$ & 0.06$_{-0.01}^{+0.01}$ & $\dagger$0.96$_{-0.21}^{+0.26}$ \\ [2pt]
WR125 & WC6--7+O & KI & \phantom{0}5.9$_{-0.6}^{+0.7}$   & 6.0$_{-0.6}^{+0.6}$ & 6.28$_{-0.25}^{+0.26}$    & a         & --11.38$_{-0.05}^{+0.04}$ & 2800    &     29.7$_{-12.7}^{+20.5}$     & 0.40$_{-0.05}^{+0.07}$ & $\cdots$ & \multicolumn{2}{c}{--- 2.98$_{-0.46}^{+0.52}$ ---}  & 1.03$_{-0.01}^{+0.01}$ & 0.29 & 0.21$_{-0.01}^{+0.02}$ & 0.23$_{-0.02}^{+0.02}$ & $\cdots$ & $\cdots$ \\ [2pt]
{\bf WR132} & WC6--7      & KI  & \phantom{0}4.2$_{-0.2}^{+0.3}$   & 4.3$_{-0.4}^{+0.4}$ & 5.34$_{-0.18}^{+0.18}$     & a        & --10.87$_{-0.04}^{+0.04}$ & 2000          &   11.8$_{-3.9}^{+5.5}$      &  1.41$_{-0.09}^{+0.10}$     &   0.80$_{-0.18}^{+0.23}$      & 2.56$_{-0.29}^{+0.22}$ & 0.64$_{-0.07}^{+0.08}$ & 0.19 & 0.10 & 0.11$_{-0.01}^{+0.00}$ & 0.16$_{-0.01}^{+0.01}$ & $\cdots$  & $\cdots$ \\ [2pt]
{\bf WR135} & WC8--9     & WI02   & \phantom{0}2.4$_{-0.2}^{+0.1}$ & 1.5$_{-0.1}^{+0.2}$  & 5.54$_{-0.07}^{+0.08}$ & a          & --9.45$_{-0.04}^{+0.04}$ & 1400         &   \phantom{0}8.6$_{-1.4}^{+1.5}$  & 0.64$_{-0.04}^{+0.04}$         &1.35$_{-0.09}^{+0.13}$  & 3.34$_{-0.14}^{+0.14}$ & 0.85$_{-0.03}^{+0.03}$ & 1.06 & 0.23 & 0.31$_{-0.01}^{+0.00}$ & 0.28$_{-0.01}^{+0.00}$ & 0.13$_{-0.01}^{+0.00}$ & $\dagger$0.59$_{-0.04}^{+0.04}$ \\ [2pt]
{\bf WR137} & WC6--7+O     & KI   & \phantom{0}2.0$\pm$0.1 & 2.4$_{-0.2}^{+0.3}$  & 5.89$_{-0.10}^{+0.10}$        & a            & --9.54$_{-0.05}^{+0.04}$ & 1700          &     11.1$_{-2.4}^{+2.9}$    & 0.37$_{-0.01}^{+0.02}$          & 0.98$_{-0.13}^{+0.15}$      & 3.38$_{-0.22}^{+0.24}$ & 0.75$_{-0.04}^{+0.05}$ & 0.59 & 0.13 & 0.26$_{-0.01}^{+0.01}$ & 0.22$_{-0.00}^{+0.01}$ & $\cdots$  & $\dagger$0.36$_{-0.03}^{+0.05}$ \\ [2pt]
{\bf WR140} & WC6--7+O     & II13   & \phantom{0}1.8$_{-0.1}^{+0.0}$ & 2.1$_{-0.2}^{+0.3}$  &$\cdots$ & b & --8.90$_{-0.09}^{+0.08}$  & 2900          &    29.3$_{-7.7}^{+8.4}$  & $\cdots$    & 0.36$_{-0.04}^{+0.05}$ & \multicolumn{2}{c}{--- 2.24$_{-0.12}^{+0.13}$ ---} & 0.34 & 0.08 & 0.11$_{-0.00}^{+0.01}$ & 0.12$_{-0.01}^{+0.00}$ & 0.10$_{-0.01}^{+0.01}$ & $\dagger$0.30$_{-0.02}^{+0.04}$ \\ [2pt]
{\bf WR143} & WC4--5+O     & KI   & \phantom{0}1.8$\pm$0.1  & 4.4$_{-0.5}^{+0.4}$ & 5.45$_{-0.18}^{+0.18}$        & a          & --10.31$_{-0.04}^{+0.04}$ & 2900         &    \phantom{0}7.3$_{-2.4}^{+3.3}$   & 0.66$_{-0.03}^{+0.04}$   & 0.37$_{-0.09}^{+0.16}$       & \multicolumn{2}{c}{--- 2.36$_{-0.35}^{+0.41}$ ---} & 0.00 & 0.00 & 0.04 & 0.08$_{-0.01}^{+0.01}$ & $\cdots$ & $\cdots$ \\ [2pt]
{\bf WR144} & WC4--5         & WI02  & \phantom{0}1.7$_{-0.1}^{+0.0}$   & 7.0$_{-0.7}^{+0.7}$ & 5.15$_{-0.28}^{+0.28}$       & a          & --11.14$_{-0.04}^{+0.04}$  & 2800         &    10.3$_{-4.8}^{+8.6}$   & 1.89$_{-0.05}^{+0.06}$           & $\cdots$ & \multicolumn{2}{c}{--- 1.13$_{-0.17}^{+0.21}$ ---} & 0.00 & 0.00 & 0.05$_{-0.01}^{+0.00}$ & 0.12$_{-0.01}^{+0.01}$ & 0.12$_{-0.01}^{+0.03}$ & $\cdots$ \\  [2pt]
WR150 & WC4--5         &  KI  & \phantom{0}6.8$_{-0.6}^{+0.8}$ & 3.5$_{-0.4}^{+0.3}$ & 5.54$_{-0.16}^{+0.17}$         & a          & --10.70$_{-0.04}^{+0.04}$ & 2900         &         23.1$_{-6.9}^{+8.9}$    & 1.73$_{-0.20}^{+0.24}$                         & 0.28$_{-0.04}^{+0.05}$       & \multicolumn{2}{c}{--- 2.29$_{-0.19}^{+0.20}$ ---} & 0.00 & 0.09 & 0.11$_{-0.01}^{+0.00}$ & 0.14$_{-0.01}^{+0.01}$ & $\cdots$ & $\cdots$ \\ [2pt]
{\bf WR154} & WC6--7         & WI02  & \phantom{0}4.7$\pm$0.3 & 2.9$_{-0.3}^{+0.3}$   & 5.76$_{-0.14}^{+0.13}$        & a          & --10.18$_{-0.04}^{+0.05}$ & 2100      &   25.9$_{-6.8}^{+8.6}$     & 1.17$_{-0.08}^{+0.10}$                          & 0.70$_{-0.11}^{+0.13}$      & 2.37$_{-0.18}^{+0.20}$ & 0.64$_{-0.05}^{+0.05}$ & 0.18 & 0.12 & 0.10$_{-0.00}^{+0.01}$ & 0.16$_{-0.00}^{+0.01}$ &  0.11$_{-0.01}^{+0.01}$ & $\dagger$0.22$_{-0.02}^{+0.04}$ \\ [2pt]
{\bf BAT99-8} & WC4--5       & MM & 49.6$\pm$1.0             & 0.4$_{-0.0}^{+0.0}$   & 5.47$_{-0.02}^{+0.03}$        &  c        & --11.31$_{-0.05}^{+0.04}$ & 2600       &            20.2$_{-2.2}^{+2.2}$     & 1.76$_{-0.03}^{+0.04}$   & 0.52$_{-0.01}^{+0.02}$ & \multicolumn{2}{c}{--- 1.71$_{-0.01}^{+0.02}$ ---} & 0.03 & 0.08 & 0.04 & 0.07 & 0.08 & 0.07 \\ [2pt]
{\bf BAT99-9} & WC4--5        & MM  & 49.6$\pm$1.0            & 0.5$_{-0.1}^{+0.0}$   & 5.47$_{-0.02}^{+0.03}$       & c          & --11.20$_{-0.04}^{+0.05}$ & 2600      &             28.7$_{-3.2}^{+3.2}$     & 2.51$_{-0.05}^{+0.06}$  & 0.25$_{-0.00}^{+0.01}$  & \multicolumn{2}{c}{--- 1.05$_{-0.02}^{+0.01}$  ---} & 0.01 & 0.06 & 0.04 & 0.03 & 0.05 & 0.02\\ [2pt]
{\bf BAT99-10} & WC4--5+O& AD    & 49.6$\pm$1.0             & 0.1$_{-0.0}^{+0.0}$   &$\cdots$& b &--11.09$_{-0.05}^{+0.04}$ & 3200     &            25.4$_{-2.6}^{+2.6}$ & $\cdots$&$\cdots$&\multicolumn{2}{c}{--- 1.26 ---}&0.00 & 0.00&$\cdots$&0.09& 0.06 & $\cdots$ \\   [2pt]
{\bf BAT99-11}&WC4--5       & MM & 49.6$\pm$1.0            &  0.4$_{-0.0}^{+0.1}$    & 5.85$_{-0.02}^{+0.03}$       & c          &--10.81$_{-0.04}^{+0.04}$ & 3500      &            66.7$_{-7.2}^{+7.2}$     & 2.43$_{-0.05}^{+0.05}$  & 0.40$_{-0.02}^{+0.01}$   & \multicolumn{2}{c}{--- 1.73$_{-0.02}^{+0.02}$ ---} & 0.01 & 0.10 & 0.05 & 0.08 & 0.06 & 0.06  \\ [2pt]
{\bf BAT99-20} &WC4--5+O& AD     & 49.6$\pm$1.0            & 1.2$_{-0.2}^{+0.1}$ & $\cdots$& b &--11.72$_{-0.05}^{+0.04}$ & 3900     &            15.2$_{-2.1}^{+2.2}$ & $\cdots$ &$\cdots$&\multicolumn{2}{c}{--- 1.81$_{-0.06}^{+0.05}$  ---}&0.00 & 0.00&$\cdots$&0.05 &0.05 & $\cdots$ \\ [2pt]
{\bf BAT99-28} & WC4--5+O& AD  & 49.6$\pm$1.0             & 0.2$_{-0.1}^{+0.0}$  & $\cdots$& b &--11.10$_{-0.04}^{+0.04}$ & 3400     &            26.8$_{-2.8}^{+2.7}$ & $\cdots$ &$\cdots$&\multicolumn{2}{c}{--- 2.20$_{-0.01}^{+0.00}$ ---}&0.13 & 0.00& 0.09 &0.09 & 0.07 & $\cdots$ \\ [2pt]
{\bf BAT99-34} & WC4--5+O& AD  & 49.6$\pm$1.0             & 0.1$_{-0.0}^{+0.0}$   & $\cdots$& b &--11.19$_{-0.04}^{+0.04}$ & 2700     &  20.4$_{-2.1}^{+2.1}$ & $\cdots$ &$\cdots$&\multicolumn{2}{c}{--- 1.41$_{-0.00}^{+0.01}$  ---}&0.01 & 0.00& 0.05 &0.05 & 0.05 & $\cdots$ \\ [2pt]
{\bf BAT99-38} & WC4--5+O& AD  & 49.6$\pm$1.0             & 0.2$_{-0.1}^{+0.0}$   & $\cdots$& b &--11.14$_{-0.04}^{+0.04}$ & 3600     &            24.4$_{-2.5}^{+2.6}$ & $\cdots$ &$\cdots$&\multicolumn{2}{c}{--- 1.26$_{-0.01}^{+0.00}$  ---}&0.01 & 0.00& $\cdots$ &0.03 & 0.06 & $\cdots$ \\ [2pt]
{\bf BAT99-39} & WC4--5+O& AD  & 49.6$\pm$1.0             & 0.1$_{-0.0}^{+0.0}$   & $\cdots$& b &--11.18$_{-0.05}^{+0.04}$ & 3900     &            21.4$_{-2.2}^{+2.2}$ & $\cdots$ &$\cdots$&\multicolumn{2}{c}{--- 1.45$_{-0.01}^{+0.00}$ ---}&0.02 & 0.00& $\cdots$ &0.04 & 0.05 & $\cdots$ \\ [2pt]
{\bf BAT99-52}&WC4--5       & MM  & 49.6$\pm$1.0            &  0.3$_{-0.0}^{+0.1}$  & 5.64$_{-0.02}^{+0.02}$       & c          &--11.06$_{-0.05}^{+0.04}$ & 3000      &   34.3$_{-3.6}^{+3.7}$     & 2.03$_{-0.04}^{+0.04}$  & 0.68$_{-0.02}^{+0.02}$   & \multicolumn{2}{c}{--- 2.05$_{-0.02}^{+0.01}$ ---} & 0.03 & 0.09 & 0.04 & 0.08 & 0.07 & 0.07 \\ [2pt]
{\bf BAT99-53} & WC4--5+O& AD  & 49.6$\pm$1.0             & 0.4$_{-0.0}^{+0.1}$   & $\cdots$& b & --11.07$_{-0.05}^{+0.04}$ & 3700     &    35.7$_{-3.9}^{+3.8}$ & $\cdots$ &$\cdots$&\multicolumn{2}{c}{--- 1.60$_{-0.02}^{+0.01}$ ---}&0.00 & 0.00& $\cdots$ &0.05 & 0.06 & $\cdots$ \\ [2pt]
{\bf BAT99-61}&WC4--5       & MM  & 49.6$\pm$1.0            &  0.5$_{-0.1}^{+0.0}$    & 5.87$_{-0.02}^{+0.03}$       & d          &--10.85$_{-0.05}^{+0.04}$ & 3400      &            63.5$_{-7.0}^{+7.1}$     & 2.21$_{-0.05}^{+0.05}$  & 0.31$_{-0.01}^{+0.01}$   & \multicolumn{2}{c}{--- 1.68$_{-0.02}^{+0.02}$ ---} & 0.01 & 0.14 & 0.06 & 0.07 & 0.05 & 0.06 \\ [2pt]
{\bf BAT99-70} & WC4--5+O& AD  & 49.6$\pm$1.0             & 1.5$_{-0.1}^{+0.2}$   & $\cdots$& b & --11.09$_{-0.05}^{+0.04}$ & 4300     &            89.4$_{-14.3}^{+15.6}$ & $\cdots$ &$\cdots$&\multicolumn{2}{c}{--- 1.03$_{-0.04}^{+0.04}$  ---}&0.00 & 0.00& 0.05 &0.03 & 0.04 & $\cdots$ \\  [2pt]
{\bf BAT99-84} & WC4--5+O& AD  & 49.6$\pm$1.0             & 0.5$_{-0.1}^{+0.0}$   & $\cdots$& b & --11.09$_{-0.04}^{+0.05}$ & 3200     &            36.7$_{-4.0}^{+4.0}$ & $\cdots$ &$\cdots$&\multicolumn{2}{c}{--- 1.96$_{-0.02}^{+0.03}$ ---}&0.01 & 0.00& 0.05 &0.09 & 0.07 & $\cdots$ \\  [2pt]
{\bf BAT99-85} & WC4--5+O& AD  & 49.6$\pm$1.0             & 1.1$_{-0.1}^{+0.1}$   & $\cdots$& b & --11.34$_{-0.04}^{+0.04}$ & 3700     &            35.6$_{-4.9}^{+5.1}$ & $\cdots$ &$\cdots$&\multicolumn{2}{c}{--- 1.25$_{-0.04}^{+0.04}$ ---}&0.00 & 0.00&$\cdots$ &0.10 & 0.06 & $\cdots$ \\  [2pt]
{\bf BAT99-90}&WC4--5       & MM  & 49.6$\pm$1.0            &  1.1$_{-0.1}^{+0.1}$   & 5.48$_{-0.04}^{+0.05}$       & d          &--11.46$_{-0.05}^{+0.04}$ & 2800      &            25.9$_{-3.5}^{+3.7}$     & 2.22$_{-0.06}^{+0.05}$  & 0.50$_{-0.03}^{+0.05}$   & \multicolumn{2}{c}{--- 1.60$_{-0.05}^{+0.04}$ ---} & 0.02 & 0.08 & 0.04 & 0.06 & 0.06 & 0.06 \\ [2pt]
BAT99-115      & WC4--5      & MM     & 49.6$\pm$1.0  & 1.6$_{-0.2}^{+0.1}$  & $\cdots$ & d         & --11.56$_{-0.10}^{+0.08}$  & 4000 & 30.7$_{-7.3}^{+7.5}$ & $\cdots$ & $\sharp$1.07$_{-0.11}^{+0.13}$ & \multicolumn{2}{c}{--- 2.49$_{-0.10}^{+0.10}$ ---} & 0.03 & 0.02 & 0.04$_{-0.00}^{+0.01}$  &  0.07$_{-0.00}^{+0.01}$ & 0.08$_{-0.00}^{+0.01}$ & $\cdots$ \\ [2pt]
{\bf BAT99-121}& WC4--5    & MM  & 49.6$\pm$1.0            & 2.2$_{-0.2}^{+0.2}$   & $\cdots$ & b & --12.16$_{-0.05}^{+0.04}$ & 3000    &           13.6$_{-2.8}^{+3.1}$ & $\cdots$ & 0.70$_{-0.10}^{+0.12}$   & \multicolumn{2}{c}{--- 1.85$_{-0.10}^{+0.11}$ ---} & 0.02 & 0.10 & 0.05 & 0.06 & 0.07$_{-0.00}^{+0.01}$ & 0.12$_{-0.01}^{+0.01}$ \\  [2pt]
\hline
\end{tabular}
\end{footnotesize}
\end{center}
\footnotesize{a: \citet{2012A&A...540A.144S}; b $E_{\rm B-V}$ = 1.21 $E_{b-v}$ and $(b-v)_{0} = -0.30$ mag adopted; c: \citet{2022ApJ...924...44A}; d: \citet{2022ApJ...931..157A} \\
 d: \citet{2001MNRAS.324...18B} \\ 
Note: $\dagger$ IRTF/SpeX observations \citep{2006MNRAS.372.1407C}; $\sharp$ FOS observations \citep{1998ApJ...493..180M}
}
\end{table}
\end{landscape}

\begin{landscape}
\begin{table}
\caption{Luminosities of prominent optical emission lines of Galactic (WR), LMC (BAT99, LMC, LH) and SMC (AB) WO stars. C\,{\sc iv} $\lambda\lambda$5801,12 FWHM are also provided in km\,s$^{-1}$ ($\pm$200 km\,s$^{-1}$). Stars included in spectral templates are indicated in bold (the remainder are excluded owing to limited spectral coverage)}
\label{WO-all}
\begin{center}
\begin{footnotesize}
 \begin{tabular}{l@{\hspace{1mm}}l@{\hspace{1.5mm}}l@{\hspace{1.5mm}}l@{\hspace{1.5mm}}l@{\hspace{1.5mm}}l@{\hspace{0mm}}c@{\hspace{1.5mm}}c@{\hspace{1mm}}c@{\hspace{1mm}}c@{\hspace{0mm}}c
 @{\hspace{1.5mm}}l@{\hspace{1mm}}l@{\hspace{1mm}}l@{\hspace{1mm}}l@{\hspace{1mm}}l@{\hspace{1mm}}l@{\hspace{1mm}}c@{\hspace{1mm}}c@{\hspace{0mm}}c}   %
\hline
Star & Category & Data & $d$ & $A_{\rm V}$  & $\log$  & Ref & $\log F_{\rm CIV~5801,12}$ & CIV~5801,12 & \multicolumn{2}{c}{$L_{\rm CIV~5801,12}$}  & $\underline{L_{\rm OIV~3403,13}}$ & $\underline{L_{\rm OVI~3811,34}}$  & $\underline{L_{\rm CIV~4658+HeII~4686}}$ & $\underline{L_{\rm OV~5572,5607}}$ & $\underline{L_{\rm HeII~6560}}$ & $\underline{L_{\rm CIV~7725}}$  \\  
        &    &     ID         &   kpc     & mag              & $L_{\rm Bol}/L_{\odot}$        &        & erg\,s$^{-1}$\,cm$^{-2}$ & FWHM & 10$^{35}$ erg\,s$^{-1}$ &  $10^{-3} L_{\rm Bol}$   & $L_{\rm CIV~5801,12}$ &  $L_{\rm CIV~5801,12}$  & $L_{\rm CIV~5801,12}$ & $L_{\rm CIV~5801,12}$  & $L_{\rm CIV~5801,12}$ &  $L_{\rm CIV~5801,12}$ \\
\hline
{\bf WR30a} &  WO+O  & AD         & \phantom{0}8.0$_{-0.5}^{+0.7}$ & 4.1$\pm$0.4           & $\cdots$ & a & --11.47$_{-0.05}^{+0.04}$ & 5200 & \phantom{0}8.7$_{-2.8}^{+3.8}$ & $\cdots$ & 0.66$_{-0.14}^{+0.18}$ & \phantom{0}0.35$_{-0.06}^{+0.08}$ & 0.81$_{-0.09}^{+0.09}$ & 0.14 & $\cdots$ & 0.05$_{-0.01}^{+0.00}$  \\ [2pt]
{\bf WR93b} & WO         & VX  & \phantom{0}2.1$_{-0.1}^{+0.1}$ & 6.3$\pm$0.6 & 4.88$_{-0.26}^{+0.26}$        & b & --11.90$_{-0.04}^{+0.05}$ &  8400  & \phantom{0}1.5$_{-0.6}^{+1.1}$ & 0.52$_{-0.03}^{+0.04}$   & 1.6$_{-0.5}^{+0.6}$ & \phantom{0}1.7$_{-0.4}^{+0.5}$ & 0.71$_{-0.10}^{+0.13}$ & 0.12$_{-0.00}^{+0.01}$ & 0.06       & 0.14$_{-0.02}^{+0.02}$  \\  [2pt]
{\bf WR102} & WO         & WI02       & \phantom{0}2.6$_{-0.1}^{+0.1}$ &   3.9$\pm$0.4          & 4.95$_{-0.16}^{+0.17}$        & b & --12.10$_{-0.04}^{+0.04}$ & 6600   & \phantom{0}0.2$_{-0.1}^{+0.1}$ & 0.05$_{-0.00}^{+0.01}$   & 3.1$_{-0.6}^{+0.8}$ & 25$_{-4}^{+5}$ & 3.2$_{-0.3}^{+0.3}$ & 1.31$_{-0.02}^{+0.02}$ & 0.38$_{-0.02}^{+0.02}$      & 1.10$_{-0.09}^{+0.13}$  \\ [2pt]
{\bf WR142} & WO        & WI02         & \phantom{0}1.6$_{-0.0}^{+0.1}$  & 4.9$\pm$0.5 &  5.33$_{-0.20}^{+0.20}$       & b  & --11.16$_{-0.04}^{+0.04}$ & 7000    & \phantom{0}1.5$_{-0.6}^{+0.7}$ & 0.18$_{-0.00}^{+0.00}$   & 2.0$_{-0.5}^{+0.7}$ & \phantom{0}7.1$_{-1.6}^{+2.1}$ & 2.2$_{-0.3}^{+0.3}$ & 0.43$_{-0.01}^{+0.01}$ & 0.22$_{-0.02}^{+0.01}$     & 0.53$_{-0.07}^{+0.09}$  \\ [2pt]
{\bf LH41-1042} & WO   & VX  & 49.6$\pm$1.0               &   0.4$\pm$0.0      & 5.25$_{-0.02}^{+0.02}$         & b & --11.52$_{-0.04}^{+0.04}$    & 5600     &  12.7$_{-1.4}^{+1.3}$ & 1.85$_{-0.04}^{+0.03}$    & 0.78$_{-0.02}^{+0.02}$ & \phantom{0}0.33$_{-0.01}^{+0.00}$ & 0.78$_{-0.01}^{+0.01}$ & 0.10 & 0.03     & 0.06 \\ [2pt]
{\bf LMC195--1} & WO & MM        & 49.6$\pm$1.0                & 0.4$\pm$0.0         & 5.40$_{-0.02}^{+0.03}$         & c & --12.26$_{-0.04}^{+0.04}$     & 4900  &  \phantom{0}2.3$_{-0.2}^{+0.2}$  & 0.24$_{-0.1}^{+0.0}$   & 4.3$_{-0.2}^{+0.1}$ & \phantom{0}5.8$_{-0.1}^{+0.1}$ & 1.5$_{-0.0}^{+0.1}$ & 0.32 & 0.17      & 0.29  \\ [2pt]
{\bf BAT99-123} & WO  & MM        &  49.6$\pm$1.0               & 0.5$_{-0.1}^{+0.0}$         & 5.40$_{-0.02}^{+0.03}$         & c & --11.50$_{-0.05}^{+0.04}$    & 5400 & 14.2$_{-1.5}^{+1.6}$ & 1.46$_{-0.03}^{+0.04}$   & 0.64$_{-0.02}^{+0.03}$ & \phantom{0}0.40$_{-0.01}^{+0.02}$ & 0.45$_{-0.01}^{+0.00}$ & 0.05 & 0.02$_{-0.00}^{+0.01}$     & 0.06  \\  [2pt]
{\bf AB 8}    & WO+O    & AD      & 61.2$\pm$1.2                 & 0.2$\pm$0.0         & 6.16$_{-0.02}^{+0.02}$           & d & --11.22$_{-0.04}^{+0.05}$ &  5300  & 32.9$_{-3.4}^{+3.4}$  & 0.59$_{-0.01}^{+0.01}$  & 0.96$_{-0.02}^{+0.01}$ & \phantom{0}0.70$_{-0.01}^{+0.01}$ & 0.65$_{-0.00}^{+0.01}$ & 0.16  & 0.04 & 0.19  \\ [2pt]
\hline
\end{tabular}
\end{footnotesize}
\end{center}
\footnotesize{a: \citet{2020MNRAS.493.1512R} b: \citet{2015A&A...581A.110T}; c: \citet{2022ApJ...931..157A}, d:  \citet{2016A&A...591A..22S} }
\end{table}
\end{landscape}

\section{Templates}

\begin{figure}
\centering
\includegraphics[width=0.68\linewidth,bb=30 65 530 780,angle=-90]{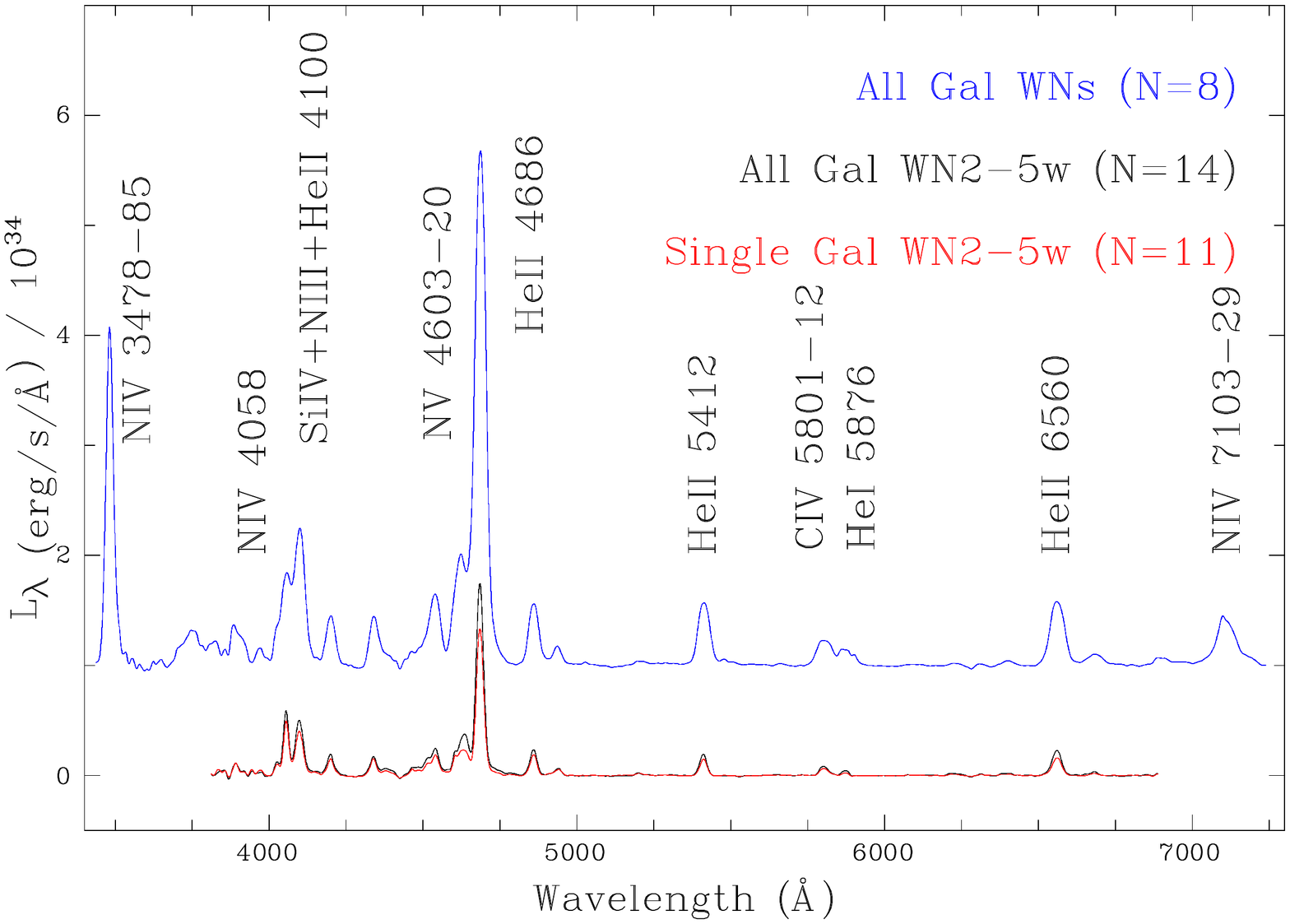}
\includegraphics[width=0.68\linewidth,bb=30 65 530 780,angle=-90]{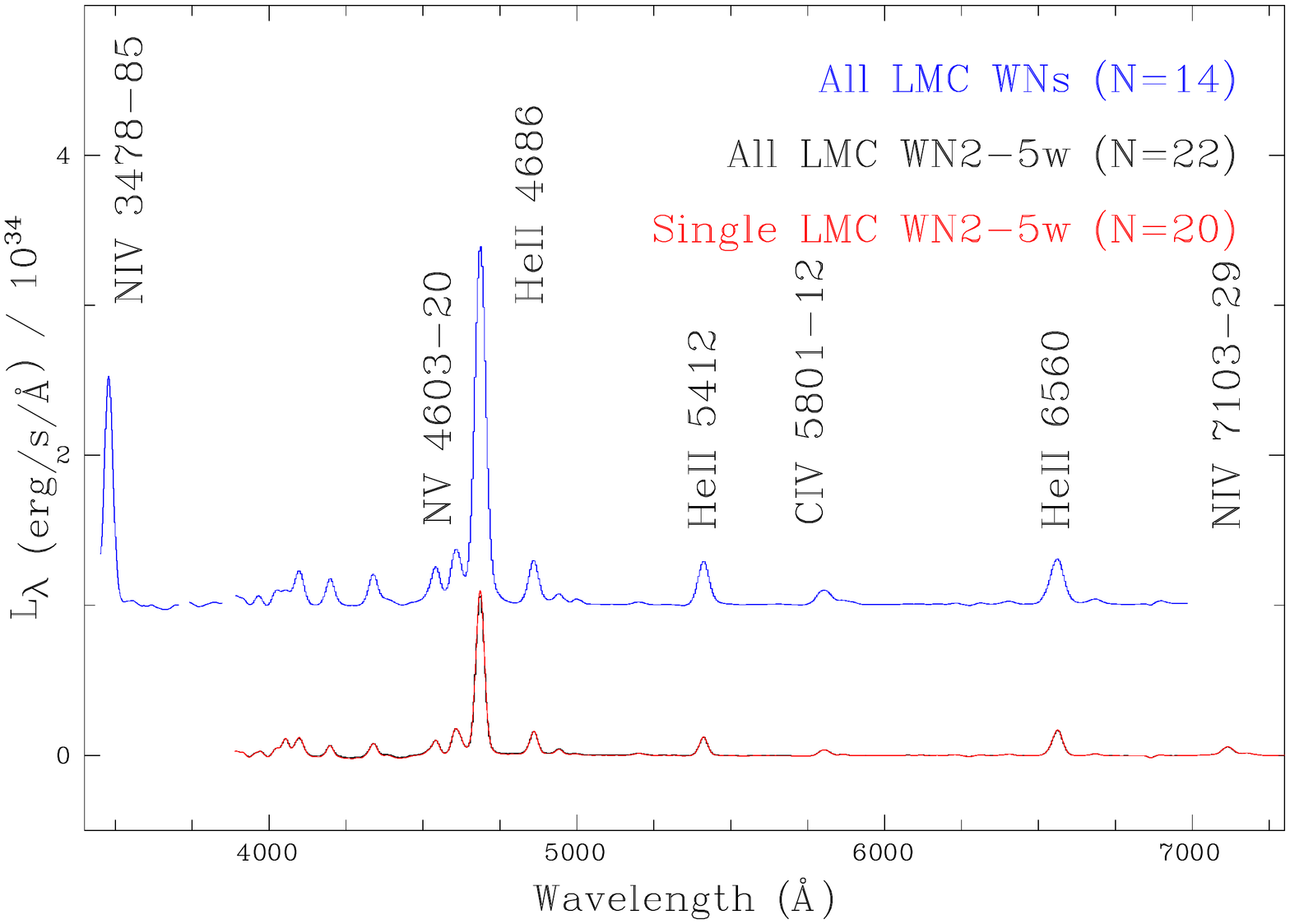}
\includegraphics[width=0.68\linewidth,bb=30 65 530 780,angle=-90]{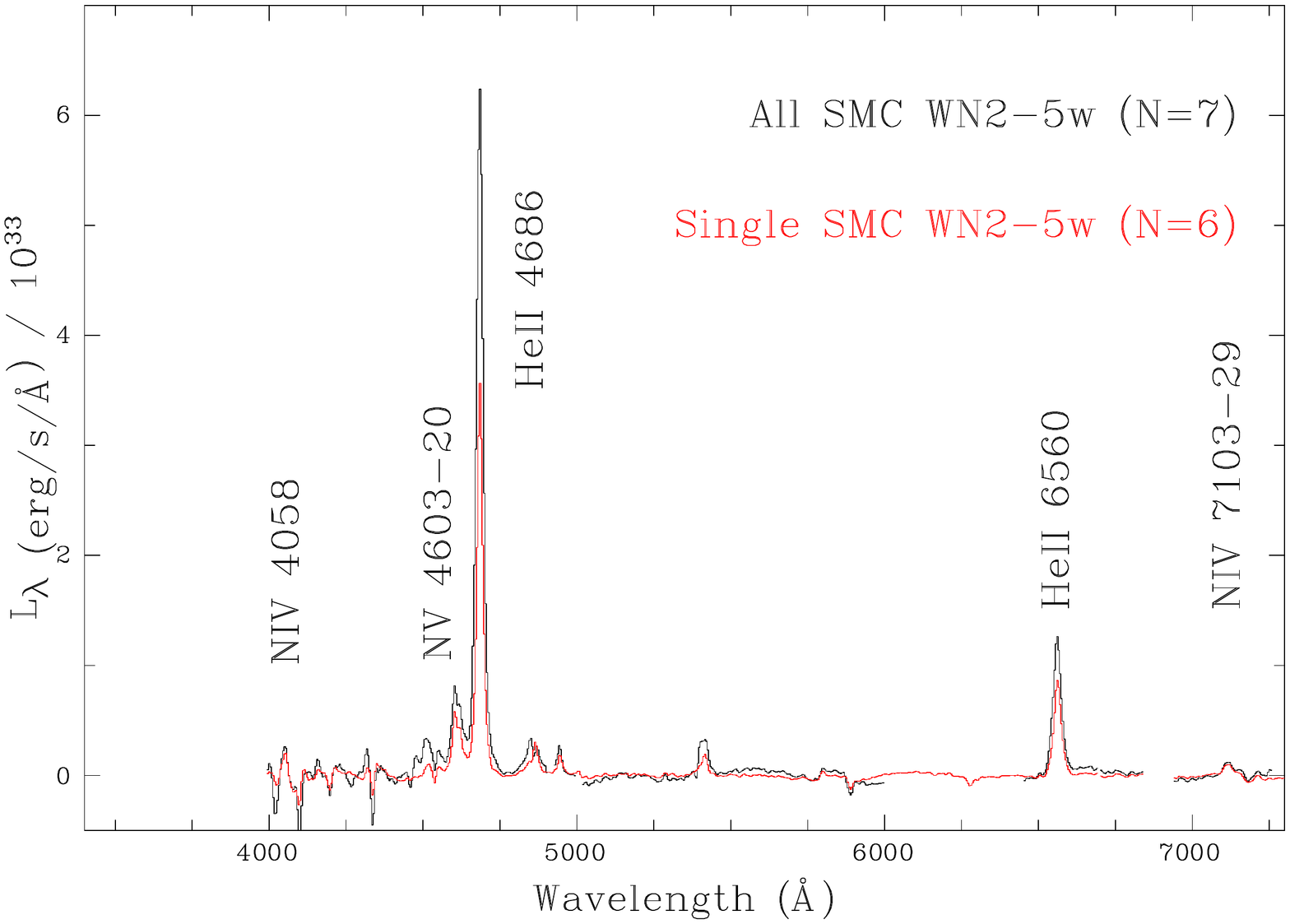}
	\centering
  \caption{Upper panel: Galactic WN2--5w emission line templates based on single (red) and all (black) stars, plus WN3--7s templates (blue, offset by 10$^{34}$ erg\,s$^{-1}$\,\AA$^{-1}$); Middle panel: LMC emission line templates; Lower panel: SMC emission line templates (no strong-lined WN stars are known).}
	\label{WNE-lum} 
\end{figure}

\begin{figure}
\centering
\includegraphics[width=0.68\linewidth,bb=30 65 530 780,angle=-90]{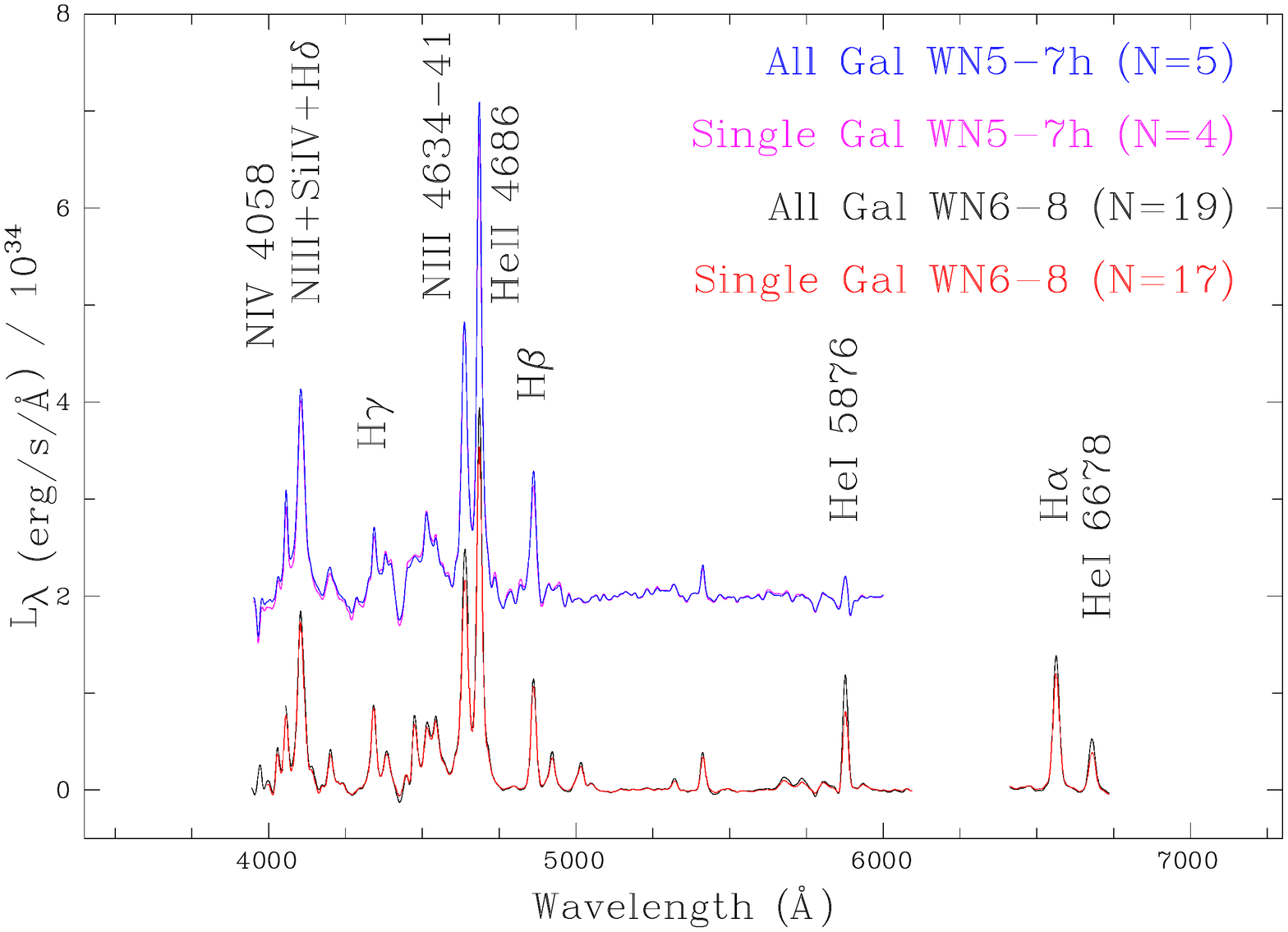}
\includegraphics[width=0.68\linewidth,bb=30 65 530 780,angle=-90]{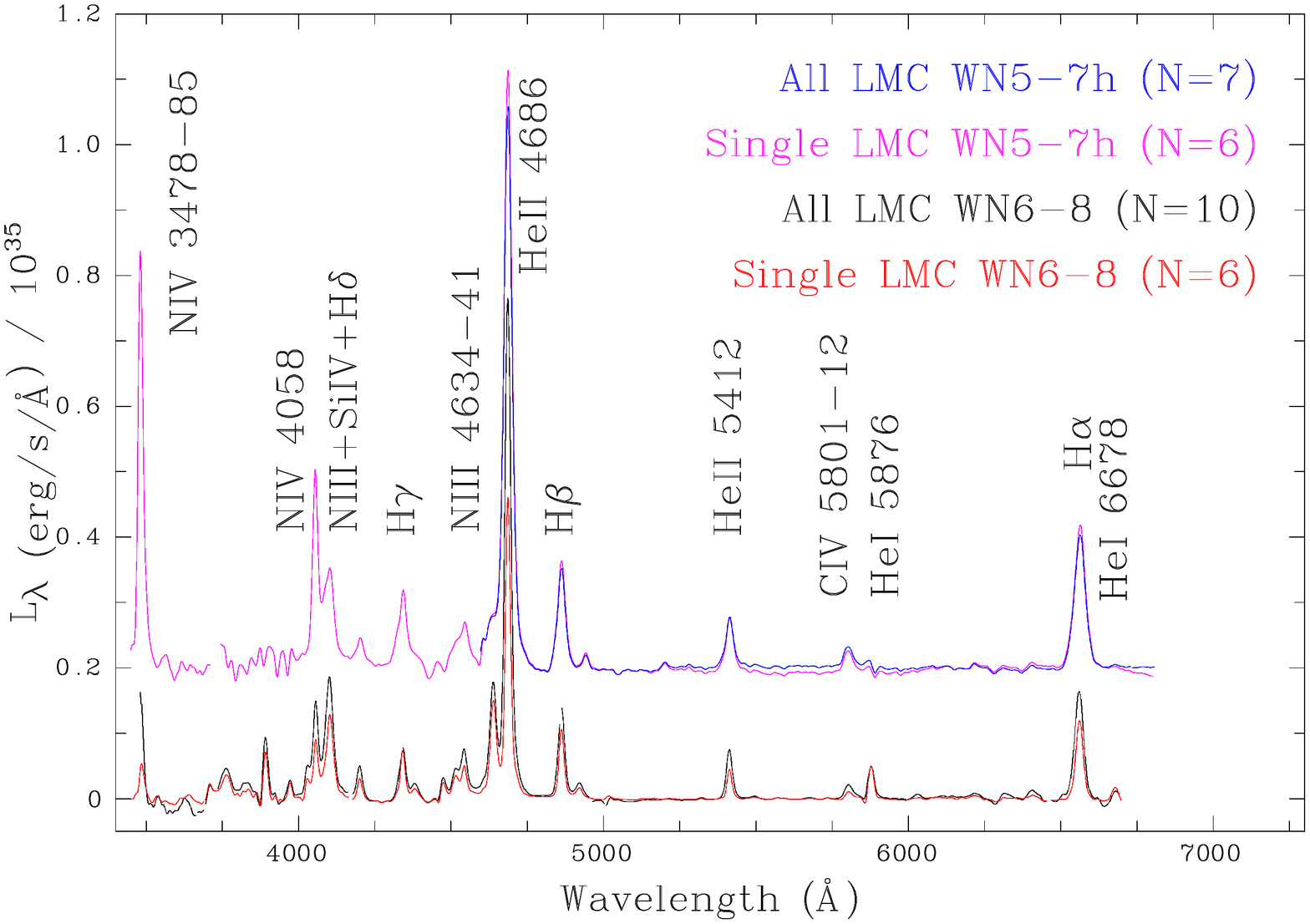}
\includegraphics[width=0.68\linewidth,bb=30 65 530 780,angle=-90]{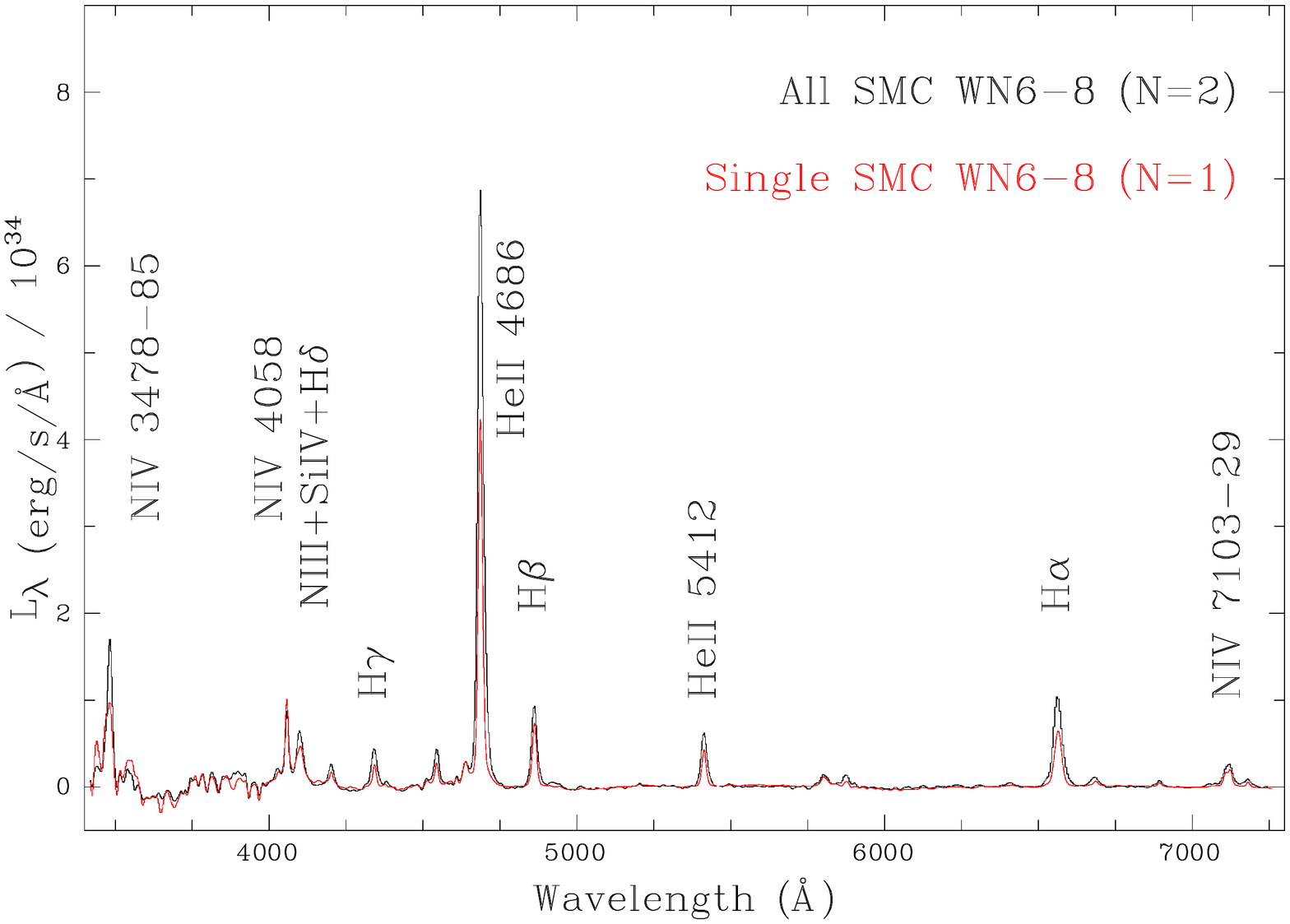}
	\centering
  \caption{Upper panel: Galactic WN6--8 emission line templates based on single (red) and all (black) stars, plus WN5--7h templates (pink and blue, offset by $2 \times 10^{34}$ erg\,s$^{-1}$\,\AA$^{-1}$); Middle panel: LMC WN6--8 and WN5--7h emission line templates; Lower panel: SMC WN6--8 emission line templates. LMC WN5--7h templates exclude the region shortward of $\lambda$4600 owing to the use of VLT/MUSE datasets \citep{2018A&A...614A.147C}.}
	\label{WNL-lum} 
\end{figure}

\begin{figure}
\centering
\includegraphics[width=0.68\linewidth,bb=30 65 530 780,angle=-90]{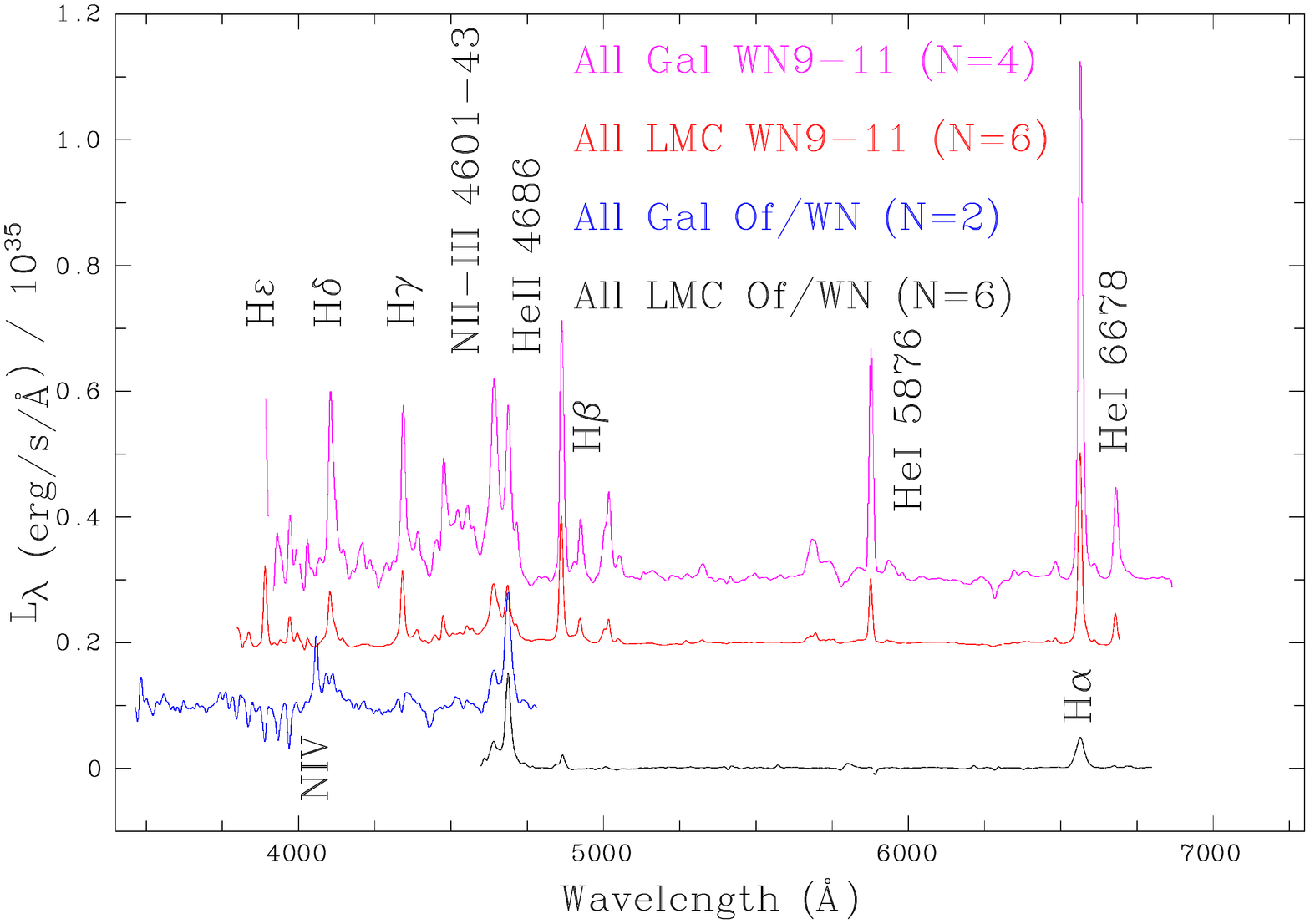}
	\centering
  \caption{Emission line templates for LMC (black) and Milky Way (blue, offset by 10$^{34}$ erg/s/\AA) Of/WN stars, plus LMC (red, offset by $2 \times 10^{34}$ erg\,s$^{-1}$\,\AA$^{-1}$) and Milky Way (pink, offset by $3 \times 10^{34}$ erg/s/\AA) WN9--11 stars. LMC Of/WN templates exclude the region shortward of $\lambda$4600 owing to the use of VLT/MUSE datasets \citep{2018A&A...614A.147C}}
	\label{OfWN-lum} 
\end{figure}

\begin{figure}
\centering
\includegraphics[width=0.68\linewidth,bb=30 65 530 780,angle=-90]{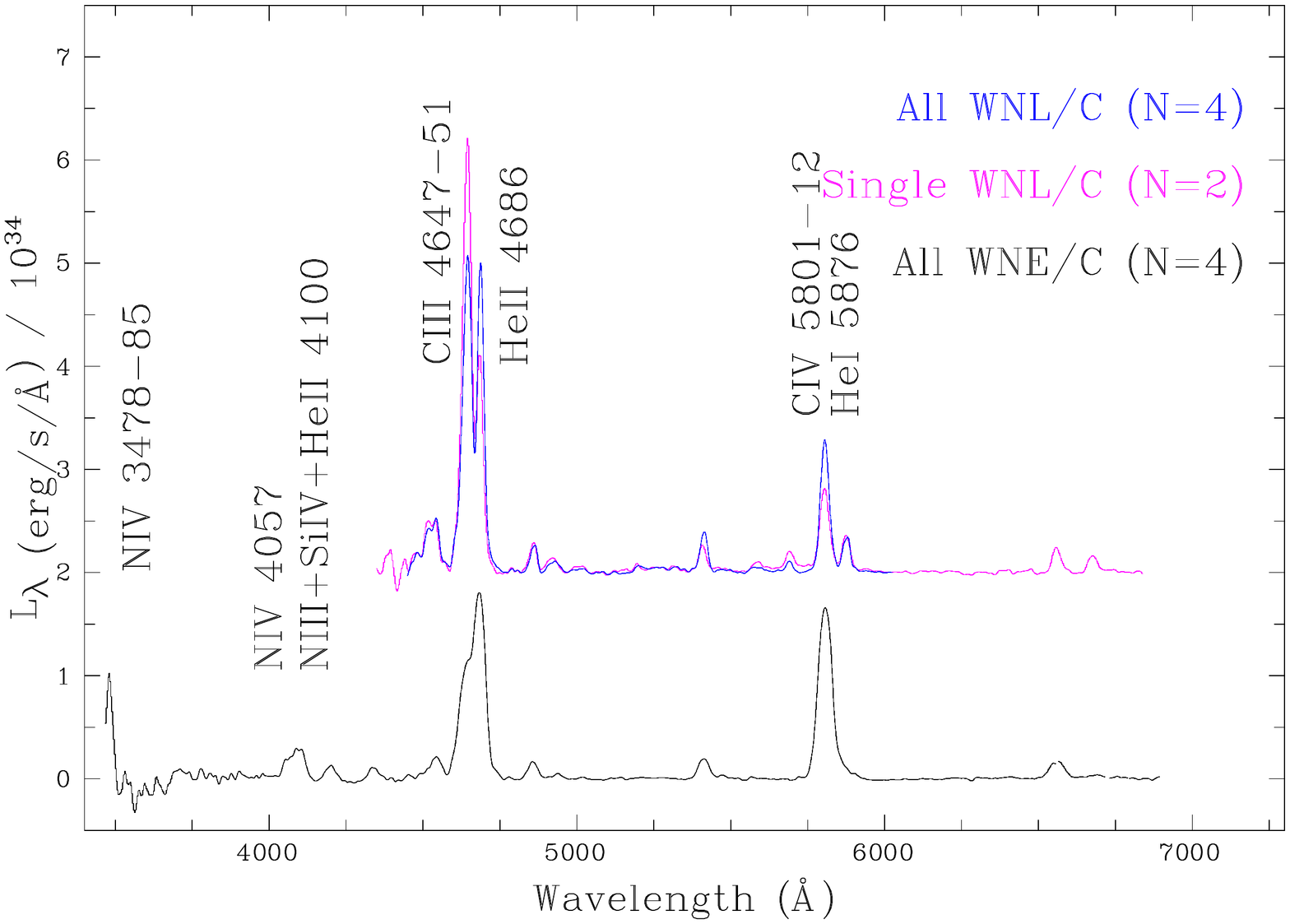}
	\centering
  \caption{WN/C emission line templates for single WNE/C (black, 3 Milky Way and 2 LMC), single WNL/C (2 Milky Way, pink) and all WNL/C (4 Milky Way, blue) the latter group offset by $2 \times 10^{34}$ erg\,s$^{-1}$\,\AA$^{-1}$.}
	\label{WNC-lum} 
\end{figure}

\begin{figure}
\centering
\includegraphics[width=0.68\linewidth,bb=30 65 530 780,angle=-90]{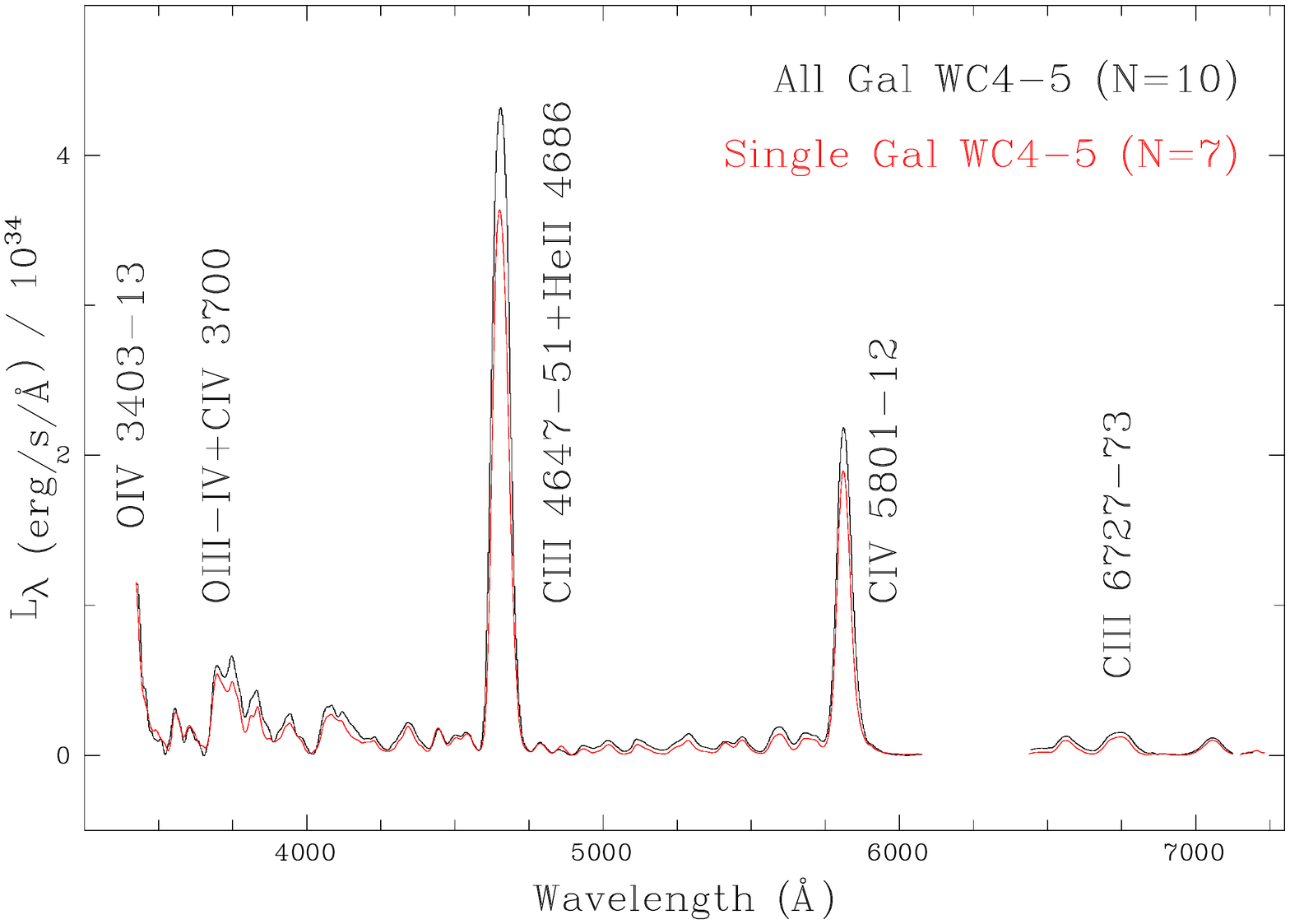}
\includegraphics[width=0.68\linewidth,bb=30 65 530 780,angle=-90]{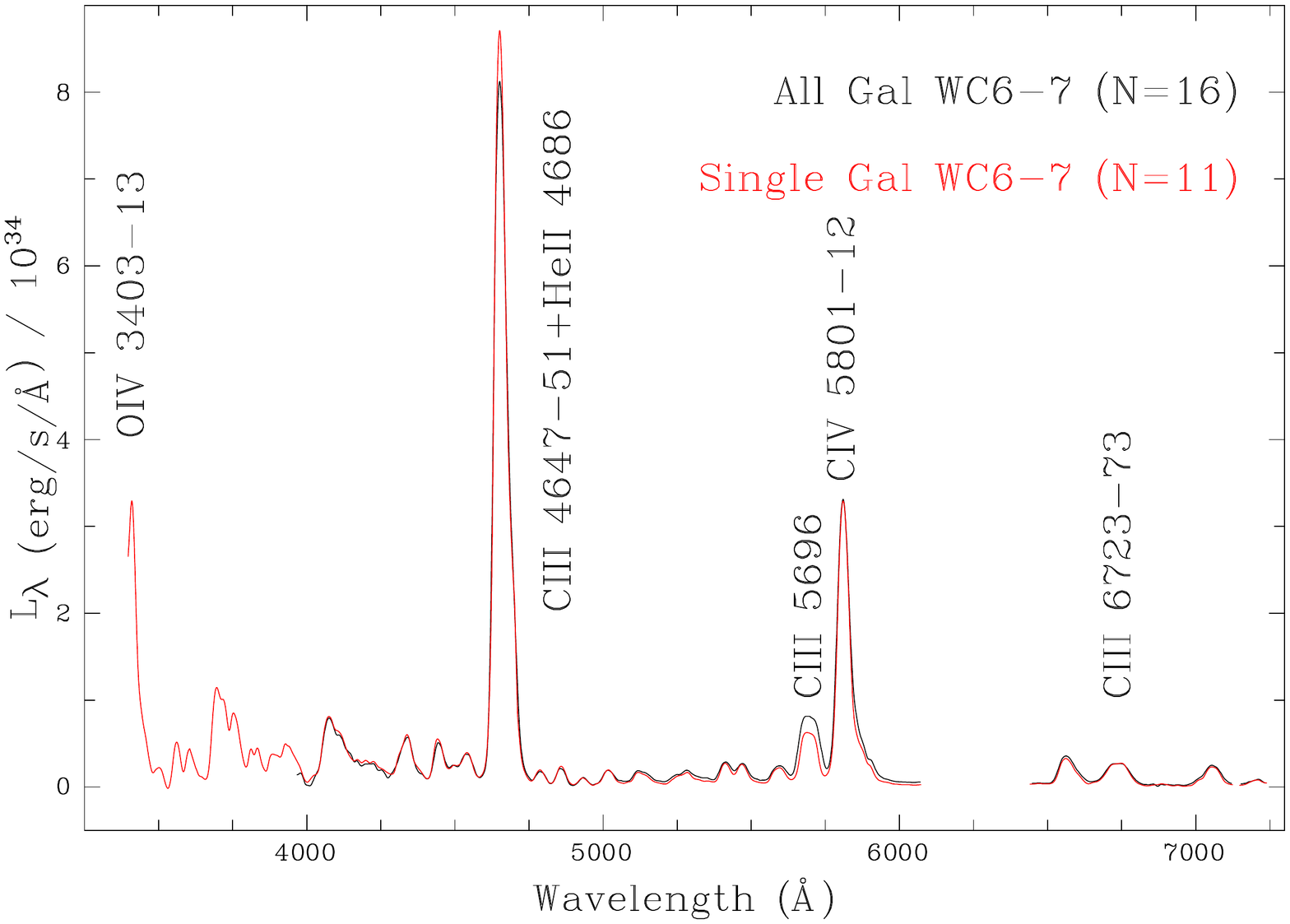}
\includegraphics[width=0.68\linewidth,bb=30 65 530 780,angle=-90]{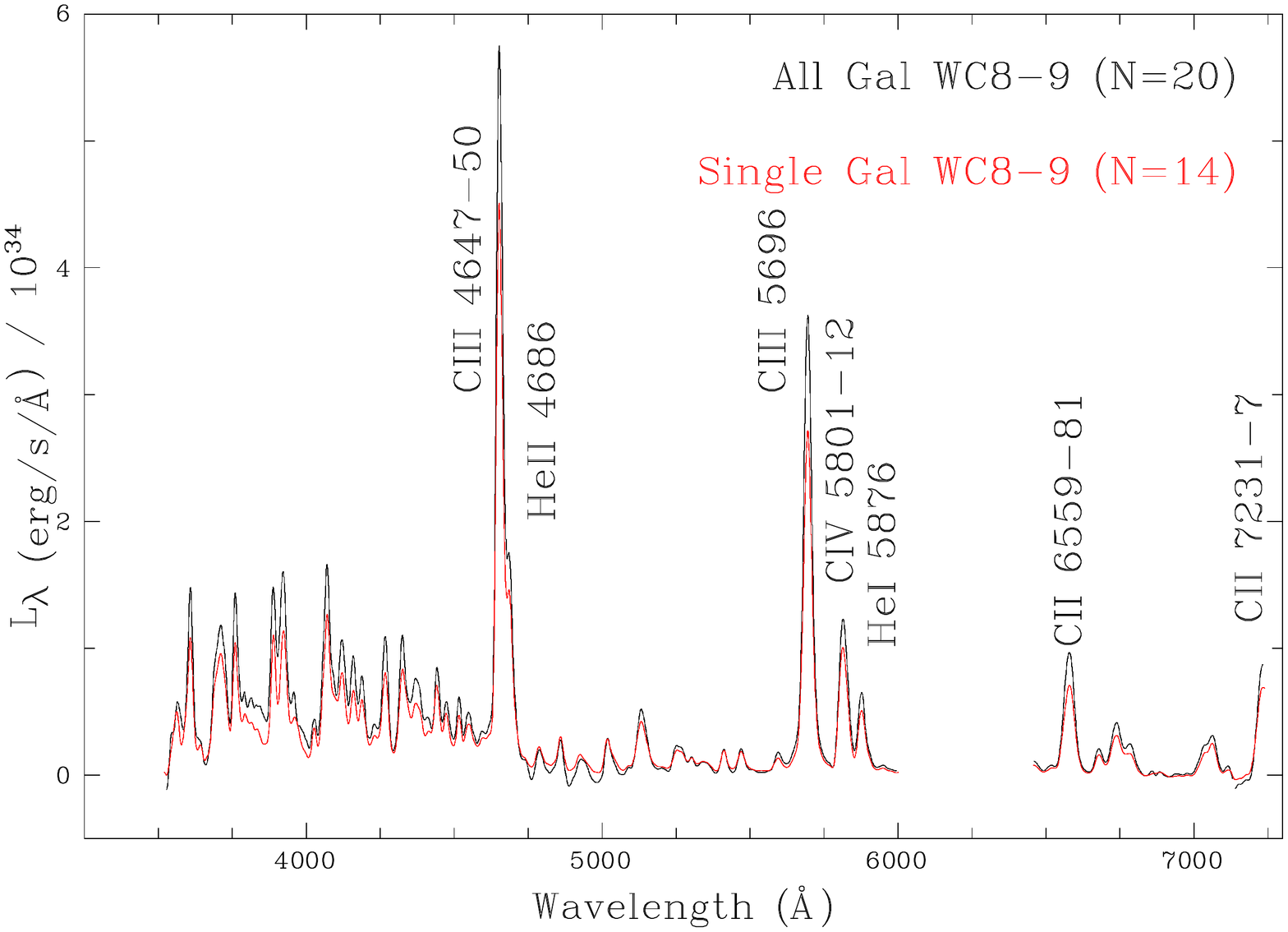}
	\centering
  \caption{Upper panel: Galactic WC4--5 emission line templates based on single (red) and all (black) stars; Middle panel: Galactic WC6--7 emission line templates. Lower panel: Galactic WC8--9 emission line templates. The forest of blue features in WC8--9 stars primarily involve C\,{\sc ii-iii} \citep{2006ApJ...636.1033C}. }
	\label{WC-gal-lum} 
\end{figure}

\begin{figure}
\centering
	\includegraphics[width=0.68\linewidth,bb=30 65 530 780,angle=-90]{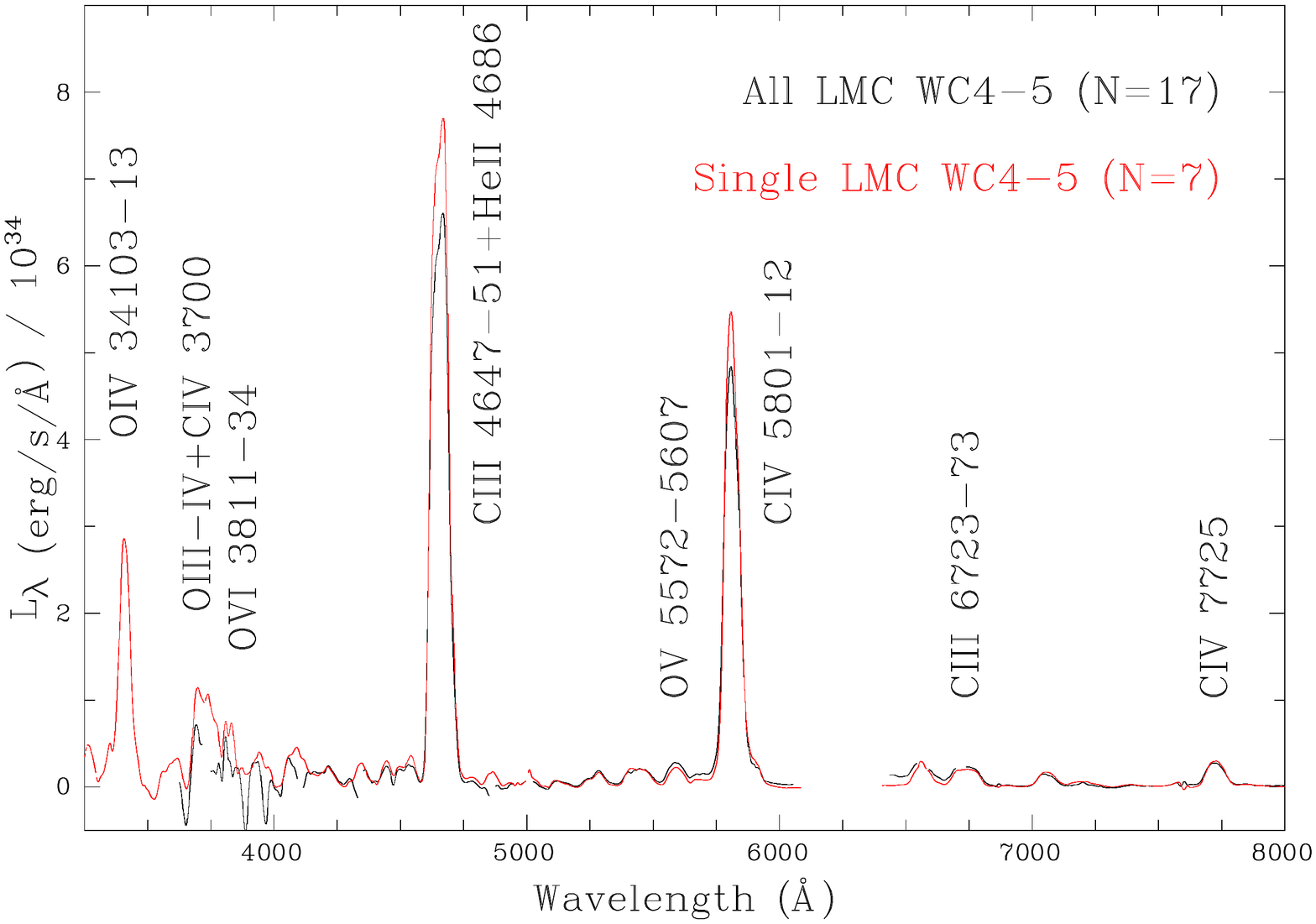}
	\includegraphics[width=0.68\linewidth,bb=30 65 530 780,angle=-90]{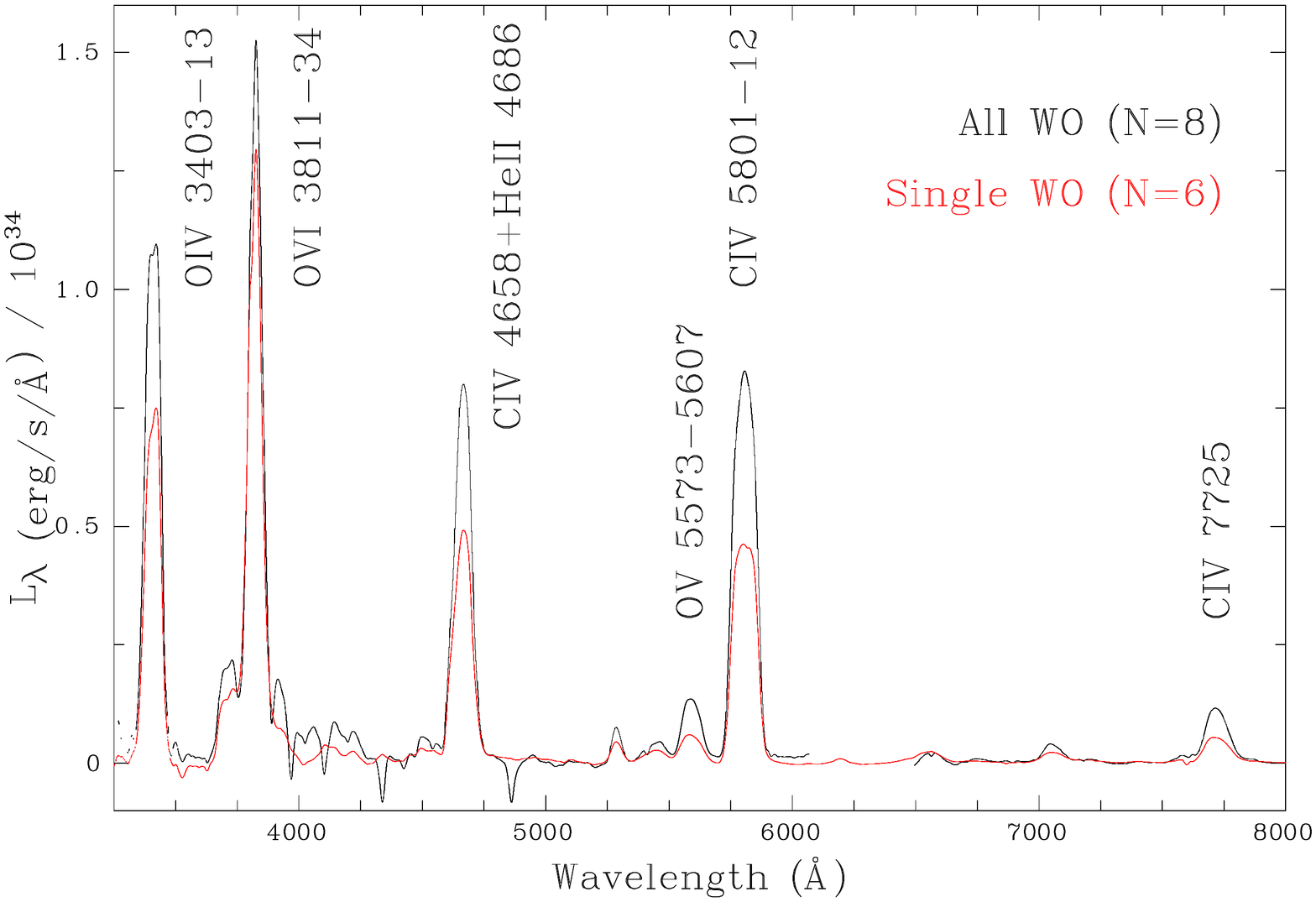}
	\centering
  \caption{Upper panel: LMC WC4--5 emission line templates based on single (red) and all (black) stars; Lower panel: WO emission line templates based on single (red) and all (black) stars, incorporating all Milky Way (4), LMC (3) and SMC (1) stars.}
	\label{WC4-WO-lum} 
\end{figure}

Continuum subtracted WR emission line templates are provided for the Milky Way, LMC and SMC in Figs.~\ref{WNE-lum}--\ref{WC4-WO-lum}. Templates are degraded to a uniform spectra resolution of 10\AA, and are provided from single and single+binary WR stars, since the latter are often contaminated by (Balmer) absorption lines from companion OB stars. Average velocity corrections of 284 km\,s$^{-1}$ and 162 km\,s$^{-1}$ have been applied for the LMC and SMC, respectively \citep{2016AJ....152...50T}. 

\bsp	
\label{lastpage}
\end{document}